\documentclass[reprint,rmp,amsmath,amssymb]{revtex4-2}
\usepackage{mathrsfs}
\usepackage{graphicx}  % 图片支持
\usepackage{dcolumn}   % 表格小数点对齐
\usepackage{bm}        % 粗体数学符号
\usepackage{braket}    % 量子力学符号 < | > (来自 Chapter 6)

\usepackage{multirow}
\usepackage{tabularx}
\usepackage{longtable}
\usepackage{makecell}
\renewcommand{\arraystretch}{2} % 表格行高设置 (来自 Chapter 3，如觉得表格太稀疏可注释掉此行)

\usepackage{times}     % Times New Roman 字体风格
\usepackage{color}
\usepackage{xcolor}    % 颜色支持 (包含了 color 包的功能)

\usepackage{lipsum}    % 生成填充文本
\usepackage[mathlines]{lineno} % 行号支持

\usepackage[colorlinks,
    citecolor=blue,
    linkcolor=blue,
    urlcolor=blue
]{hyperref}
\DeclareUnicodeCharacter{2212}{\ensuremath{-}}

\graphicspath{{chap2/}{chap3/}{chap4/}{chap5/}{chap6/}}

\begin{document}

\title{Topological Phononics}

\author{Zeguo Chen}
\email{These authors contributed equally to this work.}
\affiliation{National Laboratory of Solid State Microstructures, Nanjing University, Nanjing 210093, China}
\affiliation{School of Advanced Manufacturing Engineering, Nanjing University, Suzhou 215163, China}

\author{Tiantian Zhang}
\email{These authors contributed equally to this work.}
\affiliation{Institute of Theoretical Physics, Chinese Academy of Sciences, Beijing 100190, China}
% \affiliation{University of Chinese Academy of Sciences, Beijing 100049, China}

\author{Xulong Wang}
\email{These authors contributed equally to this work.}
\affiliation{Department of Physics, Hong Kong Baptist University, Kowloon Tong, Hong Kong, China}

\author{Jiangxu Li}
\email{These authors contributed equally to this work.}
\affiliation{Shenyang National Laboratory for Materials Science, Institute of Metal Research, Chinese Academy of Sciences, Shenyang 110016, China}

\author{Zhi-Kang Lin}
\email{These authors contributed equally to this work.}
\affiliation{New Cornerstone Science Laboratory, Department of Physics, The University of Hong Kong, Hong Kong 999077, China}

\author{Feng Gao}
%\email{These authors contributed equally to this work.}
\affiliation{School of Physical Science and Technology, Soochow University, Suzhou 215006, China}

\author{Li-Wei Wang}
\affiliation{School of Physical Sciences, University of Science and Technology of China, Hefei 230026, China}

\author{Yizhou Liu}
\affiliation{School of Physics Science and Engineering, Tongji University, 1239 Siping Road, Shanghai 200092, China}

\author{Qi Wang}
\affiliation{Suzhou Institute for Advanced Research, University of Science and Technology of China, Suzhou 215123, China}

\author{Xiujuan Zhang}
\affiliation{National Laboratory of Solid State Microstructures and Department of Materials Science and Engineering, Nanjing University, Nanjing, China}

\author{Guancong Ma}
\email{phgcma@hkbu.edu.hk}
\affiliation{Department of Physics, Hong Kong Baptist University, Kowloon Tong, Hong Kong, China}

\author{Xingqiu Chen}
\email{xingqiu.chen@imr.ac.cn}
\affiliation{Shenyang National Laboratory for Materials Science, Institute of Metal Research, Chinese Academy of Sciences, Shenyang 110016, China}

\author{Minghui Lu}
%\email{luminghui@nju.edu.cn}
\affiliation{National Laboratory of Solid State Microstructures and Department of Materials Science and Engineering, Nanjing University, Nanjing, China}
\affiliation{Collaborative Innovation Center of Advanced Microstructures, Nanjing University, Nanjing, China}

\author{Yanfeng Chen}
\email{yfchen@nju.edu.cn}
\affiliation{National Laboratory of Solid State Microstructures and Department of Materials Science and Engineering, Nanjing University, Nanjing, China}
\affiliation{Collaborative Innovation Center of Advanced Microstructures, Nanjing University, Nanjing, China}

\author{Jian-Hua Jiang}
\email{jhjiang3@ustc.edu.cn}
\affiliation{State Key Laboratory of Bioinspired Interfacial Materials Science, Suzhou Institute for Advanced Research, University of Science and Technology of China, Suzhou 215123, China}
\affiliation{School of Physical Sciences, University of Science and Technology of China, Hefei 230026, China}
\affiliation{School of Physical Science and Technology, Soochow University, Suzhou 215006, China}
\affiliation{Information Materials Research Department, Suzhou Laboratory, Suzhou 215123, China}

\begin{abstract}

Topological phononics extends the foundational concepts of topological condensed matter physics to the realm of lattice vibrations and classical mechanical waves, unlocking robust, defect-immune states and phenomena beyond the reach of conventional phononic engineering. This review provides a unified, systematic framework for understanding topological phonons across natural and artificial systems, spanning solid-state materials, acoustic/mechanical metamaterials, and non-Hermitian platforms. We cover the core theoretical principles---from Berry curvature and symmetry-protected topological invariants to bulk-boundary correspondence---alongside experimental advances in probing topological phonon states via inelastic scattering and momentum-resolved techniques for solid-state phonons as well as pump-probe measurements in acoustic/mechanical metamaterials. Key topics include Weyl/Dirac/nodal-line phonons in crystalline solids, symmetry-engineered topological phases in metamaterials, non-Hermitian effects (exceptional points, skin effect), and emergent directions such as Floquet engineering, synthetic dimensions, and real-space topological textures (skyrmions, merons). We also highlight technological applications in robust waveguides, on-chip surface-acoustic-wave devices, and acoustofluidics, while outlining future challenges and opportunities in quantum phononics, nonlinear topological phenomena, and interdisciplinary integration with photonics and electronics. This review serves as a comprehensive guide across physics, materials science, and engineering, bridging fundamental theory with cutting-edge experiments and innovations in topological phononics.

\end{abstract}

\maketitle

\tableofcontents

\section{Introduction}
Phonons, as the collective modes of mechanical dynamics, permeate every corner of the physical world—from the sub-nanometer-scale atomic vibrations in solids that govern thermal conductivity to the kilometer-scale seismic waves shaping planetary geology~\citetext{\citealp{10.1093/oso/9780192670083.001.0001}; \citealp{PhysRev.95.954}; \citealp{RevModPhys.84.1045}; \citealp{Rev.Geophys.8.1}}. Beyond their fundamental role in fundamental sciences, phonons are the backbone of numerous technologies that define modern life. Mechanical metamaterials leverage phononic engineering for advanced sound and phonon manipulation~\citetext{\citealp{Science.289.1734}; \citealp{MaterToday.12.34}; \citealp{NatRevMater.1.16001}}, enabling noise cancellation and medical imaging devices~\citetext{\citealp{Science.342.323}; \citealp{AdvMaterTechnol.7.2100698}}. On-chip surface-acoustic-wave (SAW) devices, rooted in phonon propagation, are indispensable for wireless communications, serving as state-of-the-art radio-frequency filters and resonators in 5G/6G networks~\citetext{\citealp{App.Phys.Lett.7.314}}. Phonon thermal conduction dictates heat dissipation in high-performance semiconductor systems, a critical bottleneck in advancing chip design toward smaller feature sizes and higher power densities. In biomedical engineering, acoustic tweezers use phononic fields for non-invasive manipulation of cells, tissues, and other particles, revolutionizing drug delivery and tissue engineering~\citetext{\citealp{RevModPhys.83.647}; \citealp{LabChip.13.3626}; \citealp{NatMethods.15.1021}}. Even in quantum science, phonons emerge as important carrier for quantum coherence and correlation, bridging classical and quantum technologies through their ability to couple with photons and electrons~\citetext{\citealp{NatPhys.11.37}; \citealp{Science.364.368}}. The ubiquity and versatility of phonons underscore their central importance in advancing science and technology across different disciplines.

At the core of these technological innovations lies the challenge of controlling phonon propagation with unprecedented precision and robustness. In practical devices, defects, fabrication imperfections, and environmental perturbations are unavoidable, often degrading performance by scattering phonons or altering their intended paths. Topology, a branch of mathematics concerned with robust, global, invariant properties of geometry and spaces, provides a transformative solution to this challenge. Unlike conventional materials where properties are dictated by local symmetry or composition, topological systems exhibit robust modes protected by global topological invariants—quantities that remain unchanged under smooth deformations of the system~\citetext{\citealp{RevModPhys.82.3045}; \citealp{RevModPhys.83.1057}; \citealp{RevModPhys.90.015001}}. This topological protection ensures that edge or surface states of phonons persist even in the presence of defects, disorder, or sharp bends, offering a level of stability unattainable with traditional phononic engineering. Over the past decade, the fusion of topology and phononics has given birth to the vibrant field of topological phononics~\citetext{\citealp{PhysRevLett.103.248101}; \citealp{PhysRevLett.105.225901}; \citealp{PhysRevLett.114.114301}; \citealp{Science.349.47}; \citealp{NatCommun.6.8682}; \citealp{PhysRevB.93.205158}; \citealp{NatPhys.12.1124}; \citealp{NatPhys.13.369}; \citealp{AdvFunctMater.30.1904784}; \citealp{RepProgPhys.86.106501}}, unlocking a wealth of unconventional phonon states and effects that have redefined our understanding of mechanical wave propagation.

The study of topological phonons naturally divides into two interconnected fronts: crystalline solids and artificial metamaterials. Crystalline solids, with their inherent periodic atomic lattices, provide a natural playground for applying topological band theory---originally developed for electrons---to phononic systems. These materials host diverse phonon band structures, giving rise to topological states such as Weyl phonons~\citetext{\citealp{PhysRevB.97.054305}; \citealp{PhysRevLett.120.016401}; \citealp{PhysRevLett.121.035302}}, Dirac phonons~\citetext{\citealp{PhysRevB.96.064106}; \citealp{NanoLett.18.7755}; \citealp{PhysRevB.101.081403}; \citealp{PhysRevLett.126.185301}}, and nodal-line phonons~\citetext{\citealp{PhysRevLett.123.245302}; \citealp{App.Phys.Rev.9.041304}}, each characterized by unique topological invariants. A particularly intriguing emergent phenomena in these systems is the topological surface phonons, which are confined to material interfaces and exhibit exotic physical properties. These surface phonons hold profound implications for surface physics, chemistry, and device functions, as they can notably affect interface transport of charge and energy. They could also trigger surface sensitive optical processes and be connected with quantum correlated phases at interfaces and low-dimensions as these surface phonons with large local density-of-states can drive superconductivity and charge density waves via electron-phonon interactions. In practical terms, topological surface phonons could play an active role in interfacial thermal transport---a critical mechanism for managing heat in multi-layered electronic devices and heterostructures. They also play pivotal roles in surface chemistry, catalysis, and erosion, as their localized nature amplifies solid-molecule interactions at material interfaces.

Historically, the exploration of topological phonons in solid-state materials emerged only in the past decade, notably lagging behind electronic topological systems due to experimental challenges. Phonon measurements require specialized techniques such as inelastic neutron or X-ray scattering and momentum-resolved electron energy-loss spectroscopy, which are technically demanding and less accessible than electronic probes. Nevertheless, this situation is rapidly changing. Recent experimental breakthroughs have directly observed Weyl phonons in chiral crystals like FeSi ~\citetext{\citealp{PhysRevLett.121.035302}}, nodal-line phonons in MoB$_2$~\citetext{\citealp{PhysRevLett.123.245302}}, and Dirac phonons in graphene~\citetext{\citealp{PhysRevLett.131.116602}}, validating theoretical predictions and sparking renewed interest in the field. Moreover, the scope of topological phononics in solids is expanding beyond 3D perfect crystals to amorphous materials, alloys, and nanostructures, with emerging topological phonon states that are absent in three-dimensional crystalline lattices. Given phonons’ central role in driving quantum correlated phases and governing energy/charge transport, the study of topological phonons in solids promises to unlock new avenues for designing materials with emergent quantum correlated phases or tailored thermal, electrical, optical and surface chemical properties.

Alongside with solid-state systems, artificial materials---acoustic and mechanical metamaterials~\citetext{\citealp{MaterToday.12.34}; \citealp{NatRevMater.1.16001}; \citealp{NatRevMater.3.460}}---have emerged as pioneering platforms for topological phononics. Unlike natural crystals, metamaterials offer unparalleled design freedom, allowing researchers to engineer periodic structures with precisely tailored band structures, symmetries, and non-Hermiticity. These systems provided the first experimental demonstrations of topological phonon states, laying the groundwork for the field. Over the past decade, acoustic metamaterials have enabled the observation of a wide array of topological phenomena: geometric phases in phononic waveguides~\citetext{\citealp{NatPhys.11.240}}, non-Hermitian topological effects~\citetext{\citealp{NatCommun.12.6297}; \citealp{Nature.597.655}; \citealp{NatRevPhys.6.11}}, helical and chiral edge states that propagate unidirectionally without backscattering~\citetext{\citealp{Science.349.47}; \citealp{NatPhys.12.1124}; \citealp{NatPhys.13.369}; \citealp{PhysRevLett.122.014302}}, and topological bulk-defect correspondence that links real-space topological defects to observable topological effects~\citetext{\citealp{NatPhys.11.153}; \citealp{AnnuRevCondensMatterPhys.8.211}; \citealp{PhysRevLett.127.214301}; \citealp{NatCommun.12.3654}; \citealp{NatMater.21.430}}. Notably, metamaterials have pushed the boundaries of topological physics by realizing states that were first predicted theoretically but remained elusive in electronic systems. Higher-order topological insulators, where gapless states are confined to corners or hinges rather than edges or surfaces, were first observed in phononic metamaterials~\citetext{\citealp{Nature.555.342}; \citealp{NatMater.18.108}; \citealp{NatMater.18.113}; \citealp{NatPhys.15.582}}, as were fragile topological phases that lose their nontrivial topological character when coupled to trivial bands~\citetext{\citealp{Science.367.797}; \citealp{SciBull.69.1653}}. Non-Abelian phononic topological states~\citetext{\citealp{NatPhys.17.1239}}, characterized by non-commuting braiding of band nodes, and Weyl points with large Chern numbers have also been experimentally realized first in phononic metamaterials~\citetext{\citealp{NatPhys.15.645}}, demonstrating the field’s capacity and vigor for scientific innovation.

Acoustic and mechanical systems offer unique advantages that make them ideal for exploring topological phononics. Their macroscopic scale allows for direct visualization of phonon fields using microphones or vibrometers, providing direct access to topological edge states and bulk modes. The ability to engineer non-Hermitian mechanisms---through controlled gain, loss, or nonreciprocal couplings---has opened new frontiers in topological physics, leading to phenomena like the non-Hermitian skin effect, where bulk modes localize at boundaries~\citetext{\citealp{NatCommun.12.6297}; \citealp{NatCommun.12.5377}}, and enriching the scope of topological physics~\citetext{\citealp{PhysRevLett.125.226402}; \citealp{NatRevPhys.4.745}}. Versatile pump-probe techniques at ambient conditions enable real-time tracking of the dynamics of phonon fields, facilitating the measurement of geometric phases and topological invariants~\citetext{\citealp{NatPhys.11.240}; \citealp{NatCommun.15.1601}}. These advantages have fueled the rapid growth of topological phononics in metamaterials, with numerous states predicted, realized, and characterized, establishing a solid foundation for future research. Today, the field of topological phononics stands at the intersection of solid-state physics and material engineering, with both fronts complementing each other to drive fundamental discoveries and technological innovations.

Given the rapid growth of topological phononics over the past decade, a comprehensive and systematic review is timely and necessary. Existing reviews focus on narrow subfields---such as topological phonons in solids or metamaterials in isolation~\citetext{\citealp{NatRevMater.1.281}; \citealp{NatRevMater.7.974}; \citealp{RepProgPhys.86.106501}; \citealp{NatRevPhys.5.483}; \citealp{Nature.618.687}; \citealp{RevModPhys.96.021002}; \citealp{NatCommun.16.3560}}---lacking a unified framework that connects these diverse systems and the seemingly different underlying physics. This gap hinders the ability of researchers, particularly those  new to the field, to grasp the full scope of the field of topological phononics and its interdisciplinary implications. The goal of this review is to fill this void by providing a pedagogical and comprehensive overview of the field, spanning Hermitian and non-Hermitian systems, natural and artificial materials, and from fundamental physics to practical applications.

In this article, we begin with a foundational introduction to the theoretical framework of topological phononics, including basic concepts such as Berry curvature, topological invariants (Chern numbers, $\mathbb{Z}_2$ indices), and bulk-boundary correspondence. We then systematically explore topological phonons in solid-state materials, covering experimental techniques, key material systems, and emerging phenomena like topological surface phonons and disorder-induced topological phases. Next, we delve into artificial metamaterials, detailing the design principles, realized topological states, and unique advantages of acoustic and mechanical systems. A dedicated section is devoted to non-Hermitian topological phononics, a rapidly growing subfield that explores the interplay between topology and non-Hermitian effects. We also cover emerging directions that are reshaping the field: synthetic dimensions, which extend topological phenomena to higher dimensions using parameter spaces; hidden symmetries that stabilize novel topological states; spatiotemporal modulations enabling Floquet topological phases; and nonlinearity, which introduces interactions between topological phonons. Finally, we provide an outlook on future challenges and opportunities, including the search for topological phonons in quantum systems (such as cold atoms and molecules in optical lattices), the development of topological phononic devices for information processing and related applications, and the integration of topological phononics with other fields such as photonics and electronics.

By unifying the study of topological phonons in natural and artificial materials, this review aims to provide a holistic perspective on the field’s development, highlight its interdisciplinary nature, and inspire future research. Whether for researchers seeking to enter the field, experts looking to expand their research horizons, or engineers exploring technological applications, this article offers a comprehensive guide to the exciting world of topological phononics---where mathematics, physics, and engineering converge to unlock the next generation of robust, high-performance phononic technologies.

%\documentclass[%
%reprint,
%superscriptaddress,
%groupedaddress,
%unsortedaddress,
%runinaddress,
%frontmatterverbose, 
%preprint,
%preprintnumbers,
%nofootinbib,
%nobibnotes,
%bibnotes,
%amsmath,amssymb,
%aps,
%pra,
%prb,
%rmp,
%prstab,
%prstper,
%floatfix,
%]{revtex4-1}

%\usepackage{graphicx}% Include figure files
%\usepackage{dcolumn}% Align table columns on decimal point
%\usepackage{bm}% bold math
%\usepackage{multirow}
%\usepackage{tabularx} % 
%\usepackage{longtable} % 
%\usepackage{lipsum}
%\usepackage{color}
%\usepackage{xcolor}
%\usepackage{makecell}
%\usepackage{braket} % for \braket command
%\renewcommand{\arraystretch}{2}
%\usepackage[colorlinks,
%citecolor=blue,
%linkcolor=blue,
%urlcolor=blue]{hyperref}
%\usepackage{times}

%\begin{document}

%\preprint{APS/123-QED}

\section{Basic concepts}

This section introduces only the minimum conceptual framework needed for the discussions that follow. 
We briefly recall the basic language of band topology and then emphasize those directions that are especially relevant to phononic, acoustic, and artificial systems. 
In particular, after summarizing several canonical single-gap examples, we focus on symmetry-engineered Dirac masses, real multigap topology, and dynamical extensions such as Floquet phases and topological pumping.

Although topological band theory was originally developed mainly in electronic settings, its central ideas apply much more broadly. 
Once a wave problem is cast into an appropriate band-eigenvalue form, one can discuss Berry phase, Berry curvature, topological invariants, and boundary responses in close analogy with electronic bands. 
This broader viewpoint is especially useful in phononic and acoustic systems, where symmetry breaking, effective gauge fields, and time-dependent couplings can often be engineered in a direct and highly controllable manner.

\subsection{Minimal phonon-band formulation}\label{subsec:phonon_schro}

A key step in topological phononics is to reformulate linear lattice dynamics as a band-eigenvalue problem~\citetext{\citealp{PhysRevB.96.064106}; \citealp{AdvFunctMater.30.1904784}; \citealp{PNAS.113.E4767}}. 
Starting from the equations of motion
\begin{equation}
\mathbf M\,\ddot{\mathbf u}(t)+\mathbf G\,\dot{\mathbf u}(t)+\mathbf K\,\mathbf u(t)=0,
\label{eq:eom_real}
\end{equation}
where \(\mathbf M\) is the mass matrix, \(\mathbf K\) the force-constant matrix, and \(\mathbf G\) a velocity-coupling term, one obtains after Bloch transformation
\begin{equation}
\mathbf M(\mathbf k)\,\ddot{\mathbf u}(\mathbf k,t)+\mathbf G(\mathbf k)\,\dot{\mathbf u}(\mathbf k,t)+\mathbf K(\mathbf k)\,\mathbf u(\mathbf k,t)=0.
\label{eq:eom_k}
\end{equation}
Introducing mass-weighted coordinates yields the dynamical matrix
\begin{equation}
\mathbf D_{\mathbf k}
=
\mathbf M^{-1/2}(\mathbf k)\,\mathbf K(\mathbf k)\,\mathbf M^{-1/2}(\mathbf k),
\label{eq:D_def}
\end{equation}
together with the normalized antisymmetric term
\begin{equation}
\bm{\eta}_{\mathbf k}
=
\frac{1}{2}\,\mathbf M^{-1/2}(\mathbf k)\,\mathbf G(\mathbf k)\,\mathbf M^{-1/2}(\mathbf k),
\qquad
\bm{\eta}_{\mathbf k}^{T}=-\bm{\eta}_{\mathbf k}.
\label{eq:eta_def}
\end{equation}
The quadratic phonon problem can then be written in the Schr\"odinger-like first-order form~\citetext{\citealp{PhysRevB.96.064106}; \citealp{AdvFunctMater.30.1904784}}
\begin{equation}
\mathbf H_{\mathbf k}\,\bm{\psi}_{n\mathbf k}
=
\omega_n(\mathbf k)\,\bm{\psi}_{n\mathbf k},
\label{eq:sch_phonon}
\end{equation}
with
\begin{equation}
\mathbf H_{\mathbf k} =
\begin{pmatrix}
\mathbf 0 & i\,\mathbf D_{\mathbf k}^{1/2}\\[2pt]
-\,i\,\mathbf D_{\mathbf k}^{1/2} & -\,2i\,\bm{\eta}_{\mathbf k}
\end{pmatrix}.
\end{equation}
For lossless systems, this Hermitian formulation allows Berry curvature and topological invariants to be defined for positive-frequency phonon bands in close analogy with electronic band theory~\citetext{\citealp{PhysRevB.96.064106}; \citealp{AdvFunctMater.30.1904784}; \citealp{PNAS.113.E4767}}.

This form also makes the role of time-reversal-symmetry breaking transparent. 
When \(\bm{\eta}_{\mathbf k}=0\), the Berry curvature is odd in momentum and the Chern number vanishes for an isolated nondegenerate two-dimensional band. 
A nonzero gyroscopic or Coriolis-type term \(\bm{\eta}_{\mathbf k}\neq 0\) removes this constraint and enables Chern-type phononic phases~\citetext{\citealp{PhysRevB.96.064106}; \citealp{PNAS.112.14495}; \citealp{PhysRevLett.115.104302}}. 
For the purposes of this review, this minimal formulation is sufficient; below we focus mainly on the resulting topological structures rather than on the detailed derivation of the phonon eigenproblem.

With this minimal phonon-band formulation in place, it is useful to briefly recall several canonical examples of single-gap topology that provide the basic vocabulary for later discussions.

\subsection{Canonical single-gap examples}\label{sec:basic_examples}

We briefly summarize several standard single-gap topological phases that provide a common language across electronic, phononic, and acoustic systems. 
The purpose here is not to review these well-known models in detail, but to fix the basic relation between dimensionality, symmetry, topological invariant, and boundary response.

\subsubsection{Su--Schrieffer--Heeger model}\label{subsec:ssh}

The SSH model~\citetext{\citealp{PhysRevLett.42.1698}; \citealp{PhysRevB.22.2099}; \citealp{RevModPhys.60.781}} is the canonical one-dimensional example of a chiral-symmetry-protected topological phase. 
For a bipartite chain with intracell hopping \(t_1\) and intercell hopping \(t_2\),
\begin{equation}
H(k)=d_x(k)\sigma_x+d_y(k)\sigma_y,
\end{equation}
with
$d_x=t_1+t_2\cos k, d_y=t_2\sin k$.
Because \(H(k)\) anticommutes with \(\mathcal C=\sigma_z\), its topology is characterized by an integer winding number.
The phase with \(|t_2|>|t_1|\) is topological and hosts zero-energy end states under open boundary conditions. 
In phononic and mechanical lattices, the same dimerization principle can be implemented through alternating couplings or effective spring constants, producing localized boundary vibrations.

\subsubsection{Chern phases in two dimensions}\label{subsec:chern2d}

In two dimensions, broken time-reversal symmetry allows an integer Chern number,
\begin{equation}
C=\frac{1}{2\pi}\int_{\mathrm{BZ}} d^2k\,\Omega_{xy}(\mathbf k),
\end{equation}
which determines the quantized Hall response and the number of chiral edge modes. 
The integer quantum Hall effect provides the original example, while the Haldane model~\citetext{\citealp{PhysRevLett.61.2015}} shows that a lattice system can realize a Chern insulator even with zero net magnetic flux through the unit cell. 
Near the valleys, the problem reduces to massive Dirac fermions, and the total Chern number is fixed by the relative signs of the Dirac masses. 
In phononic and acoustic settings, analogous Chern phases can be generated by gyroscopic bias, circulating flow, or spatiotemporal modulation, which act as synthetic time-reversal-symmetry-breaking fields.

\subsubsection{Time-reversal-invariant \texorpdfstring{$\mathbb Z_2$}{Z2} phases}\label{subsec:qshe}

A distinct two-dimensional topology arises when time-reversal symmetry is preserved and \(\mathcal T^2=-1\). 
The BHZ model~\citetext{\citealp{science.314.1757}} provides the minimal prototype: it consists of two time-reversed massive Dirac sectors and enters a nontrivial \(\mathbb Z_2\) phase when band inversion occurs. 
Its defining boundary signature is a pair of helical edge states related by Kramers symmetry. 
In inversion-symmetric crystals, the \(\mathbb Z_2\) index can be obtained from parity eigenvalues; more generally, it can be diagnosed from Wilson-loop spectra~\citetext{\citealp{PhysRevLett.95.146802}; \citealp{PhysRevB.89.155114}}. 
Phononic analogues of this physics rely on pseudo-spin degrees of freedom together with an effective antiunitary symmetry that protects counter-propagating edge transport. 
Three-dimensional time-reversal-invariant topological insulators provide the corresponding higher-dimensional extension, but in the present review we focus mainly on lower-dimensional and engineered bosonic platforms, where the connection to phononic and acoustic realizations is more direct.

\begin{figure}[!htbp]
  \centering
\includegraphics[width=1.0\linewidth]{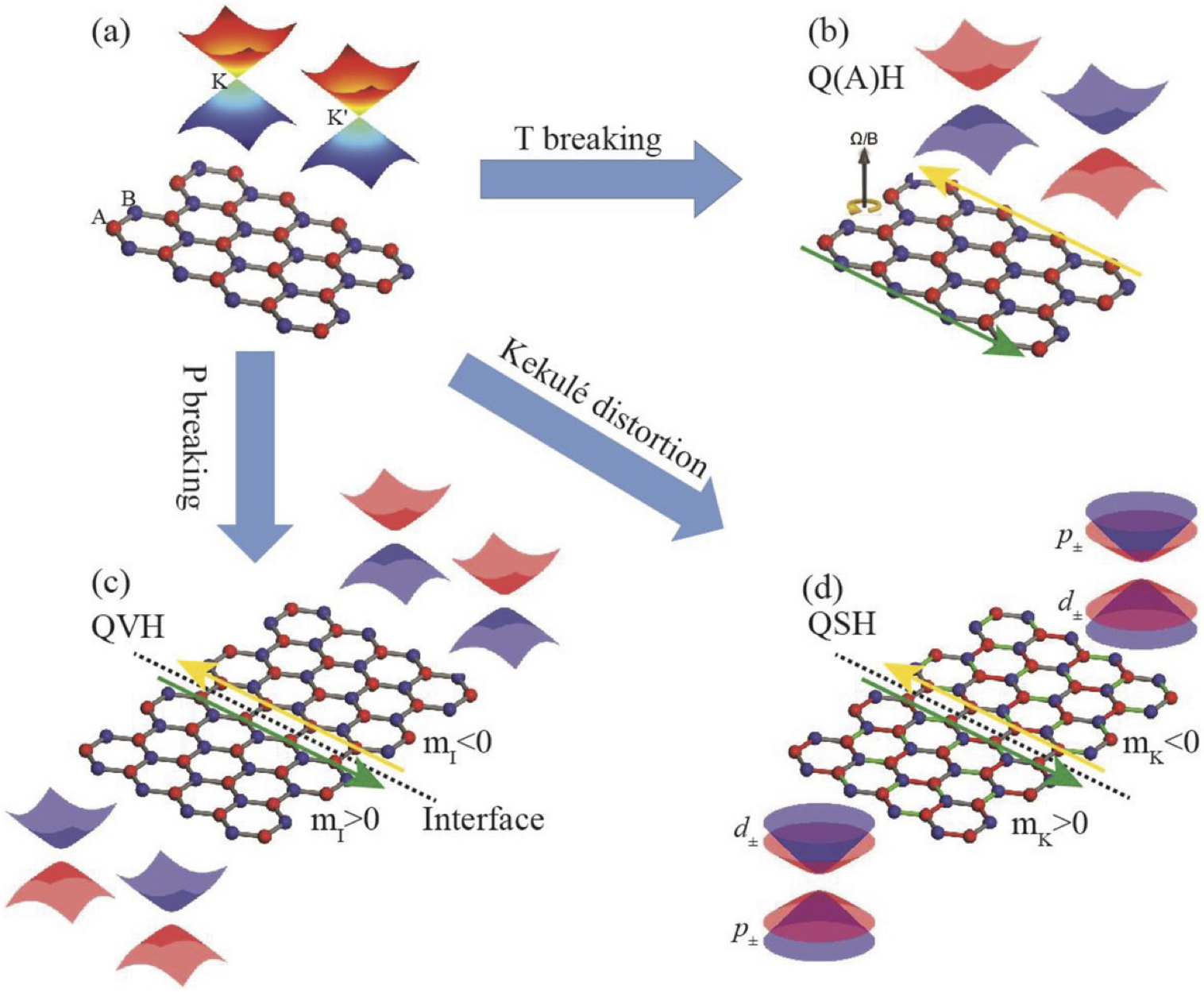}
\caption{Symmetry-breaking routes to Hall-type topological phases for phonons in a honeycomb lattice.
(a) Gapless phononic Dirac cones at the valleys $K$ and $K'$ for a $\mathcal{P}\mathcal{T}$-symmetric lattice (A/B sublattices indicated by distinct colors).
(b) Breaking time-reversal symmetry $\mathcal{T}$ (e.g., via rotation/Coriolis or gyroscopic bias) gaps the Dirac cones into a phononic quantum (anomalous) Hall phase with chiral one-way edge modes.
(c) Breaking inversion symmetry $\mathcal{P}$ (e.g., by an A/B sublattice bias) yields a quantum valley Hall phase with valley-contrasting Berry curvature and valley-polarized interface states at a domain wall where the inversion-breaking mass changes sign.
(d) A Kekul\'e-type distortion (valley-mixing superlattice) gaps the Dirac cones into a QSH-like phase supporting helical edge transport protected by an appropriate pseudo-time-reversal symmetry in the effective pseudospin sector.
Adapted from Ref.~\citetext{\citealp{AdvFunctMater.30.1904784}}.}
\label{fig:phonon_hall_dirac}
\end{figure}

\subsection{Hall-type phononic phases from gapped Dirac cones}\label{subsec:phonon_hall}

While these standard models establish the basic language of single-gap topology, phononic systems often admit a more direct symmetry-based description in terms of Dirac cones and the mass terms that gap them~\citetext{\citealp{AdvFunctMater.30.1904784}}.

A particularly transparent route to topological phononic phases is to begin with a two-dimensional Dirac spectrum and gap it by symmetry-breaking perturbations. As illustrated in Fig.~\ref{fig:phonon_hall_dirac}, this viewpoint provides a unified description of phononic quantum (anomalous) Hall, quantum valley Hall, and QSH-like phases~\citetext{\citealp{AdvFunctMater.30.1904784}; \citealp{PNAS.112.14495}; \citealp{PhysRevLett.115.104302}}. The essential idea is that distinct symmetry-allowed Dirac mass terms generate distinct Berry curvatures and, correspondingly, distinct boundary excitations.

Near the two valleys \(K\) and \(K'\), a broad class of honeycomb-based phononic lattices is described by the minimal valley--sublattice Hamiltonian
\begin{equation}
H_0(\mathbf q)=v\left(\tau_z\sigma_x q_x+\sigma_y q_y\right),
\label{eq:dirac_phonon}
\end{equation}
where \(\tau\) labels the valleys and \(\sigma\) acts in the two-mode Dirac subspace. When both inversion \(\mathcal P\) and time-reversal \(\mathcal T\) are present, the two valleys form symmetry-related gapless Dirac cones. Topological phases arise once symmetry breaking permits a mass term that gaps this spectrum~\citetext{\citealp{AdvFunctMater.30.1904784}; \citealp{PhysRevApplied.11.014040}; \citealp{SciRep.8.13814}}.

Breaking \(\mathcal T\) yields the Haldane-type mass
\begin{equation}
H_{\mathcal T}=m_{\mathcal T}\sigma_z\tau_z,
\label{eq:mass_qah}
\end{equation}
which produces a nonzero Chern number and hence chiral edge states. In phononic systems, such a term can be generated by gyroscopic or Coriolis couplings, active nonreciprocity, or spatiotemporal modulation, which act as effective magnetic fields for bosonic waves~\citetext{\citealp{PNAS.112.14495}; \citealp{PhysRevLett.115.104302}; \citealp{AdvFunctMater.30.1904784}}.

Breaking \(\mathcal P\) while preserving \(\mathcal T\) gives the Semenoff-type mass
\begin{equation}
H_{\mathcal P}=m_{\mathcal P}\sigma_z,
\label{eq:mass_qvh}
\end{equation}
corresponding to an A/B sublattice bias. In this case the total Chern number vanishes, but the two valleys acquire opposite Berry curvature, leading to a quantum valley Hall phase with valley-polarized interface states when \(m_{\mathcal P}\) changes sign across a domain wall~\citetext{\citealp{PhysRevApplied.11.014040}; \citealp{SciRep.8.13814}; \citealp{AdvFunctMater.30.1904784}}.

A third possibility is to introduce valley mixing through a Kekul\'e-type distortion,
\begin{equation}
H_{K}=m_{K}\left(\tau_x\sigma_x-\tau_y\sigma_y\right),
\label{eq:mass_kekule}
\end{equation}
which gaps the Dirac cones by hybridizing the valleys. When the resulting two-component sector is protected by an appropriate pseudo-time-reversal symmetry, the gapped phase supports helical edge transport and provides a bosonic analogue of the quantum spin Hall effect~\citetext{\citealp{SciRep.6.32752}; \citealp{PhysRevApplied.11.044086}; \citealp{AdvFunctMater.30.1904784}}.

These three mass terms furnish a compact taxonomy of Hall-type phononic phases. Their boundary excitations follow from the same domain-wall principle: whenever the relevant mass changes sign in space, the Dirac theory predicts localized interface modes whose chirality, valley polarization, or helicity is fixed by the underlying symmetry and topological index~\citetext{\citealp{PhysRevLett.115.104302}; \citealp{PhysRevApplied.11.014040}; \citealp{PhysRevApplied.11.044086}}. This Dirac-mass viewpoint offers a simple bridge between symmetry breaking and the more general band-topology tools introduced below.

\subsection{Real multigap topology and Euler class}\label{subsec:euler}
The canonical examples above mainly belong to the paradigm of \emph{single-gap} topology, where an isolated band or an occupied-band subspace is characterized by invariants such as the Chern number or the \(\mathbb Z_2\) index. 
A conceptually distinct situation arises in \emph{real multigap topology}, where the relevant topological object is an \emph{oriented real two-band subspace} in two dimensions rather than a single isolated band~\citetext{ 
\citealp{PhysRevB.92.081201}; \citealp{PhysRevLett.121.106403}; \citealp{Science.365.1273}; \citealp{ChinesePhysB.28.117101}; \citealp{PhysRevX.9.021013}; \citealp{PhysRevB.100.195135}; \citealp{PhysRevB.100.205126}; \citealp{NatPhys.16.1137}; \citealp{PhysRevB.102.115135};\citealp{PhysRevResearch.4.023188};
}. 
This setting is especially relevant to phononic, acoustic, and mechanical systems, because in the absence of explicit time-reversal-symmetry breaking the Bloch eigenvectors are often naturally real, or are constrained by an effective antiunitary symmetry such as \(\mathcal{PT}\) or \(C_{2z}\mathcal{T}\)~\citetext{\citealp{PhysRevB.105.064301}; \citealp{Science.383.844}}.

The Euler class is an integer invariant defined for two real bands that remain separated from all other bands throughout the Brillouin zone~\citetext{\citealp{PhysRevX.9.021013}; \citealp{PhysRevB.102.115135}}. 
Accordingly, the minimal \emph{global insulating} realization is a three-band model: two bands form the real two-band subspace of interest, while the third band isolates this subspace from the rest of the spectrum~\citetext{\citealp{PhysRevLett.125.053601}; \citealp{PhysRevResearch.4.023188}}. 
A convenient minimal form is
\begin{equation}
H_{3\mathrm{b}}(\mathbf k)
=
\Delta\Big(2\,|n(\mathbf k)\rangle\langle n(\mathbf k)|-\mathbb I_3\Big),
\label{eq:H3b_euler}
\end{equation}
where \(|n(\mathbf k)\rangle\in\mathbb R^3\) is a smooth normalized real vector. 
The corresponding isolated two-band subspace is described by the projector
\begin{equation}
P_{12}(\mathbf k)=\mathbb I_3-|n(\mathbf k)\rangle\langle n(\mathbf k)|.
\label{eq:proj_euler}
\end{equation}
When this rank-two real bundle is orientable, one may choose a smooth real orthonormal frame
\(\{|u_1(\mathbf k)\rangle,|u_2(\mathbf k)\rangle\}\),
and define the non-Abelian Berry connection
\begin{equation}
[A_\mu(\mathbf k)]_{ab}
=
\langle u_a(\mathbf k)|\partial_{k_\mu}u_b(\mathbf k)\rangle,
\qquad a,b=1,2.
\label{eq:NA_connection}
\end{equation}
In a real frame, \(A_\mu\) takes values in \(\mathfrak{so}(2)\), and the associated Euler class characterizes the global topology of the two-band subspace~\citetext{\citealp{PhysRevB.102.115135}; \citealp{PhysRevLett.125.053601}; \citealp{PhysRevResearch.4.023188}}.

The physical meaning of the Euler class becomes particularly transparent near isolated band nodes. 
Although the global invariant requires a three-band insulating setting, the local structure near a node can always be projected to an effective real two-band Hamiltonian,
\begin{equation}
H_{\mathrm{eff}}(\mathbf q)
=
a(\mathbf q)\,\mathbb I
+
r(\mathbf q)\Big[\cos\theta(\mathbf q)\,\sigma_x+\sin\theta(\mathbf q)\,\sigma_z\Big],
\label{eq:Heff_euler}
\end{equation}
where \(\mathbf q\) is measured relative to the node. 
Because the Hamiltonian is real, only \(\sigma_x\) and \(\sigma_z\) appear. 
The node then carries an integer winding number
\begin{equation}
N_v
=
\frac{1}{2\pi}
\oint_{\partial D} d\mathbf q\cdot \nabla \theta(\mathbf q),
\label{eq:winding_euler}
\end{equation}
with \(\partial D\) a small loop encircling the degeneracy. 
For a real two-band subspace, the total winding of all nodes is tied to the Euler class through
\begin{equation}
e_2=-\frac12\sum_i N_i,
\label{eq:euler_winding}
\end{equation}
up to an overall sign convention fixed by the choice of orientation~\citetext{\citealp{PhysRevX.9.021013}; \citealp{PhysRevB.102.115135}; \citealp{NatCommun.14.1261}}. 
This relation shows that Euler topology is not merely a geometric abstraction: it controls the global accounting of node charges in momentum space, and underlies a variety of non-Abelian nodal phenomena and their dynamical manifestations~\citetext{\citealp{PhysRevB.105.214108}; \citealp{PhysRevResearch.4.023188}; \citealp{PhysRevB.108.075129}}.

From the viewpoint of topological phononics, this real multigap topology is particularly natural. 
Phononic and acoustic lattices often provide real-symmetric dynamical matrices or effective real band structures, and engineered metamaterials offer fine control over isolated two-band subspaces and their symmetry-protected nodes~\citetext{\citealp{PhysRevB.105.064301}; \citealp{PhysRevB.105.214108}}. 
As a result, Euler and related non-Abelian topological structures have become especially visible in artificial wave systems, including acoustic semimetals, ideal metamaterials, and acoustic Euler insulators~\citetext{\citealp{NatPhys.17.1239}; \citealp{NatCommun.14.1261}; \citealp{SciBull.69.1653}}. 
More broadly, recent progress has highlighted real multigap topology as an important extension of the modern language of topological waves beyond the familiar Chern- and \(\mathbb Z_2\)-band paradigms~\citetext{\citealp{Science.383.844}; \citealp{PhysRevB.108.075129}}.

Having introduced both single-gap phases and real multigap topology, we now briefly summarize the most commonly used diagnostic tools for identifying these structures in crystalline band systems.

\begin{figure}[!htbp]
  \centering
  \includegraphics[width=\linewidth]{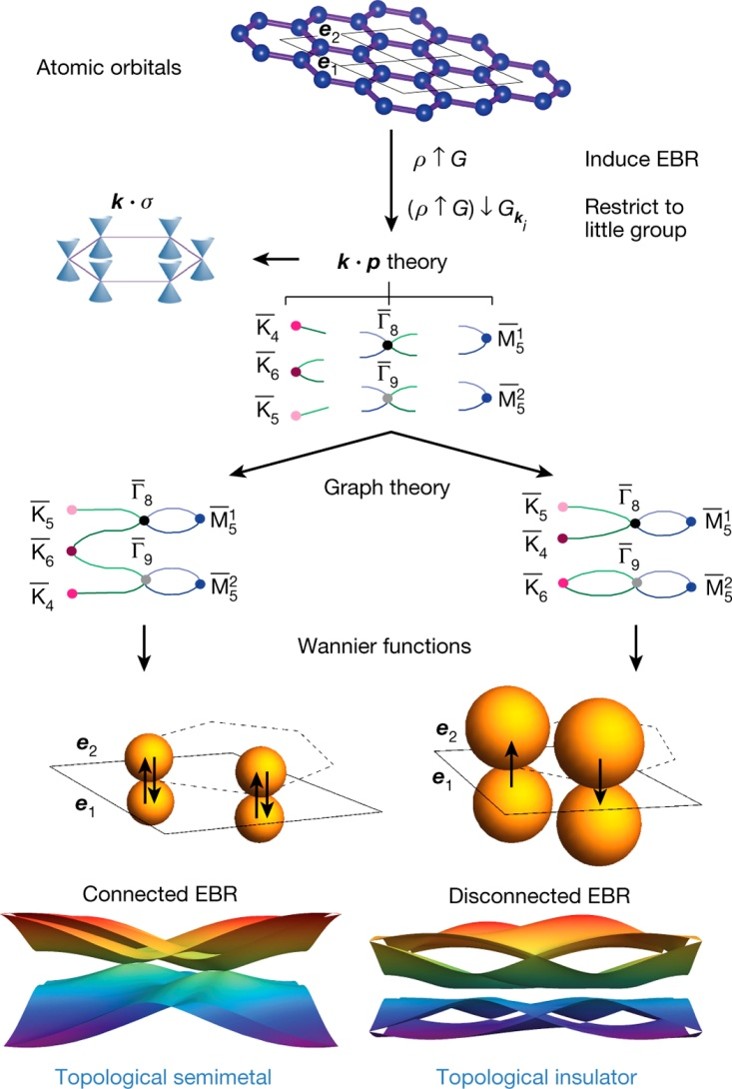}
  \caption{Schematic of the topological classification for graphene with SOC. Atomic orbitals  ($\left | p_z \uparrow  \right \rangle$ and $\left | p_z \downarrow  \right \rangle$ )  induce an elementary band representation (EBR), whose compatibility relations are analyzed via graph theory. The graph connectivity diagnoses the phase: a single connected component (left) corresponds to a symmetry-protected semimetal with atomic-like Wannier functions, while two disconnected components (right) indicate a topological insulator with symmetry-breaking Wannier centers and a gapped spectrum. Adapted from~\citetext{\citealp{nature.547.298}}.}
  \label{fig:TQC}
\end{figure}

\begin{figure*}[!htbp]
  \centering
  \includegraphics[width=0.9\textwidth]{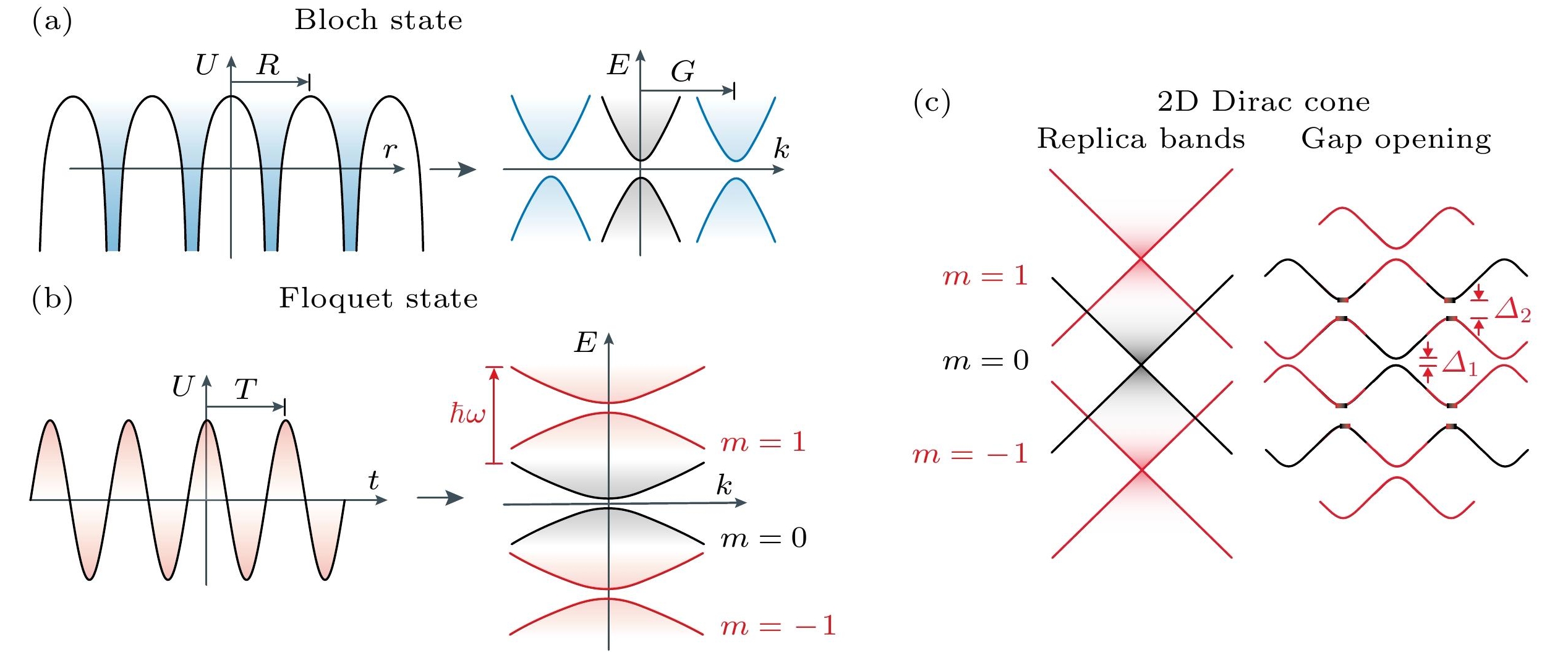}
  \caption{(a) Spatially periodic potential and Bloch bands in the k-space; (b) time-periodic potential and Floquet bands in energy; (c) schematics for Floquet engineering. Adapted from~\citetext{\citealp{Acta.Phys.Sin72.234202-1},\citealp{Nat.Rev.Phys.4.33}}.}
  \label{fig:floquet}
\end{figure*}

\subsection{Topological band diagnostics}\label{subsec:band_diagnostics}

Beyond canonical model Hamiltonians, topological phases are often identified through a small set of general diagnostic tools. 
For the purposes of this review, we only summarize the most commonly used ideas. 
The emphasis is not on a complete classification of all band-topological possibilities, but on a minimal framework for diagnosing topology in crystalline wave systems, including phonons.

\subsubsection{Symmetry-based classification}

A general starting point is the Altland--Zirnbauer tenfold way~\citetext{\citealp{PhysRevB.55.1142}}, which classifies noninteracting gapped Hamiltonians according to the presence or absence of time-reversal, particle-hole, and chiral symmetries. 
For each symmetry class and spatial dimension, the allowed topological phases are characterized by invariants valued in \(\mathbb Z\), \(\mathbb Z_2\), or \(0\). 
Although originally formulated for electronic insulators and superconductors, this symmetry-based viewpoint remains a useful organizing principle for bosonic band structures once the relevant wave problem has been cast into an appropriate Hermitian eigenvalue form.

\subsubsection{Wilson loops and \texorpdfstring{$\mathbb Z_2$}{Z2} diagnosis}

For time-reversal-invariant bands, one of the most useful practical diagnostics is the Wilson loop~\citetext{\citealp{PhysRevB.89.155114}}. 
Given an isolated set of bands, one defines the non-Abelian Berry connection
\begin{equation}
[\mathcal A_\mu(\mathbf k)]_{mn}
=
i\langle u_{m\mathbf k}|\partial_{k_\mu}u_{n\mathbf k}\rangle,
\end{equation}
from which the Wilson loop operator is constructed along a closed path in momentum space. 
For example, in two dimensions one may write
\begin{equation}
\mathcal W(k_x)
=
\overline{\exp}\!\left[
- \int_{-\pi}^{\pi} dk_y\, \mathcal A_y(k_x,k_y)
\right],
\end{equation}
whose eigenphases define hybrid Wannier centers. 
Their winding, or partner switching between time-reversal-invariant momenta, gives a direct characterization of the \(\mathbb Z_2\) topology. 
In inversion-symmetric systems, the same invariant can often be obtained more simply from the Fu--Kane parity criterion~\citetext{\citealp{PhysRevB.76.045302}}, but the Wilson-loop formulation is more general and provides a more geometric picture of the band topology.

\subsubsection{Crystalline-symmetry diagnostics}

When crystalline symmetries are important, topology can often be diagnosed directly from symmetry data at high-symmetry momenta. 
This idea underlies symmetry indicators~\citetext{\citealp{Nat.Commun.8.50}}, which compare the symmetry representations of the bands with those realizable by atomic insulators, and topological quantum chemistry (TQC)~\citetext{\citealp{nature.547.298}; \citealp{PhysRevB.26.3010}}, which asks whether a set of isolated bands can be expressed as a sum of elementary band representations induced from localized Wannier orbitals. 
A mismatch signals a Wannier obstruction and hence nontrivial topology. 
These crystalline diagnostics are especially valuable in materials searches, because they connect momentum-space symmetry data to real-space orbital pictures and provide an efficient route to identifying topological bands without explicitly computing Berry curvature over the full Brillouin zone.

Taken together, these methods form a compact diagnostic toolbox for modern band topology. 
With suitable adaptations, the same ideas are widely used in topological phononics, where symmetry, Berry geometry, and band connectivity play roles closely analogous to those in electronic systems.

The diagnostics summarized above are formulated primarily for static band structures. 
Periodic driving extends these ideas into the time domain and gives rise to genuinely dynamical topological phases.

\subsection{Floquet topological states}\label{subsec:floquet}

Periodic driving provides a powerful route to topological phases that have no static counterpart~\citetext{\citealp{PhysRevX.3.031005}; \citealp{PhysRevB.96.155118}; \citealp{NatCommun.7.11744}}. 
For a $T$-periodic Hamiltonian, $H(t+T)=H(t)$, the evolution over one cycle is encoded in the Floquet operator
\begin{equation}
U_F \equiv \mathcal{T}\exp\!\Big[-i\!\int_{0}^{T} H(t)\,dt\Big],
\end{equation}
whose eigenphases define the quasienergies modulo the driving frequency \(\Omega=2\pi/T\). 
Equivalently, one introduces an effective stroboscopic Hamiltonian through
\begin{equation}
U_F=e^{-iTH_{\mathrm{eff}}}.
\end{equation}
In contrast to static band theory, quasienergy is periodic, so both the \(0\) and \(\pi\equiv \Omega/2\) gaps may carry distinct topology~\citetext{\citealp{PhysRevX.3.031005}; \citealp{PhysRevB.96.155118}}. 
As a result, periodically driven systems can host anomalous edge states even when the Chern numbers of the Floquet bands themselves vanish~\citetext{\citealp{PhysRevX.3.031005}; \citealp{PhysRevB.96.155118}; \citealp{NatCommun.7.13368}}. 
This feature makes Floquet engineering especially attractive in phononic and acoustic systems, where time-dependent couplings, active elements, and gyroscopic bias can be implemented with high tunability~\citetext{\citealp{NatCommun.7.11744}; \citealp{SciAdv.6.eaba8656}; \citealp{PhysRevApplied.11.044029}}.

\subsubsection{Minimal Floquet framework}

The basic topological distinction from static systems is that Floquet phases are classified gap by gap in quasienergy rather than solely band by band~\citetext{\citealp{PhysRevX.3.031005}; \citealp{PhysRevB.96.155118}}. 
In one dimension, a periodically driven chiral chain such as the Floquet SSH model can support protected end modes in either the \(0\) or \(\pi\) gap, providing the simplest example of anomalous Floquet topology. 
In two dimensions, periodic driving can generate Floquet Chern insulators with chiral edge modes, and may even produce anomalous phases in which edge transport survives although the band Chern numbers vanish. 
More generally, these phases are diagnosed by gap-resolved winding invariants constructed from the full time-evolution operator, reflecting the fact that the topology of driven systems is encoded not only in \(H_{\mathrm{eff}}\) but in the micromotion over an entire period.

\subsubsection{Representative driven phases in phononic and acoustic systems}

Two representative examples illustrate the basic ideas. 
First, in one-dimensional modulated dimer chains, time-dependent intracell and intercell couplings realize a Floquet SSH phase with end modes pinned at quasienergy \(0\) or \(\pi\)~\citetext{\citealp{PhysRevB.96.155118}; \citealp{PhysRevResearch.1.033069}}. 
Second, in driven hexagonal lattices, periodic modulation can gap Dirac points and generate either conventional Floquet Chern bands or anomalous chiral edge states~\citetext{\citealp{NatCommun.7.11744}; \citealp{SciAdv.6.eaba8656}; \citealp{PhysRevApplied.11.044029}}. 
In both cases, the physically relevant topological information is attached to quasienergy gaps rather than to static bands alone~\citetext{\citealp{PhysRevX.3.031005}; \citealp{PhysRevB.96.155118}}.

Phononic and acoustic metamaterials provide particularly natural platforms for these phenomena. 
Time-programmable couplers, electroacoustic shunts, piezoelectric networks, and gyroscopic elements enable direct control of the drive protocol, while stroboscopic measurements make it possible to resolve edge transport in the \(0\) and \(\pi\) gaps~\citetext{\citealp{NatCommun.7.11744}; \citealp{NatCommun.7.13368}; \citealp{SciAdv.6.eaba8656}}. 
For this reason, Floquet engineering has become an important extension of topological phononics, complementing static symmetry-based routes with genuinely dynamical band topology~\citetext{\citealp{PhysRevApplied.11.044029}; \citealp{PhysRevResearch.1.033069}}.

While Floquet topology concerns quasienergy bands and edge states in periodically driven systems, adiabatic pumping highlights a complementary dynamical phenomenon in which the topology of the cycle itself produces quantized transport.

\subsection{Topological pumping}\label{subsec:pumping}

Topological pumping is the adiabatic counterpart of band topology in which a cyclic parameter modulation produces quantized bulk transport over one period. 
Rather than being characterized by edge states in a static gap, the pump is classified by the topology of the cycle itself in the extended \((k,\lambda)\) space, where \(\lambda(t)\) is the slowly varying periodic parameter. 
This provides a particularly transparent link between Berry geometry, Chern numbers, and measurable transport, and it also offers a natural route to topological control in phononic and acoustic metamaterials.

\subsubsection{Adiabatic pumping and the Rice--Mele paradigm}

The paradigmatic example is the one-dimensional Thouless pump~\citetext{\citealp{PhysRevB.27.6083}; \citealp{PhysRevA.111.053308}}, in which a gapped Bloch Hamiltonian \(H(k,\lambda)\) evolves adiabatically along a closed cycle \(\lambda(t+T_p)=\lambda(t)\). 
If the gap remains open throughout the cycle, the transported charge (or, more generally, the transported wave-packet center) over one period is quantized as
\begin{equation}
Q
=
\frac{1}{2\pi}\int_0^{2\pi} d\lambda \int_{\mathrm{BZ}} dk\, \Omega(k,\lambda),
\label{eq:pump_chern}
\end{equation}
where \(\Omega(k,\lambda)=\partial_k A_\lambda-\partial_\lambda A_k\) is the Berry curvature in the extended parameter space. 
Thus, \(Q\) is the first Chern number of the adiabatic cycle and remains invariant under smooth deformations that do not close the bulk gap.

The canonical realization is the Rice--Mele model~\citetext{\citealp{PhysRevLett.49.1455}},
\begin{equation}
H(k,t)
=
\bigl[t_1(t)+t_2(t)\cos k\bigr]\sigma_x
+
t_2(t)\sin k\,\sigma_y
+
\delta(t)\sigma_z,
\label{eq:rice_mele_hamiltonian}
\end{equation}
where the staggered hoppings and onsite potential are modulated cyclically so that the path in \((t_1,t_2,\delta)\) space encloses the gapless SSH point. 
Its geometric interpretation is especially transparent in terms of the Wannier center,
\begin{equation}
\bar{x}(\lambda)=\frac{a}{2\pi}\gamma(\lambda),
\qquad
\gamma(\lambda)= i\int_{\mathrm{BZ}} dk\, \langle u(k,\lambda)|\partial_k u(k,\lambda)\rangle,
\label{eq:wannier_pump}
\end{equation}
which shifts by an integer number of unit cells over one cycle. 
Under open boundary conditions, the same topology appears as spectral flow of edge states across the bulk gap, providing a direct boundary signature of the pump.

\begin{figure}[!htbp]
  \centering
  \includegraphics[width=\linewidth]{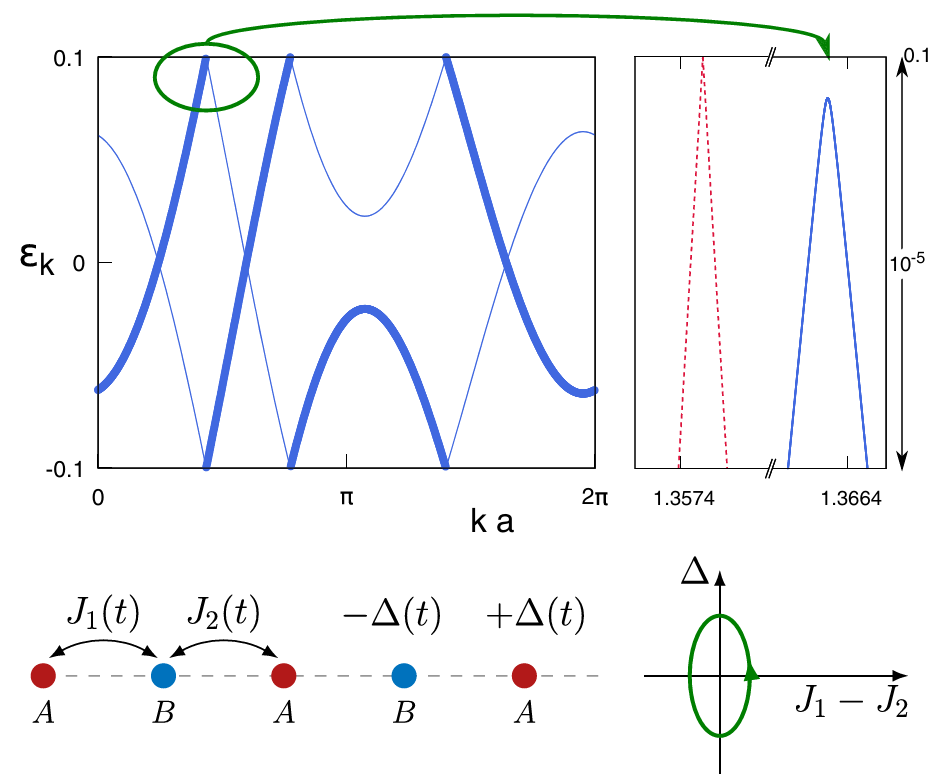}
  \caption{Top left: quasienergy spectrum of the Rice-Mele Model. The thick band is the lowest-energy Floquet band. Top right: (solid line) zoom of the previous gure close to the upper border of the Floquet Brillouin zone around $ka$ = 13664; the dashed line denotes the quasienergies in the adiabatic limit $\epsilon_{\alpha,k}$. The gap is of order $10^{-6}$; Bottom: A cartoon of the Rice-Mele model (left) and a path in parameter space. Adapted from~\citetext{\citealp{PhysRevLett.120.106601}}.}
  \label{fig:RM}
\end{figure}

\subsubsection{Pumping in phononic and acoustic systems}

The same adiabatic principle carries over naturally to bosonic wave systems. 
In phononic and acoustic lattices, cyclic modulation of couplings, onsite resonant frequencies, or effective stiffness parameters can realize Thouless-type pumps in which the center of a localized wave packet is displaced in a quantized and robust manner. 
Equivalently, one may observe adiabatic spectral flow of boundary or interface modes across a frequency gap as the pumping parameter is varied.

This makes topological pumping particularly attractive for engineered metamaterials. Compared with electronic systems, phononic and acoustic platforms offer direct control over the modulation protocol and straightforward real-space detection of transported intensity, displacement, or pressure fields. 
As a result, topological pumping has become an important dynamical route to robust transport, mode transfer, and frequency-space control in classical wave systems, complementing the static band-topology mechanisms discussed above.

Taken together, these static and dynamical concepts provide the basic language needed to discuss topological phonons in both equilibrium and driven settings.

\section{\label{sec:level1}Topological phonons in crystalline solids}

%\begin{abstract}
% {T.T. Zhang} 
% {Y.Z. Liu}
% {Qi Wang}

As quantized normal modes of atomic motion, phonons not only determine the thermodynamic and transport properties of solids but can also exhibit nontrivial topological characteristics, giving rise to Weyl, Dirac, and nodal-line phonons as well as Berry-curvature-driven transport phenomena. This section provides an overview of the theoretical foundations, first-principles computational frameworks, and experimental developments in the study of topological phonons. Finally, we discuss how topology manifests in lattice dynamics and highlight its profound implications for emergent quantum phenomena in condensed matter systems.

%\end{abstract}

%\keywords{Suggested keywords}%Use showkeys class option if keyword
                              %display desired
\maketitle

%\tableofcontents

\subsection{\label{sec:level2}An introduction of topological phonons in solids}

Phonons are the quantized normal modes of lattice vibrations~\citetext{\citealp{grosso2013solid}; \citealp{hook2013solid}; \citealp{srivastava2022physics}}, offering a powerful framework for understanding how atoms in a crystalline solid collectively oscillate around their equilibrium positions. Within the harmonic approximation, the lattice Hamiltonian can be written as
\begin{equation}
H = \sum_{l\kappa} \frac{\mathbf{p}_{l\kappa}^2}{2M_{\kappa}} + \frac{1}{2}\sum_{l\kappa,l'\kappa'} \Phi_{\kappa\kappa'}(l - l') \mathbf{u}_{l\kappa}\cdot\mathbf{u}_{l'\kappa'},    
\end{equation}
where $\mathbf{u}_{l\kappa}$ and $\mathbf{p}_{l\kappa}$ denote the displacement and momentum of atom $\kappa$ in unit cell $l$, $M_\kappa$ is its mass, and $\Phi_{\kappa\kappa'}$ are the interatomic force constants (IFCs). Fourier transforming to momentum space yields the dynamical matrix

\begin{equation}
D_{\kappa\alpha, \kappa'\beta}(\mathbf{q}) = \frac{1}{\sqrt{M_\kappa M_{\kappa'}}}\sum_{l'} \Phi_{\kappa\alpha,\kappa'\beta}(l - l') e^{i\mathbf{q}_(l-l')}.    
\end{equation}

Solving the eigenvalue problem
\begin{equation}
D(\mathbf{q}) \mathbf{e}_{n\mathbf{q}} = \omega_{n}^2(\mathbf{q}) \mathbf{e}_{n\mathbf{q}}    
\end{equation}
gives the phonon dispersion $\omega_{n}(\mathbf{q})$ and corresponding normalized eigenvectors $\mathbf{e}_{n\mathbf{q}}$, describing the normal modes. These solutions form phonon bands in momentum space, directly analogous to electronic Bloch bands, though phonons are neutral bosons without Pauli exclusion constraints or charge-based couplings.

Traditionally, phonons were regarded as topologically trivial because they lack spin and a Fermi level. However, phonon eigenmodes can still acquire Berry phase or Berry curvature due to geometric phases associated with internal vibrational motion~\citetext{\citealp{PhysRevLett.105.225901}; \citealp{PhysRevB.86.104305}; \citealp{PhysRevB.96.064106}; \citealp{PhysRevLett.119.255901}; \citealp{PhysRevMater.2.114204}; \citealp{PhysRevB.98.220103}; \citealp{PhysRevLett.121.035302}; \citealp{NatlSciRev.5.314}; \citealp{PhysRevLett.120.016401}; \citealp{PhysRevB.97.054305}}. For example, the Berry curvature of a phonon band $n$ is defined as
\begin{equation*}
\Omega_{n}(\mathbf{q}) = i\sum_{m\neq n}\frac{\langle \mathbf{e}_{n\mathbf{q}} | \partial_{q_x} D | \mathbf{e}_{m\mathbf{q}} \rangle \langle \mathbf{e}_{m\mathbf{q}} | \partial_{q_y} D | \mathbf{e}_{n\mathbf{q}} \rangle - \text{c.c.}}{(\omega_n^2(\mathbf{q}) - \omega_m^2(\mathbf{q}))^2},    
\end{equation*}
which enables the definition of topological invariants, such as Chern numbers and monopole charges associated with Weyl points~\citetext{\citealp{PhysRevB.97.054305};
\citealp{PhysRevLett.120.016401}; \citealp{PhysRevB.99.174306};
\citealp{PhysRevB.100.081204}; \citealp{PhysRevB.102.125148}; \citealp{npjComputMater.6.95}; \citealp{PhysRevB.103.104101}; \citealp{PhysRevB.103.094306}; \citealp{PhysRevB.103.184301}; \citealp{PhysRevB.103.L161303}; \citealp{PhysRevB.106.214309}; \citealp{PhysRevB.106.214308}; \citealp{PhysRevMater.6.084201}; \citealp{AdvSci.10.2207508}; \citealp{NanoLett.23.7561}; \citealp{PhysRevB.108.054305}; \citealp{PhysRevB.108.235302}; \citealp{PhysRevB.109.045203}; \citealp{ChinesePhysB.33.100301};
\citealp{NatCommun.16.3560};
\citealp{arXiv.2025.}}. This recognition has led to the development of topological phonon band theory~\citetext{\citealp{PhysRevX.8.031069}; \citealp{PhysRevRes.2.022066}; \citealp{JPhysDApplPhys.54.414002}; \citealp{arXiv.2025.}}, revealing that vibrational excitations can host Dirac points, nodal lines, and even Weyl nodes endowed with topological chirality.

{\subsection{Computational Framework}}

\begin{figure*}
    \centering
    \includegraphics[width=0.7\linewidth]{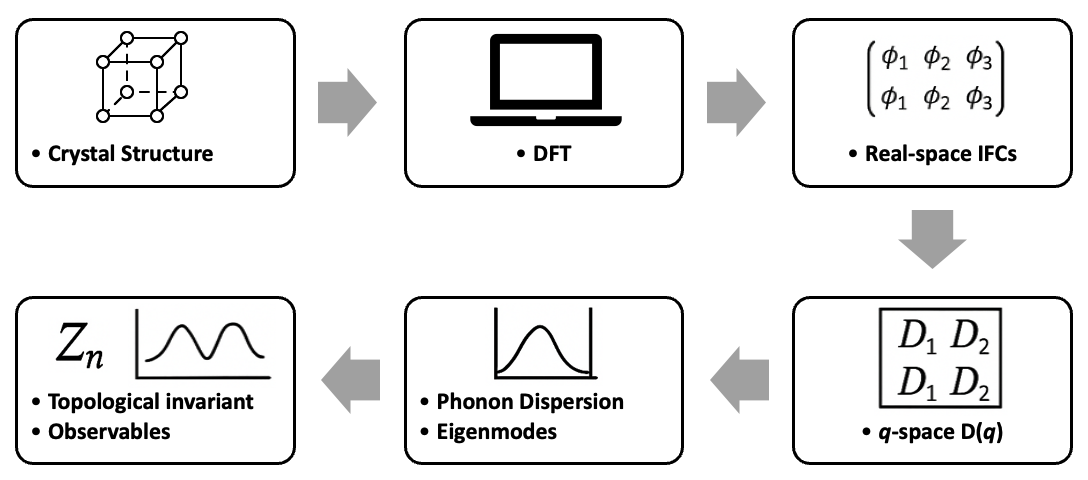}
    \caption{Workflow for obtaining and analyzing topological phonons from first-principles calculations. Starting from the crystalline structure, density functional theory (DFT) calculations yield the real-space interatomic force constants (IFCs). After performing a Fourier transform, the dynamical matrix D($\mathbf{q}$) in momentum space is obtained. Diagonalizing D($\mathbf{q}$) provides the phonon dispersion and eigenmodes, which enable the evaluation of topological invariants and the computation of experimental observables (such as dynamical structure factors in the neutron scattering process) for direct comparison with experiments.}
    \label{fig:flowchart}
\end{figure*}

The theoretical identification of topological phonons generally follows a systematic workflow that bridges \textit{ab initio} lattice dynamics with topological band theory~\citetext{\citealp{PhysRevLett.120.016401}; \citealp{PhysRevB.97.054305}; \citealp{PhysRevLett.123.245302}}. The starting point is the calculation of interatomic force constants (IFCs) within density functional theory (DFT)~\citetext{\citealp{PhotosynthRes.102.443}; \citealp{engel2011density}}, as shown in Fig.~\ref{fig:flowchart}. Using either density functional perturbation theory (DFPT) or finite-displacement methods, one obtains the real-space IFCs
\begin{equation}
\Phi_{\kappa\alpha, \kappa'\beta}(l - l') = \frac{\partial^2 E_{\text{tot}}}{\partial u_{l\kappa,\alpha} \partial u_{l'\kappa',\beta}},    
\end{equation}
where $E_{\text{tot}}$ is the total electronic ground-state energy, $\kappa,\kappa'$ label atoms inside the unit cell, and $\alpha,\beta$ denote Cartesian components. These IFCs form the basis of the dynamical matrix $D_{\kappa\alpha, \kappa'\beta}(\mathbf{q}) $,
whose diagonalization yields phonon eigenfrequencies $\omega_{n}(\mathbf{q})$ and normalized eigenvectors $\mathbf{e}_{n}(\mathbf{q})$.

To efficiently interpolate phonon dispersions throughout the Brillouin zone as well as the topological surface states canculation under the open boundary condition, Zhang \textit{et al.}, constructs a phonon tight-binding (TB) or Wannier-like model~\citetext{\citealp{PhysRevLett.120.016401}}. In this representation, the IFCs play the role of effective hopping amplitudes, analogous to orbital hoppings in electronic tight-binding Hamiltonians. Explicitly, the phonon Hamiltonian in reciprocal space can be written as
\begin{equation}
H_{\text{ph}}(\mathbf{q}) = \frac{1}{2}
\begin{pmatrix}
\mathbf{p}^\dagger(\mathbf{q}) & \mathbf{u}^\dagger(\mathbf{q})
\end{pmatrix}
\begin{pmatrix}
M^{-1} & 0 \\
0 & D(\mathbf{q})
\end{pmatrix}
\begin{pmatrix}
\mathbf{p}(\mathbf{q}) \
\mathbf{u}(\mathbf{q})
\end{pmatrix},
\end{equation}

where $\mathbf{u}(\mathbf{q})$ and $\mathbf{p}(\mathbf{q})$ denote displacement and momentum operators, $M$ is the mass matrix. In practice, Fourier transformation of IFCs allows the dynamical matrix to be smoothly interpolated, avoiding expensive DFT calculations at every $\mathbf{q}$-point. This interpolation framework is directly analogous to electronic Wannier interpolation, and is implemented in opensource codes.

The final step is the calculation of topological invariants. The normalized phonon eigenvectors ${\mathbf{e}_{n}(\mathbf{q})}$ define a fiber bundle over the Brillouin zone, allowing one to compute the Berry connection~\citetext{\citealp{RevModPhys.82.1959}},
\begin{equation}
\mathbf{A}_{n}(\mathbf{q}) = i\langle \mathbf{e}_{n}(\mathbf{q}) | \nabla_{\mathbf{q}} \mathbf{e}_{n}(\mathbf{q}) \rangle,    
\end{equation}
and the corresponding Berry curvature,

\begin{equation}
    \boldsymbol{\Omega}_{n}(\mathbf{q}) = \nabla_{\mathbf{q}} \times \mathbf{A}_{n}(\mathbf{q}).
\end{equation}

Integrating the Berry curvature yields topological invariants, such as the phonon Chern number,

\begin{equation}
C_{n} = \frac{1}{2\pi}\int_{\mathrm{BZ}} \boldsymbol{\Omega}_{n}(\mathbf{q}) \cdot d^2\mathbf{q},    
\end{equation}

or the chiral charge of a Weyl point,
\begin{equation}
\chi = \frac{1}{2\pi} \oint_{\mathcal{S}} \boldsymbol{\Omega}_{n}(\mathbf{q}) \cdot d\mathbf{S}.
\end{equation}

Here we note that, for most crystalline materials, phonon topology can also be efficiently diagnosed through symmetry data without explicitly computing the Berry curvature~\citetext{\citealp{NatCommun.8.50}; \citealp{PhysRevX.8.031069}; \citealp{JPhysDApplPhys.54.414002}; \citealp{PhysRevResearch.4.033170}; \citealp{Sci.Rep.13.9239}}. In particular, the irreducible representations (irreps) of the normalized phonon eigenvectors at high-symmetry points in the Brillouin zone provide a powerful input to the symmetry-based indicator or topological quantum chemistry (TQC) frameworks. By comparing these irreps and symmetry-based indicator analysis, one can directly infer the configuration of phonon band crossings, including: the type of node (Weyl point, Dirac point or nodal line), the location in momentum space (high-symmetry point, high-symmetry line, or generic k-point), the number of such nodes imposed by the crystal space group, and the corresponding topological charge ($\mathbb{Z}_2$ charge).
This approach is particularly advantageous for high-throughput or database-assisted screening~\citetext{\citealp{Innovation.2.100134}; \citealp{Science.384.eadf8458}}, as it relies primarily on symmetry properties of the phonon eigenvectors and does not require full topological invariant integration.

% \begin{tabularx}{\linewidth}{|c|X|}
% \hline
% \textbf{Probe} & \textbf{Physical quantity measured} \\
% \hline
% IXS & Phonon dispersion \newline Dynamical structure factor $S(\mathbf{q},\omega)$ \\
% \hline
% \end{tabularx}

\begin{table*}[t]
\centering
\caption{Representative experimental techniques for probing phonon topology. The first column illustrates the principle of each technique, the second column summarizes the experimental observables or information accessed, and the last column highlights their significance from the topological perspective.}
%\begin{tabular}{p{4cm} | p{5.5cm} | p{8.5cm}}
\setlength{\tabcolsep}{2pt}
\renewcommand{\arraystretch}{1.05}
\begin{tabular*}{\textwidth}{@{\extracolsep{\fill}}l l l}
\hline
\parbox[t]{3.0cm}{\raggedright\textbf{Technique}} & \parbox[t]{7.0cm}{\raggedright\textbf{Information Accessed}} & \parbox[t]{5.6cm}{\raggedright\textbf{Significance}} \\ 
\hline
 \parbox[t]{3.0cm}{\raggedright INS / IXS}
& \parbox[t]{7.0cm}{\raggedright Full phonon dispersion $\omega_{n} (\mathbf{q})$; \newline Dynamical structural factor $S(\mathbf{q},\omega)$.}
& \parbox[t]{5.6cm}{\raggedright Direct mapping of topological phonon bands; \newline Extract the wavefunction information.} \\
\hline
 \parbox[t]{3.0cm}{\raggedright Raman Spectroscopy / Infrared Spectroscopy}
& \parbox[t]{7.0cm}{\raggedright Zone-center vibrational symmetries; \newline Mode splitting; \newline Selection rules.}
& \parbox[t]{5.6cm}{\raggedright Identify rotational chirality; \newline Identify degeneracy lifting; \newline Identify symmetry-enforced phonons.} \\
\hline
 \parbox[t]{3.0cm}{\raggedright Ultrafast Pump-Probe Spectroscopy}
& \parbox[t]{7.0cm}{\raggedright Coherent phonon phase evolution; \newline Circularly polarized phonon dynamics.}
& \parbox[t]{5.6cm}{\raggedright Detect Berry-curvature-induced effects; \newline Detect phonon angular momentum.} \\
\hline
 \parbox[t]{3.0cm}{\raggedright Momentum-resolved EELS}
& \parbox[t]{7.0cm}{\raggedright Full phonon dispersion $\omega_{n} (\mathbf{q})$; \newline Energy-loss function $L(\mathbf{q}, \omega)$.}
& \parbox[t]{5.6cm}{\raggedright Toward direct observation of topological surface phonons; \newline Direct mapping of topological phonons in 2D materials.} \\
\hline
\end{tabular*}
\end{table*}

{\subsection{Experimental probes of phonon topology}}

A range of experimental techniques can reveal phonon topology, either by directly resolving the phonon band dispersion $\omega_{n}(\mathbf{q})$ or by measuring the dynamical structure factor, which characterizes the momentum- and energy-resolved response of the lattice. In general, the measurable quantity is the dynamical structure factor
\begin{equation}
S(\mathbf{q},\omega) = \sum_{n} \frac{|\mathbf{F}_{n}(\mathbf{q})|^{2}}{2\omega_{n}(\mathbf{q})}
\left[ \delta(\omega - \omega_{n}(\mathbf{q})) - \delta(\omega + \omega_{n}(\mathbf{q})) \right],    
\end{equation}
where $\mathbf{F}_{n}(\mathbf{q})$ denotes the mode-dependent scattering form factor. The intensity profile of $S(\mathbf{q},\omega)$ directly maps how individual phonon modes are distributed in momentum-energy space, enabling access to band crossings, degeneracy lifting, and mode chirality.

\subsubsection{Inelastic Neutron Scattering (INS)}

INS measures the dynamical structure factor using neutron momentum and energy transfer~\citetext{\citealp{PhysRev.87.366}; \citealp{RepProgPhys.39.911}; \citealp{NeutronNews.3.26}}:
\begin{equation}
\mathbf{q} = \mathbf{k}_{\text{in}} - \mathbf{k}_{\text{out}}, \quad
\omega = \frac{\hbar}{2m_n}\left(k_{\text{in}}^{2} - k_{\text{out}}^{2}\right).    
\end{equation}

Because neutrons couple to nuclear displacements, INS provides full Brillouin-zone mapping of phonon dispersions, including longitudinal and transverse branches. INS has been crucial in identifying Weyl phonons.

\subsubsection{Inelastic X-ray Scattering (IXS)}

IXS probes lattice vibrations through electron density modulation~\citetext{\citealp{RevModPhys.73.203}; \citealp{RevModPhys.83.705}}. The scattering cross-section is proportional to
\begin{equation}
I_{\text{IXS}}(\mathbf{q},\omega) \propto | \mathbf{q}\cdot \mathbf{e}_{n}(\mathbf{q}) |^{2} S(\mathbf{q},\omega),    
\end{equation}
making it particularly sensitive to optical phonon modes and high-energy dispersions. Due to higher momentum resolution and small sample requirements, IXS is powerful for detecting Weyl phonon nodes and node-line phonons.

\subsubsection{Raman and Infrared (IR) Spectroscopy}

Raman~\citetext{\citealp{clark1977raman}; \citealp{Springer.10.978}; \citealp{smith2019modern}} and IR spectroscopy~\citetext{\citealp{HandbInstrumTechAnalChem.249.1}; \citealp{Kirk-OthmerEncyclChemTechnol.2000.1}; \citealp{alpert2012ir}} probe zone-center phonons $(\mathbf{q}=0)$ via light--matter coupling with symmetry-dependent selection rules. Raman intensity is proportional to
\begin{equation}
I_{\text{Raman}} \propto \left| \frac{\partial \chi_{\alpha\beta}}{\partial u_{n}} \right|^{2},    
\end{equation}
where $\chi_{\alpha\beta}$ is the optical susceptibility tensor. Degeneracy lifting of doubly degenerate modes (e.g., $E_g \rightarrow E_1 + E_2)$ provides a key signature of broken inversion, time-reversal, or circularly polarized phonon polarization. Circularly polarized Raman can further detect phonon pseudo-angular momentum and topology.

\subsubsection{Ultrafast Pump-probe Spectroscopy}

Ultrafast pump-probe experiments excite coherent phonons and track the time-dependent phase of the oscillation~\citetext{\citealp{ChemPhys.193.211}; \citealp{ChemRev.104.1781}; \citealp{JAmChemSoc.142.3}; \citealp{AdvSci.8.2102488}}:
\begin{equation}
\phi_{n}(t) = \arg( u_{n}(t) ).    
\end{equation}

Berry curvature in momentum space induces phase shifts and polarization rotation of coherent phonon wave packets, resulting in observable chiral oscillation dynamics. This method allows direct observation of circularly polarized phonons.

\subsubsection{Momentum-Resolved EELS Techniques}

Momentum-resolved electron energy-loss spectroscopy (EELS) measures vibrational excitations with nanometer spatial resolution~\citetext{\citealp{RepProgPhys.72.016502}; \citealp{egerton2011electron}; \citealp{ibach2013electron}; \citealp{hofer2016fundamentals}; \citealp{PhysRevB.99.094105}; \citealp{MicroscMicroanal.27.136}}, enabling detection of surface phonon arcs and edge-localized topological vibrational states. In experiments, a focused electron beam interacts with the sample, and the energy loss of the transmitted electrons reflects the excitation spectrum of lattice vibrations, encoded in the energy-loss function
    $L(\mathbf{q}, \omega) = -\operatorname{Im}\left[\frac{1}{\varepsilon(\mathbf{q}, \omega)}\right]$,
where $\varepsilon(\mathbf{q}, \omega)$ is the complex dielectric function. The peaks in $L(\mathbf{q}, \omega)$ correspond to the phonon excitation energies at momentum $\mathbf{q}$, allowing reconstruction of the phonon band structure with high spatial and momentum resolution.

\begin{figure*}
    \centering
    \includegraphics[width=1\linewidth]{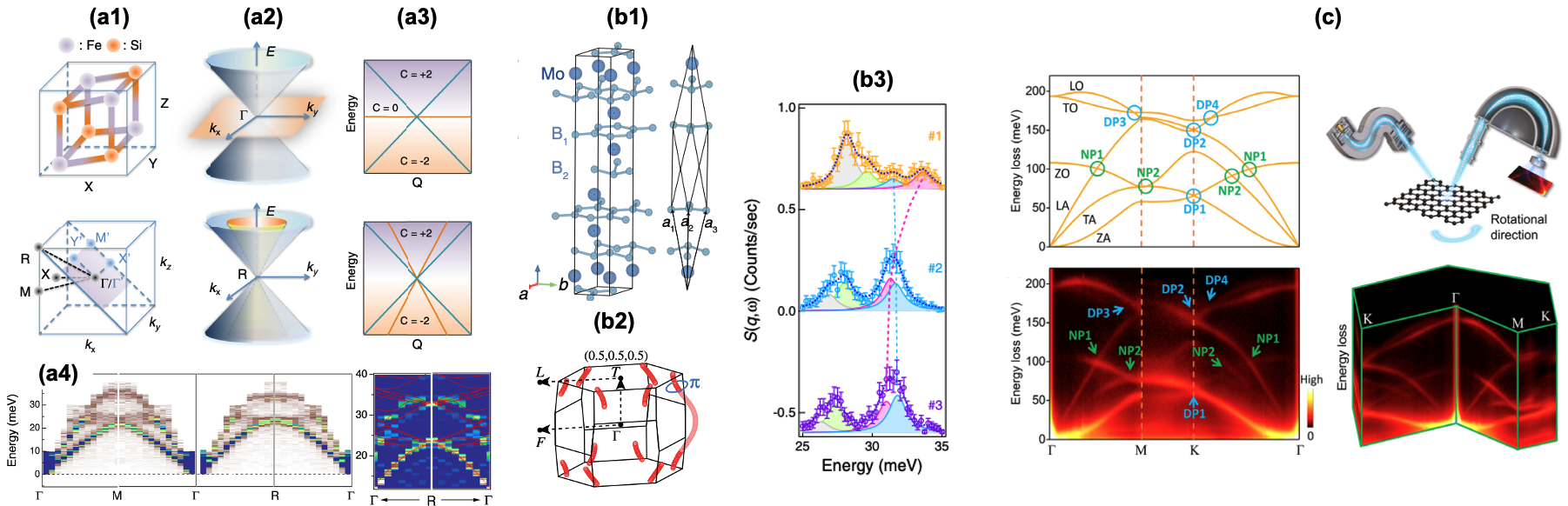}
    \caption{Experimentally verified topological phonons in three representative materials: (a) The FeSi family, (b) MoB$_2$, and (c) the two-dimensional material graphene. (a1) shows the crystal structure and Brillouin zone (BZ) of the FeSi family. (a2) and (a3) present the calculated phonon band dispersions featuring the charge-2 Dirac phonon and spin-1 Weyl phonon, respectively, while (a4) displays the corresponding experimental phonon dispersions measured by inelastic X-ray scattering (IXS). (b1) and (b2) show the crystal structure and BZ of MoB$_2$, where the red curved lines indicate the $\mathcal{PT}$-symmetry-protected nodal lines; (b3) shows the experimentally observed phonon nodal-line crossing from IXS measurements. (c) presents the phonon band dispersion of graphene obtained from DFT calculations (upper left) and electron energy-loss spectroscopy (EELS) experiments (lower pannels). This figure is adapted from \citetext{\citealp{PhysRevLett.121.035302}; \citealp{PhysRevLett.123.245302}; \citealp{PhysRevLett.131.116602}}.}
    \label{fig:exp_mat}
\end{figure*}

{\subsection{Experimentally observed topological phonons}}

The experimental discovery of topological phonons has opened a new frontier in condensed matter physics, bridging lattice dynamics with concepts traditionally associated with electronic topology. In this section, we highlight three representative material systems, the double Weyl phonons in FeSi family, the twofold quadruple Weyl phonons in BaPtGe, nodal line phonons in MoB$_2$, and Dirac phonons in 2D graphene, among which distinct types of topological phonons have been theoretically predicted~\citetext{\citealp{PhysRevLett.120.016401}; \citealp{PhysRevLett.123.245302}; \citealp{PhysRevB.102.125148}; \citealp{PhysRevB.102.125148,npjComputMater.6.95}} and experimentally verified~\citetext{\citealp{PhysRevLett.121.035302}; \citealp{PhysRevLett.123.245302}; \citealp{PhysRevB.103.184301}; \citealp{PhysRevB.103.184301}; \citealp{PhysRevB.106.224304}; \citealp{PhysRevLett.131.116602}}, primarily through inelastic X-ray scattering (IXS), inelastic neutron scattering (INS) and electron energy-loss spectroscopy (EELS).

\subsubsection{Weyl phonons in B20-type chiral crystals}

The FeSi-type compounds (including FeSi, CoSi, BaPtGe and other B20-type chiral crystals) crystallize in a non-centrosymmetric cubic structure belonging to the space group ( P$2_13$ ), as shown in Fig.~\ref{fig:exp_mat}. The lack of both inversion and mirror symmetry allows for nontrivial phonon band degeneracies with quantized topological charges.
DFT calculations for FeSi reveal two types of topological phonon quasiparticles: a charge-2 Dirac phonon at the Brillouin zone corner, characterized by a fourfold degeneracy with total Chern number C=$\pm$2, and a spin-1 Weyl phonon at the zone center, corresponding to a threefold degeneracy with quantized monopole charge C=0, $\pm$ 2, as shown in Figs.~\ref{fig:exp_mat} (a2)-(a3).

Using high-resolution IXS, two kinds of Weyl phonons have been experimentally observed in transition-metal monosilicides such as FeSi, CoSi, and MnSi. The linear crossings measured by IXS agree well with \textit{ab initio} predictions, confirming the presence of chiral phonon modes and topological phonons.

We note that another class of topological phonon quasiparticles, twofold quadruple Weyl phonons, has also been theoretically predicted in B20-type chiral crystals~\citetext{\citealp{PhysRevB.102.125148}}. These excitations correspond to a twofold band degeneracy carrying a topological charge of C = $\pm$4, representing higher-order Weyl points in the phonon spectrum. Remarkably, such charge-4 Weyl phonons were subsequently experimentally confirmed in BaPtGe~\citetext{\citealp{PhysRevB.103.184301}} via IXS and in CoSi/MnSi via INS~\citetext{\citealp{PhysRevB.106.224304}}.

\begin{figure}
    \centering
    \includegraphics[width=0.95\linewidth]{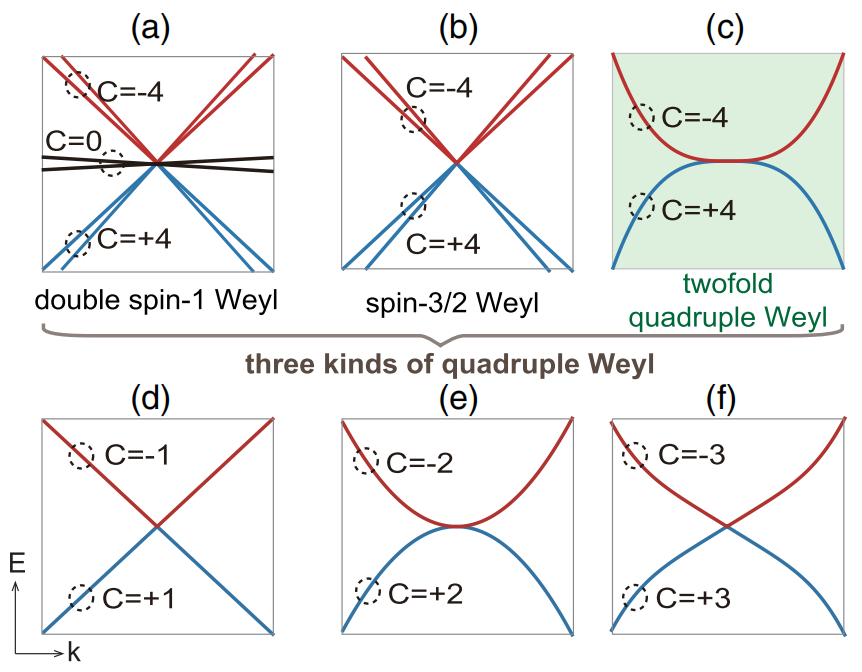}
    \caption{Different types of Weyl quasiparticles carrying different monopole charge. (a) A sixfold double spin-1 Weyl node, composed of two identical spin-1 Weyl nodes, which can be only obtained in fermionic and spinful systems. (b) A fourfold spin-3/2 Weyl node, with Chern numbers +1, +3, $−1$, and $−3$ assigned to the respective bands. This kind of Weyl quasiparticle also can be only obtained in fermionic and spinful systems. (c) A twofold quadruple Weyl node, with band Chern numbers of +4 and $−4$, which can be only obtained in Bosonic systems. (d–f) Three additional types of twofold Weyl nodes realizable in crystalline materials, namely single, double, and triple Weyl nodes, respectively, with the monopole charge of 1, 2 and 3. This figure is adapted from \citetext{\citealp{PhysRevB.102.125148}}.}
    \label{fig:tqw}
\end{figure}

\subsubsection{\textcolor{blue}{Twofold quadruple Weyl phonons in BaPtGe}}

In addition to the two types of double Weyl phonons, unconventional Weyl phonons with higher Chern numbers (e.g., $C = 4$) can in principle be realized, although the required conditions are rather stringent. In fermionic systems, multifold Weyl fermions, such as sixfold double spin-1 Weyl nodes or fourfold spin-3/2 Weyl nodes, can emerge in spinful case, giving rise to a total monopole charge of 4, as illustrated in Figs.~\ref{fig:tqw}(a,b).
By contrast, in phonon spectra, the realization of such high monopole charges is far more constrained and typically requires the combined protection of time-reversal symmetry and specific crystalline symmetries. Previous studies have shown that in systems preserving time-reversal symmetry and chiral cubic symmetry~\citetext{\citealp{PhysRevB.102.125148,npjComputMater.6.95}}, twofold Weyl phonons can host a monopole charge of 4, as depicted in Fig.~\ref{fig:tqw}(c).
Beyond the twofold quadruple Weyl~(TQW) phonon, twofold Weyl phonons can also exhibit lower Chern numbers, such as $C = 1$ and $C = 2$, as shown in Figs.~\ref{fig:tqw}(d,e). In contrast, the $C = 3$ case is generally inaccessible in phononic systems and can only be realized in fermionic systems when spin–orbit coupling is taken into account.

\begin{figure*}
    \centering
    \includegraphics[width=0.9\linewidth]{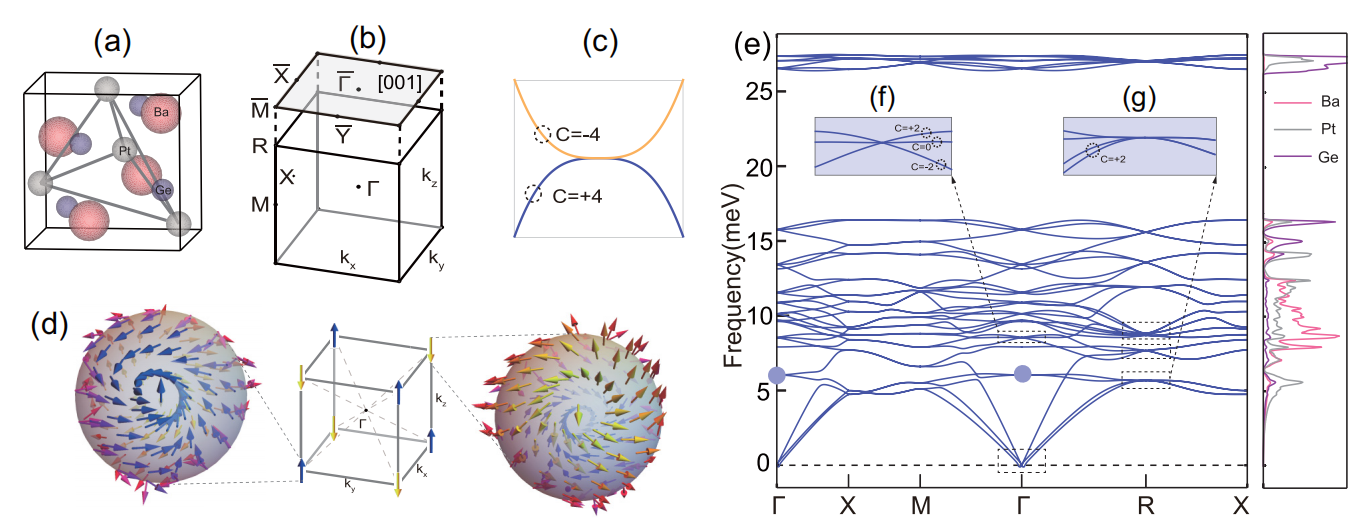}
    \caption{Twofold quadruple Weyl (TQW) phonons in BaPtGe. (a), (b) Crystal structure and Brillouin zone of BaPtGe, respectively. (c) TQW node characterized by band Chern numbers $C = +4$ and $-4$. (d) Pseudospin textures for the TQW at the eight symmetry-related R points, where colored arrows denote the pseudospin orientations. Enlarged views of the pseudospin textures near the time-reversal-related momenta reveal opposite chiralities. (e) Phonon spectra of BaPtGe obtained from first-principles calculations. Solid circles at the $\Gamma$ point mark the predicted TQW nodes. The dashed rectangles around the $\Gamma$ and R points highlight threefold and fourfold double Weyl nodes, respectively, which are further magnified in (f) and (g). This figure is adapted from \citetext{\citealp{PhysRevB.102.125148}}.}
    \label{fig:tqw-dft}
\end{figure*}

\begin{figure*}
    \centering
    \includegraphics[width=0.9\linewidth]{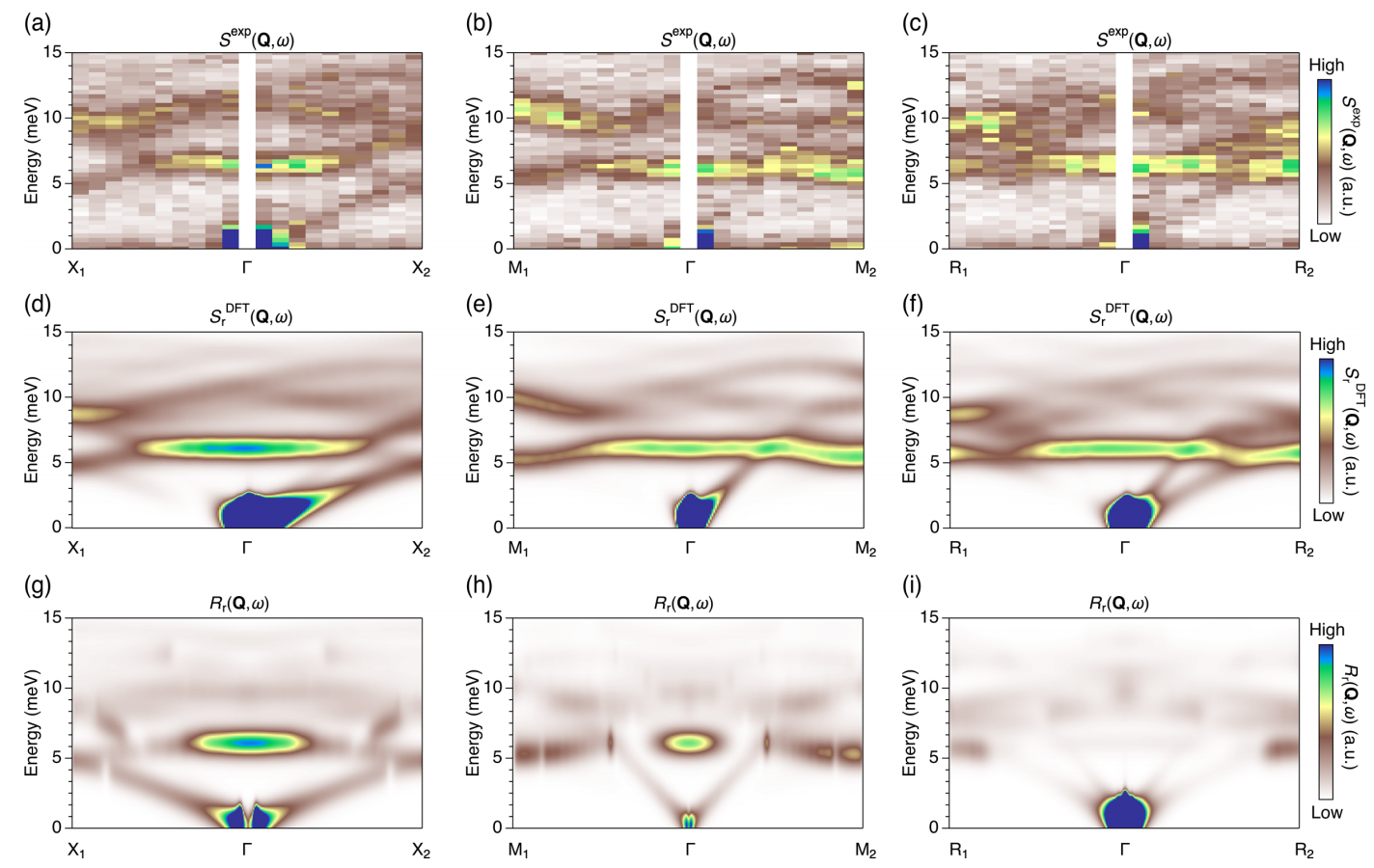}
    \caption{Momentum-resolved chiral wavefunction associated with the TQW. (a)–(c) Measured dynamic structure factor $S^{\mathrm{exp}}(\mathbf{Q}, \omega)$ along the high-symmetry directions [001], [011], and [111] by inelastic x-ray scattering. The selected momentum points are defined as $X_1 = (0, 3, 2.5)$, $X_2 = (0, 3, 3.5)$, $M_1 = (0, 2.5, 2.5)$, $M_2 = (0, 3.5, 3.5)$, $R_1 = (-0.5, 2.5, 2.5)$, and $R_2 = (0.5, 3.5, 3.5)$. (d)–(f) Corresponding DFT-derived spectra $S^{\mathrm{DFT}}_{r}(\mathbf{Q}, \omega)$ after convolution with the experimental resolution function, while (g)–(i) present the similarly processed response $R_{r}(\mathbf{Q}, \omega)$. $R_{r}(\mathbf{Q}, \omega)$ considers only the real part of the phonon wavefunction. Panels (d)–(i) share a common color scale.  This figure is adapted from \citetext{\citealp{PhysRevB.103.184301}}.}
    \label{fig:tqw-exp}
\end{figure*}

First-principles calculations predict that BaPtGe hosts a TQW phonon in its bulk phonon spectra~\citetext{\citealp{PhysRevB.102.125148}}. As shown in Fig. \ref{fig:tqw-dft}, the crystal structure and corresponding Brillouin zone support symmetry-protected band crossings at high-symmetry points. In particular, a TQW node emerges at the $\Gamma$ point, characterized by band Chern numbers $C=+4$ and $-4$ displayed in Fig. \ref{fig:tqw-dft}(c). The associated pseudospin textures further reveal the chiral nature of the phonon wave functions, with opposite chirality at time-reversal-related momenta, as shown in Fig. \ref{fig:tqw-dft}(d). In addition to the TQW node, Figure \ref{fig:tqw-dft}(e) shows that the phonon dispersion also hosts multiple types of Weyl phonons, including threefold and fourfold double Weyl phonons, highlighting the richness of topological phonon states in this material.

Experimentally, the topological chiral phonon wave functions are probed using high-resolution inelastic X-ray scattering~\citetext{\citealp{PhysRevB.103.184301}}, as presented in Figs. \ref{fig:tqw-exp} (a-c). The measured dynamical structure factor $S^{\mathrm{exp}}(\mathbf{Q}, \omega)$ along several high-symmetry directions exhibits clear signatures of the TQW, including characteristic spectral features near the band crossing points. By directly comparing the experimental results with resolution-convoluted first-principles calculations, $S^{\mathrm{DFT}}_{r}(\mathbf{Q}, \omega)$ and $R_{r}(\mathbf{Q}, \omega)$, an excellent agreement is observed. This consistency provides strong evidence for the existence of TQW phonons in BaPtGe.

\subsubsection{Node-line phonons in MoB$_2$}

IXS has been applied to resolve nodal-line phonon degeneracies.
Transition-metal diborides such as MoB$_2$ crystallize in a centrosymmetric hexagonal structure (space group P6/mmm), as shown in Fig.~\ref{fig:exp_mat} (b1), which preserves both inversion $(\mathcal{P})$ and time-reversal $(\mathcal{T})$ symmetries. The coexistence of $\mathcal{P}$ and $\mathcal{T}$ enforces the quantization condition: $\boldsymbol{\Omega}_n(\mathbf{q}) = -\boldsymbol{\Omega}_n(-\mathbf{q})$,
ensuring that the total Berry curvature vanishes globally but allowing for topological nodal-line phonons where two phonon branches cross and remain degenerate along closed loops in the Brillouin zone, as shown in Fig.~\ref{fig:exp_mat} (b2).

Theoretical calculations show that in MoB$_2$, phonon branches intersect along extended one-dimensional nodal rings protected by combined $\mathcal{PT}$ symmetry~\citetext{\citealp{PhysRevLett.123.245302}}. These nodal lines carry a quantized Berry phase of $\pi$:
$\gamma = \oint_C \mathbf{A}(\mathbf{q}) \cdot d\mathbf{q} = \pi$,
where  $\mathbf{A}(\mathbf{q}) = i \langle u(\mathbf{q}) | \nabla_{\mathbf{q}} | u(\mathbf{q}) \rangle$  is the phonon Berry connection.
Inelastic X-ray scattering experiments have directly resolved these band crossings in MoB$_2$~\citetext{\citealp{PhysRevLett.123.245302}}, with the measured phonon dispersion showing linear crossings consistent with the theoretical nodal-ring topology, as shown in Fig.~\ref{fig:exp_mat} (b3). 

\subsubsection{Dirac phonons in 2D graphene}

In addition to topological phonon in 3D materials, recently, momentum-resolved electron energy-loss spectroscopy (EELS) has been successfully implemented to directly probe the phonon dispersion of graphene, providing compelling evidence for the existence of topological phonon modes in two-dimensional materials~\citetext{\citealp{PhysRevLett.131.116602}}, as shown in Fig.~\ref{fig:exp_mat} (c). 

In graphene, high-resolution EELS measurements have mapped out the acoustic and optical phonon branches across the Brillouin zone, revealing degeneracy points and band crossings consistent with DFT predictions of Dirac-type topological phonons at the high-symmetry $K$ and $K'$ valleys. These degeneracies, protected by the underlying honeycomb lattice symmetry and time-reversal symmetry, correspond to quantized phonon Berry phases of $\pi$, analogous to the spinless Dirac cones in graphene electronic band structure.

% \begin{tabular}{|c|c|c|c|}
% % \begin{tabularx}{\textwidth}{|X|X|X|X|}
% \hline
% Type & classification & prediction & experiments \\ 
% \hline
% point  & Weyl & Fesi & PRL2018 \\
% \hline
% point  & conventional Dirac & ... & ... \\
% \hline
% point  & charge-2 Dirac & ... & ... \\
% \hline
% point  & Z2 Dirac & ... & ... \\
% \hline
% line  & nodal line & MoB2 & PRL2019 \\
% \hline
% line  & nodal chain & ...& ... \\
% \hline
% plane  & nodal plane & ... & ... \\
% \hline
% gapped & QHE & ... & ... \\
% \hline
% ...
% \label{tab:3d}
% \end{tabular}

% zhang2025natcommun, zhang2025thechirality

{\subsection{Theoretically predicted topological phonons and database}}

%This section provides a systematic review of topological phonons in two-dimensional crystalline systems, structured through three interconnected dimensions: (i) model studies on the Hall family, (ii) prototypical material realizations by first-principles predictions spanning from graphene analogs to hexagnoal magnets (EuSi2, etc.), and (iii) synergistic advances bridging first-principles predictions with EELS measurements.

%\begin{center}
%\begin{tabular}{cccc}
%\hline
%Type & classification & prediction & experiments \\ 
%\hline
%Model & QAHE & liu & none \\
%material study & graphene & ... & jiade li \\
%\hline
%\end{tabular}
%\end{center}

Topological phonons were theoretically predicted to exist among a large number of crystalline solids but only a few of them are experimentally verified possibly due to the complexity of phonon dispersion in realistic solids. This section provides a introductory review of the theoretically proposed models and representative material candidates. 

\subsubsection{Examples of 2D topological phononic models}

%\begin{itemize}
%    \item 2D Dirac phonon, Zak phase, polarization
%    \item The phononic Chern insulator, time-reversal symmetry breaking, phonon Chern number in 2D
%    \item Valley Chern number, 
%    \item 3D Weyl
%\end{itemize}

Honeycomb lattices represent typical model systems that can support various topological states in 2D. The $C_{3v}$ symmetry at Brillouin zone corners $K$ and $K'$ can support two-dimensional irreducible representations which will form cone-like band dispersion relations. 
%Similarly, the inplane and out-of-plane modes of graphene also exhibit phononic Dirac cones at the same high-symmetry momentum \citetext{\citealp{PhysRevB.101.081403}; \citealp{PhysRevB.106.L121401}}. 
Based on symmetry analysis \citetext{\citealp{PhysRevB.96.064106}}, the effective model of phononic Dirac cones can be described by 
\begin{equation}\label{Heff_2D_Dirac}
    H^{\textrm{Dirac}}_{\textrm{2D}} = \omega_D + v_D (k_y \tau_z \sigma_x - k_x \sigma_y),
\end{equation}
with $\sigma_z=\pm$ referring to the two degenerate orbitals, and $\tau_z=\pm$ referrring to the $K/K'$ valley, respectively; $\omega_D$ and $v_D$ refer to the frequency and group velocity at the Dirac points, respectively. For such effective model, the Berry curvature vanishes everywhere except at the Dirac points where the Berry curvature becomes singularly large due to the band degeneracy. If we integrate the Berry flux around each valley or calculate the Berry phase acquired along a closed loop enclosing each valley, we get a quantized $\pm\pi$ flux or phase indicating its topologically nontrivial character [Figs. \ref{fig_theory_materials}(a1)-(a4)]. The topologically nontrivial character is also manifested in the existence of edge states which have been unveiled in various hexagonal materials including CrI$_3$ and YGaI \citetext{\citealp{NanoLett.18.7755}}, graphene \citetext{\citealp{PhysRevB.101.081403}; \citealp{PhysRevB.106.L121401}} and other related materials \citetext{\citealp{PhysRevB.107.144307}}. 

The 2D Dirac phonons are protected by the combined inversion and time-reversal symmetry $\mathcal{PT}$. By further introducing symmetry breaking terms, the Dirac phonons can be gapped out. Upon the different mass terms introduced, the band topology can be driven into distinct phases leading to the quantum-Hall family of topological phonon states. 

For example, the 2D Dirac phonon can be gapped by the inversion symmetry breaking Semenoff-type mass term $H_I = m_I \sigma_z$. Such a mass term combined with the Dirac effective model in Eq. \eqref{Heff_2D_Dirac} leads to a valley contrasting Berry curvature 
\begin{equation}
    \Omega^I_z(\mathbf{k}) = \tau_z \frac{m_I v^2_D}{\left( v^2_D \mathbf{k}^2 + m^2_I \right)^{\frac{3}{2}} }.
\end{equation}
In the limit of $m_I \rightarrow 0$, the Berry curvature mainly distributes around the $K/K'$ valleys leading to nearly quantized valley-polarized Chern number $C_K = -C_{K'} = \frac{1}{2}\mathrm{sgn}(m_I) + O(m_I)$ if we integrate the Berry curvature around each valley. Such system belong to quantum valley Hall systems. Materials like monolayer boron nitride and transition metal dichalcogenides belong to this kind. 

The 2D Dirac phonon can also be gapped by time-reversal symmetry breaking Haldane-type mass term $H_T = m_T \sigma_z \tau_z$ \citetext{\citealp{PhysRevB.96.064106}}. In this case the effective model exhibits Berry curvature 
\begin{equation}
    \Omega^T_z(\mathbf{k}) = \tau^2_z \frac{m_T v^2_D}{\left( v^2_D \mathbf{k}^2 + m^2_T \right)^{\frac{3}{2}}} = \frac{m_T v^2_D}{\left( v^2_D \mathbf{k}^2 + m^2_T \right)^{\frac{3}{2}}}.
\end{equation}
Such system is characterized by nonzero total Chern number $C=\mathrm{sgn}(m_T)$ which belong to the quantum (anomalous) Hall system. In crystalline solids, the time-reversal symmetry breaking effects can be induced by Raman-type spin-lattice interactions \citetext{\citealp{PhysRevLett.96.155901}; \citealp{PhysRevLett.100.145902}; \citealp{PhysRevLett.105.225901}}, Coriolis force \citetext{\citealp{PhysRevLett.115.104302}; \citealp{NewJPhys.17.073031}}, or molecular Berry curvature effects \citetext{\citealp{PhysRevB.105.064303}}. 

The microscopic lattice Hamiltonian of phonon with time-reversal symmetry breaking can be expressed as \citetext{\citealp{PhysRevLett.105.225901}; \citealp{PhysRevB.96.064106}}
\begin{equation}
    \begin{split}
        H =& \frac{1}{2} \dot{u}^2_i + \frac{1}{2} D_{ij} u_i u_j \\
        =& \frac{1}{2} (p_i - \eta_{ij} u_j)^2 + \frac{1}{2} D_{ij} u_i u_j,
    \end{split}
\end{equation}
where $u_i = \sqrt{m_i} x_i$ refers to the mass-weighted atomic displacement for $i$-th microscopic degree of freedom. Einstein's convention of summation over repeated index is assumed. $\dot{u}_i = p_i - \eta_{ij} u_j$ is the atomic velocity, where $p_i$ refers to the canonical momentum and $\eta_{ij}$ is an antisymmetric tensor which play the role of gauge potential for phonons. Since the atomic velocity (or canonical momentum) is coupled with displacement, the lattice dynamics can only be correctly described in the extended velocity-displacement space. By making use of the canonical Hamiltonian equation of motion, the lattice dynamics can be expressed into a Schr\"odinger-like form as \citetext{\citealp{PhysRevB.96.064106}; \citealp{NatlSciRev.5.314}}
\begin{equation}
    \begin{split}
        & H_\mathbf{k} |\psi_{n\mathbf{k}} \rangle = \omega_{n\mathbf{k}} |\psi_{n\mathbf{k}} \rangle, \\
        & H_\mathbf{k} = \left(
        \begin{array}{cc}
            0 & iD^{1/2}_\mathbf{k} \\
            -iD^{1/2}_\mathbf{k} & -2i\eta_\mathbf{k}
        \end{array}
        \right), |\psi_{n\mathbf{k}} \rangle = \left(
        \begin{array}{c}
            D^{1/2}_\mathbf{k} |\mathbf{u_k} \rangle \\
            | \mathbf{\dot{u}_k} \rangle
        \end{array}
        \right).
    \end{split}
\end{equation}
When $\eta_{\mathbf{k}}=0$, i.e., when the time-reversal symmetry is present, the effective Hamiltonian $H_\mathbf{k}$ can be block-diagonalized into two independent blocks, which reduces to the ordinary normal-mode equation $D_\mathbf{k} |\mathbf{u_k} \rangle = \omega^2_{n\mathbf{k}} |\mathbf{u_k} \rangle $.

\subsubsection{Examples of 2D topological phononic materials}

\begin{figure*}
    \includegraphics[width=\linewidth]{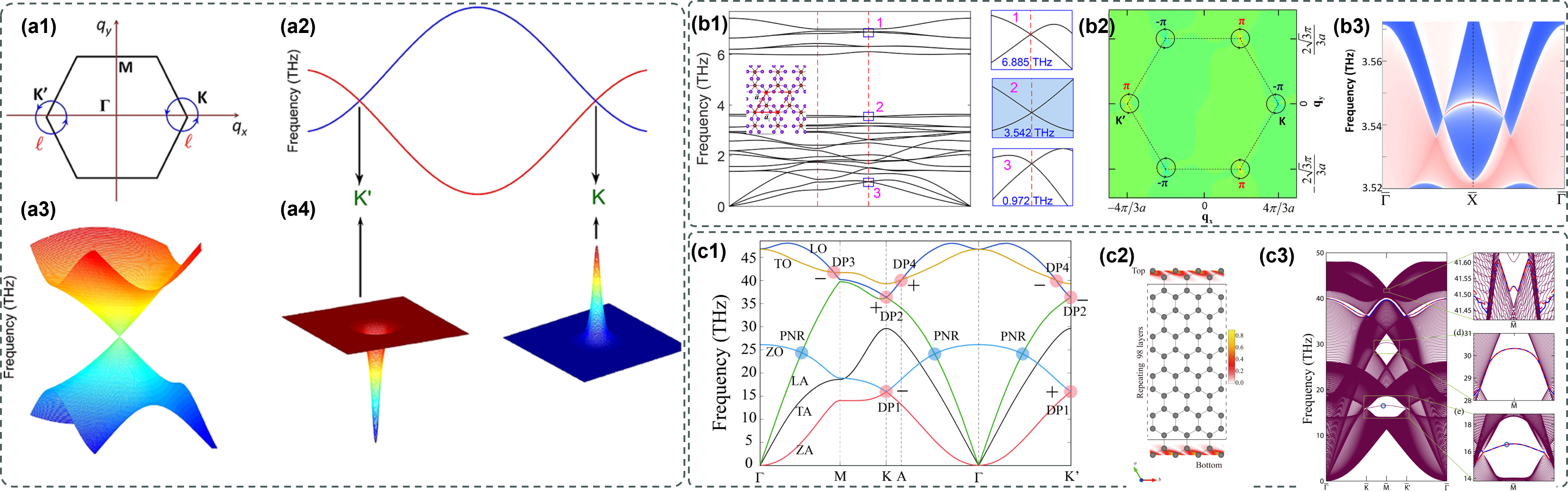}
    \caption{Topological Dirac phonon materials. (a1)-(a4) Schematic topological Dirac phonons in 2D hexagonal lattices. (b1)-(b3) Topological phonons in CrI$_3$. (c1)-(c3) Topological Dirac phonons in graphene. Adapted from \citetext{\citealp{NanoLett.18.7755}} and \citetext{\citealp{PhysRevB.101.081403}}. }
    \label{fig_theory_materials}
\end{figure*}

The 2D Dirac phonons have been proposed in several kinds of hexagonal materials \citetext{\citealp{NanoLett.18.7755}; \citealp{PhysRevB.101.081403}; \citealp{PhysRevB.106.L121401}; \citealp{PhysRevB.107.144307}} including CrI$_3$, YGaI, and graphene as shown in Figs. \ref{fig_theory_materials}(b1)-(b3) and (c1)-(c3). All the two-band degenerate points at $K/K'$ are phononic Dirac points characterized by quantized Berry flux $\pm\pi$. 

Graphene is a representative candidate supporting 2D Dirac phonons. Both its out-of-plane and in-plane modes can support Dirac points at $K/K'$ points as shown by the ``DP1'' and ``DP2'' respectively in Fig. \ref{fig_theory_materials}(c1). Besides, graphene also has other two Dirac points at the mirror symmetric lines $\Gamma M$ and $\Gamma K$ as shown by ``DP3'' and ``DP4''. All the Dirac points carry quantized Berry phase of $\pm\pi$ when integrating along a small loop enclosing the Dirac point. This leads to the in-gap edge states for a zigzag nanoribbon as shown in Figs. \ref{fig_theory_materials}(c2) and (c3). The topological phonons in graphene was further experimentally studied by \citetext{\citealp{PhysRevLett.131.116602}} using HR-EELS.

\begin{figure}
    \centering
    \includegraphics[width=\linewidth]{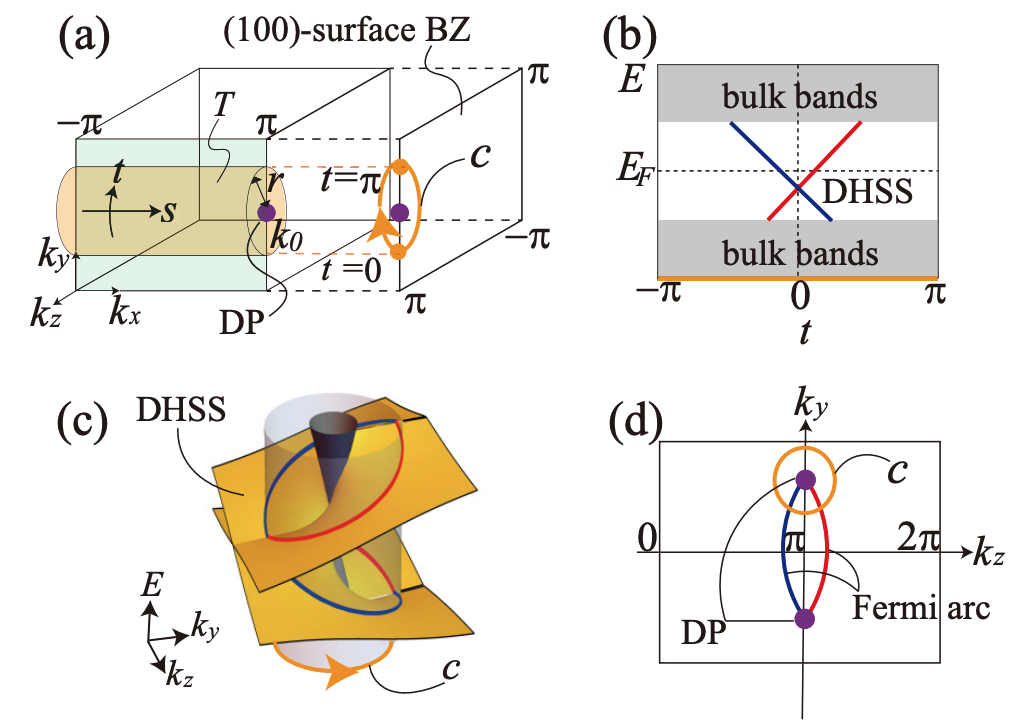}
    \caption{Bulk-surface correspondence associated with the $\mathbb{Z}_2$ Dirac point (DP), which is different from the conventional Dirac point carrying the $\mathbb{Z}$-type monopole charge of zero. (a) A torus $T$ defined in the bulk Brillouin zone for analyzing the topological correspondence, which projects onto a closed loop $c$ in the (100) surface BZ. (b) Surface band dispersion along $c$. When the torus $T$ encloses a Dirac point, the surface states exhibit a helical dispersion, reflecting a nontrivial $\mathbb{Z}_2$ invariant. (c) Double-helicoid surface states (DHSSs), which can be continuously revealed by tuning the radius $r$ of the torus $T$. (d) Double Fermi arcs in the (100) surface BZ, connecting the projected Dirac points. For conventional Dirac points with a $\mathbb{Z}$-type monopole charge of zero, Fermi arcs are absent. This figure is adapted from \citetext{\citealp{{PhysRevB.112.045137}}}. }
    \label{fig:dhss}
\end{figure}

\begin{figure*}
    \centering
    \includegraphics[width=0.7\linewidth]{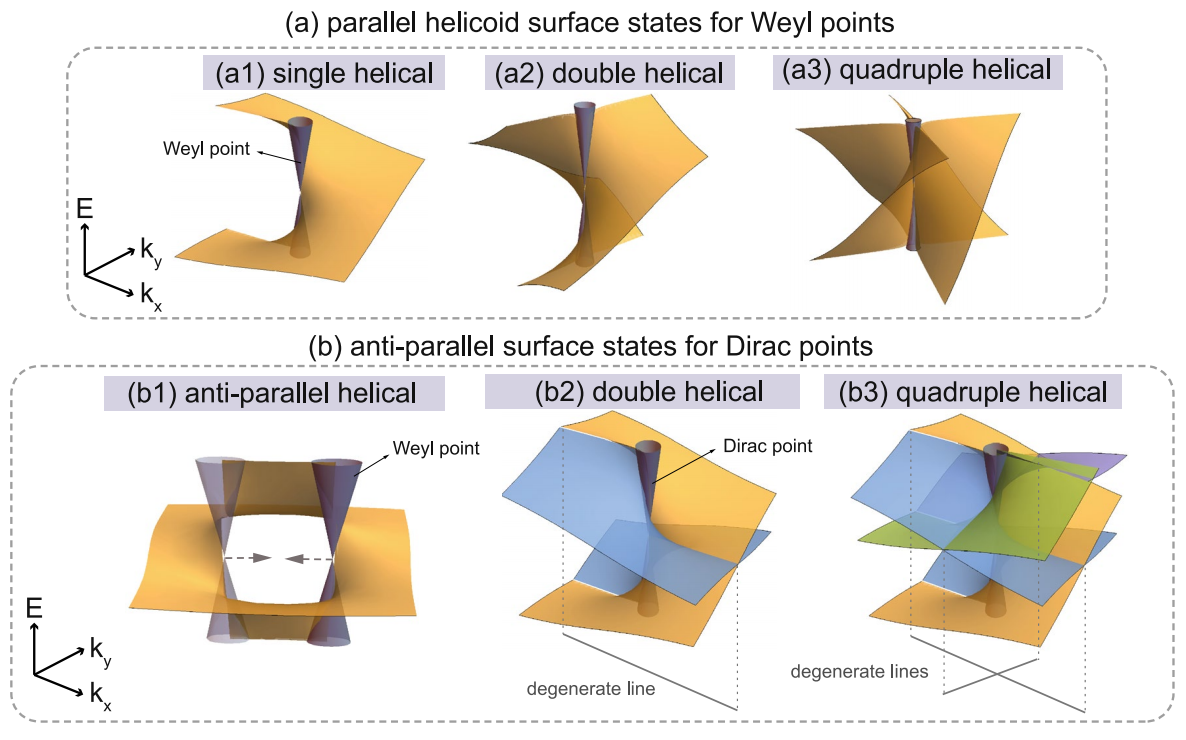}
    \caption{Bulk-surface correspondence for Weyl phonons and Dirac phonons, i.e., parallel and antiparallel helical surface states. (a) Parallel helical surface state for Weyl phonons carrying $\mathbb{Z}$-type monopole charge of $C = +1,+2$ and $C = +4$. (b1) Surface states for a pair of Weyl phonons with monopole charges of $C = \pm 1$. (b2) Antiparallel double-helical surface states for a Dirac point carrying a nonzero $\mathbb{Z}_2$-type monopole charge, where the crossing of surface states forms a degeneracy line along the boundary of the surface Brillouin zone. (b3) Antiparallel quad-helical surface states for Dirac points with a nonzero $\mathbb{Z}_2$-type monopole charge, where the intersections occur along the lines protected by two time-reversal-glide-mirror symmetries. This figure is adapted from \citetext{\citealp{Sci.Rep.13.9239}}. }
    \label{fig:hss}
\end{figure*}

\subsubsection{\textcolor{blue}{$\mathbb{Z}_2$ Dirac phonons and material prediction}}

In addition to conventional Dirac phonons with a trivial $\mathbb{Z}$-type monopole charge and the $\mathbb{Z}$-type charge-2 Dirac phonons identified in the B20 structure discussed above, Dirac points may also carry a $\mathbb{Z}_2$-type monopole charge~\citetext{\citealp{PhysRevResearch.4.033170}; \citealp{PhysRevB.112.045137}}.
As illustrated in Fig.~\ref{fig:dhss}, the bulk–surface correspondence associated with a $\mathbb{Z}_2$ Dirac point exhibits fundamentally different behavior from that of conventional Dirac points. In general, a Dirac point can be viewed as the superposition of two Weyl nodes with opposite topological charges, resulting in a net $\mathbb{Z}$-type monopole charge of zero and, consequently, the absence of topologically protected surface arcs. However, when additional symmetries are present, i.e., time-reversal symmetry and glide mirror symmetry, the topological characterization can be enriched beyond the conventional $\mathbb{Z}$ classification.

In particular, by introducing a torus $T$ in the bulk Brillouin zone that encloses the Dirac point, as shown in Figs.~\ref{fig:dhss}(a,c), one can define a $\mathbb{Z}_2$ topological invariant associated with the enclosed region. Upon projection onto the (100) surface, this torus maps onto a closed loop $c$, along which the surface band structure exhibits a helical dispersion when the $\mathbb{Z}_2$ invariant is nontrivial. This leads to the emergence of double-helicoid surface states (DHSSs) in Fig.~\ref{fig:dhss}(c), which can be continuously visualized by tuning the radius of the torus. Similarly, these nontrivial surface states give rise to double surface arcs that connect the projected Dirac points in the surface Brillouin zone, as shown in Fig.~\ref{fig:dhss}(d). Unlike the trivial case, these features are protected by symmetry and remain robust as long as the underlying symmetries are preserved, providing a clear manifestation of the nontrivial $\mathbb{Z}_2$ topology in Dirac semimetals~\citetext{\citealp{Sci.Rep.13.9239}}.

Figure~\ref{fig:hss} summarizes the bulk–surface correspondence for both Weyl and Dirac phonons in terms of their helical surface states. It illustrates the distinction between parallel helical surface states associated with Weyl phonons carrying $\mathbb{Z}$-type monopole charges and antiparallel helical surface states emerging from Dirac phonons with nontrivial $\mathbb{Z}_2$ topology~\citetext{\citealp{Sci.Rep.13.9239}}.

\begin{figure*}
    \centering
    \includegraphics[width=0.8\linewidth]{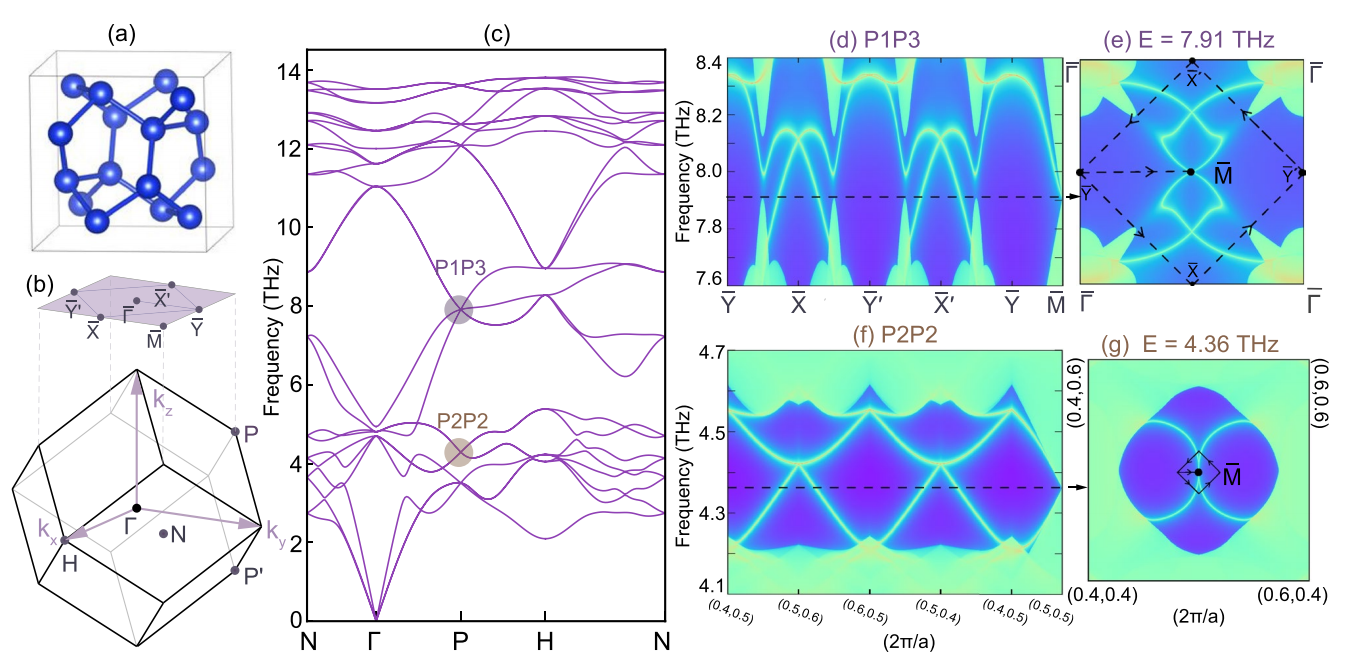}
    \caption{$\mathbb{Z}_2$-type Dirac phonons in silicon with space group $Ia\bar{3}$. (a) Crystal structure. (b) Bulk Brillouin zone and the projected (001) surface BZ. (c) Phonon band structure of Si, where the points $P_1P_3$ and $P_2P_2$, highlighted by colored markers, both correspond to Dirac phonons with distinct irreducible representations. (d,e) Calculated surface states and surface arcs associated with the pair of Dirac phonons at $P$ and $P'$ characterized by the $P_1P_3$ irreducible representation. (f,g) Corresponding surface states and surface arcs for the Dirac phonons at $P$ and $P'$ with the $P_2P_2$ irreducible representation. This figure is adapted from \citetext{\citealp{Sci.Rep.13.9239}}. }
    \label{fig:qhss}
\end{figure*}

\citetext{\citealp{Sci.Rep.13.9239}} proposes silicon in space group \#206 ($Ia\bar{3}$) as an experimentally accessible platform hosting Dirac phonons with nontrivial $\mathbb{Z}_2$ topology. As shown in Figs.~\ref{fig:qhss}(a-c), the body-centered structure with two perpendicular glide mirrors $G_x$ and $G_y$ gives rise to two non-TRIM points $P$ and $P'$ related by time-reversal symmetry, where fourfold-degenerate phonon bands form Dirac phonons with irreps $P_2P_2$ and $P_1P_3$, both carrying a nonzero $\mathbb{Z}_2$-type monopole charge.

Surface state calculations on the (001) surface reveal anti-parallel quad-helical surface states for both types of Dirac phonons. As shown in Figs.~\ref{fig:qhss}(d-g), the projected Dirac points coincide at $\bar{M}$, and the presence of those two time-reversal-glide-mirror symmetries enforces Kramers-like degeneracies along the surface BZ boundaries. Consequently, symmetry-protected crossings of multiple helical surface states emerge, demonstrating that anti-parallel quad-HSSs originate from both $P_1P_3$ and $P_2P_2$ Dirac phonons.

\subsubsection{Examples of 3D topological phonon materials and topological phononic database}

The three-dimensional topological phases of phonon can be roughly classified into topological ``insulators'' with topological gaps and ``semimetals'' with gapless nodes. The ``insulators'' can be fully classified based on symmetry representations of the ``occupied'' states below the band gap. The topologically nontrivial bands can be defined by the impossibility to be connected to the ``atomic insulator (AI)'' without symmetry breaking or gap closing. Such ``insulators'' are usually named as ``obstructed atomic insulator (OAI)''. Recently, Xu et al. developed a systemtic way to classify phonon bands into OAIs and AIs \cite{Science.384.eadf8458} based on the method of topological quantum chemistry \cite{nature.547.298}. 

%The topological gapless phonons include Dirac \cite{PhysRevLett.126.185}, Weyl, and nodal line phonons.

Recently, topological phononic material database has been built by \citetext{\citealp{NatCommun.12.1204}; \citealp{Science.384.eadf8458}} based on high-throughput first-principles calculations. \citetext{\citealp{NatCommun.12.1204}} filted out topological phononic nodal lines and Weyl phonons in currently avaible phononic database. \citetext{\citealp{Science.384.eadf8458}} further identified all topological nontrivial band structures based on the theory of topological quantum chemistry. These studies provide versatile potential candidate materials for further experimental studies. 

\begin{figure*}
    \centering
    \includegraphics[width=\linewidth]{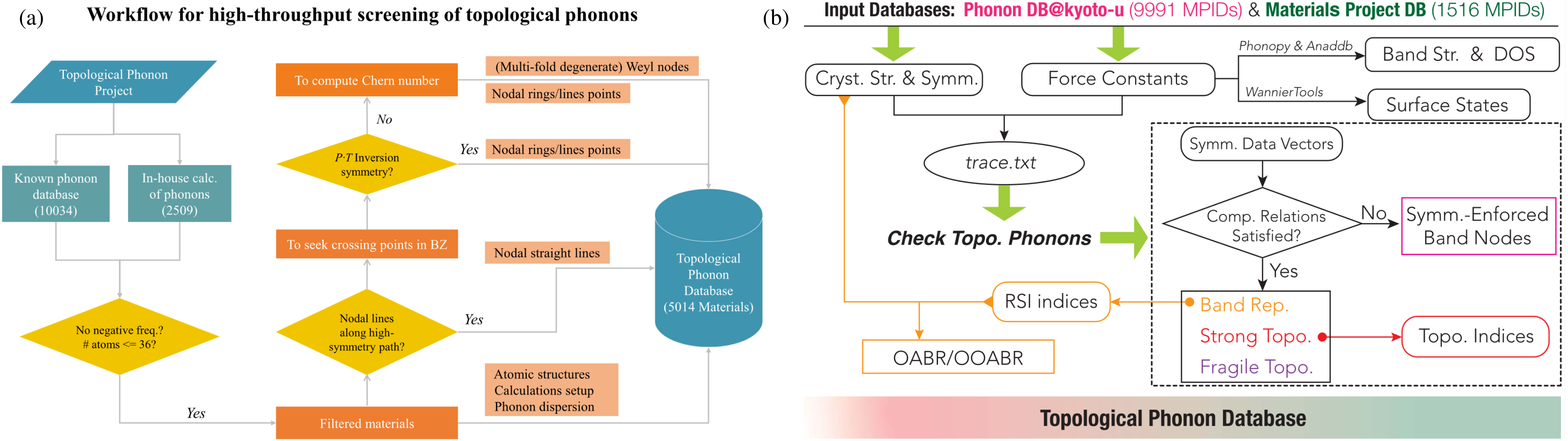}
    \caption{(a) Schematic flow of high-throughput calculation of topological phonon materials \cite{NatCommun.12.1204}. (b) Workflow of topological phononic OAIs discovery \cite{Science.384.eadf8458}. }
    \label{fig_database}
\end{figure*}

We first discuss the band topology formed by two-band crossings. A generic two-band model can be described by the Hamiltonian:
\begin{equation}\label{H_two-band}
    H_{\textrm{two-band}} = \varepsilon_0(\bm{k}) \sigma_0 + h_x(\bm{k}) \sigma_x + h_y (\bm{k}) \sigma_y + h_z(\bm{k}) \sigma_z,
\end{equation}
where $\sigma_0$ is the $2\times2$ identity matrix, and $\sigma_{x,y,z}$ refer to the Pauli matrices. $\varepsilon_0 (\bm{k})$ represents a band-bending term and does not affect the band topology. The topological properties of the effective model are determined by the nodes of $h_{x,y,z}$ in Eq. \eqref{H_two-band}. Symmetries place fundamental restrictions on the structure and topological properties of nodes. For example, in the presence of $\mathcal{PT}$ symmetry, the effective model Hamiltonian should be real-valued up to a gauge transformation, i.e. $h_y = 0$ for arbitrary wave vector $\bm{k}$. In 3D momentum space, the node condition $h_x(\bm{k}) = h_z (\bm{k}) = 0$ typically gives rise to a continuous line manifold leading to various topological nodal-line structures including nodal lines, rings, chains, and cages. If $\mathcal{PT}$ symmetry is broken, which typically exists in noncentrosymmetric nonmagnetic materials, all the coefficients $h_x$, $h_y$, and $h_z$ are generally nonzero, which makes the band nodes discrete points in the momentum space. Such nodes are generally Weyl phonons characterized by nonzero Chern number. This symmetry-breaking process has been investigated in the MgB$_2$ system \citetext{\citealp{PhysRevB.101.024301}}.
In 2D space, however, the $\mathcal{PT}$ symmetry allows the discrete nodes which is 2D Dirac phonons, while the breaking of $\mathcal{PT}$ give rises to gapped bands. 

%\subsubsection{Theoretically predicted material candidates}

%\begin{itemize}
%    \item Graphene (2D Dirac)
%    \item CrI$_3$ (2D Dirac)
%    \item BN (2D valley Hall, 2D massive Dirac)
%    \item Te, SiO$_2$ (3D Weyl)
%    \item Database
%\end{itemize}

% \subsubsection{Experimental verification}

%\begin{figure*}
%    \includegraphics[height=4cm]{Topo_1D_2D_3D.png}
%    \caption{Topological phononic models and their corresponding realization in (a) 1D (SSH) (b) 2D (massless and massive Dirac), and (c) 3D (Weyl) systems respectively.}
%\end{figure*}

{\subsubsection{High-order phonon topology in solids}}

Compared to the well-established field of high-order band topology in electronic systems, the investigation of their phononic counterparts in solids remains in its infancy. Current research on higher-order phonon topology in solids primarily focuses on identifying and calculating topological invariants that extend beyond conventional first-order indices, enabling the characterization of boundary states residing in dimensions at least two levels lower than the bulk. To date, first-principles calculations have successfully predicted and realized various phenomena, including high-order nodal points~\cite{PhysRevB.105.035429}, two-dimensional second-order phononic topological insulators~\cite{acs.nanolett.1c04239}, higher-order topology characterized by quadrupole and bulk dipole polarization~\cite{PhysRevB.109.115422}, and the integration of higher-order topology with unconventional Weyl dipole phonons~\cite{advs.202504812}. Recently, the modulation of high-order phonon topology~\cite{PhysRevB.111.115407} and the transport behaviors of phononic topological corner states~\cite{PhysRevB.112.075415} have also attracted considerable attention.

\begin{figure*}
    \includegraphics[width=\linewidth]{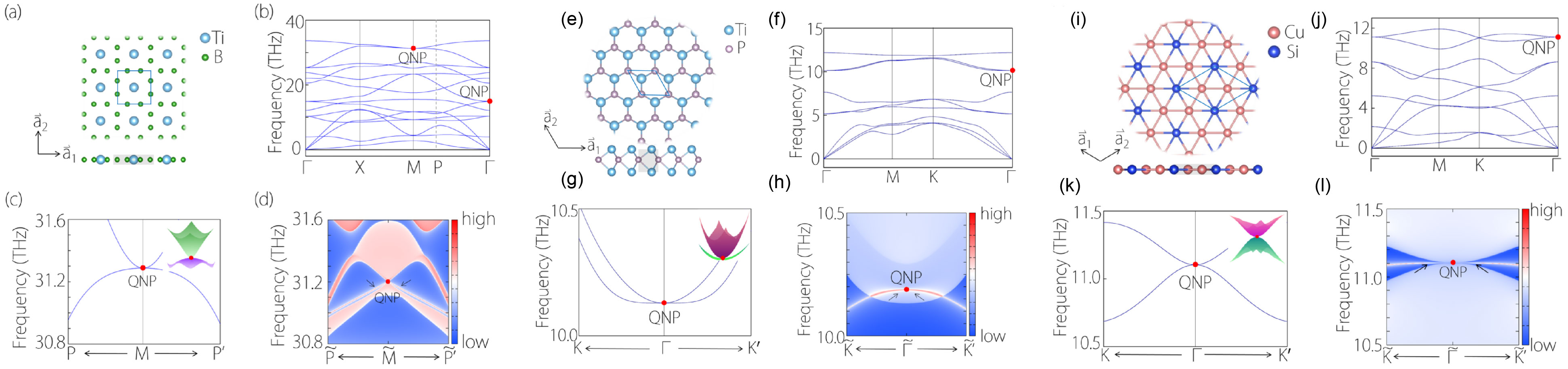}
    \caption{Lattice structures of (a) TiB$_4$, (e) Ti$_2$P, and (i) Cu$_2$Si. Phonon spectrum along high symmetry path for (b) TiB$_4$, (f) Ti$_2$P and (j) Cu$_2$Si monolayer. Enlarged view for QNP at $\Gamma$ points and its corresponding 3D band structure of (c) TiB$_4$, (g) Ti$_2$P, and (k) Cu$_2$Si is shown in the inset. Surface spectrum along [100], [110], and [100] directions for (d) TiB$_4$, (h), and (l) Cu$_2$Si, respectively. Adapted from \citetext{\citealp{PhysRevB.105.035429}}.}
    \label{fig:high-order-qnps}
\end{figure*}

TiB$_4$, Ti$_2$P, and Cu$_2$Si host higher-order quadratic nodal points (QNPs)~\cite{PhysRevB.105.035429}, which are protected by the combination of rotation symmetry ($C_{nz}$ for n=3,4,6) and time-reversal symmetry. Analysis  based on $\textbf{k·p}$ theory reveals that these QNPs possess a quadratic dispersion and are associated with distinct edge states spanning the Brillouin zone boundary. The lattice structures of TiB$_4$, Ti$_2$P, and Cu$_2$Si, phonon spectra along high symmetry paths, and the corresponding surface states are displayed in Fig.~\ref{fig:high-order-qnps}, which also presents the 3D view of the quadratic dispersion.

\begin{figure*}
    \includegraphics[width=\linewidth]{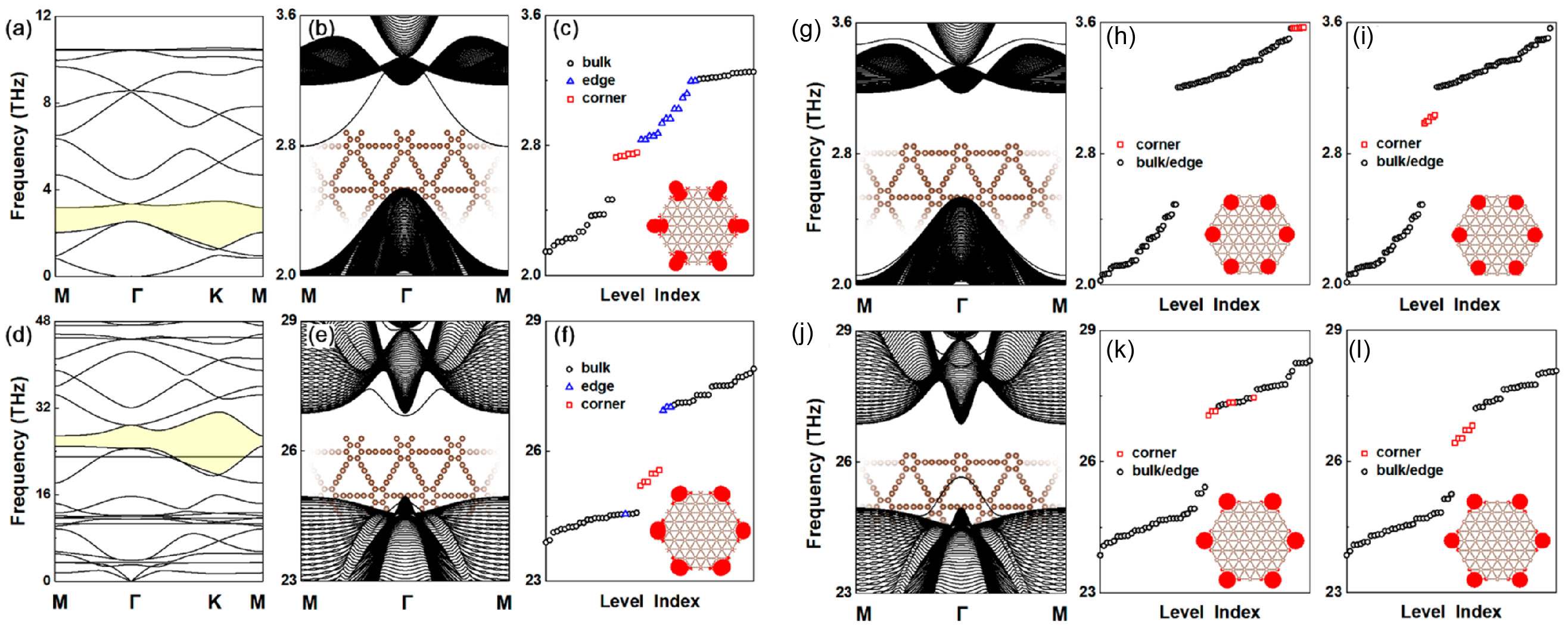}
    \caption{Numerical simulation of phononic SOTI (a-f) and  robust topological corner states (g-l) in graphdiyne. (a) Out-of-plane phonon bands of graphdiyne with the SOTI gap highlighted by a light-yellow color. (b) Out-of-plane phonon bands of graphdiyne nanoribbon with edge termination shown in the inset. (c) Discrete out-of-plane phonon levels of hexagonal graphdiyne cluster with the spatial distribution of six corner states shown in the inset. The radii of red circles denote the intensity of phonon mode. (d−f) Same as in panels a−c, but for the in-plane modes of graphdiyne. (g) Out-of-plane phonon bands of graphdiyne nanoribbon with edge termination shown in the inset. (h) Discrete out-ofplane phonon levels of hexagonal graphdiyne cluster with the spatial distribution of six corner states shown in the inset. (i) Same as panel b but with an additional local corner potential that can tune the frequencies of six corner states into bulk gap. (j−l) Same as in panels g−i, but for the in-plane modes of graphdiyne. Adapted from \citetext{\citealp{acs.nanolett.1c04239}}.}
    \label{fig:high-order-2dsoti}
\end{figure*}

The Kekulé lattice, graphdiyne, has been proposed as a two-dimensional second-order topological insulator (SOTI) for both phononic and electronic systems, manifesting in both out-of-plane and in-plane phonon modes~\cite{acs.nanolett.1c04239}, as shown in Fig.~\ref{fig:high-order-2dsoti}. It exhibits a nontrivial quadrupole moment of $q_{12} = 1/2$ and hosts spatially localized topological corner states. Notably, these topological corner states can reside either inside or outside the bulk gap and are tunable by the local corner potential. Further analysis reveals that the electron–phonon coupling between the localized phononic and electronic topological corner states is significantly stronger and more robust than that between nontrivial (trivial) electronic states and trivial (nontrivial) phononic states in other topological systems, providing an exotic platform to explore the interplay between higher-order topological phonons and electrons.

\begin{figure*}
    \includegraphics[width=0.5\linewidth]{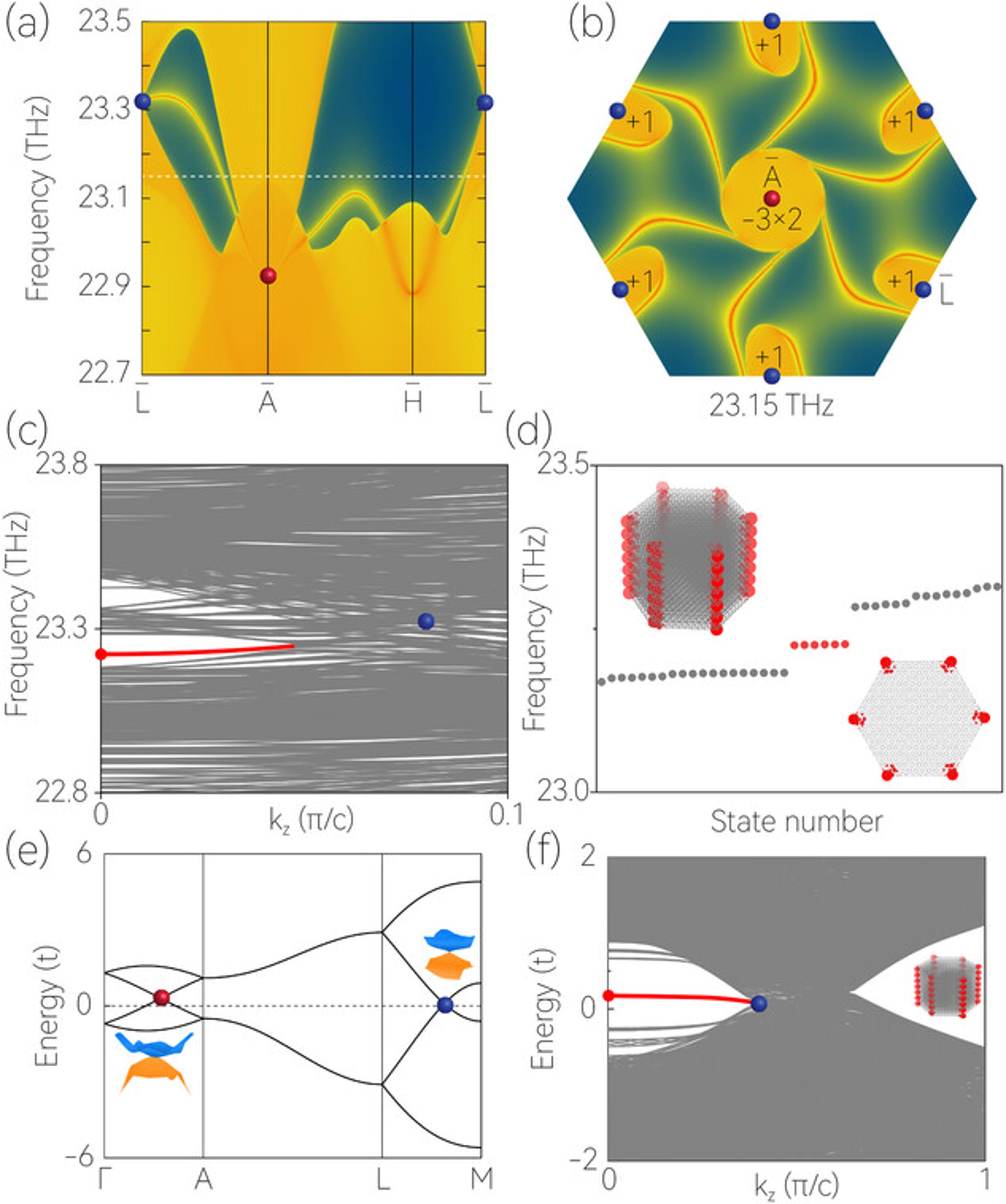}
    \caption{(a) Projected spectrum of Y(OH)$_3$ on the (001) surface. (b) Constant-frequency slice at 23.15 THz. (c) Enlarged phonon spectrum (in the $k_z$ direction) within the frequency range from 22.8 to 23.8 THz for a sample of Y(OH)$_3$ with a tube-like geometry. The hinge states are highlighted by the red curves. (d) Frequency spectrum for a sixfold degenerate state (red dot) at $k_z$ = 0. These insets are the spatial distributions for the sixfold degenerate state. (e) Bulk band structure of the tight-binding model. These insets are 3D plots of bands around the charge-3 WP and charge-1 WP. (f) The spectrum of a 1D tube geometry (in the $k_z$ direction). The inset is the spatial distribution for the state at the $k_z$ = 0. Adapted from \citetext{\citealp{advs.202504812}}. }
    \label{fig:high-order-WPs}
\end{figure*}

In the $P6_3-$type Y(OH)$_3$, a unique unconventional Weyl dipole phase exists in the phonon spectrum~\cite{advs.202504812}. This dipole consists of an unconventional charge-3 Weyl point (WP) and three conventional charge-1 WPs, collectively carrying a quantized quadrupole moment of $Q_{WD} = 1/2$. This state is distinguished by its multidimensional boundary modes, including 2D sextuple-helicoid Fermi-arc states on the surfaces and 1D hinge states connecting the dipoles, as demonstrated in Fig.~\ref{fig:high-order-WPs}.

\begin{figure*}
    \includegraphics[width=\linewidth]{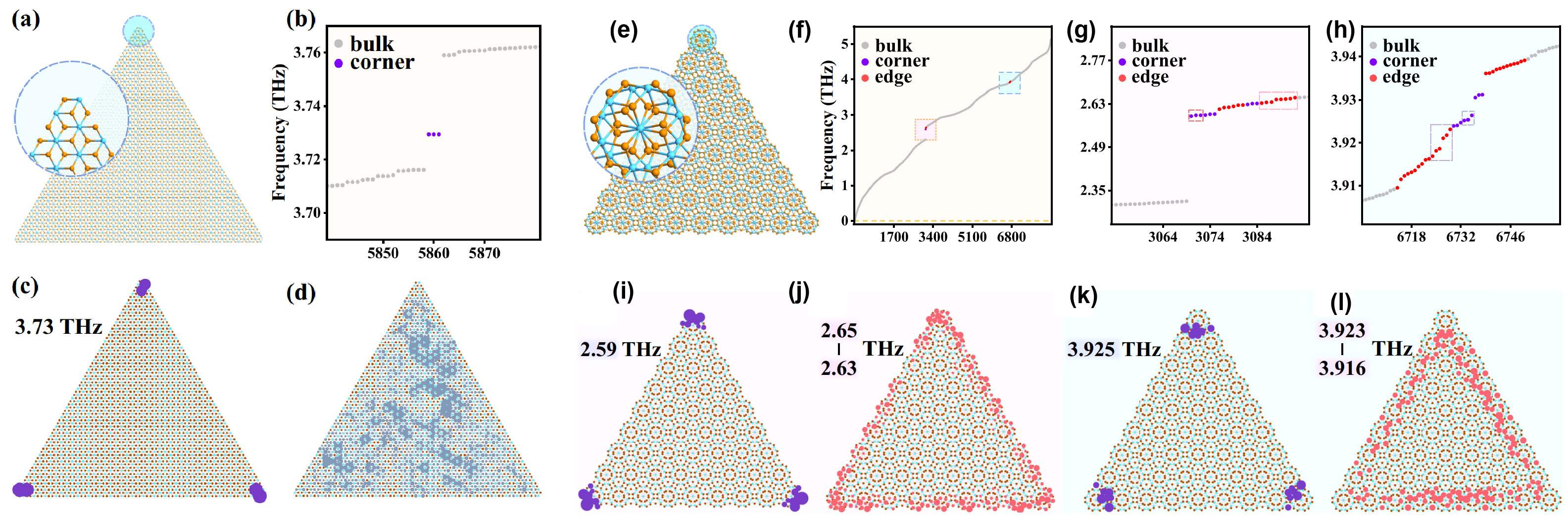}
    \caption{The triangular nanodisk of 1T-HfTe$_2$ monolayer and corresponding distributions of topological modes in both momentum space and real space. (a) Top view of the 1T-HfTe$_2$ double armchairedged triangular nanodisk, with the circular area on the left showing an enlarged view of the top corner. (b) The phonon spectrum within the frequency range that encompasses the phonon topological modes of the nanodisk in (a). The gray and purple dots represent the bulk and topological corner modes, respectively. (c) Spatial distribution of phonon topological modes in the double armchair-edged nanodisk of the 1T-HfTe$_2$ monolayer, with purple circles representing the phonon topological corner modes. (d) Similar to (c), but corresponding to the spatial distribution of phonon bulk modes, with blue circles indicating the bulk modes. The triangular nanodisk of the HfTe$_2$ twisted smallest moiré inversion lattice (TSMIL) and corresponding distribution of phonon topological modes in momentum space and real space. (e) The top view of the HfTe$_2$ TSMIL triangular nanodisk. The circle area on the left corresponds to the enlarged view of the region marked by the small blue circle at the top corner of the nanodisk. (f) The phonon spectrum corresponding to the HfTe$_2$ TSMIL triangular nanodisk in (e). The light pink and light blue boxes in (f) encompass phonon topological modes and part of the bulk modes. Gray, purple, and red dots represent bulk modes, topological corner modes, and topological edge modes, respectively. (g) and (h) The magnified views corresponding to the light pink and light blue boxes in (f), respectively. (i) and (j) The distribution of phonon topological corner and edge modes in the HfTe$_2$ TSMIL nanodisk corresponding to the topological modes within the dashed boxes in (g); purple and red dots represent corner and edge modes, respectively. The frequency of the topological corner modes in (i) and the frequency range of the topological edge modes in (j) are indicated on the left side of the respective nanodisks. (k) and (l) Similar to (i) and (j), but corresponding to the topological modes in (h). Adapted from \citetext{\citealp{PhysRevB.111.115407}}. }
    \label{fig:high-order-twist}
\end{figure*}

Beyond the realization of higher-order phononic band topology in materials, the modulation of such topology has also attracted recent attention~\cite{PhysRevB.111.115407}. The 1T-HfTe$_2$ monolayer and its twisted trilayer moiré lattice have been demonstrated to undergo transitions from trivial to nontrivial phases upon the application of external stimuli. In the strained monolayer, the system manifests as a second-order topological insulator characterized by a fractional quadrupole moment, whereas the twisted moiré lattice features fractional bulk dipole polarization. Both configurations support topological edge modes as well as corner modes, as shown in Fig. \ref{fig:high-order-twist}. 

{\subsubsection{Non-Abelian and Euler topology in solids}}
Phononic systems have proven exceptionally suited for the experimental and computational exploration of real band topology protected by $PT$ or $C_2T$ symmetry (see Secs. IID and IVC3 for a more detailed introduction). In solids, a seminal work established a first-principles methodology to compute the Euler class and non-Abelian frame charges from phonon eigenvectors~\citetext{\citealp{PRB.105.085115}}, applying it to monolayer Al$_2$O$_3$. By simulating electrostatic doping, they demonstrated a symmetry-constrained braiding process within a three-band subspace, wherein band inversions at high-symmetry points led to the transfer of stable nodal pairs between adjacent frequency gaps. The accompanying evolution of edge states and the $Z_2$-quantized Zak phase provided a clear bulk-boundary signature of the underlying multigap topology. Concurrently, another work~\citetext{\citealp{NatCommun.13.423}} identified layered silicates, such as monolayer Si$_2$O$_3$ with its characteristic kagome lattice, as a versatile material platform. Under experimentally feasible strain and electric fields, its phonon spectrum exhibits braiding processes involving more than three bands—a significant extension beyond the three-band paradigm—where triply and quadruply degenerate points mediate the conversion of non-Abelian charges. Notably, these band inversions were shown to be directly observable through the evolution of Raman-active modes, offering a clear experimental fingerprint for the detection of non-Abelian braiding in real materials.

A complementary and comprehensive perspective on real band topology in phonons is provided by the study of real triple points (RTPs)~\citetext{\citealp{PRB.110.174111}}. Tang et al. performed a systematic symmetry analysis across all 1651 magnetic space groups containing PTPT symmetry, deriving the $k\cdot p$ models for RTPs pinned at high-symmetry points or emerging as the Nambu-Goldstone mode of acoustic phonons~\citetext{\citealp{PRB.110.174111}}. They demonstrated that the Euler number serves as the topological invariant for these triply degenerate points, even though their Chern number vanishes by symmetry. By integrating first-principles calculations with large-scale materials databases (PhononDB@kyoto-u and ICSD), they cataloged 386 materials hosting RTPs with a non-vanishing Euler number. Taking KM$_g$F$_3$ as a representative example, they further showed that applying external strain to break the cubic symmetry transforms the RTP into a linked nodal structure—a direct manifestation of the underlying Euler topology. This work not only provides a unified symmetry-based picture of RTPs but also underscores the ubiquity of multigap topological phonons in naturally occurring crystalline solids.

From an experimental standpoint, the verification of non-Abelian phonon topology relies on a suite of established spectroscopic techniques. While Raman spectroscopy offers a convenient route to track band inversions at the Brillouin zone center~\citetext{\citealp{PRB.105.085115}}, the full momentum-resolved phonon dispersion can be mapped using inelastic neutron scattering, inelastic x-ray scattering, or high-resolution electron energy-loss spectroscopy. The predicted linked nodal structures and topological edge states await direct imaging and provide compelling targets for future measurements.

{\subsection{Topological phonon induced physical phenomena}}

% \subsection{Experimental approaches}

% NOTE: the existing ones and potential developments, (wide figure illustration, e.g., EELS, IXS, RIXS, Neutron scattering, Raman scattering, infrared absorption)

% NOTE: The figure should include information of \textbf{Experimental principle, conservation rule, obeservable physical quantity, etc.}

Topological phonons in solids induce rich physical phenomena which can provide a new dimension for investigating electrical, thermal and optical interaction mechanisms and offer an opportunity to detect themselves in turn. Inelastic x-ray (IXS), resonant inelastic x-ray scattering (RIXS), and Raman and infrared (IR) spectroscopy use the photon-phonon interaction mechanism to detect topological phonons in multiple frequency and momentum ranges, energy and space resolutions, interaction strengths, etc. 

From 2018, IXS measurement has become a useful tool for identifying topological phonons~\cite{PhysRevLett.121.035302}. IXS probes the coherent motion of the electronic cloud around atoms (Thomson scattering) and under the adiabatic approximation, phonon information can be extracted. Using this method, spin-1 Weyl phonons at the $\Gamma$ point and charge-2 Dirac phonons at the R point in FeSi were firstly experimentally observed and achieved good agreement with DFT results~\cite{PhysRevLett.121.035302}. Later on, the \textit{PT}-symmetry protected helical phonon nodal lines in MoB$_2$ were also confirmed~\cite{PhysRevLett.123.245302}. A Dirac cone approximant at the \textit{K} point in graphite was clearly demonstrated by IXS, and its intensity winding can also reveal the winding of quasiparticle eigenvector~\cite{PhysRevLett.131.246601}. Inelastic neutron scattering is another method which can obtain all possible phonons with full momentum and frequency ranges by probing the nuclear motion. ~\cite{PhysRevB.106.224304} determined the large Chern number of topological phonon band-crossing nodes using inelastic neutron scattering.

Compared with the scarcity and expense of INS and IXS spectrometers, HR-EELS has been another choice to probe the phonon spectrum with nanometer spatial resolution including topological phonons in recent years. Topological phonon nodal rings in monolayer graphene and BN, and Dirac phonons in graphene have been observed in HR-EELS experiments recently~\cite{PhysRevLett.131.116602,ChinPhysLett.42.027405}.

Depending on sample conditions, resolution requirements, and research objectives, different detection techniques are better suited for specific studies. Inelastic X-ray scattering (IXS) offers high energy resolution—down to a few meV—and requires only small sample volumes, but it is primarily sensitive to bulk topological phonons. Raman and infrared (IR) spectroscopies exhibit high sensitivity to phonons; however, they are limited to probing modes near the $\Gamma$ point and can only access optical phonons within a restricted frequency range. Electron energy-loss spectroscopy (EELS) demands high-quality sample surfaces for detecting topological surface states and typically suffers from relatively low energy resolution for phonon measurements. Therefore, it is essential to select an appropriate technique tailored to the specific research goals of each topological phonon study.

Thermal Hall effect is another featured phenomenon which is related to topological phonons. In 2012, ~\cite{PhysRevB.86.104305} established the theory between the phonon Hall effect and the intrinsic phonon Berry curvature, and predicted that certain topological phonon materials with strong spin-orbit coupling which can reorganize the phonon bands may display phonon quantum Hall effect. However, since many experimental studies on thermal Hall effect have been observed in various crystals without strong spin-orbit coupling such as BP~\cite{NatCommun.14.1027}, Si and Ge~\cite{PhysRevLett.135.196302}, the mechanism behind thermal Hall effect remains varied (e.g., chiral phonons are also considered to induce thermal Hall effect~\cite{NatPhys.21.1532,PhysRevLett.105.225901,PhysRevLett.131.236301,NatPhys.16.1108}) and has yet to reach a consensus. In spite of this, the phonon Hall effect induced by phonon Berry curvature is still a solid way to reach thermal Hall effect, and it remains a hot topic in experimental measurements.

Phonons can also possess pseudo-angular momentum due to their atomic circular or elliptical rotation, which is known as phonon chirality. Wely phonons have been demonstrated to be both topological and chiral, offering a new freedom to detect and utilize the Wely phonons and the corresponding topological surface states. ~\cite{NanoLett.23.7561} found that in certain systems, Wely phonons own chirality and can be detected by the circularly polarized Raman scattering, which is demonstrated by the following experiments in Te ~\cite{NanoLett.23.7561}and $\alpha$-HgS~\cite{NatPhys.19.35}. 

\subsubsection{\label{sec:level3}Current challenges and future directions}

Compared with other topological systems, topological phonons in solids remain insufficiently explored, presenting numerous open questions that require urgent investigation. Unlike electronic systems, where topological surface states and their transport behaviors have been comprehensively characterized, direct experimental evidence of topological surface or edge phonon states in solids remains scarce. Surface-sensitive techniques—including HR-EELS and IXS for probing topological phonons across the entire Brillouin zone, as well as Raman and infrared spectroscopy for detecting optical topological surface phonons near the $\Gamma$ point—offer promising avenues for experimental verification.

In addition, a class of strongly anharmonic crystalline materials is often overlooked in the study of phonon topology, largely due to the presence of imaginary frequencies in their harmonic phonon dispersions. By incorporating temperature renormalization effects and higher-order anharmonicity~\cite{PhysRevB.87.104111, JPhysCondensMatter.33.363001}, these materials typically exhibit stable phonon dispersions at finite temperatures—albeit at the cost of substantial computational expense. Nevertheless, investigating the phonon topology of such materials holds strong application potential: they are often relevant for thermoelectric conversion, and understanding their topological phonon properties may facilitate the development of thermoelectric materials~\cite{AdvFunctMater.34.2401684}.

Furthermore, despite the critical importance of topological surface phonons in nanoscale thermal transport across surfaces and interfaces, their transport properties remain far less understood than those of their electronic counterparts. Experimental quantification of these transport behaviors remains challenging because topological surface phonons participate in thermal transport through complex multiple-phonon scattering processes. To address this limitation, machine learning force fields trained on first-principles data combined with the Boltzmann transport equation (BTE) have recently been employed to isolate and quantify the contribution of topological surface phonons to thermal transport with first-principles accuracy in semiconductor thin films~\cite{arXiv.2025.su}, specifically for Si, 4\textit{H}-SiC, and \textit{c}-BN. Using machine learning potentials, the phonon local density of states of semiconductor thin films with the reconstructed surfaces can be accurately calculated. The transport properties of bulk and topological surface state phonon modes can also be distinguished, as well as their contributions to the total lattice thermal conductivity. Further investigations are essential to elucidate the role of topological surface phonons in other functional contexts, including thermoelectric materials and thermal interfacial conductance.

Another natural question that arises is the study of phonon coupling with other topological quasi-particles. For instance, it is believed that the coupling between topological surface phonons and electrons has potential significance in Weyl semimetals like CoSi—where both Weyl phonons and Weyl fermions govern transport~\cite{PhysRevB.102.245116}. Such interactions can drive phenomena including the nonlinear Hall effect~\cite{PhysRevB.102.245116,PhysRevLett.123.016801}, charge/spin density waves~\cite{PhysRevX.14.011053,npjComputMater.10.264}, quantum Hall states~\cite{Research.2018.6793752}, and superconductivity~\cite{NatPhys.16.1108,PhysRevB.108.064510}. Other applications such as thermoelectric enhancement~\cite{PhysRevMater.2.114204,PhysRevLett.112.226801} can also benefit from these studies.

Beyond crystalline solids, the topological properties of phonons in non-periodic structures—such as quasicrystals, amorphous materials, alloys, and nanostructures—have also emerged as a compelling research direction. As demonstrated in the field of topological acoustics, quasicrystalline and amorphous systems can host rich topological phenomena, including the quantum Hall effect, quantum spin Hall (QSH) effect, and topological crystalline insulators (TCIs). Despite these advances, the phonon topology in non-periodic solids remains largely unexplored, with limited theoretical derivation or experimental observation. Given the broad applications of non-periodic materials—such as alloys in thermoelectrics, high-entropy alloys in aerospace and radiation-resistant devices, and amorphous materials in transformers and catalysis—the systematic investigation of phonon topology in these real-world systems is of both fundamental and practical importance. Comprehensive theoretical and experimental efforts are urgently needed to uncover their potential.

\nocite{*}

%\bibliography{1}% Produces the bibliography via BibTeX.

%
%  End of file apssamp.tex 

%\documentclass[%
%reprint,
%superscriptaddress,
%groupedaddress,
%unsortedaddress,
%runinaddress,
%frontmatterverbose, 
%preprint,
%preprintnumbers,
%nofootinbib,
%nobibnotes,
%bibnotes,
%amsmath,amssymb,
%aps,
%pra,
%prb,
%rmp,
%prstab,
%prstper,
%floatfix,
%]{revtex4-2}

%\usepackage{braket} 
%\usepackage{graphicx}% Include figure files
%\usepackage{dcolumn}% Align table columns on decimal point
%\usepackage{bm}% bold math
%\usepackage{color}
%\usepackage{amssymb} 
%\usepackage{amsmath}
%\usepackage{hyperref}% add hypertext capabilities
%\usepackage[mathlines]{lineno}% Enable numbering of text and display math
%\linenumbers\relax % Commence numbering lines

%\usepackage[showframe,%Uncomment any one of the following lines to test 
%%scale=0.7, marginratio={1:1, 2:3}, ignoreall,% default settings
%%text={7in,10in},centering,
%%margin=1.5in,
%%total={6.5in,8.75in}, top=1.2in, left=0.9in, includefoot,
%%height=10in,a5paper,hmargin={3cm,0.8in},
%]{geometry}

%\preprint{APS/123-QED}
\section{\label{sec:level1}Topological phonons in artificial metamaterials}
%\title{Chapter 4: Topological phonons in artificial materials}% Force line breaks with \\
%\thanks{A footnote to the article title}%

%\author{Ann Author}
%\altaffiliation[Also at ]{Physics Department, XYZ University.}%Lines break automatically or can be forced with \\
%\author{Second Author}%
%\email{Second.Author@institution.edu}
%\affiliation{%
%Authors' institution and/or address\\
%This line break forced with \textbackslash\textbackslash
%}%

%\collaboration{MUSO Collaboration}%\noaffiliation

%\author{Charlie Author}
%\homepage{http://www.Second.institution.edu/~Charlie.Author}
%\affiliation{
%Second institution and/or address\\
%This line break forced% with \\
%}%
%\affiliation{
%Third institution, the second for Charlie Author
%}%
%\author{Delta Author}
%\affiliation{%
%Authors' institution and/or address\\
%This line break forced with \textbackslash\textbackslash
%}%

%\collaboration{CLEO Collaboration}%\noaffiliation

%\date{\today}% It is always \today, today,
             %  but any date may be explicitly specified

%\begin{abstract}
%An article usually includes an abstract, a concise summary of the work
%covered at length in the main body of the article. 
%\begin{description}
%\item[Usage]
%Secondary publications and information retrieval purposes.
%\item[Structure]
%You may use the \texttt{description} environment to structure your abstract;
%use the optional argument of the \verb+\item+ command to give the category of each item. 
%\end{description}
%\end{abstract}

%\keywords{Suggested keywords}%Use showkeys class option if keyword
                              %display desired
%\maketitle

%\tableofcontents
\subsection{\label{sec:level2}Introduction}
Having surveyed the manifestations of topological phonons in natural crystals where lattice symmetries, atomic interactions intrinsic constraints dictate the accessible phases, we now turn to artificial phononic structures. By this term we mean engineered platforms that emulate or extend lattice dynamics for sound and vibration, encompassing acoustic metamaterials, elastic resonator arrays, water-wave analogues, and surface-bound waves such as Rayleigh or Love modes. The rationale for moving to such structures parallels the motivation behind photonic crystals~\citetext{\citealp{PhysRevLett.135.080001}}: natural systems impose severe constraints, while artificial ones offer unprecedented design freedom.For instance, artificial lattices can be freely designed to realize any of the 230 crystallographic space groups without being bound to the atomic motifs of real materials. This opens the way to explore symmetry-protected topological phases in a deliberately engineered setting. Beyond symmetry, their programmable degrees of freedom-on-site resonances, intersite couplings, and even spatiotemporal modulations make it possible to emulate synthetic magnetic fluxes. Moreover, although phonons cannot couple directly to electromagnetic fields as photons do, they admit multi-physics integrations (piezoelectric, electroacoustic, hydrodynamic) and dissipative elements, turning loss, feedback, and nonlinearity into knobs for accessing phases unavailable in optics. This complexity provides an expansion of the accessible topological landscape.

Artificial phononic structures also provide unmatched experimental accessibility. In contrast to atomic phonons, whose band structures are fixed by chemistry and confined to the terahertz regime, artificial phononic lattices operate in the kHz–GHz range where fabrication is versatile and measurements are direct. Their macroscopic wavelengths allow direct field mapping with microphones or vibrometers, not only in amplitude but also in phase, enabling reconstruction of Berry phases, Wilson loops, and even full wavefunctions-quantities that in photonics often require elaborate interferometry. The slower timescales make it possible to track state evolution in real time, offering a privileged window onto geometric responses that in solids remain hidden.

In what follows, we show how artificial phononic lattices elevate topology from a fixed attribute of atomic crystals to a tunable and reconfigurable resource. The discussion begins with spatial-symmetry engineering, where designed structures can realize, reinterpret, or even surpass the catalog of crystallographic topological phases. It then moves to the programming of local resonances and couplings, which enables synthetic gauge fields, projective symmetry realizations, and non-Abelian band structures. We next turn to the temporal dimension, where deliberate modulation gives rise to pumping protocols, Floquet phases, and braiding dynamics, and then to real-space textures, where patterned resonances emulate skyrmions and related quasiparticles. Finally, we consider mechanical and optomechanical extensions, including zero-frequency topological mechanics and light-controlled topological phononics, which illustrate how topology can be embedded into hybrid platforms with active control and readout. Taken together, these developments show that artificial phononic structures are not merely analog simulators of topological phases known from condensed matter, but versatile architectures in which topological phonons can be designed, programmed, and dynamically manipulated.

\subsection{\label{Spatial-Symmetry Engineering}Spatial-Symmetry Engineering: From Crystalline Topology to Artificial Gauge Fields}
\subsubsection{\label{Chern Insulator Phases}Chern Insulator Phases in Phononic Systems via Time-Reversal Symmetry Breaking}

Chern insulator phases-originally discovered in the context of the quantum Hall effect-require the \textit{breaking of time-reversal symmetry} ($\mathcal{T}$) to support nonzero Chern numbers and associated unidirectional edge states. In classical acoustic systems, however, the governing equations are typically second-order in time and fully symmetric under time reversal, making such topological phases inaccessible in static media. A key advance came with the realization that embedding a moving background flow breaks this symmetry at the level of the wave operator, thereby enabling nonreciprocal topological transport.

Specifically, in a barotropic, inviscid fluid with steady background velocity $\mathbf{v}_0(\mathbf{r})$, the acoustic perturbations are most naturally described using the velocity potential $\phi(\mathbf{r}, t)$, which satisfies the convected wave equation:
\begin{equation}
\left( \frac{\partial}{\partial t} + \mathbf{v}_0 \cdot \nabla \right)^2 \phi = c^2 \nabla^2 \phi.
\end{equation}

 Under time reversal $t \rightarrow -t$, the time derivative flips sign, but the background flow $\mathbf{v}_0$ remains fixed, resulting in a broken symmetry of the entire operator. This explicit $\mathcal{T}$-symmetry breaking lifts the degeneracy between forward and backward propagating waves, enabling the emergence of Chern-type topological edge states.
These principles have inspired several designs, where steady flow introduces synthetic gauge fluxes that mimic magnetic fields, with the background velocity field $\mathbf{v}_0$ playing the role of an effective vector potential analogous to $\mathbf{A}$ in electromagnetic systems.

\begin{figure}[!htbp]
  \centering
  \includegraphics[width=\linewidth]{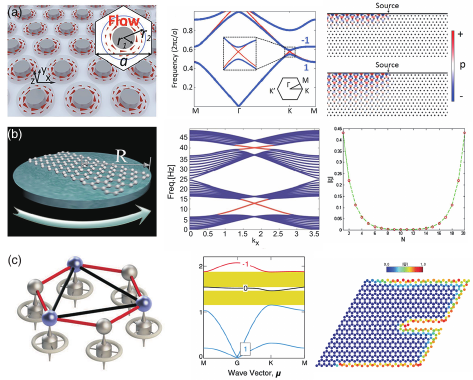}
  \caption{Designs in artificial phononic system that breaks the time-reversal symmetry to achieve Chern insulator phase. (a) Phononic crystal with surrounding airflow. (b) Masses and springs system placed on a constantly rotational coordinate system. (c) Two dimensional gyroscopic lattices. (a) Adapted from~\citetext{\citealp{PhysRevLett.114.114301}}. (b) Adapted from~\citetext{\citealp{PhysRevLett.115.104302}}. (c) Adapted from~\citetext{\citealp{NewJPhys.17.073031}}.}
  \label{fig:Chern phases}
\end{figure}

To ensure that this symmetry breaking opens a full band gap rather than merely shifting degeneracies, it is advantageous to implement the airflow in lattices that naturally host Dirac cones protected by time-reversal symmetry. Hexagonal (C$_{6v}$) lattices, in particular, host symmetry-protected Dirac points at the $K$ and $K'$ points of the Brillouin zone, which can be gapped by introducing appropriate flow fields. This strategy has been proposed~\citetext{\citealp{NewJPhys.17.053016}; \citealp{PhysRevLett.114.114301}} and experimentally implemented in coupled ring-resonator arrays with azimuthal airflow~\citetext{\citealp{PhysRevLett.122.014302}}, where robust unidirectional edge modes emerge, in agreement with the presence of nonzero Chern numbers, as shown in Fig.~\ref{fig:Chern phases}(a).

Another powerful strategy for breaking time-reversal symmetry in classical mechanical systems is to introduce velocity-dependent forces that emulate the behavior of charged particles in a magnetic field. This can be achieved either by placing the system in a rotating reference frame, where \textit{Coriolis forces} arise naturally, or by embedding \textit{gyroscopic elements} that exert lateral forces proportional to the velocity of motion.

In both cases, the governing equation for the displacement field $\mathbf{u}$ acquires a first-order-in-time term of the form:
\begin{equation}
\rho\, \ddot{\mathbf{u}} + \gamma\, \mathbf{G} \cdot \dot{\mathbf{u}} = \mathbf{F}_{\mathrm{elastic}},
\end{equation}
where $\gamma$ is the coupling strength and $\mathbf{G}$ is an antisymmetric tensor that encodes the angular momentum bias. For rotating systems, $\mathbf{G}$ is set by the angular velocity $\boldsymbol{\Omega}$, giving rise to the Coriolis term $2\rho\, \boldsymbol{\Omega} \times \dot{\mathbf{u}}$~\citetext{\citealp{NewJPhys.17.073031}}. In gyroscopic metamaterials, as realized in~\citetext{\citealp{PhysRevLett.115.104302}}, each mass is mounted on a spinning rotor that produces an effective Lorentz-like force of the form $\eta\, \hat{z} \times \dot{\mathbf{u}}$, with $\eta$ determined by the spin angular momentum of the gyroscope, see Fig.~\ref{fig:Chern phases}(c). In both scenarios, the added term explicitly breaks time-reversal symmetry: under $t \rightarrow -t$, the velocity $\dot{\mathbf{u}}$ reverses sign while $\mathbf{G}$ remains fixed, rendering the dynamics non-invariant under $\mathcal{T}$. This allows the phononic band structure to host nonzero Chern numbers and support robust, unidirectional edge modes, as demonstrated in~\citetext{\citealp{PhysRevLett.115.104302}; \citealp{NewJPhys.17.073031}}, also see Fig.~\ref{fig:Chern phases}(b,c).

\subsubsection{\label{sec:level3}Phononic Quantum Spin Hall-Like Phase}

Moving from Chern insulators to quantum spin Hall insulators, $\mathcal {T}$ becomes necessary and spin degrees of freedom come to play. Recall that the spin-$\frac{1}{2}$ of electrons and spin-orbit interactions let $\mathcal {T}^{2}=-1$, leaving the topological classification into the AII class that admits the $Z_2$-classified topological phases in both two and three dimensions, that is, the 2D QSH insulator and the 3D TI, respectively. However, both phonons and the other classical waves are intrinsically Bosons, lacking the fractional spins, the resultant $\mathcal {T}^{2}=1$ falls forever beyond the AII symmetry class. Such a fundamental disparity suggests an impossible mission for realizing these phases in Bosonic systems. Nevertheless, considering that the crucial topological phenomenon of the QSH phase is the emergence of counterpropagating (or helical) edge states that differ by spin-up and spin-down spaces, there are several ways to mimic such analogs, where the spatial symmetries and gauge fields take the most central roles. 

\begin{figure}[!htbp]
  \centering
  \includegraphics[width=\linewidth]{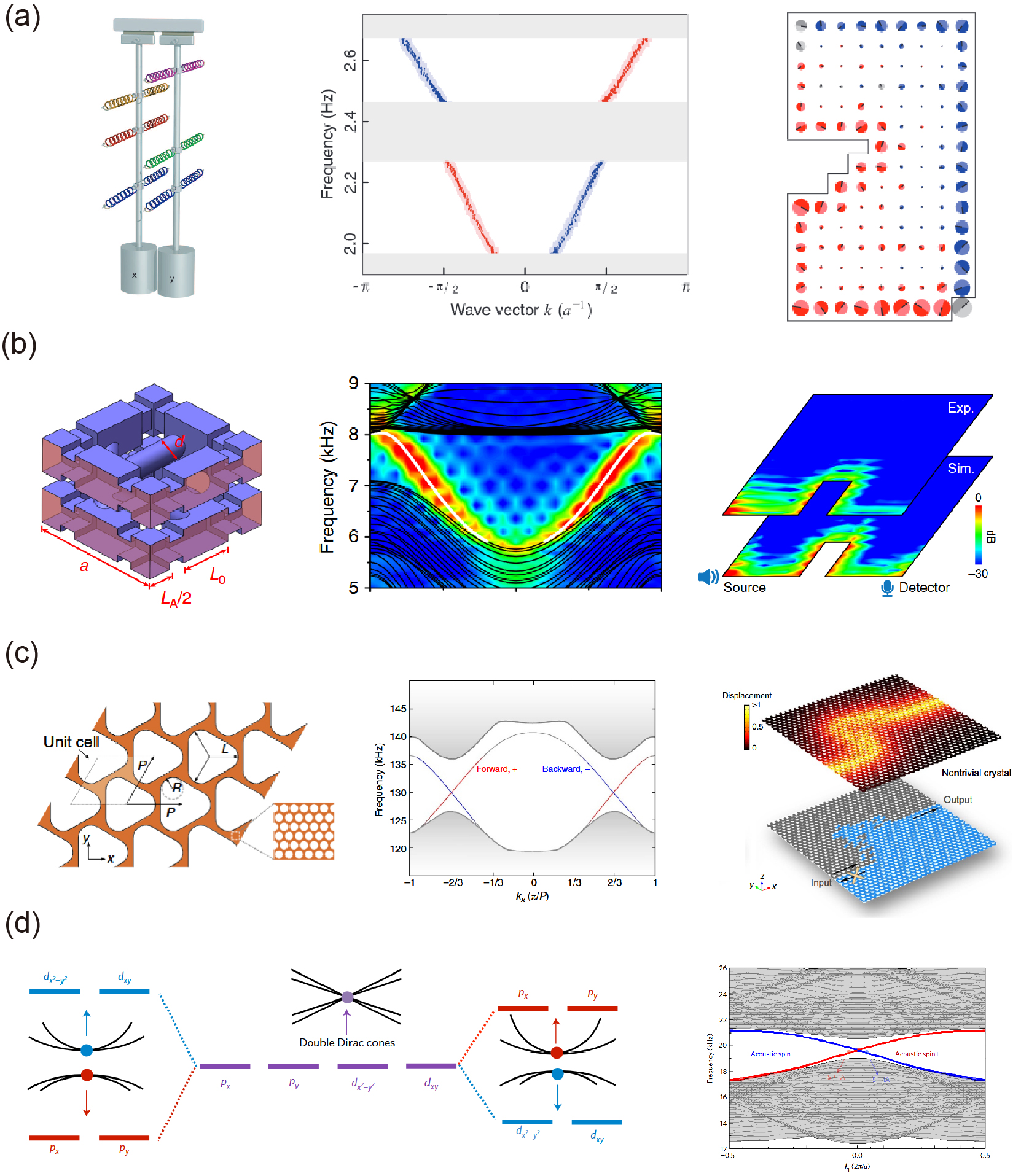}
  \caption{Designs in artificial phononic systems mimicking QSH-like phases. (a) Left panel: Two pendula are taken as the effective site. Springs denote couplings; Middle panel: Dispersion of phononic helical edge states. Right panel: Robustness of topological edge state. (b) Left panel: Illustration of the unit cell for the acoustic spin Chern number; Middle panel: Phononic dispersion of edge states; Right panel: Robustness against the defect of edge states. (c) Left panel: Illustration of an elastic spin Chern insulator; Middle panel: Dispersion of elastic edge states; Right panel: Robustness of edge states along the S-shaped interface. (d) Left panel:  Illustration of $pd$ inversion. Right panel: Nearly gapless edge state. (a) Adapted from~\citetext{\citealp{PNAS.113.E4767}}. (b) Adapted from~\citetext{\citealp{NatCommun.11.3227}}. (c) Adapted from~\citetext{\citealp{NatCommun.6.8682}}.(d) Adapted from~\citetext{\citealp{NatPhys.12.1124}}}
  \label{fig:QSH phases}
\end{figure}

The phononic helical edge states were first experimentally observed in a mechanical lattice system composed of classical harmonic oscillators [see Fig.~\ref{fig:QSH phases}(a)], of which the dynamics is described by Newton’s equation of motion~\citetext{\citealp{PNAS.113.E4767}},
\begin{equation}
\ddot{x}_i=-D_{ij}x_j,
\end{equation}
where $x_i$ denotes the coordinates of pendulas. The dynamic matrix $D_{ij}$ including couplings between pendulas is analogous to the Hamiltonian in quantum mechanics,  but only real-valued. To obtain a purely real $D$ with nontrivial topology, two independent copies of the Hofstadter model $H_+$ and $H_-$ but with opposite Chern numbers were acted on by a unitary transformation,
$U^{\dagger}HU \equiv D,
$
where 
$
H=diag(H_+, H_-)$, and
\begin{equation}
    U=\frac{\sqrt2}{2}\begin{pmatrix}
1 & -i \\
1 & i 
\end{pmatrix}\bigotimes 1_3.
\end{equation}
The sublattices $A_+$ and $A_-$ at two different layers are mixed to 
$
U^{-1}[A_+, A_-]^T 
=[A_++iA_-, A_+-iA_-]^T$
to denote $\mathcal{T}$-symmetric pseudo-spins. The resultant $D$ contains solely positive and negative couplings, complying with the design of passive phononic lattices. Meanwhile, the unitary transformation does not alter the band structure of $H$, such that the two pairs of gapless chiral edge states remain in the projected edge spectrum of $D$ [see Fig.~\ref{fig:QSH phases}(a)].  This work first introduced the bilayer degrees of freedom and negative couplings to construct spin-resolved band topology, which has further inspired the mapping of all ten-fold classified topological phases to real Hamiltonians and has enriched the topological classification with gauge fields (see Sec. III for more details). In comparison, Deng \emph{et al.}, constructed a phononic bilayer system with only positive interlayer couplings to realize the so-called spin Chern insulator that does not demand any symmetries~\citetext{\citealp{NatCommun.11.3227}}[see Fig.~\ref{fig:QSH phases}(b)]. Although the absence of negative couplings leads to the breakdown of pseudo-spin conservation, the edge states lack the pseudo-spin-momentum locking, yet remain gapless. The nontrivial spin Chern number defined in pseudo-spin space, and the gaps of both bulk and edge spin spectra, together protect the gapless feature of edge states.

The above-mentioned QSH-like edge states are self-guided without surrounding trivial or topologically distinct phases. Indeed, similar topology can also be realized at interfaces, where two adjacent insulator phases are derived from gapping a double Dirac cone into two different topology. The most representative work is based on the vector characteristics of elastic waves~\citetext{\citealp{NatCommun.6.8682}} [see Fig.~\ref{fig:QSH phases}(c)]. With certain geometric parameters, an accidental double Dirac cone formed by two degenerate symmetric ($S$) and two degenerate asymmetric ($A$) modes can be found at two $K$ points in a triangular elastic lattice. Once breaking the mirror symmetry along the $z$ direction, the two mode pairs are hybridized as $(A+S)/\sqrt{2}$ and $(A-S)/\sqrt{2}$, acting as two pseudo-spins. Meanwhile, the symmetry-breaking operations introduce an effective gauge field emulating SOC, leaving two topological BHZ-like models captured by reversed spin Chern insulators.  Once the bulk phases at opposite sides of the domain wall possess opposite spin Chern numbers, two pairs of helical interface states emerge, according to the bulk-boundary correspondence. Specifically, the domain wall satisfies the $C_2$ symmetry along the interface, which protects the crossing of edge states with opposite $C_2$ eigenvalues and thus keeps their gapless feature.

Another strategy to generate the accidental double Dirac cone is based on two degenerate $p$- and two $d$-orbitals at the $\Gamma$ point in a $C_6$-symmetric lattice, which can be obtained in scalar wave systems~\citetext{\citealp{PhysRevLett.114.223901}; \citealp{NatPhys.12.1124}}, as this mechanism relies solely on the spatial symmetry. By tuning the geometric parameter without breaking the $C_6$ symmetry, the accidental Dirac cone can be gapped out to form a topological phase when the $d$-orbitals lie below the $p$-orbitals [see Fig.~\ref{fig:QSH phases}(d)], vice versa for the trivial phase. The effective Hamiltonian in the topological phase derived from four orbitals also resembles the celebrated BHZ model. In addition, the pseudo-spins are formed by the hybridization of $p$- and $d$- orbitals, the pseudo-time reversal symmetry $(U\mathcal{T})^2=-1$ can also be well fulfilled with $U=\frac{(C_6+{C_6^2})}{\sqrt3}=-i\sigma_y$.  However, as the pseudo-spins are only well defined at the $\Gamma$ point with $C_6$ symmetry, the resultant topological edge states always exhibit a band gap, whose band width has a positive correlation to the symmetry breaking around the boundaries or interfaces. A later theoretical work reveals that another global pseudo-spin can be defined by utilizing the so-called Berry bands, where a global nontrivial spin Chern number can be defined in the whole BZ as long as the $C_2\mathcal{T}$ symmetry is present~\citetext{\citealp{PhysRevResearch.3.L022013}}. He \emph{et al.}, first implemented the phononic experiments and observed acoustic spin-dependent topological interface states [see Fig.~\ref{fig:QSH phases}(d)]. Later, they introduced both the layer degrees and dimension extension to the 2D symmetric lattice and developed several novel phononic topological phases, including the 2D ideal QSH insulator without the tiny gap in the edge dispersion~\citetext{\citealp{NatCommun.14.952}}, 3D acoustic QSH insulator with 2D surface Dirac cones~\citetext{\citealp{PhysRevLett.123.195503}; \citealp{NatCommun.11.2318}}, higher-order topological states~\citetext{\citealp{PhysRevLett.128.115701}; \citealp{Research.6.0235}}, and topological fiber protected by second spin-Chern numbers~\citetext{\citealp{PhysRevLett.133.226602}}, greatly enriching the phononic TIs. Additionally, the pseudo-spins and pseudo-$(U\mathcal{T})^2=-1$ can also be supported in the boundary of BZ, with $U$ being nonsymmorphic symmetries including screw rotations and glide mirrors~\citetext{\citealp{NatPhys.15.582}; \citealp{NatCommun.11.65}}. The nonsymmorphic symmetries may protect the gapless feature of edge states or surface Dirac cones, stabilizing the emergence of boundary states in the bulk band gap.  Properly speaking, most of the above-mentioned phononic analogs are classified as obstructed atomic insulators, fragile phases, or Euler insulators. However, as long as the protecting spatial symmetries and pseudo-spins are not destroyed, and the nontrivial distributions of Berry curvatures in the whole Brillouin zone in different pseudo-spin spaces are still well defined, the band topology and the associated bulk-boundary correspondence ought to be robust. 
To date, the analogs of 3D  TIs in phononic systems with single or paired Dirac cones lying at all the surfaces of a sample are still absent, which may demand the nonsymmorphic symmetries for symmetry protection for all the surfaces, or the global pseudo-$(U\mathcal{T})^2=-1$ that may be possible with the equipment of gauge fields.  
\subsubsection{\label{sec:level3}Phononic Valley Hall Insulators}

\begin{figure}[!htbp]
  \centering
  \includegraphics[width=\linewidth]{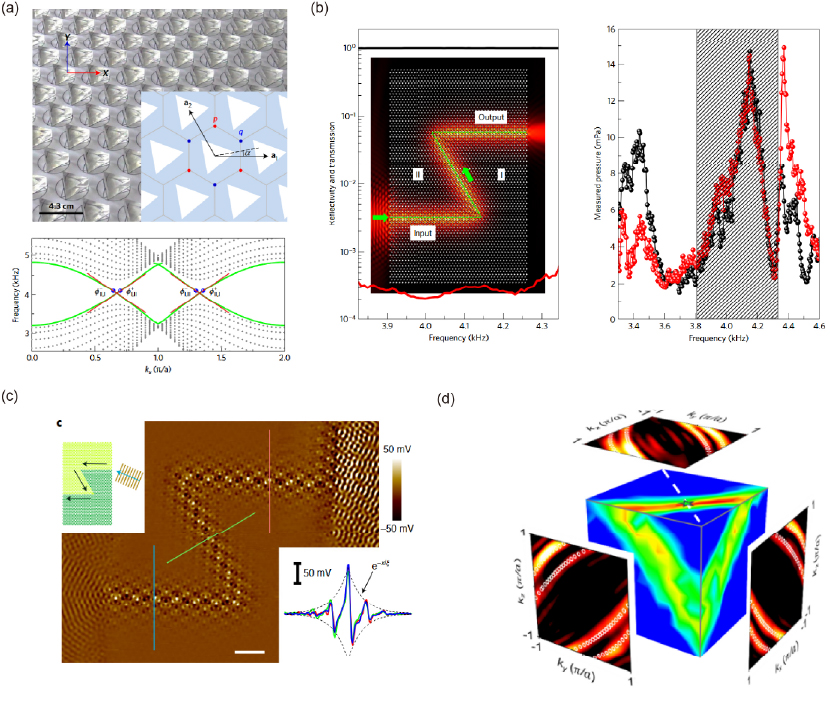}
  \caption{Phononic Valley Hall phases. (a) Illustration of the phononic valley Hall system and its gapless edge states. (b) Simulated edge state transmission along the Z-shaped interface and experimentally obtained edge state singularities. (c) Valley Hall edge state transmission at gigahertz frequencies. (d) Edge states of a 3D phononic valley Hall insulator. (a) and (b) Adapted from~\citetext{\citealp{NatPhys.13.369}}. (c) Adapted from~\citetext{\citealp{NatElectron.5.157}}.(d) Adapted from~\citetext{\citealp{PhysRevLett.134.216601}}}
  \label{fig:Valley phases}
\end{figure}

The topological valley Hall insulators also derive from the nontrivial distributions of Berry curvatures, but only around local valleys. Valleys are energy extremums in the band structures, which can also be regarded as a pseudo-spin degree of freedom and are rather easily obtained in phononic systems, usually by gapping out the Dirac points at $K$ points in triangular lattices through breaking the inversion or mirror symmetries. The simplest valley Hall model can be described by the effective Hamiltonian 
\begin{equation}
H_{\mathrm{V}}(\mathbf{q})=v_x q_x \sigma_x+\tau v_y q_y \sigma_y+\tau \Delta q_z \sigma_z,
\end{equation}
with $\mathbf{q}=\mathbf{k}-\mathbf{k}_\mathrm{V}$ and the Fermi velocity $\mathbf{v}$. $\tau=\pm1$ represents two different valleys and $\Delta$  the mass term that breaks the inversion symmetry. $\mathbf{\sigma}$ is the Pauli matrix denoting the sublattice degrees.
The integration of the Berry curvature in the vicinity of one valley gives rise to an effective half-quantized valley Chern number $C_v=\frac{\tau}{2}sgn(\Delta)$, which is equal to $\pm0.5$ at two nonequivalent $K$ points. Such nontrivial valley Chern numbers also lead to topological valley interface states. Fig.~\ref{fig:Valley phases}(a) shows a typical phononic valley Hall insulator~\citetext{\citealp{NatPhys.13.369}}, where, by altering the rotation angle of the acoustic scatterers that plays the role of $\Delta$, the symmetry of the triangular lattice is reduced from $C_{3v}$ to $C_3$ symmetry, lifting the degeneracy of the Dirac point at both $K$ points, where two valleys carry opposite valley Chern numbers. By constructing an interface formed by two valley insulators with opposite rotation angles, the topological interface states emerge, whose propagation directions are locked by different valleys with opposite vortex polarizations. It was meanwhile experimentally demonstrated that valley interface states can pass through sharp corners and disorders, almost without backscattering [see Fig.~\ref{fig:Valley phases}(b)]. The degree of layers has also been introduced to the phononic valley insulator to obtain layer-mixed and layer-polarized interface states~\citetext{\citealp{PhysRevLett.120.116802}}. Moreover, large-area and valley-locked interface states were obtained with surface acoustic waves~\citetext{\citealp{NatCommun.13.1324}}, which may be useful for integrated devices with high-energy sound transport.  The valley Hall topology is further reported in nanoelectromechanical aluminium nitride membranes at gigahertz frequencies~\citetext{\citealp{NatElectron.5.157}}, also exhibiting great transmission in a $Z$-shape interface [see Fig.~\ref{fig:Valley phases}(c)]. Moreover, it is combined with the recently introduced method of soft clamping, enabling the on-chip phononic propagation with extremely low loss~\citetext{\citealp{Nature.2025.1}}. 

From the perspective of global topology, as the valley topology is effective, when the band gap becomes wider, the valley Chern numbers largely deviate from $\pm0.5$, the valley interface states tend to be gapped or merge into the bulk continuum, and the vortex-locked transport is no longer well-protected. While if an additional synthetic dimension is involved, the valley topology can be connected to strictly quantized Weyl topological charges, which have been validated in phononic experiments~\citetext{\citealp{PhysRevLett.128.216403}}, further demonstrating the topological origin of valley topology. Beyond the condensed matter systems, recent work in acoustic systems exhibits that the valley topology can also be well defined in 3D~\citetext{\citealp{SciAdv.10.eadp0377}; \citealp{PhysRevLett.134.216601}}, where the valley Chern numbers can take a 3-tuple vectorial form $(C_{v,x},C_{v,y}, C_{v,z})$ with all elements nonvanishing, and the resulant valley-locked interface states can circularly propagate around the corner of the sample, exhibiting novel negative refraction [see Fig.~\ref{fig:Valley phases}(d)].

\subsubsection{\label{sec:level3}Gapless Topological Phases in Three dimensional Artificial Phononic Systems}
%\subsection{Gapless Topological Phases in Three dimensional Artificial Phononic Systems}

While Chern insulators and quantum spin Hall phases are typically associated with gapped two-dimensional systems, gapless topological phases in three dimensions offer a distinct avenue for realizing robust band topology. In these systems, stable degeneracies-points, lines, or even surfaces in momentum space-can carry quantized topological charges and give rise to protected boundary modes and anomalous transport, despite the absence of a full bulk gap. Unlike 2D topological phases that often require external  fields or fine-tuned structural asymmetries, many gapless topological features in three dimensions arise purely from crystalline symmetries. This makes them particularly well suited to implementation in artificial phononic systems, whose geometry can be precisely tailored to realize the desired space group symmetries. Among the various possibilities, \textit{Weyl points} are the most elementary and robust topological singularities in 3D band structures. They appear as linear, twofold-degenerate crossings that act as monopoles of Berry curvature and are quantified by integer-valued Chern numbers. Other symmetry-protected band degeneracies include multi-Weyl nodes, Dirac points, threefold and sixfold crossings, nodal lines, and nodal surfaces-all of which can, in principle, be systematically realized in architected acoustic lattices. In what follows, we focus on Weyl points as a prototypical example, and show how their existence, location, and topological charge can be fully predicted from symmetry principles via group-theoretic constraints on the effective Hamiltonian.

The presence and nature of Weyl points in a band structure are fundamentally dictated by the spatial symmetries of the lattice. At any high-symmetry momentum point $\mathbf{k}_1$, the local band topology is constrained by the \textit{little group} of $\mathbf{k}_1$-that is, the subgroup of space group operations that leave $\mathbf{k}_1$ invariant up to a reciprocal lattice vector. These operations, denoted $O_i$, include unitary symmetries such as rotations, mirror reflections, or screw axes, as well as possible antiunitary symmetries like time reversal. Together, they constrain the form of the effective $k \cdot p$ Hamiltonian through the covariance relation:
\begin{equation}
D(O_i)\, H(\mathbf{q})\, D(O_i)^{-1} = H(O_i \mathbf{q}),
\end{equation}
where $D(O_i)$ is the matrix representation of the symmetry operation $O_i$, and $\mathbf{q} = \mathbf{k} - \mathbf{k}_1$ is the local momentum deviation. Solving these constraints yields a symmetry-allowed Hamiltonian whose degeneracy, dispersion, and topological charge are all fixed by the representation structure of the little group.

This framework provides a direct route for topological band design: by tailoring the space group symmetries (e.g., breaking inversion, embedding rotation or screw symmetries), one can enforce desired degeneracies and even prescribe the topological charge of emergent Weyl points. Acoustic metamaterials, with their high geometric tunability, offer an ideal platform to realize such symmetry-engineered Weyl phases in both predicted and previously unexplored configurations. 

\begin{figure}[!htbp]
  \centering
  \includegraphics[width=\linewidth]{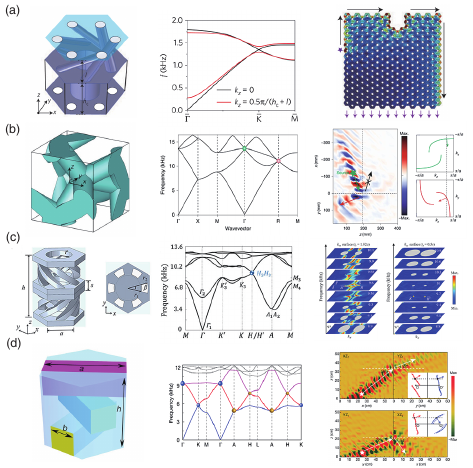}
  \caption{Designs in 3D artificial phononic system that holds gap less topological phases. (a) Charge-1 Weyl point in layered Chiral structure. (b) Phononic crystal with space symmetry No. 198 that supports triple degeneracy (Spin-1 weyl point). (c) 3D Dirac acoustic crystal. (d) Reflectionless negative refraction of surface states in acoustic Weyl crystals. (a) Adapted from~\citetext{\citealp{NatPhys.11.920}}. (b) Adapted from~\citetext{\citealp{NatPhys.15.645}} (c) Adapted from~\citetext{\citealp{PhysRevLett.128.115701}} (d) Adapted from~\citetext{\citealp{Nature.560.61}}.}
  \label{fig:Weyl phases}
\end{figure}

We introduce the concept from the acoustic construction of the simplest charge-1 Weyl node, which is captured by the two-band $k\!\cdot\!p$ Hamiltonian
\begin{equation}
H_{\mathrm{W}}(\mathbf{q})=v_x q_x \sigma_x+v_y q_y \sigma_y+v_z q_z \sigma_z,
\end{equation}
with $\mathbf{q}=\mathbf{k}-\mathbf{k}_\mathrm{W}$ and band velocities $(v_x,v_y,v_z)$. The spectrum
$E_{\pm}(\mathbf{q})=\pm\sqrt{(v_x q_x)^2+(v_y q_y)^2+(v_z q_z)^2}$ is linear in all three directions, and the Berry flux through a small enclosing sphere yields a unit Chern charge. Practically, this model implies a simple design route:  start from a 2D lattice hosting a linear Dirac-like crossing and endow it with a symmetry-compatible interlayer coupling to promote the dispersion into the third direction, thereby splitting the Dirac crossing into a pair of Weyl points. This stacking philosophy underpinned one of the earliest acoustic realizations of Weyl physics, achieved prior to modern space-group classifications: ~\textcite {NatPhys.11.920} engineered a synthetic gauge flux in an architected acoustic crystal to generate symmetry-allowed Weyl points and mapped their linear bulk dispersions [see Fig.~\ref{fig:Weyl phases}(a)]. Beyond the untilted (type-I) case, a Weyl node becomes \emph{type-II} when a symmetry-allowed tilt term overwhelms the isotropic cone, so that the linear $k\!\cdot\!p$ Hamiltonian reads
\begin{equation}
H_{\text{II}}(\mathbf{q})=\mathbf{t}\!\cdot\!\mathbf{q}\;\sigma_0+\sum_{i=x,y,z}v_i q_i \sigma_i,
\end{equation}
and there exists a direction $\hat{\mathbf{n}}$ with $|\mathbf{t}\!\cdot\!\hat{\mathbf{n}}|>\sqrt{\sum_i (v_i \hat n_i)^2}$.
At the nodal frequency, the equifrequency surface is then an open electron–hole (acoustic) pocket intersection rather than a point, while the bulk topological charge and associated Fermi-arc surface states persist. A constructive route in acoustics is to stack dimerized 1D chains (e.g., SSH-like layers) with symmetry-compatible interlayer couplings that generate a strong, directional tilt in the effective 3D dispersion ~\citetext{\citealp{PhysRevLett.117.224301}}.

Distinct space groups can enforce band crossings with higher topological charge. In a hexagonal lattice of space group No.\,181 (P6mm), ~\textcite{NatCommun.11.1820} observed quadratic Weyl points (Chern number $\pm 2$) protected by $C_{6v}$ symmetry, together with the characteristic \emph{double-helicoid} surface arcs that wind twice around the projected nodes. In a chiral cubic lattice of space group No.\,207 (P432), ~\textcite{PhysRevB.106.134108} realized quadruple Weyl points (Chern number $\pm 4$) in a hybrid-Weyl phononic crystal, where cubic crystal symmetries stabilize higher-charge monopoles and shape the connectivity of the surface arcs accordingly.

Similar symmetry-guided strategies also apply to engineer higher-multiplicity degeneracies, such as triple and fourfold crossings. For triple degeneracies, a chiral phononic crystal in space group No. 198 exhibits a topological triply-degenerate point accompanied by double Fermi arcs ~\citetext{\citealp{NatPhys.15.645}}, also shown in Fig.~\ref{fig:Weyl phases}(b). In contrast, symmetry-enforced Dirac points have been demonstrated in a nonsymmorphic body-centered-cubic lattice of space group (No. 230), yielding fourfold degeneracies pinned at BZ corners and quad-helicoid surface states ~\citetext{\citealp{LightSciAppl.9.38}}. Related phenomena around Dirac phases-including intrinsic topological surface states protected by glide symmetries ~\citetext{\citealp{PhysRevLett.124.104301}} and a bulk-hinge correspondence measured in a 3D Dirac acoustic crystal ~\citetext{\citealp{PhysRevLett.128.115701}}-underscore how explicit space-group control fixes the allowed degeneracy type and dictates the protected boundary connectivity in artificial phononic platforms, as shown in  Fig.~\ref{fig:Weyl phases}(c).

Beyond reproducing known symmetry-protected degeneracies, architected phononic structures offer two decisive advantages afforded by their large design freedom. First, they enable "ideal" band-crossing nodes. Symmetry-enforced degeneracies in natural crystals are often non-ideal-energetically misaligned, densely clustered in momentum space, or entangled with parasitic bands. By tailoring lattice symmetry, metric (cell shape and aspect), and interlayer couplings, one can realize ideal Weyl points that are well separated in $\mathbf{k}$-space and isolated within a clean frequency window, free from nearby trivial bands-conditions that sharpen bulk responses and simplify surface phenomenology ~\citetext{\citealp{PhysRevLett.124.206802}}. Second, they permit deliberate engineering of the surface-state dispersion. Because Fermi-arc connectivity, curvature, and group velocity follow from both bulk topology and boundary termination, artificial lattices can co-design bulk symmetry and surface microgeometry to sculpt arc dispersions for targeted functionalities-e.g., reflectionless, topology-protected negative refraction, directive arc-guided beam steering, or robust surface routing across complex interfaces ~\citetext{\citealp{Nature.560.61}}, also see Fig.~\ref{fig:Weyl phases}(d). Together, these capabilities elevate artificial phononic crystals from "demonstration platforms" to band-topology foundries, where both bulk nodes and their surface manifestations can be specified to order.

\subsubsection{\label{sec:level3}Topological-Defect Correspondence}

\begin{figure}[!htbp]
  \centering
  \includegraphics[width=\linewidth]{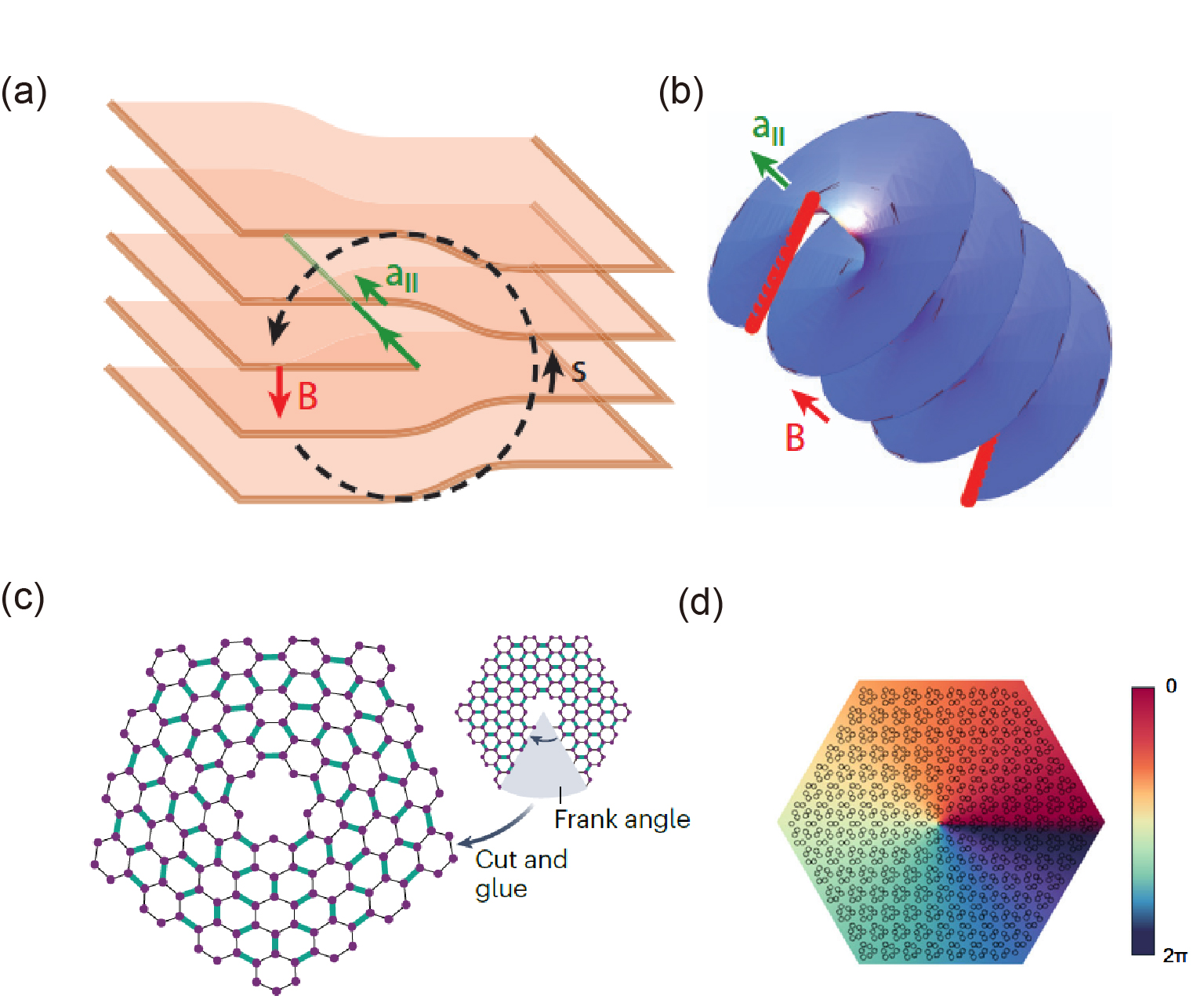}
  \caption{Illustration of various topological defects. (a) and (b) Edge and screw dislocations with the Burger's vectors perpendicular and parallel to the dislocation line, respectively. (c) Disclination with a Frank angle of $\pi/3$. (d) Dirac vortex, the color denotes the effective phase of Dirac mass terms. (a) and (b) Adapted from~\citetext{\citealp{AnnuRevCondensMatterPhys.8.211}}. (c)  Adapted from~\citetext{\citealp{NatRevPhys.5.483}}. (d)  Adapted from~\citetext{\citealp{NatCommun.15.4431}}.}
  \label{fig:Topo_defect}
\end{figure}
Nodal degeneracies carrying topological charges are widely recognized as topological defects (TDs) in momentum space, while the study of TDs originated earlier in real space. Defects are ubiquitous in nature and can be more precisely engineered in phononic systems. To some extent, such defects represent intrinsic boundaries or interfaces embedded within the bulk. Unquestionably, the conventional topological bulk-boundary correspondence can be naturally extended to a more general topological-defect correspondence, particularly in the context of TDs. Common types of TDs include dislocations, disclinations, and vortices in an order parameter field (see Fig.~\ref{fig:Topo_defect}), where symmetry breaking or singularities in the order parameter cannot be eliminated by local perturbations and are characterized by quantized real-space topological invariants. For instance, both edge and screw dislocations break translational lattice symmetry [see Figs.~\ref{fig:Topo_defect} (a) and (b)]. In the case of an edge dislocation, the removal of certain atomic layers leads to localized lattice compression, described by a Burger's vector perpendicular to the dislocation line. In a screw dislocation, lattice planes wind helically around a central axis, analogous to a screw thread, with the Burger's vector aligned parallel to the dislocation line.  A disclination breaks rotational symmetry and is characterized by a Frank angle [see Fig.~\ref{fig:Topo_defect} (c)]. The interplay between real-space topology and momentum-space band topology in the host material can give rise to emergent phenomena such as defect-localized states and fractional charges, which are jointly protected by both real- and momentum-space topological invariants. A comprehensive classification of topological defect-induced band topology has been established based on the Altland-Zirnbauer tenfold classification, using generalized Bloch Hamiltonian $H(\textbf{k},\textbf{r})$, where $\textbf{k}$ resides in a $d_1$-D BZ, $\textbf{r}$ spans a $d_2$-D surface enclosing the defect~\citetext{\citealp{PhysRevB.82.115120}}. The first experimental validation of this theory is based on a mechanical metamaterial, where topological bound states at edge dislocations can result from the interplay between the momentum-space nontrivial Berry phase and the real-space Burger's vector~\citetext{\citealp{NatPhys.11.153}} [see Fig. {\ref{fig:Topodislocation} (a)]. Prior to this formal framework~\citetext{\citealp{PhysRevB.82.115120}}, an early seminal example demonstrated that screw dislocations can support 1D robust gapless helical modes with spin-momentum locking in 3D weak TIs, justifying the strong side of weak TIs~\citetext{\citealp{NatPhys.5.298}}. The presence of such dislocation-bound states is governed by a $Z_2$ topological number
\begin{equation}
\nu=\frac{1}{2\pi}\textbf{B}\cdot\textbf{G} \quad mod\quad 2.
\end{equation}
Here $\textbf{B}$ is the Burger's vector. $\textbf{G}=\sum_i{\nu_i\textbf{b}_i}$ with $\textbf{b}_i$ denoting the primitive reciprocal lattice vectors and $\nu_i$ the weak topological invariants of occupied bands. Such long-sought bulk-dislocation correspondence was recently experimentally verified in acoustic systems~\citetext{\citealp{PhysRevLett.127.214301}; \citealp{NatCommun.13.508}}, benefiting from the mature 3D printing technology. One is based on the 3D stacking of 2D QSH-like phononic insulators with $Z_2$ gauge fields [see Fig. {\ref{Fig-site_coupling} (d)], which mimics a 3D weak TI. By applying the cutting and glue procedure onto the stacked 2D phononic lattice as illustrated in Fig. {\ref{fig:Topodislocation} (b), a screw dislocation can be obtained. The spin-dependent localized gapless topological dislocation states were then experimentally detected. Such gapless states are not only protected by both the real- and momentum-space topology, but also by the projective crystal symmetry that ensures the pseudo-spin conservation, which is attainable in acoustic systems. The other one study also implemented a 3D weak TI analog and a similar dislocation structure, but based on acoustic ring waveguides. Both these two work demonstrated in experiments the pseudo-spin-resolved unidirectional dislocation states. 
\begin{figure}[!htbp]
  \centering
  \includegraphics[width=\linewidth]{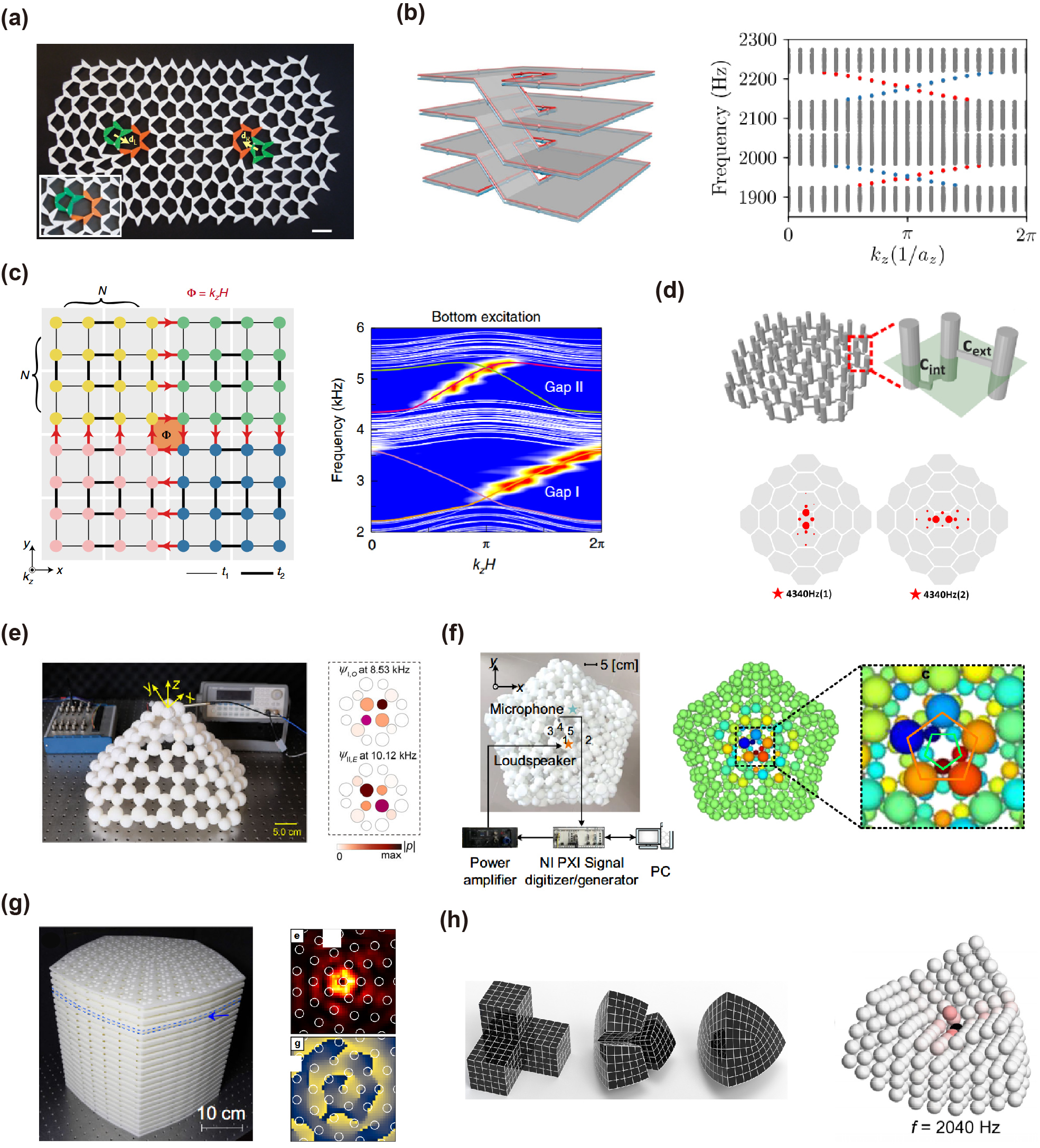}
  \caption{Bulk-defect correspondences in phononic systems. (a) Topological bound states at edge dislocations in a mechanical metamaterial. (b) Left panel:  Illustration of a screw dislocation in phononic systems; Right panel: Dispersion of dislocation-localized states. (c) Illustration of the $k$-dependent local gauge flux and experimentally measured dispersions of localized dislocation states. (d) Illustration of the chiral-symmetric acoustic crystal, and localized disclination states. (e) Phononic non-Euclidean conic structure supporting localized states. (f) Phononic fullerene-like structure with localized disclination states. (g) Phononic Weyl system with a disclination structure that captures localized states with vortex patterns. (h) Illustration of a 3D disclination and localized states. (a) Adapted from~\citetext{\citealp{NatPhys.11.153}} (b) Adapted from~\citetext{\citealp{PhysRevLett.127.214301}}. (c) Adapted from~\citetext{\citealp{NatMater.21.430}}. (d)  Adapted from~\citetext{\citealp{PhysRevLett.128.174301}}. (e)  Adapted from~\citetext{\citealp{PhysRevLett.129.154301}}. (f)  Adapted from~\citetext{\citealp{NatCommun.15.9644}}. (g)  Adapted from~\citetext{\citealp{NatCommun.12.3654}}. (h)  Adapted from~\citetext{\citealp{arXiv.2505.12330}}.}
  \label{fig:Topodislocation}
\end{figure}

When moving to spinless crystalline TIs, screw dislocations play a more significant role in probing topological responses. As the protecting spatial symmetries tend to be broken around boundaries, the robust bulk-boundary correspondence is, in general, absent. However, the screw dislocation may inherit the bulk spatial protecting symmetries. For instance, the screw dislocation can take the role of the $k_z$-dependent local gauge flux~\citetext{\citealp{NatMater.21.430}}, which, designed in a symmetric shape, does not break the rotation symmetry of the original lattice. Such local gauge flux insertion induces in-gap spectral flows localized at dislocation lines for the rotation symmetry-protected higher-order topological insulators with corner states, as illustrated in Fig. {\ref{fig:Topodislocation} (c). To explain it briefly, the nontrivial band topology leaves the imbalance of different symmetry eigenvalues within each bulk continuum, characterized by nonzero real-space topological invariants. These imbalanced bulk states evolve into each other when the gauge flux $k_z$ goes around a loop. Meanwhile, to keep the invariant of the band structures at $k_z=0$ and $2\pi$, they must traverse the band gaps and form the gapless spectral flows, which have been detected in acoustic stacked SSH lattices with a screw dislocation [see Fig. {\ref{fig:Topodislocation} (c)]. 

Another alternative way to probe the higher-order topology is through the fractional charges located at disclinations. Disclinations originate from the breaking of rotation symmetries via a cut-and-glue process, characterized by the Frank angle $\Omega$. The fractional disclination charges are determined by both the real- and momentum-space topology as well, 
\begin{equation}
Q=\frac{\Omega}{2\pi}\xi,
\end{equation}
where $\xi$ is an integer topological index depending on the symmetry eigenvalues of Bloch eigenvalues at high-symmetry points~\citetext{\citealp{PhysRevB.101.115115}}. In classical wave systems, the fractional charges can be mimicked by the fractional mode numbers below the band gap, integrated from the local density of states in one unit cell over a concerned frequency range, which are much more easier to observe than the real electron charges in solids. Although they were first verified in photonics and transmission lines~\citetext{\citealp{Nature.589.376}; \citealp{Nature.589.381}}, the disclination defects have been widely extended to phononics, as the lattice distortion for a disclination is more easily implemented in phononic designs. These systems include topological elastic plates~\citetext{\citealp{ActaMechSin.38.521459}}, surface acoustic waves~\citetext{\citealp{PhysRevB.106.174112}}, and chiral-symmetric acoustic crystals~\citetext{\citealp{PhysRevLett.128.174301}}. The additional chiral symmetry in phononics can be well-maintained in a subtle design [see Fig. {\ref{fig:Topodislocation} (d)], as elaborated in the following section, which further stabilizes the degenerate zero modes localized at disclinations. Furthermore, 3D acoustic metamaterials also support topological in-gap disclination states, including  non-Euclidean conic (hyperbolic) surfaces enriched by
the $p$-orbital physics~\citetext{\citealp{PhysRevLett.129.154301}} [see Fig. {\ref{fig:Topodislocation} (e)], the fullerene-like structure~\citetext{\citealp{NatCommun.15.9644}} [see Fig. {\ref{fig:Topodislocation} (f)], and the 3D acoustic Weyl semimetal~\citetext{\citealp{NatCommun.12.3654}}, which also host fractional charges and localized states even with vortex distributions [see Fig. {\ref{fig:Topodislocation} (g)], originating from the effective gauge flux response with nontrivial Chern topology. More innovatively, a recent acoustic experimental work first brings forth the 3D extension of the disclination without translational lattice symmetry along all directions,  as illustrated in Fig. {\ref{fig:Topodislocation} (h) via a 3D cutting and glue procedure, which also sustains 0D ingap disclination states~\citetext{\citealp{arXiv.2505.12330}}. 
\begin{figure}[!htbp]
  \centering
  \includegraphics[width=\linewidth]{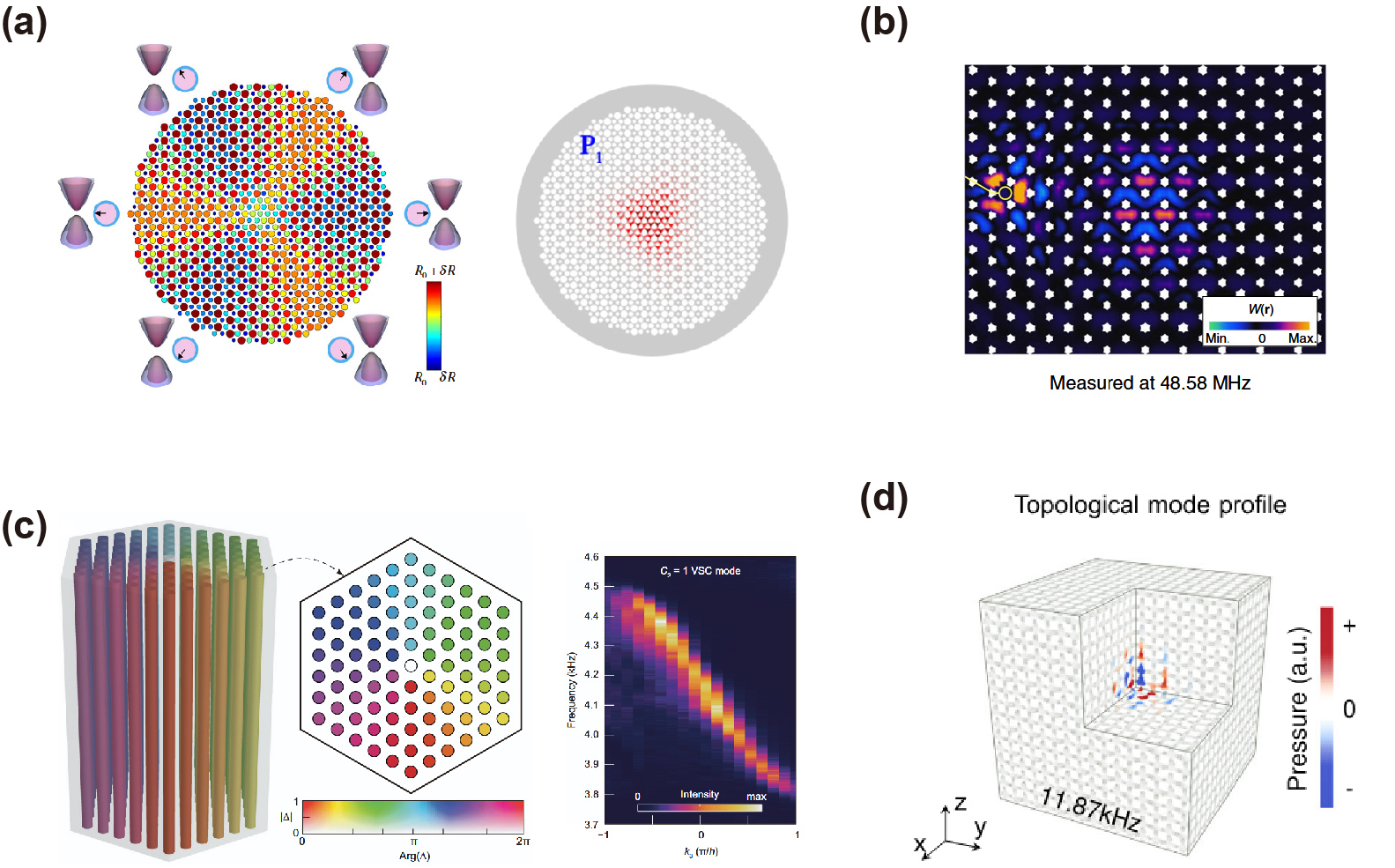}
  \caption{Topological Dirac vortex states in phononic systems. (a) Acoustic Dirac vortex structure formed by rigid cylinders with varying radii and Dirac vortex states. (b) Optical microscope image of Dirac vortex localized states. (c) Illustration of Dirac vortex string and resultant acoustic chiral modes. (d) Localized acoustic states induced by a 3D extension of the Dirac vortex. (a) Adapted from~\citetext{\citealp{PhysRevLett.123.196601}}. (b) Adapted from~\citetext{\citealp{NatNanotechnol.16.576}}. (c)  Adapted from~\citetext{\citealp{PhysRevLett.132.066601}}. (d)  Adapted from~\citetext{\citealp{NatCommun.15.7327}}.}
  \label{fig:Topovortex}
  \end{figure}

Turning to Vortex defects, the Dirac vortex, characterized by the vortex-like texture of the 2D Dirac mass term around a singular point, was first simulated in honeycomb lattices with Kekul\'e-distortion ~\citetext{\citealp{PhysRevLett.98.186809}}. It can be effectively described by the Jackiw–Rossi model~\citetext{\citealp{NuclPhysB.190.681}}, a 2D extension of the 1D Jackiw–Rebbi model, which predicts the existence of topological zero modes localized at the vortex core.  In phononic systems,  Kekul\'e-distortion can be readily achieved by continuously tuning the distances or radii of acoustic cavities and scatterers. Fig. {\ref{fig:Topovortex} (a) illustrates the first experimental realization of a Dirac vortex texture in a phononic system~\citetext{\citealp{PhysRevLett.123.196601}}, realized through spatially varying radii of rigid cylinders arranged in a triangular lattice. A topologically protected acoustic mode is localized around the vortex center, exhibiting a robust frequency immune to defects that preserve chiral symmetry. This framework was quickly extended to mechanical systems using a similar design~\citetext{\citealp{AdvMater.31.1904386}; \citealp{NatNanotechnol.16.576}}, particularly enabling implementation in nano-mechanical platforms operating at megahertz frequencies, where additional orbital degrees of freedom and nonlinear effects were involved, and the vortex state localization was observed as well [see Fig. {\ref{fig:Topovortex} (b)]. Similar to the disclination mentioned above, the 3D extension of the Dirac vortex texture was also developed. One variant is the Dirac string structure, which exhibits lattice periodicity perpendicular to the Dirac vortex plane and can support unidirectional chiral acoustic modes, enabling the realization of topological phononic fibers~\citetext{\citealp{PhysRevLett.133.226602}; \citealp{PhysRevLett.132.066601}} [see Fig. {\ref{fig:Topovortex} (c)]. The other 3D configuration requires lattice symmetry breaking along all spatial directions, hosting 0D localized states within 3D materials, which is also easily modulated and has been first verified in acoustic experiments~\citetext{\citealp{NatCommun.15.7327}} [see Fig. {\ref{fig:Topovortex} (d)]. Indeed, through effective field theory, the dislocations, disclinations, and singular vortexes, can be characterized by  nontrivial Abelian or non-Abelian effective gauge fluxes, demonstrating the power of gauge fields for various topological responses and their facilitation for realizations in phononics.

\subsection{\label{sec:level2} Programmable On-Site and Coupling Parameters with artificial materials.}

While spatial-symmetry engineering provides a natural entry point to topological phonons---mirroring how crystalline point and space groups enforce degeneracies and constrain band representations---it is not, by itself, sufficient to span the landscape of mechanical topological phases. Symmetry fixes admissible universality classes, but realizing and navigating among them in practice requires control over the effective parameters of the phononic Hamiltonian. This logic is formalized by \textcite{PNAS.113.E4767}, who mapped linear mechanical metamaterials onto the ten-fold way, and extended by \textcite{PhysRevB.98.094310} to comprehensive tables for passive metamaterials. Once the class is identified, the concrete on-site terms (local resonances, effective masses, stiffnesses) and hopping terms (coupling magnitudes and phases) determine which topological invariant is actually realized. Without tunability, spatial symmetry is a static scaffold; with programmability, it becomes a platform for writing Hamiltonians.

A constructive route to full topological control is to go beyond merely setting the sign of couplings and to synthesize complex hopping $J e^{i\phi}$ with a tunable phase. Positive/negative couplings already enable band shaping and effective $\pi$ inversions along selected links, but a continuous phase $\phi$ is needed to write synthetic fluxes and, in the presence of degeneracies, non-Abelian holonomies. Practically, this is achieved by assigning to each site two real, quadrature-like degrees of freedom  and replacing every target complex edge by a local $2\times2$ real block,
\begin{equation}
\label{eq:block_coupler}
J e^{i\phi}\;\longmapsto\;
J\!\begin{pmatrix}
\cos\phi & -\sin\phi \\
\sin\phi & \phantom{-}\cos\phi
\end{pmatrix},
\end{equation}
so that sign-programmable links synthesize $\cos\phi$ while cross-couplings synthesize $\pm\sin\phi$. 
Let $H\in\mathbb{C}^{N\times N}$ be the target Hermitian tight-binding Hamiltonian with complex hoppings. Define its real embedding $H_{\mathbb{R}}\in\mathbb{R}^{2N\times 2N}$ by the block map
\begin{equation}
\label{eq:real_embedding}
H\;\longmapsto\;
H_{\mathbb{R}}=
\begin{pmatrix}
\Re H & -\,\Im H\\
\Im H & \phantom{-}\Re H
\end{pmatrix},
\qquad
\Psi=\Re\psi\oplus\Im\psi\in\mathbb{R}^{2N}.
\end{equation}
Then the following isospectral equivalence and invariant preservation holds: For Hermitian $H(k)$,
(i) \(\mathrm{spec}\,H_{\mathbb{R}}(k)=\mathrm{spec}\,H(k)\) with duplicated multiplicity (isospectral).
(ii) Abelian Berry connection/curvature from $\Psi$ equals that from $\psi$.
(iii) For an isolated $m$-fold multiplet, the non-Abelian Berry connection and Wilson loops computed from an orthonormal frame are \emph{identical} after the embedding (up to a trivial $2\times 2$ identity on the doubled real fiber).

\paragraph*{Reaching the full ten-fold way constructively.}
Equation~\eqref{eq:block_coupler} makes complex hopping available in passive mechanical media (from sign-reversal links plus quadrature cross-links). Unit\-ary classes (A/AIII) follow immediately. Imposing appropriate real constraints on $H_{\mathbb{R}}$ recovers time-reversal--even/odd classes (AI/AII with $T^2=\pm1$), and adding particle--hole--type constraints yields the remaining real classes (BDI/D/DIII/C/CI/CII). In this constructive view, the \emph{choice of symmetries} (chiral, TRS, PHS) selects the invariant ($\mathbb{Z}$, $\mathbb{Z}_2$, or trivial) predicted by the ten-fold way, while the complex phases---implemented via \eqref{eq:block_coupler}---write the requisite synthetic gauge structures.

\begin{figure}[!htbp]
		\centering
		\includegraphics[width=\linewidth]{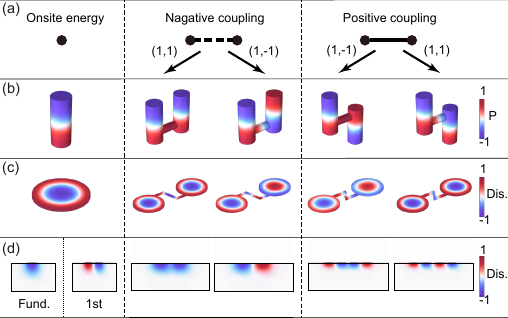}
		\caption{\label{Programmable onsite energy and coupling in various phononic system}
			Programmable onsite energy and coupling. (a) Eigenmode distribution in the basic Hamiltonian model. For the simplest $2\times2$ Hamiltonian, the coupling $(\pm t)$ between two sites shifts the eigenfrequencies from the onsite value $f_0$ to $f_0\pm t$. The sign of the coupling determines the symmetry of the eigenmodes: symmetric $(1,1)$ and antisymmetric $(1,-1)$. In various phononic crystal frameworks, both onsite energy and coupling can be individually controlled: (b) Acoustic cavity system. The onsite energy and coupling correspond to Helmholtz resonators and the tubular connections between them, respectively. By varying the length or connection position, the sign of the coupling can be inverted. (c) Elastic wave system. The onsite energy and coupling correspond to steel disks and their zigzag connections. Adjusting the shape, length, or position of the connections enables coupling-sign inversion. (d) Surface acoustic wave system. The onsite energy and coupling correspond to surface mass blocks and their surface-mediated interactions. By exploring higher-order modes, the coupling sign between modes can be reversed.}
\end{figure}

Distinct phononic platforms realize these controls in complementary ways (Fig.~\ref{Programmable onsite energy and coupling in various phononic system}). In acoustic cavity lattices, on-site resonances shift via cavity-volume adjustments or compliant diaphragms, while asymmetric channels or time-modulated apertures imprint phases on inter-cavity coupling. In elastic lattices, piezoelectric shunts continuously tune stiffness/damping (amplitude channel), and feedback circuits impose phase delays, effectively realizing complex hopping. Surface-acoustic-wave devices push this further: lithographically defined electrodes interfaced with programmable electronics digitally address both on-site frequencies and coupling phases, enabling rapid and reversible reconfiguration. Across platforms, the pattern is uniform: amplitude programmability sculpts where and how gaps open; phase programmability writes the synthetic gauge fields that establish topological order; and multi-parameter programmability makes phase switching on demand routine rather than exceptional. This subsection will show how programmable on-site terms and complex couplings enable a constructive realization of topological phases across the Altland–Zirnbauer ten-fold way in acoustic platforms. After that, we will move beyond the ten-fold way to higher-order topological phases, projective-symmetry–protected phases and to non-Abelian topology characterized by the Euler class, including multi-band nodal braiding.

\subsubsection{\label{sec:level3}Constructing AZ Ten-Fold Topological Phases using programmable onsite and coupling designs}

\label{subsec:AZ_construct_acoustics}

 As outlined above, artificial phononic crystals and metamaterials have become a programmable laboratory for exploring exotic topological phases. By engineering the onsite energy and the inter-site coupling, these systems can faithfully emulate the tight-binding Hamiltonians. We assume programmable on-site terms $\varepsilon_{i}$ and complex hoppings $J_{ij}e^{i\phi_{ij}}$ implemented via real $2\times2$ quadrature blocks. Below we introduce several basic real-space Hamiltonians from different AZ classes and their realizations in phononic systems.

\begin{figure*}[t]
		\centering
		\includegraphics[width=1\linewidth]{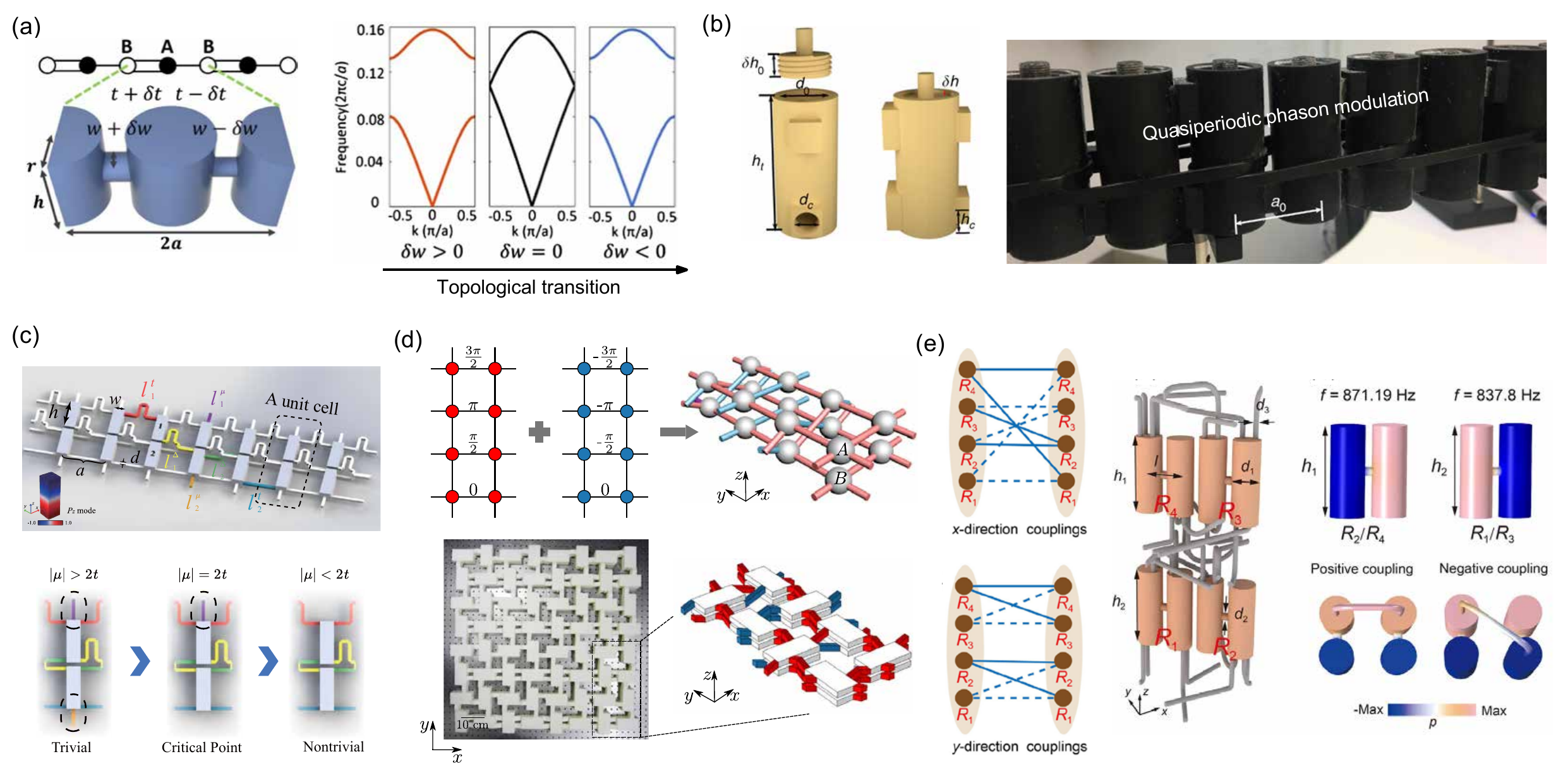}
		\caption{\label{Fig-site_coupling}
			Topological phases based on programmable onsite energy and coupling in phononic lattices. (a) Acoustic cavity-tube design to emulate the SSH model. Varying the tube cross-section drives the topological phase transition. (b) In-situ tuning of cavity height imprints a quasiperiodic phason modulation, realizing an acoustic AAH chain. (c) Positive and negative couplings engineered by cross and straight connections between adjacent cavities, respectively. (d) Positive and negative couplings engineered by cross and straight connections between adjacent cavities, respectively. (e) Complex couplings realized by real couplings. (a) Adapted from~\citetext{\citealp{PhysRevLett.117.224301}}. (b) Adapted from~\citetext{\citealp{CommunPhys.2.55}}. (c) Adapted from~\citetext{\citealp{PhysRevB.107.134107}}. (d) Adapted from~\citetext{\citealp{PhysRevLett.127.214301}}. (e) Adapted from~\citetext{\citealp{SciBull.69.893}}}
\end{figure*}

 The SSH model is one of the most celebrated examples. Two-sublattice chain with alternating bonds:
\begin{align}
H_{\mathrm{SSH}}
&=\sum_{n}\Big[
\varepsilon_A\, a_{n}^{\dagger} a_{n}+\varepsilon_B\, b_{n}^{\dagger} b_{n}
+J_{1}\, b_{n}^{\dagger} a_{n}\nonumber\\
&
+J_{2}\, a_{n+1}^{\dagger} b_{n}
+\mathrm{h.c.}
\Big].
\end{align}

When the sublattices are strictly balanced (\(\varepsilon_A=\varepsilon_B=0\)) and only inter-sublattice hoppings are present, the SSH Hamiltonian takes an off-diagonal (block–off-diagonal) form, obeying the chiral (sublattice) symmetry \(S H S^{-1}=-H\) with \(S=\sigma_z\). 
With generic complex hopping phases this symmetry is the only one enforced (no simple TRS or PHS), placing the model in class AIII. 
If, in addition, all couplings are strictly real (\(\phi_{ij}\in\{0,\pi\}\)), the Hamiltonian is invariant under complex conjugation \(T=\mathcal{K}\) with \(T^2=+1\), and, together with chiral symmetry \(S\), this induces a particle–hole symmetry \(C\equiv S T\) with \(C^2=+1\); the model is then promoted to class BDI. (Any finite sublattice detuning \(\varepsilon_A-\varepsilon_B\neq 0\) breaks \(S\) and drives the system out of AIII/BDI.) Using phononic structures, it can be realized through spatial dimerization of coupled resonators, such as the cavity-tube hybrids or corrugated waveguides for airborne sound waves~\citetext{\citealp{PhysRevLett.117.224301}; \citealp{ApplPhysLett.113.20}; \citealp{PhysRevB.103.224309}}, the rod-beam designs for elastic/mechanical waves~\citetext{\citealp{ProcNatlAcadSciUSA.111.13004}; \citealp{NatPhys.10.39}; \citealp{PhysRevAppl.14.054035}}, and the coupled on-chip waveguides for surface acoustic waves~\citetext{\citealp{AdvMater.36.2312861}}. Resonance or waveguide-mode frequencies set the onsite energy (\(\varepsilon_A , \varepsilon_B\))and resonator spacing or connecting can conveniently tune the intracell versus intercell coupling strengths (\(J_1 , J_2\)), enabling on-demand transitions between trivial and topological phases. 

Starting from the SSH chain with dimerized couplings $(J_1,J_2)$ on sublattices $A/B$, add a staggered detuning
$\Delta(t)\!\equiv\![\varepsilon_A(t)-\varepsilon_B(t)]/2$ (i.e., $A/B$ shifted by $\pm\Delta$):
\begin{align}
H_{\mathrm{RM}}(t)=\sum_{n}\big[
&\,\Delta(t)\, a_{n}^\dagger a_{n}-\Delta(t)\, b_{n}^\dagger b_{n}\nonumber\\
&+J_{1}(t)\, b_{n}^\dagger a_{n}
+J_{2}(t)\, a_{n+1}^\dagger b_{n}
+\mathrm{h.c.}\big].
\end{align}

Here $\Delta=0$ recovers SSH (AIII/BDI), while any $\Delta\neq 0$ breaks chiral symmetry and the instantaneous 1D model is class~A (AI if all couplings are real). 
If $(J_1,J_2,\Delta)$ are varied adiabatically, time acts as a synthetic dimension for Thouless pumping (details deferred to Sec. II-F).

With finer, on-demand control of the on-site magnitudes-for example by imposing a quasiperiodic modulation-we can construct the Aubry–André–Harper (AAH) model, a 1D toy model with sinusoidally varying onsite energy~\citetext{\citealp{AnnIsrPhysSoc.3.18}}:

\begin{align}
H_{\mathrm{AAH}}
&=\sum_{n}\Big[
\lambda\cos(2\pi\alpha n+\phi)\, a_{n}^{\dagger} a_{n}
+J\, a_{n+1}^{\dagger} a_{n}
+\mathrm{h.c.}\Big].
\end{align}

It belongs to class A and is topologically trivial; viewing the phase $\phi$ as a synthetic dimension promotes it to an effective 2D class-A system with nonzero Chern numbers (i.e., topologicsally nontrivial)~\citetext{\citealp{PhysToday.56.38}; \citealp{CommunPhys.2.55}}. Using acoustic cavities, whose heights are varied in a quasiperiodic sequence to produce the required phason modulation, as illustrated in Fig.~\ref{Fig-site_coupling}(b), a complete Hofstadter's butterfly has been successfully demonstrated without ever subjecting sound to an actual magnetic field~\citetext{\citealp{CommunPhys.2.55}}. Extending the modulation to two orthogonal directions generates two independent phasons, promoting a 2D lattice into an effective 4D system~\citetext{\citealp{PhysRevX.11.011016}}. This phononic structure presents 3D topological hypersurface states, which are otherwise unattainable in real physical space, given the dimension reduction of topological phases. The modulation of onsite energy, or in a broader scope, the creation of synthetic dimensions has also facilitated other interesting topological phases that are typically difficult in real and static space. For example, combining lower spatial dimensions with synthetic dimensions, one can studies the Weyl points and semimetals in 1D or 2D systems~\citetext{\citealp{PhysRevLett.122.136802}; \citealp{PhysRevB.102.064309}; \citealp{PhysRevLett.133.256602}}, as a supplement to discussion in Sec. IV-B. Likewise, a synthetic“time" dimension can be encoded by a spatially varying phase profile, allowing static structures to emulate Floquet dynamics, which will be detailed in Sec. IV-D.

With simultaneous programmability of the on-site term  and the couplings, a site-wise Nambu doubling/basis transformation---treating two internal modes as a particle--hole pair---maps the real-space acoustic chain to the BdG form of the 1D Kitaev chain. This construction enforces particle--hole symmetry with \(C^{2}=+1\) without imposing time-reversal symmetry, thereby placing the system in class D, as realized by~\citetext{\citealp{PhysRevB.107.134107}}.

Let us move to 2D and start with class A. Unlike the previous subsection-where an external, magnetic-field–like bias was used to lift degeneracies-here we take a tight-binding viewpoint and consider a Hamiltonian:

\begin{equation}
H_{C} =
- \sum_{m,n}
\Big[
J_{1}\, a_{m+1,n}^{\dagger} a_{m,n}
+ J_{2}\, e^{\,i\,2\pi \phi\, m}\, a_{m,n+1}^{\dagger} a_{m,n}
+ \text{h.c.}
\Big],
\label{2DChern}
\end{equation}

in which complex hoppings generate a uniform flux per plaquette, producing bands with a nonzero Chern number-a minimal acoustic Chern insulator. Practically, such complex bonds can be realized using the same recipe as Eq. \eqref{eq:block_coupler}: combine positive/negative links with quadrature cross-links to synthesize the desired bond phase, thereby implementing a 2D class-A topology~\citetext{\citealp{PhysRevLett.127.214301}}. Duplicating the lattice into two decoupled copies with opposite Chern numbers yields an emergent fermionic time-reversal symmetry with ($U_T^2=-1$), promoting the composite system to class AII (the quantum spin Hall analog): the edge hosts helical, counter-propagating acoustic channels protected by this emergent TRS, while each copy remains an individual class-A Chern sector. An entirely analogous construction leads to class D topological phases~\citetext{\citealp{SciBull.69.893}}; pairing two opposite-Chern D sectors produces an overall DIII-like phase-each D sector can be observed and controlled independently (with a chiral Majorana-like edge in isolation), yet together they exhibit helical Majorana-like edges characteristic of class DIII.

This idea readily extends to, e.g., class C or to higher-dimensional topological designs, though the architectures become substantially more intricate and, to our knowledge, remain unreported; one promising route to reduce complexity is to leverage synthetic dimensions. Three brief comments are in order. (1) Complex couplings can also be introduced by alternative means: via a pure gauge transformation one can engineer bonds of the form 
$e^{i2\pi/m}$ while keeping all physical couplings with the same sign-an approach already used to realize QSH phases~\citetext{\citealp{NatCommun.11.3227}; \citealp{NatCommun.14.952}} and Weyl points~\citetext{\citealp{NatPhys.11.920}} (see discussions in Sec. IV-B). (2) The constructions showcased here mostly perturb a single eigenmode; when multiple spatial symmetries are present, multiple eigenmodes (an "orbital" degree of freedom) naturally arise, enabling richer phenomena. (3) Nonlocal/long-range couplings can also be incorporated in a straightforward manner; we omit those details here for brevity.

%\subsection{\label{sec:level2} Constructing Higher-Order Topological Phases in Acoustics.}
\subsubsection{\label{sec:level3}Constructing Higher-Order Topological Phases in Acoustics}

HOTIs represent a profound extension of topological phases, where nontrivial topology gives rise to gapless boundary states not at the ($d–1$)-dimensional edges, but at lower-dimensional boundaries such as corners or hinges (with $d–2$ or $d–3$ dimensions)~\citetext{\citealp{NatRevPhys.3.520}}.  This hierarchy is typically enforced by spatial symmetries-such as mirror, glide, rotation, or even nonsymmorphic symmetries-which quantize topological invariants like bulk polarizations or multipole moments, leading to a nested bulk-boundary correspondence. 

The prototypical HOTI is the quantized quadrupole insulator, exemplified by a square lattice with $\pi$ flux per plaquette [see Fig. \ref{HOTI}(a), left panel]. This model is govenred by a four-band Hamiltonian ~\citetext{\citealp{Science.357.61}}
\begin{equation}
\begin{split}
H_{BBH}=[\gamma + \lambda \cos(k_x)] \Gamma_4 + \lambda \sin(k_x) \Gamma_3 + \\
[\gamma + \lambda \cos(k_y)] \Gamma_2 + \lambda \sin(k_y) \Gamma_1 + \delta \Gamma_0,
\end{split}
\end{equation}
where $\gamma$ and $\lambda$ are hopping amplitudes. $k_x$ and $k_y$ are the wave vectors along the $x$- and $y$-directions, respectively. $\Gamma_0 = \tau_3 \sigma_0$, $\Gamma_n = - \tau_2 \sigma_n$, and $\Gamma_4 = \tau_1 \sigma_0$, with $n=1,2,$ and 3 and $\tau, \sigma$ the Pauli matrices. This structure can be implemented via dimerized hopping with positive and negative amplitudes [see Fig.~\ref{HOTI}(a), right panel], which gives rise to anti-commuting mirror symmetries, i.e., $\{\hat{m}_x, \hat{m}_y\}=0$. This algebraic relation serves as the fundamental symmetry protection, quantizing a nonzero quadrupole moment. As a result, the system exhibits gapped bulk and edge spectra, within which topological corner states emerge as a direct consequence of the higher-order bulk-boundary correspondence. The topological nature of these low-dimensional boundary states is diagnosed via the general theory of Wilson loops~\citetext{\citealp{PhysRevB.96.245115}}. 

Acoustic metamaterials provide an ideal platform for engineering these complex hopping signs. In mechanical systems, this is achieved by modulating the geometry of coupling beams between vibrating plates [see  Fig. \ref{HOTI}(a), right panel]~\citetext{\citealp{Nature.555.342}}. In airborne acoustics, the negative hopping is elegantly realized by tuning the phase of acoustic dipoles or adjusting the connectivity of resonant cavities and narrow tubes, where the tube length or position modulates the coupling phase ~\citetext{\citealp{PhysRevLett.124.206601,NatCommun.11.2442}}.

\begin{figure}[t]
		\centering
		\includegraphics[width=\linewidth]{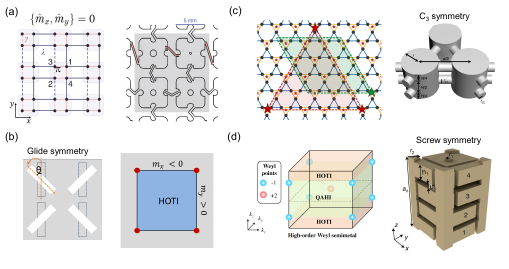}
		\caption{\label{HOTI}
			Higher-order topological insulators induced by spatial symmetries.  (a) Quadrupole topological insulator enabled by anti-commuting mirror symmetries. Left: square lattice implementation with $\pi$-flux per plaquette. Right: an elastic metamaterial design with engineered negative (shaded red) and positive couplings, generating an effective $\pi$-flux per plaquette (unit cell). (b) Nonsymmorphic glide-symmetric phononic crystal supporting Jackiw-Rebbi-type corner states as domain walls between edges with opposite Dirac masses. (c) Left: $C_3$-symmetric breathing kagome lattice hosting corner states only at specific Wyckoff positions protected by rotational symmetry. Right: an acoustic metamaterial design using cavity-tube architecture. (d) A higher-order Weyl semimetal
exhibiting concurrent hinge and surface states and fractional hinge
charges. Its acoustic realization is shown in the right. (a) Adapted from~\citetext{\citealp{Science.357.61, Nature.555.342}}. (b) Adapted from~\citetext{\citealp{NatPhys.15.582}}. (c) Adapted from~\citetext{\citealp{NatMater.18.108}}. (d) Adapted from~\citetext{\citealp{PhysRevLett.125.146401,NatMater.20.794}}.}
\end{figure}

Beyond anti-commuting symmetries, HOTIs can emerge from commuting spatial symmetries or nonsymmorphic protections. For instance, in a square lattice with glide symmetries $G_x$, $G_y$ ~\citetext{\citealp{NatPhys.15.582}}, the combined anti-unitary operator $\Theta_j=G_j T$ (with $T$ time-reversal operation) enforces Kramers-like Dirac degeneracies on the Brillouin zone boundaries. Gapping these Dirac points via symmetry-breaking perturbations leads to gapped edge states, described by 1D massive Dirac Hamiltonians
\begin{equation}
H_{edge,j}=v_j k_j s_z + m_j s_y.
\end{equation}
with $j=x,y$ for the edges along the $x$- and $y$-directions. $v_j$ is the group velocity of the edge modes, $s_z=1$ and $-1$ represent the pseudo-spin up and down states, and $s_y=1$ and $-1$ denote the odd- and even-parity edge modes, respectively. The Dirac mass $m_j$ changes sign between adjacent edges. This sign reversal, protected by mirror or glide symmetries, triggers the formation of zero-dimensional corner states via the Jackiw–Rebbi mechanism, as shown in Fig. \ref{HOTI}(b), a phenomenon demonstrated in glide-symmetric acoustic crystals ~\citetext{\citealp{NatPhys.15.582}}.    

Point-group symmetries also facilitate higher-order topology via quantized bulk polarization.  A prominent example is the breathing kagome lattice, whose $C_3$ symmetry quantizes a nonzero bulk polarization ~\citetext{\citealp{NatMater.18.108,NatMater.18.113}}. This quantized invariant dictates that corner states emerge exclusively at specific acute-angled corners—namely, those associated with symmetry-invariant Wyckoff positions where the Wannier centers become localized, as illustrated in Fig. \ref{HOTI}(c), left panel. This geometry-dependent localization highlights the profound interplay between crystal symmetry and higher-order band topology. More generally, a systematic framework based on Wannier centers has been established, showing how $C_n$-symmetric crystals host distinct corner state patterns depending on the placement of Wannier orbitals relative to the crystal's rotational axes ~\citetext{\citealp{PhysRevB.99.245151}}. Acoustic realizations again benefit from cavity–tube architectures [see Fig. \ref{HOTI}(c), right panel], which are readily fabricated via 3D printing or precision machining ~\citetext{\citealp{NatMater.18.108,NatMater.18.113}}.     

The high tunability of acoustic and mechanical metamaterials further enables HOTIs beyond idealized tight-binding limits. For example, a nonsymmorphic $p4g$-symmetric phononic crystal enables a hierarchy of topological band gaps: the first gap hosts a quantized Wannier dipole polarization (emulating the quantum spin Hall effect), while the second gap exhibits an anomalous quadrupole moment hosting higher-order corner states [see Fig. \ref{HOTI}(d)] ~\citetext{\citealp{NatCommun.11.65}}. Transitioning from $p4g$ to $C_{4v}$ symmetry via geometric deformation annihilates the quadrupole topology, directly visualizing the symmetry-topology interplay.

In addition to 2D settings, higher-order topology naturally extends to 3D systems as well as to semimetallic regimes. In 3D HOTIs, the hierarchical bulk–boundary correspondence manifests through the emergence of 1D hinge states and 0D corner states residing within the band gaps of both bulk and surface spectra ~\citetext{\citealp{NatCommun.10.5331,NatCommun.11.2442,SciAdv.6.eaay4166}}. In the semimetallic regime, higher-order topology gives rise to gapless bulk nodal points that not only connect surface Fermi arcs but also coexist with hinge-localized modes ~\citetext{\citealp{PhysRevLett.125.146401, PhysRevLett.125.266804}}, forming an enriched boundary hierarchy [see Fig. \ref{HOTI}(d)]. Realizing these exotic phases requires carefully engineered symmetries and couplings, which have been brought to reality again using acoustic metamaterial designs incorporating screw symmetries ~\citetext{\citealp{NatMater.20.794}}, chiral interlayer couplings ~\citetext{\citealp{PhysRevLett.127.146601}}, or stacked canonical 2D HOTI lattices ~\citetext{\citealp{NatMater.20.812,PhysRevLett.132.186601}}.

\subsubsection{\label{sec:level3}Designing Real Band Topology: Euler, Stiefel-Whitney, and Non-Abelian Topological Phases}
As the band topology undergoes a more refined classification, a new paradigm has shifted to the real band topology rooted in the real-valued Bloch wavefunctions~\citetext{\citealp{PhysRevB.92.081201}; \citealp{PhysRevLett.121.106403}; \citealp{Science.365.1273}; \citealp{ChinesePhysB.28.117101}; \citealp{PhysRevX.9.021013}; \citealp{PhysRevB.100.195135}; \citealp{PhysRevB.100.205126}; \citealp{NatPhys.16.1137}; \citealp{PhysRevB.102.115135};  \citealp{Science.383.844}}. The premise is the real condition of the Hamiltonian stabilized by the space-time inversion symmetry $I=I_ST$, where $I_S$ denotes the spatial inversion for both spinful and spinless lattices, or $\pi$ rotation only in two-dimensional spinless systems, $T$ is the time-reversal symmetry. Such a combined anti-unitary symmetry sends each $k$ point to itself, acting as a momentum-space onsite symmetry. Moreover, when $I^2=1$, a suitable basis can always be taken to yield
$IH(k)I^{-1}=H^{*}(K)=H(k)$, and $I\ket{u(k)}= {\ket{u(k)}}^*=\ket{u(k)}$,
 leading the eigenstates at each $k$ point to be only positive- or negative-, rather than complex-valued. The gauge freedom of eigenstates then reduces from $U(1)$ to $Z_2$. Furthermore, the real eigenstates at each $k$ can be regarded as an orthonormal rotation frame $\{u_i(k)\}$. The way the frame rotates smoothly along $k$ contributes to the real band topology. Although the Chern number has to vanish due to zero Berry curvature at each momenta, novel topological invariants can be rebuilt from real eigenstates to reinterpret the conventional topological phenomena or further give rise to unprecedented topological phases, including multigap non-Abelian topological phases, Euler insulators, and Stiefel–Whitney phases, to name a few. In addition, as the real condition is friendly to phononic designs, most of the experimental verifications of the real band topology have been accomplished in phononic systems. 

 The nontrivial first Stiefel-Whitney number denotes the nonorientation, or the real twist of Bloch wavefunctions along a 1D momentum loop. As such, the rotation frame cannot smoothly evolve to itself, but ends up with a $\pi$ rotation. The $Z_2$ classification comes from the first homotopy group of a real Hamiltonian space, which can be identified as the real version of the Berry phase, without generating additional physical consequences. In comparison, the Euler integer can be regarded as a time-reversal version of the Chern number, defined for two orientable real bands $u_1(\textbf{k})$ and $u_2(\textbf{k})$ in two dimensions. Like the Chern number integrated from the Berry curvatures, the quantized Euler number is defined by~\citetext{\citealp{NatPhys.16.1137}} 
 \begin{equation}
 \epsilon=\frac{1}{2\pi} \int_{BZ}d^2\textbf{k}F(\textbf{k}), 
 \end{equation}
 where $F(\textbf{k})=\nabla\times\bra{u_1(\textbf{k})}\nabla\ket{u_2(\textbf{k})}$ represents the Euler curvature. The $Z$-valued quantization originates from the second homotopy group of the real Hamiltonian space, i.e., $\pi_2(M_{n,l})=Z$, where $M_{n,l}=\frac{O(n+l)}{O(n)\times O(l)}$ with $n=2$ denoting the number of occupied and $l$ the unoccupied bands. The nonzero Euler number indicates the Wannier obstruction of occupied bands. However, when the number of occupied bands $n>2$, the second homotopy group of the Hamiltonian space reduces from $Z$ to $Z_2$, characterized instead by the second Stiefel-Whitney number. In other words, the Euler band topology is fragile, and the Wannier obstruction can be trivialized by an additional trivial band added upon the occupied bands. Furthermore, as the $PT$ symmetry is usually broken at boundaries, the Euler band topology lacks a robust bulk-boundary correspondence. Specifically, for a three-band Euler Hamiltonian, the two-band Euler topology can be encoded in the Bloch wavefunctions of another individual bulk band, whose orientations exhibit a meronic Skyrmion texture in the whole BZ [see Fig.~\ref{fig:real} (a)], which has been experimentally demonstrated in an acoustic kagome lattice~\citetext{\citealp{SciBull.69.1653}}. As both the phases and amplitudes of sound responses can be extracted in the air-borne acoustic system, the momentum-resolved Bloch wavefunctions can be immediately rebuilt, providing an alternative experimental approach for probing bulk responses of band topology that lacks the robust boundary signatures. Although the Euler topology is fragile, the reduced nontrivial second Stiefel-Whitney number can still exhibit higher-order topological phenomena, which has been demonstrated in two 3D acoustic nodal-line semimetals with $PT$ symmetry as well as additional projective crystal symmetries (see the following section)~\citetext{\citealp{NatCommun.14.4563}; \citealp{PhysRevLett.132.197202}}, where the emergent higher-order hinge states and the first-order surface states [see Fig.~\ref{fig:real} (b)] are simultaneously protected by the second and first Stiefel-Whitney numbers, respectively. A recent study also revealed that the broadly realized QSH-like model can also be deemed as the Euler phase or the second Stiefel-Whitney topological phase~\citetext{\citealp{PhysRevLett.131.053802}}, which supports topological higher-order corner states, rather than the robust edge states. We remark that the Euler insulator, the fragile phase~\citetext{\citealp{PhysRevLett.121.126402,PhysRevX.9.021013,PhysRevB.100.195135,PhysRevB.100.205126,PhysRevB.102.115135}}, and the QSH-like phase or the spin Chern insulator, although classified from different perspectives of band topology, may point to an identical model or system, in the presence of $PT$ or $C_2T$ symmetry~\citetext{\citealp{NatCommun.11.3227,Science.367.797}}.

\begin{figure}[!htbp]
  \centering
  \includegraphics[width=\linewidth]{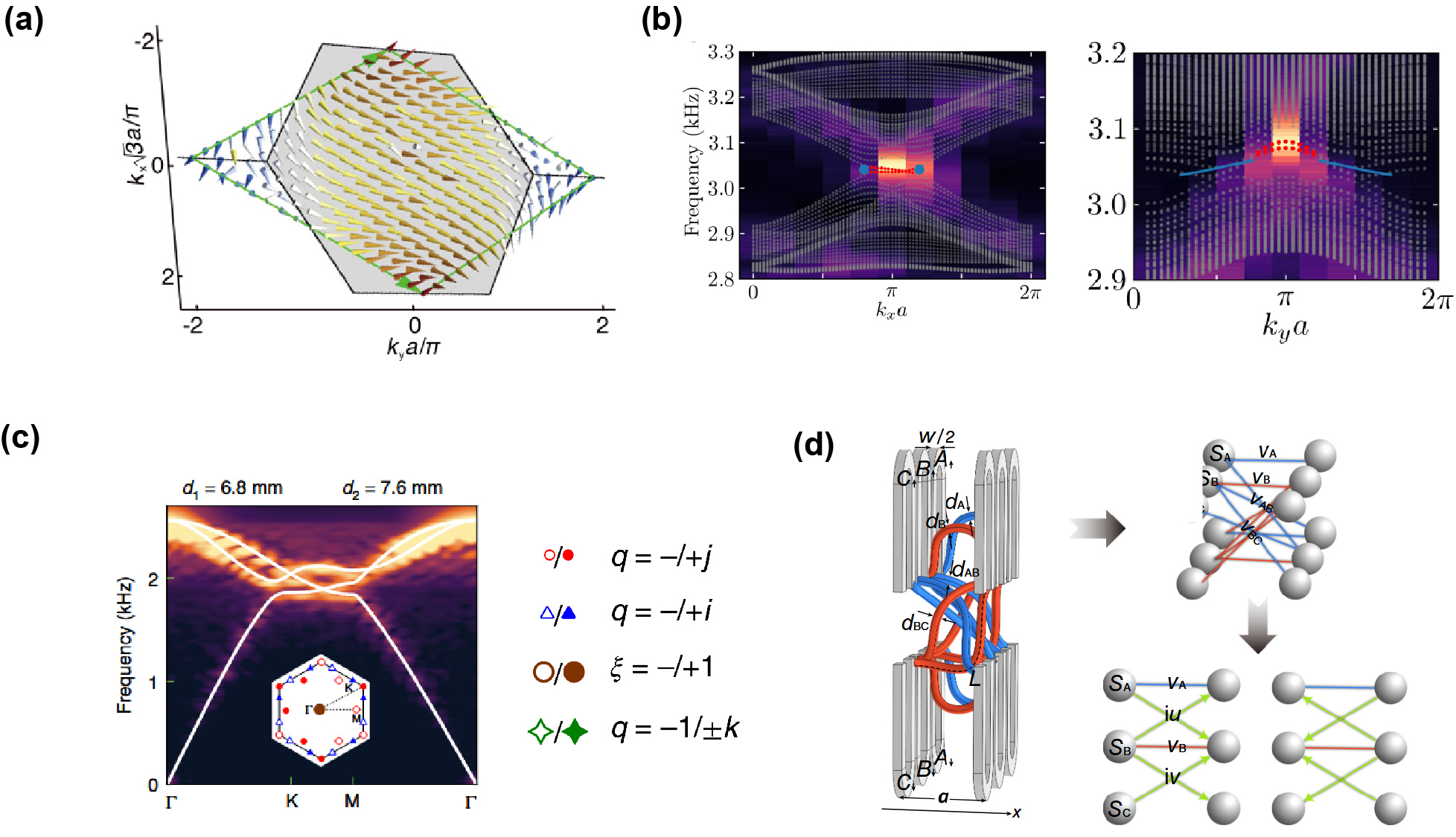}
  \caption{Real band topology in phononic systems. (a)  The half Skyrmion texture of the first band for the acoustic topological Euler insulator were detected. (b) Measured  surface and hinge states in the acoustic nodal line semimetal characterized by both the first and second Stiefel-Whitney numbers,respectively. (c) One typical evolution of non-Abelian charges in the acoustic kagome lattice. (d) Real couplings for mimicking a minimal non-ABelian model. (a) Adapted from~\citetext{\citealp{SciBull.69.1653}}. (b) Adapted from~\citetext{\citealp{NatCommun.14.4563}}. (c) Adapted from~\citetext{\citealp{NatPhys.17.1239}}. (d) Adapted from~\citetext{\citealp{PhysRevLett.132.216602}}.}
  \label{fig:real}
  \end{figure}

The conventional topological invariants, such as the Chern number and the Berry phase, are always defined within two subspaces spanned by the so-called 'occupied' and 'unoccupied' states and separated by a full band gap throughout the Brillouin zone. In this case, the topological invariants are always of Abelian character, obeying additivity and commutativity.   However, the non-Abelian band topology defined in a real multi-band system, first proposed by Wu et.al~\citetext{\citealp{Science.365.1273}}, just goes beyond the Abelian paradigm. Taking the 3-band real Hamiltonian as an example, the conventional Abelian topological invariant can usually be obtained from the homotopy group of the bipartite Hamiltonian space, such as the first Stiefel-Whitney invariant mentioned above, which is derived from $\pi_1(M_{1,2})=Z_2$, where $M_{1,2}=\frac{O(3)}{O(1)\times O(2)}$ corresponds to the space of the occupied (or unoccupied) bands. In comparison, if the binary partition for the Hamiltonian is abandoned, the Hamiltonian space is then altered to
 $M_{3}=\frac{ O(3)}{O(1)\times O(1) \times O(1)}.
$ 
 The striking resultant difference is that the first homotopy group of such a manifold is changed to
$
 \pi_1(M_{3})=Q=\{{\pm i,\pm j,\pm k,+1,-1}\},
$
 forming a non-Abelian quaternion group for the topological characterization of the whole three bands. For a 1D gapped 3-band insulator, the $\pi$ rotation of the rotation frame around the first, second, and third eigenstates induces, respectively, the non-Abelian topological charges of $+i$, $+j$, and $+k$, corresponding to the nontrivial Berry phase of each band. The conjugate ones $-i$, $-j$, and $-k$ represent the inverse rotations. $+1$ denotes the trivial phase without any rotation. The $2\pi$ rotation of the frame around each eigenstate axis constitutes the charge $-1$, which also exhibits robustness and probable boundary responses. Nevertheless, it cannot be elucidated by the conventional topological invariants. Furthermore, the non-Abelian feature of these topological charges can lead to the non-Abelian braiding process in higher dimensions, which is also absent in Abelian topological phases. Phononic systems have recently witnessed the rise of the non-Abelian band topology. The phonon spectra of layered silicates were soon predicted to host non-Abelian physics~\citetext{\citealp{NatCommun.13.423}}. The non-Abelian charges were then first observed in transmission line networks in 1D gapped systems~\citetext{\citealp{Nature.594.195}}, whereas the braiding and phase transitions of non-Abelian charged band nodes were first observed in a tunable 2D acoustic semimetal based on the kagome lattice~\citetext{\citealp{NatPhys.17.1239}}. The 3-site kagome model with nearest and next-nearest neighbor couplings naturally forms a 2D 3-band $PT$-symmetric Hamiltonian, which can be perfectly cast to an acoustic structure using acoustic cavities and tubes. Meanwhile, the band nodes in the band structure are always tangled together within three bands, upon which the non-Abelian topological charges can be defined. By tuning the geometric parameters of the acoustic system, the band nodes with different non-Abelian charges can create, convert, or annihilate, through the braiding of nodes in adjacent gaps, as exemplified by the evolution of experimentally measured band structures [see Fig.\ref{fig:real} (c)].  Specifically, the quadratic node at $\Gamma$ always remains due to the charge $-1$ and a nonzero Euler number $1$ defined in the vicinity of $\Gamma$, proving the robustness of the non-Abelian charge $-1$ within its own bands. The $-1$ charge was also observed in a 3D phononic nodal link system, where the link of the nodal line and chain cannot be severed, further experimentally indicating the topological protection of the non-Abelian charge $-1$. Another acoustic experiment based on the 2D square lattice provided more evidence of the braiding process and the stability of non-Abelian charges~\citetext{\citealp{NatCommun.14.1261}}. The non-abelian topology has also been verified in a 1D acoustic crystal with the introduction of $Z_2$ gauge fields~\citetext{\citealp{PhysRevLett.132.216602}}. The two layers of original minimal 1D 3-band models with opposite imaginary couplings can be unitarily transformed to a time-reversal 6-band model with only real couplings, which can then be constructed by meticulous layout of acoustic cavities and connecting tubes [see Fig.~\ref{fig:real} (d)]. The topological end states of $i$, $j$, and $k$ phases, and the domain wall states determined by the quaternion operation of two charges on either hand, were detected, further confirming the straightforward bulk-boundary correspondence of non-Abelian band topology. Moreover, a time-varying programmable acoustic system mimicking a synthetic 1D non-Abelian phase was also developed, which first visualized in acoustics the experimentally obtained topological charges in both the eigenstate frame sphere and $SO(3)$ space~\citetext{\citealp{PhysRevLett.134.136601}}. The diabatic evolution of bulk states is within only one unit cell, which provides tremendous convenience for obtaining real Bloch eigenstates.

\subsubsection{\label{sec:level3}Designing Projective-Crystal-Symmetry-Protected Phases}

Symmetry groups, including crystalline symmetries, are generally projectively represented under gauge fields, where the phase degrees of freedom of wavefunctions have to be considered in spatial transformations. The simplest gauge field is the time-reversal symmetric $\pi$ flux or named the $Z_2$ gauge field, where the hopping amplitudes are allowed to take $\pm$ signs, which play the most central role in projective crystal symmetry-protected phases~\citetext{\citealp{MaterTodayQuantum.2025.100055}}, and can be readily engineered in and almost perfectly cast to phononic systems, using the design scheme as elaborated in Sec. IIIA. The pioneering example, as previously mentioned, is the phononic analog of a QSH insulator realized in bilayer systems with both negative and positive interlayer couplings, which is also later referred to as the spinless mirror Chern insulator~\citetext{\citealp{PhysRevB.108.205126}}. The key there is that the $Z_2$ gauge field leads to $(MT)^2=-1$ for all momentum points with $M$ denoting mirror symmetry between two layers, which protects the pseudo-Kramers degeneracy in spinless systems. A more refined acoustic demonstration of a spinless mirror Chern insulator utilizing $p$-orbital induced $Z_2$ gauge fields was conducted recently, by further experimentally detecting the opposite Chern numbers (denoted by Wilson loop winding) within different mirror spaces~\citetext{\citealp{CommunPhys.6.268}}, as exhibited in Fig.~\ref{fig:projective} (a). Meanwhile, a 1D acoustic experimental sample with $(PT)^2=-1$ induced by the $Z_2$ gauge field also exhibits the spinful-like nontrivial topology with even winding numbers~\citetext{\citealp{PhysRevLett.130.026101}}. Another representative model is the BBH model, which, together with the $Z_2$ gauge field, inspired the birth of the M$\ddot{o}$bius insulator protected by projective translation symmetries, where the edge bands are entangled and form a M$\ddot{o}$bius twist over the edge BZ  [see Fig.~\ref{fig:projective} (b)]. Furthermore, the $Z_2$ gauge field can induce the so-called momentum-space nonsymmorphic symmetries~\citetext{\citealp{PhysRevLett.130.256601}}. One of the physical consequences of the momentum-space glide symmetry is that the torus 2D BZ $S^2$ can be resolved into two non-orientable Klein bottles $K^2$ with only one being independent. Moreover, the topological invariant can be defined only in a half of the BZ, giving a general and stable $Z_2$ classification instead of $Z$, based on the second-order cohomology group~\citetext{\citealp{NatCommun.13.2215}}. In addition, once the extra mirror symmetry $M_y$ is added, the $Z$-valued Chern number can be stabilized in a half or even a quarter of the BZ, leading to the dipole or quadrupole Chern insulators~\citetext{\citealp{arXiv.2503.09970}}. As the core thought is to involve the $Z_2$ gauge field, the M$\ddot{o}$bius and Klein bottle insulators were mainly observed in both 2D and 3D in phononic systems~\citetext{\citealp{PhysRevLett.128.116803}; \citealp{PhysRevLett.128.116802}; \citealp{PhysRevLett.130.026101}; \citealp{PhysRevB.108.L220101}; \citealp{arXiv.2305.07174}; \citealp{SciBull.69.2050}; \citealp{PhysRevLett.132.213801}; \citealp{PhysRevB.109.134107}; \citealp{PhysRevAppl.21.044002}; \citealp{PhysRevB.111.L100101}}. Fig.~\ref{fig:projective} (b) shows an acoustic M$\ddot{o}$bius insulator, whose boundary states and their $4\pi$ winding were both observed. In contrast, the multiple Chern insulator model demands complex hoppings, which were observed in phononic systems with synthetic dimensions~\citetext{\citealp{PhysRevLett.135.206603}; \citealp{NatCommun.16.9669}}, exhibiting the novel returning Thouless pumping [see Fig.~\ref{fig:projective} (c)], derived from the opposite Chern numbers in two halves of BZ. Here the synthetic dimension is denoted by varying parameters, while it can also be extended to temporal modulation in a non-static system, as elaborated in the following section D. 
\begin{figure}[!htbp]
  \centering
  \includegraphics[width=\linewidth]{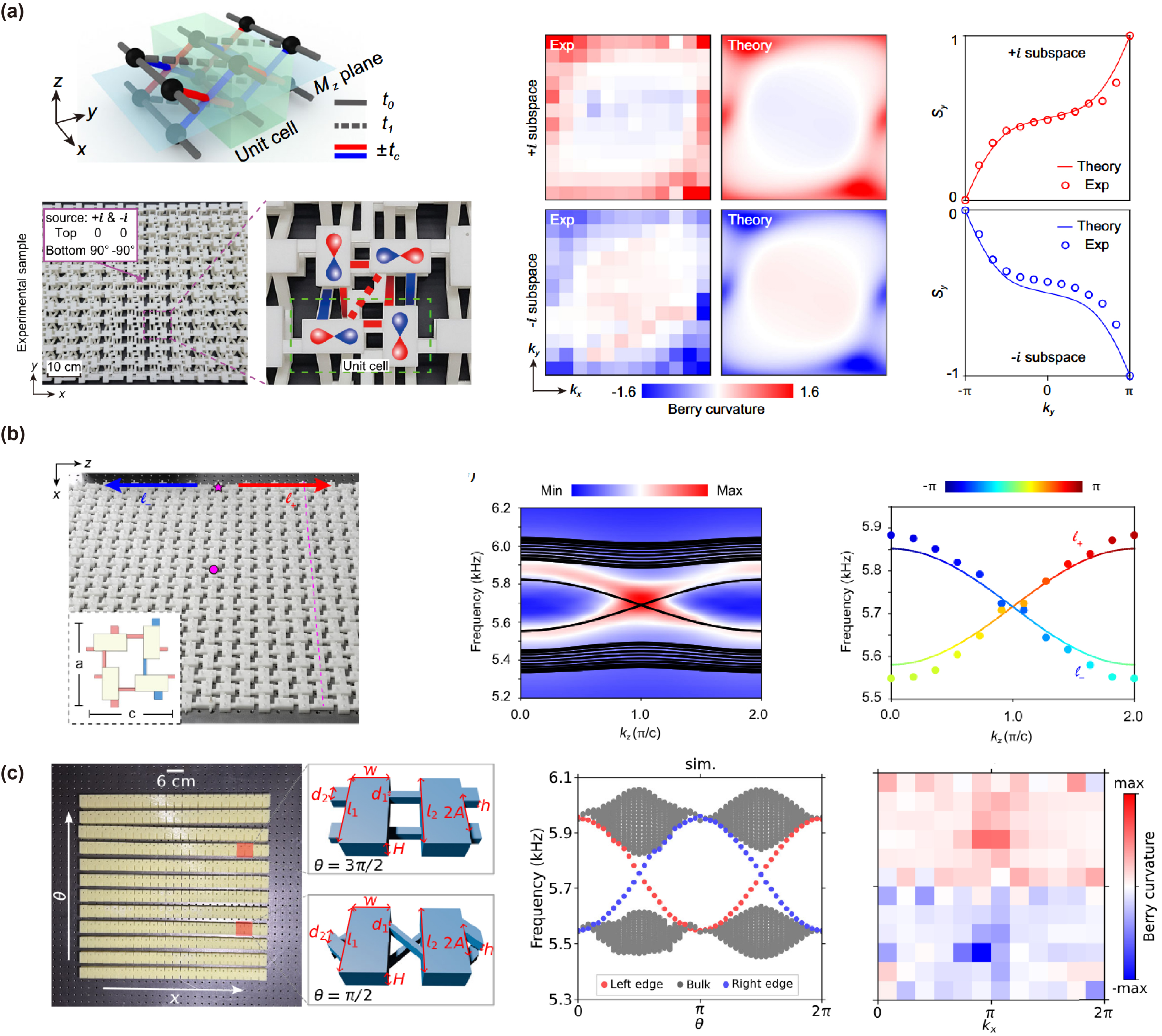}
  \caption{Projective crystal symmetry protected phases in phononic systems. (a) Schematic of a bilayer spinless mirror Chern model with its acoustic expermental sample based on the connected $p$-orbital cavities. The left panels denote the experimentally obtained Berry curvature distributions in BZ and the Wilson loop spectra in the opposite mirror spaces. (b) Acoustic M$\ddot{o}$bius insulator sample with its edge dispersion as well as phase winding measurements. (c) The 2D dipolar Chern insuator in 1D acoustic lattice with an additional parameter dimension. The projected spectra exhibit chiral edge states in each half of the BZ. The Berry curvature distributions were detected. (a) Adapted from~\citetext{\citealp{CommunPhys.6.268}}. (b) Adapted from ~\citetext{\citealp{PhysRevLett.128.116803}}. (c) Adapted from~\citetext{\citealp{NatCommun.16.9669}}.}\label{fig:projective}
  \end{figure}

\subsubsection{\label{sec:level3}Simulating Hidden Symmetry-Protected Phases}
The conventional classification of topological phases relies on explicit symmetries such as time-reversal, chiral, or reflection symmetries. However, recent researches have revealed that topological protection can arise from more subtle, hidden or latent symmetries that are not immediately apparent in the real-space structure of a system~\citetext{\citealp{PRL.130.077201,PRB.108.L220303,PRB.110.035106,Sci.Phys.18.061,PRB.111.245106,APL.126.133102,PRapplied.23.L031001}}. This concept, rooted in the mathematical framework of isospectral reduction, has recently found fertile ground in the study of phononic crystals, where structural tunability enables precise experimental realization.
The theoretical underpinning of hidden symmetry protection stems from graph theory and spectral analysis. Consider a lattice Hamiltonian $H$. For a chosen subset of vertices $S$ of the defined graph, the isospectral reduction of $H$ onto $S$ yields an energy-dependent non-linear effective Hamiltonian $H_{\text{eff}}(E)$ that shares the identical energy spectrum and satisfies
\begin{equation}
H_{\text{eff}}(E) = H_{SS} + H_{S\bar{S}} (E - H_{\bar{S}\bar{S}})^{-1} H_{\bar{S}S},
\end{equation}
where $H_{SS}$ denotes the block of $H$ restricted to $S$, and $\bar{S}$ is the complement. The key insight is that $H_{\text{eff}}(E)$ may possess symmetries—such as chiral symmetry,
\begin{equation}
\Gamma H_{\text{eff}}(E)\Gamma^{-1}=− H_{\text{eff}}(E),
\end{equation}
that are not present in the original Hamiltonian $H$. These emergent symmetries, termed hidden symmetries, can protect topological phases even when the original lattice appears trivial.

The graph theory-based Hamiltonian can be directly mapped to acoustic systems. An typical example is the acoustic trimer SSH (SSH3) model~\citetext{\citealp{PRapplied.23.L031001}}. Through isospectral reduction, the  SSH3 lattice lacking explicit chiral symmetry can be mapped onto a two-node effective Hamiltonian with exact chiral symmetry. The two pairs of edge states derived from nonzero hidden winding numbers were observed with great robustness, establishing a clear bulk-edge correspondence for systems that fall outside the conventional tenfold way classification. The interplay between hidden symmetry and flat bands~\citetext{\citealp{PRB.104.035105}}, degeneracy~\citetext{\citealp{PRL.126.180601}}, non-Hermiticity~\citetext{\citealp{PhysRevLett.131.237201}}, higher dimension, and information transfer in topological acoustic systems, to name a few, may potentially bridge the gap between classical simulators and quantum topological materials.

\subsection{\label{sec:level2}Adding the Temporal Dimension}

Topology in wave systems need not be specified by a static Hamiltonian alone; it can be encoded directly in the evolution operator. Given a possibly time-dependent generator \(H(t)\), the finite-time propagator
\begin{equation}
U(t_2,t_1)=\mathcal{T}\exp\!\Big[-i\!\int_{t_1}^{t_2} H(t)\,dt\Big],
\label{eq:U-propagator}
\end{equation}
governs all observables. Accordingly, topological information may reside in the path \(\{U(t,t_0)\}\) itself, without requiring a static band structure or even a periodic drive. For a \(d\)-dimensional Bloch lattice, the momentum-resolved propagator \(U_k\) allows topology to be defined on extended manifolds such as \((k,t)\) or \((k,\lambda)\). When \(\lambda\in S^1\) is a closed control cycle, this viewpoint reduces to standard band-geometric invariants such as the Chern number; more generally, loop unitaries and non-Abelian holonomies provide dynamical invariants that govern robust state transfer and braiding.

In artificial phononic platforms, the relevant dynamics can often be reduced, near to an effective first-order form
\begin{equation}
i\,\partial_{\tau}\psi(\tau)=H(\tau)\psi(\tau),
\qquad
\tau \in \{t,\ z,\ \text{other synthetic paths}\},
\label{eq:first-order-reduction}
\end{equation}
where \(H(\tau)\) captures the instantaneous on-site terms and couplings. This reduction naturally organizes temporal topological phenomena into three experimentally relevant regimes:
\begin{flalign}
\label{eq:dyn-regimes}
& \begin{aligned}
\text{(Adiabatic)}\;&
\bigl|\langle u_m|\partial_{\tau} u_n\rangle\bigr|
\ll \bigl|\varepsilon_m-\varepsilon_n\bigr|
; \\[2pt]
\text{(Floquet)}\;&
H(\tau{+}T)=H(\tau)
; \\[2pt]
\text{(Quench/interface)}\;&
H(\tau)=
\begin{cases}
H_1,& \tau<\tau_0,\\
H_2,& \tau\ge \tau_0.
\end{cases}
\end{aligned} &
\end{flalign}
namely, adiabatic pumping, periodic (Floquet) driving, and abrupt or finite-rate temporal control. Crucially, the relevant protection is encoded in the evolution itself, so robustness extends beyond instantaneous spectra to cycle- and path-resolved dynamics.

\noindent

Equation~\eqref{eq:first-order-reduction} is sufficient to implement the full dynamical-topology toolkit: adiabatic pumping (when a slow closed path exists), Floquet band engineering (when periodic), and quench/time-interface control (when sudden change).

\subsubsection{\label{sec:level3} Thouless Pumping within Adiabatic Limit}
In a gapped 1D band \(n\) driven adiabatically and cyclically by a parameter \(\lambda\in[0,2\pi)\), quantized transport is equivalently described by the winding of the hybrid Wannier center \(X_n(\lambda)\):
\begin{equation}
\Delta x_n \;=\; X_n(2\pi)-X_n(0)\;=\; C_n\,a ,
\label{eq:wan-shift}
\end{equation}
where \(C_n\) is the first Chern number on \(\mathrm{BZ}\times S^1_\lambda\), \(a\) is the lattice constant, and the sign is fixed by the loop orientation. Quantization persists as long as the bulk gap remains open throughout the cycle. In this sense, Thouless pumping is the dynamical manifestation of quantized polarization evolution.

Artificial phononic systems provide a particularly transparent setting for this physics. Acoustic cavity arrays first implemented the bulk viewpoint by tracking Wannier-center evolution under parameter modulation, thereby realizing a classical analogue of quantized adiabatic transport~\citetext{\citealp{JAcoustSocAm.146.742}}. Subsequent works shifted the emphasis from bulk polarization to protected boundary-state transport: in elastic and acoustic lattices, slow modulation of couplings or effective on-site parameters drives edge-localized states across the sample with high robustness against disorder~\citetext{\citealp{PhysRevLett.123.034301}; \citealp{PhysRevB.101.094307}}, as shown in Fig.~\ref{fig:waveguide-pumping}(a). This established adiabatic pumping as a genuinely dynamical topological protocol in phononics rather than a static band analogy.

The same logic extends naturally to synthetic dimensions and higher-order transport. Strongly coupled acoustic cavity systems have demonstrated edge-bulk-edge and corner-bulk-corner transfer [Fig.~\ref{fig:waveguide-pumping}(b)], providing direct probes of higher Chern numbers and 4D quantum-Hall analogues~\citetext{\citealp{NatCommun.12.5028}; \citealp{SciBull.67.1950}}. More recently, Berry-dipole systems have realized a returning Thouless pump  [Fig.~\ref{fig:waveguide-pumping}(c)], in which quantized transport acquired in the first half-cycle is reversed in the second~\citetext{\citealp{NatCommun.16.9669}; \citealp{PhysRevLett.135.206603}}. These examples show that adiabatic pumping in artificial phononics is not restricted to unidirectional edge transfer, but can be generalized to richer cycle-resolved transport.

\begin{figure}[!htbp]
  \centering
  \includegraphics[width=\linewidth]{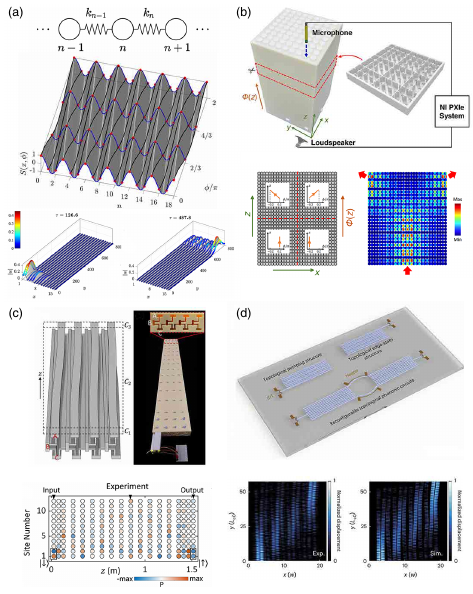}
  \caption{Topological pumping in phononics. (a) A $\phi$-modulated 1D mass–spring chain achieves edge-to-edge Thouless pumping. (b) A 3D-printed acoustic cavity array with an axial phase gradient $\phi(z)$ supports directed pumping with edge accumulation. (c) Returning Thouless pumping realized in a one-dimensional acoustic waveguide array, where adiabatic modulation along the propagation direction drives an edge state into the bulk and back to its original edge. (d) GaN on-chip phononic circuit integrating interdigitated transducers and microheater-based phase control exhibits controllable edge states and GHz-regime pumping. (a) Adapted from~\citetext{\citealp{PhysRevB.101.094307}}. (b) Adapted from~\citetext{\citealp{NatCommun.12.5028}}. (c) Adapted from~\citetext{\citealp{PhysRevLett.135.206603}}. (d) Adapted from~\citetext{\citealp{NatElectron.8.689}}.}
  \label{fig:waveguide-pumping}
\end{figure}

A further advantage of phononic platforms is that the relevant modulation rates are experimentally accessible. As a result, adiabatic pumping can be implemented not only through spatially encoded synthetic parameters, but also through direct temporal modulation. Surface-acoustic and integrated on-chip platforms have demonstrated robust pumping of Rayleigh-type or guided phononic modes under slowly varying couplers or programmable phase control [see Fig.~\ref{fig:waveguide-pumping}(d)], extending the same topological principle toward chip-scale and gigahertz architectures~\citetext{\citealp{SciAdv.9.eadh4310}; \citealp{NatElectron.8.689}}. Real-time temporal implementations based on rotating boundary elements~\citetext{\citealp{PhysRevLett.125.253901}}, magneto-mechanical modulation~\citetext{\citealp{NatCommun.11.974}}, and electromechanical waveguides~\citetext{\citealp{PhysRevLett.126.095501}} further exploit the accessible timescales of acoustic and elastic media to realize programmable topological transport in the time domain, see Fig.~\ref{fig:Temporal pumping}. Together, these studies establish adiabatic pumping as a broadly accessible route to quantized and robust state transport in artificial phononic systems.

\begin{figure}[!htbp]
  \centering
  \includegraphics[width=\linewidth]{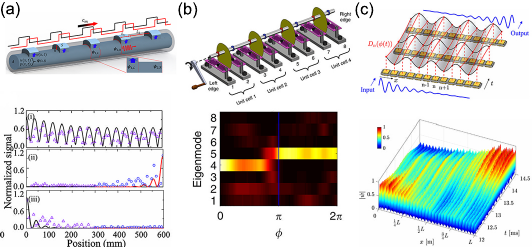}
  \caption{Topological pumping with temporal dimension. (a) An acoustic waveguide with a rotating helical sleeve produces a traveling boundary modulation, enabling robust edge-to-edge pumping. (b) A magneto--mechanical chain of torsional resonators synchronously modulates coupling and on--site frequencies with the pump phase $\phi$, exhibiting spectral flow across the gap. (c) An electromechanical waveguide with piezoelectric shunts programs a time--dependent stiffness $D_n[\phi(t)]$, adiabatically transferring an edge state from input to output. (a) Adapted from~\citetext{\citealp{PhysRevLett.125.253901}}. (b) Adapted from~\citetext{\citealp{NatCommun.11.974}}. (c) Adapted from~\citetext{\citealp{PhysRevLett.126.095501}}.}
  \label{fig:Temporal pumping}
\end{figure}

\subsubsection{\label{sec:level3} Beyond the Adiabatic Limit in Acoustic and Elastic Lattices}
When the modulation rate becomes comparable to the relevant spectral gaps, adiabatic following breaks down and the dynamics must instead be described in terms of transitions, temporal scattering, and stroboscopic evolution. In practice, this regime is usefully organized into three classes: finite rate sweeps that induce Landau-Zener tunneling, abrupt parameter changes that create temporal interfaces, and periodic drives that require a Floquet description.

Replacing a slow adiabatic passage by a finite-speed sweep converts gap traversals into Landau--Zener (LZ) problems~\citetext{\citealp{EurJPhys.31.389}}. For an avoided crossing with minimum gap \(\Delta\) and local detuning rate \(|\dot{\varepsilon}|\), the transition probability is
\begin{equation}
\label{eq:LZ}
P_{\mathrm{LZ}}
=\exp\!\Bigl(-\frac{\pi\,\Delta^{2}}{2\,|\dot{\varepsilon}|}\Bigr),
\end{equation}
which provides a practical criterion for the onset of non-adiabatic leakage. In acoustic topological pumps, this mechanism has been directly observed as boundary states are driven across finite-size minigaps, thereby delimiting the fidelity window of adiabatic transfer~\citetext{\citealp{PhysRevLett.126.054301}}, see Fig.~\ref{fig:non-adiabatic evolution}(a). More generally, reconfigurable electro-acoustic and mechanical lattices show that finite rate control need not merely degrade transport: by shaping the temporal protocol, one can relocate interfaces or accelerate state transfer while retaining substantial topological protection~\citetext{\citealp{PhysRevAppl.18.054058}; \citealp{PhysRevB.102.174312}; \citealp{PhysRevAppl.22.044009}}. In this sense, non-adiabaticity becomes a programmable resource rather than only a source of breakdown.

An abrupt parameter change defines a temporal boundary. The state remains continuous across the interface,
\begin{equation}
\label{eq:time-interface-cont}
\psi(t_0^{+})=\psi(t_0^{-}),
\end{equation}
but its modal decomposition changes because the eigenbases of \(H_-\equiv H(t_0^-)\) and \(H_+\equiv H(t_0^+)\) differ. Writing the amplitudes before and after the jump in the two instantaneous eigenbases, one obtains a temporal scattering matrix
\begin{equation}
\label{eq:time-S-matrix}
a_m^{+}=\sum_{n} S^{(t)}_{mn}\,a_n^{-},
\qquad
S^{(t)}_{mn}=\langle u_m^{+}\,|\,u_n^{-}\rangle .
\end{equation}
\textcite{PhysRevLett.133.077201} demonstrated this by abruptly altering the grounding stiffness in an elastic waveguide, observing temporal refraction (frequency translation at nearly fixed spatial wavelength) and band-to-band conversion [see Fig.~\ref{fig:non-adiabatic evolution}(b)]. In topological settings, the same mechanism provides a natural route to dynamically reroute or switch boundary-localized states between different instantaneous band structures.

\begin{figure}[!htbp]
  \centering
  \includegraphics[width=\linewidth]{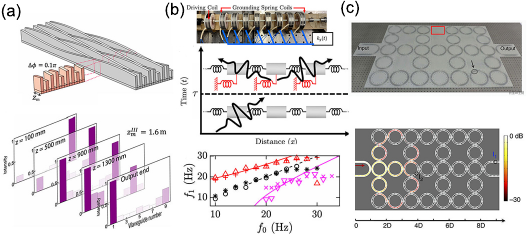}
  \caption{non-adiabatic evolution.  (a) A height-modulated acoustic waveguide array implementing a Harper modulation: adiabatic phase sweeping transfers a topological edge state across the bulk, whereas rapid sweeping induces Landau--Zener transitions. 
  (b) A magneto-mechanical phononic chain in which a step change of the grounding stiffness establishes a temporal boundary, producing temporal refraction and frequency conversion. 
  (c) A two-dimensional coupled-ring acoustic lattice (an anomalous Floquet topological insulator) with near-unitary inter-ring coupling that supports pseudo-spin-selective, backscattering-immune edge transport around bends.  (a) Adapted from~\citetext{\citealp{PhysRevLett.126.054301}}. (b) Adapted from~\citetext{\citealp{PhysRevLett.133.077201}}. (c) Adapted from~\citetext{\citealp{NatCommun.7.13368}}.}
  \label{fig:non-adiabatic evolution}
\end{figure}

The third route is periodic modulation, for which the relevant object is the Floquet operator \(U_k(T)\). Its eigenvalues define quasienergies \(\varepsilon_n(k)\) modulo \(2\pi/T\), and topology may reside in the full micromotion over \((k,t)\), even when all static Chern numbers vanish. This Floquet viewpoint has enabled several phononic manifestations of dynamical topology, including nonreciprocal chiral transport under space--time modulation, drive-induced anomalous boundary modes at \(\pi\) or fractional quasienergies, and Floquet higher-order corner states~\citetext{\citealp{SciAdv.6.eaba8656}; \citealp{PhysRevB.110.085137}; \citealp{PhysRevLett.129.254301}; \citealp{NatCommun.13.11}; \citealp{PhysRevB.109.L020302}}. Related elastic implementations based on piezo-shunted or time-modulated resonator arrays further confirm that spatiotemporal driving can open directional band gaps and realize reconfigurable, nonreciprocal wave transport~\citetext{\citealp{NewJPhys.18.083047}; \citealp{PhysRevAppl.13.031001}; \citealp{JMechPhysSolids.145.104181}}.

A particularly compact realization of Floquet topology dispenses with explicit time modulation altogether and instead uses ring-resonator scattering networks. Each loop implements a discrete propagation step, junctions are modeled by scattering matrices, and one full circulation plays the role of a Floquet period. Popularized in photonics~\citetext{\citealp{Nature.496.196}} and translated to acoustics by \textcite{NatCommun.7.11744}, such networks host anomalous Floquet topological insulators with robust chiral edge transport. Building on this framework, \textcite{NatCommun.7.13368} developed a scattering-matrix formalism for ring networks, making explicit how topological invariants emerge from the network evolution operator; Subsequent 3D-printed ring-network experiments confirmed Floquet edge modes and their immunity to disorder [see Fig.~\ref{fig:non-adiabatic evolution}(c)].  Taken together, finite rate sweeps, temporal interfaces, time-periodic drives, and scattering-network realizations show that non-adiabatic phononic dynamics provides a broad toolbox for temporal refraction, accelerated state transfer, and anomalous Floquet topology.

\subsubsection{\label{sec:level3} Geometric Phases and Multi-State Evolution}

\noindent\textbf{Geometric phases as a bulk probe.}

Beyond energy transport, programmable temporal control in artificial phononic lattices provides direct access to topology through \emph{geometric phases}. Treating time (or a protocol parameter) as a controllable coordinate, one can prepare bulk states and monitor their evolution along synthetic cycles; in a single isolated band this reduces to the Berry phase
\begin{equation}
\label{eq:berry}
\gamma_n=\oint_{\mathcal{C}} A_\lambda^{(n)}\,d\lambda,\qquad 
A_\lambda^{(n)}=i\langle u_n|\partial_\lambda u_n\rangle,
\end{equation}
so that band topology is revealed not by edge transport but by phase holonomies of \emph{bulk} trajectories.
As an antecedent in acoustics, \textcite{SciAdv.4.eaaq1475} showed that sound-vortex transport is governed by a spin-redirection geometric phase, establishing the role of geometric phases in acoustic wave control at the bulk level, see Fig.~\ref{fig:geomtric phase}(a).
Such state can be designed to be a Bloch state, \textcite{PhysRevLett.134.136601} demonstrated a complete bulk protocol: by preparing states in an electrically coupled acoustic lattice and evolving them adiabatically in time, they reconstructed Berry phases and Chern numbers of a synthetic Brillouin zone purely from transient dynamics.
This strategy highlights how programmable couplings turn geometric phases into experimentally accessible observables, providing a bulk probe complementary to boundary pumping.

\begin{figure}[!htbp]
  \centering
  \includegraphics[width=\linewidth]{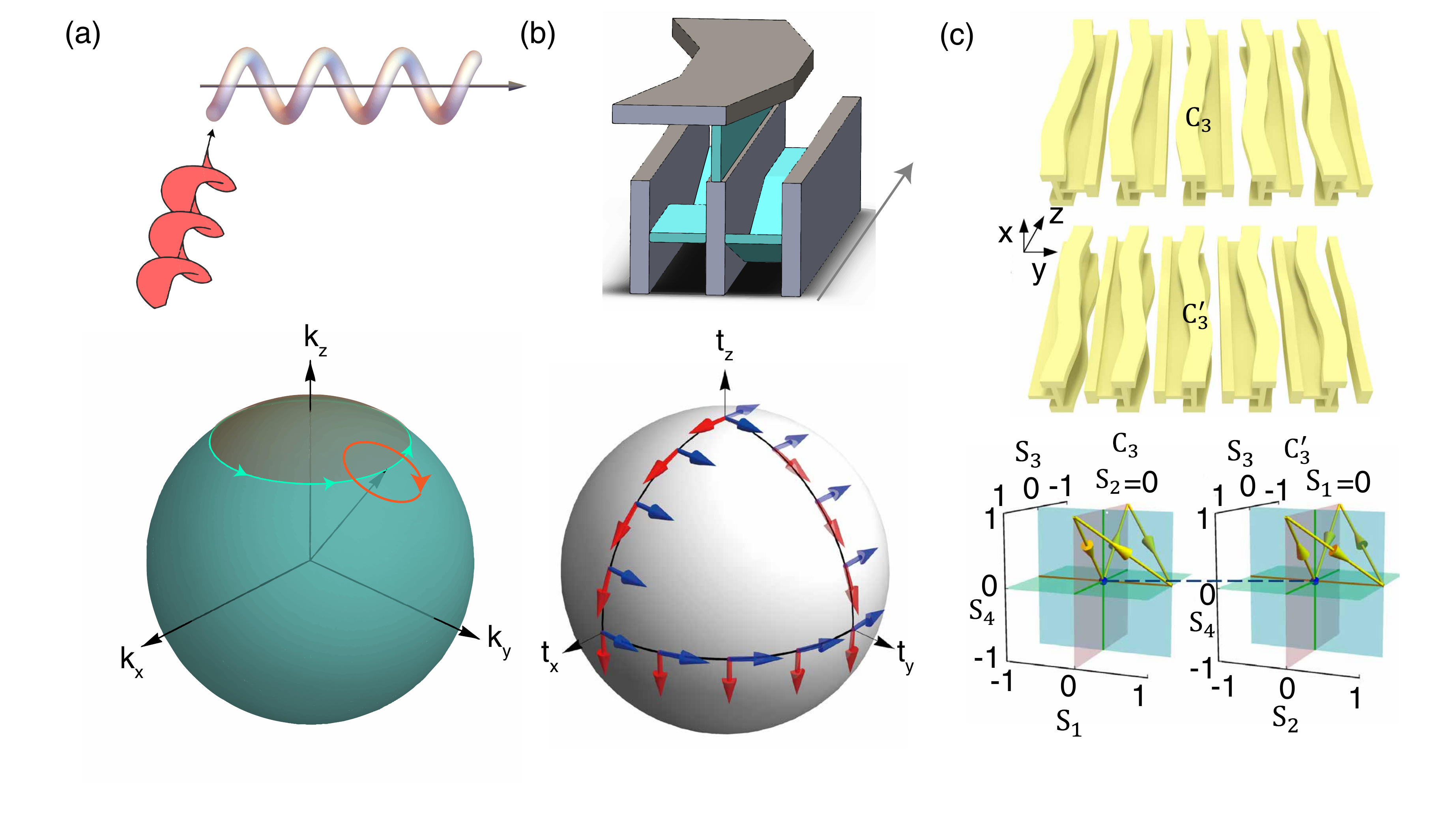}
  \caption{ (a) Scalar and (b) matrix valued Geometric phase in acoustics. (c) Non-Abelian Thouless pumping. (a) Adapted from~\citetext{\citealp{SciAdv.4.eaaq1475}}. (b) Adapted from~\citetext{\citealp{NatPhys.18.179}}. (c) Adapted from~\citetext{\citealp{PhysRevLett.128.244302}}.}
  \label{fig:geomtric phase}
\end{figure}

\medskip
\noindent\textbf{Non-Abelian generalization and braiding.}

In the presence of degeneracies, the phase becomes matrix valued: the Wilczek--Zee holonomy 
\(\mathcal{U}_{\mathcal{C}}=\mathcal{P}\exp\!\bigl(\oint_{\mathcal{C}}\!\mathcal{A}\bigr)\) with non-Abelian connection \(\mathcal{A}_{mn}=i\langle u_m|\partial_\lambda u_n\rangle\) governs how an entire multiplet is braided under a closed evolution. 
Leveraging programmable degeneracies, synthetic acoustic lattices furnish a natural arena for such effects: \textcite{NatPhys.18.179} designed a cavity--waveguide network supporting chiral-symmetry-protected degeneracies, and steered the couplings along intersecting loops in a three-parameter space, observing path-dependent mode exchange---a classical analogue of Wilczek--Zee holonomies, as illustrated in Fig.~\ref{fig:geomtric phase}(b). Fig.~\ref{fig:geomtric phase}(c) depicts a non-Abelian Thouless pump by encoding a pair of degenerate boundary modes; two spectrally indistinguishable loops executed in different orders yielded distinct final superpositions, directly evidencing non-commutativity~\citetext{\citealp{PhysRevLett.128.244302}}. These results position synthetic acoustic lattices as a controlled testbed for preparing and manipulating degenerate states---central to topological quantum control. An immediate direction is to push toward high-frequency and cryogenic regimes where intrinsic phonons act as carriers; there, transient topological control could couple directly to quantum degrees of freedom, enabling on-chip manipulation of long-coherence phonon wave packets and informing phonon-based quantum simulation and information processing.

\subsection{\label{sec:level2}Real-Space Topological Textures}

The study of real-space topological textures is fundamentally concerned with spatial configurations of continuous vector fields. A classical geometric constraint on such fields is illustrated by the Hairy Ball Theorem, which states that any continuous tangent vector field on the two-sphere $S^2$ must vanish at least at one point, implying the unavoidable emergence of singularities. Mathematically, this reflects the non-trivial topology of the tangent bundle $TS^2$, reflected in its nonvanishing Euler class, and highlights how global geometric constraints can obstruct the existence of nowhere-vanishing continuous vector fields. 

In condensed matter and acoustic systems, a closely related but conceptually distinct situation arises. The relevant vector fields, such as magnetization $\mathbf{m}(\mathbf{r})$ in chiral magnets or the nematic director $\mathbf{n}(\mathbf{r})$ in liquid crystals~\citetext{\citealp{RevModPhys.46.617}}, are not constrained to lie in the tangent space of the real-space manifold. Instead, they represent internal degrees of freedom defined independently at each spatial point, taking values in an intrinsic order-parameter space. The Hairy Ball theorem applies to tangent vector fields on the base manifold, whereas internal order-parameter fields defined as maps $M \to E$ are not subject to the same constraint, since they are not required to be tangent to $M$. From this broader perspective, the Hairy Ball theorem can be viewed as a special manifestation of geometric constraints associated with tangent bundles, whereas internal fields are naturally described within a different but parallel geometric framework. Specifically, the field configuration is no longer a section of the tangent bundle of the base manifold, but is described by a continuous mapping $\mathbf{v}: M \to E$, where the order-parameter space $E$ is regarded as a trivial fiber over $M$. Within this unified geometric language, the classification of field configurations is governed by homotopy classes of maps between the base space and the target space under appropriate boundary compactification conditions.

For localized textures or under appropriate compactification/boundary conditions, the topological classification is governed by the relevant homotopy groups of the order-parameter manifold, expressed as
\begin{equation}
    \pi_k(E) = [S^k, E].
\end{equation}
Typical examples include
\begin{equation}
\begin{aligned}
    \pi_1(S^1) &= \mathbb{Z} \quad (\text{vortices}), \\
    \pi_2(S^2) &= \mathbb{Z} \quad (\text{skyrmions}), \\
    \pi_3(S^2) &= \mathbb{Z} \quad (\text{Hopf textures}).
\end{aligned}
\end{equation}
Here, the last case corresponds to Hopf textures in three-dimensional compactified real space, where the integer invariant is associated with the linking structure of preimages under the map $S^3 \to S^2$.

Equipped with these mapping invariants, one can analyze how the real-space manifold $M$ supports different textures. For the infinite plane $M=\mathbb{R}^2$, the one-point compactification yields $S^2$, making the mapping equivalent to $S^2 \to S^2$. This provides the familiar $\pi_2(S^2)=\mathbb{Z}$ classification for skyrmions, leading to quantized integer topological charges in the continuum limit. For the disk $M=D^2$, analogous local skyrmionic textures may exist, although their stability and quantization depend sensitively on the imposed boundary conditions. Crucially, for a torus \(M=T^2\) with periodic boundary conditions, smooth maps $T^2 \to S^2$ admit an integer-valued topological invariant given by the mapping degree, which captures the global wrapping of the torus onto the target sphere, reflecting a global topological invariant defined for smooth maps between orientable closed two-dimensional manifolds. The vanishing Euler characteristic of the torus $\chi(T^2)=0$ merely implies that it admits globally non-vanishing tangent vector fields, a geometric property distinct from the topological charge associated with internal skyrmion textures. These invariants are properties of the mapping $M \to E$, and should be distinguished from the intrinsic topology of the base manifold $M$ itself. Thus, homotopy theory provides a unified mathematical framework for classifying vector-field textures \(\mathbf{v}: M \to E\) and their associated topological invariants.

These manifold-dependent constraints set the stage for defining concrete topological invariants in acoustic systems. Although the acoustic velocity field serves as the primary physical observable, evaluating a quantized topological charge requires mapping it to a normalized direction field. Specifically, through appropriate embedding/normalization of multi-component acoustic fields, one constructs a normalized pseudospin (or orientation) vector field $\mathbf{n}(\mathbf{r}) \in S^2$ within an effective three-component representation of the acoustic field from the raw velocity or displacement fields, where zeros of the field correspond to singularities of the normalized texture and hence to possible topological defects. The Skyrmion number, which quantifies the global twisting of this continuous unit vector field, is conventionally defined as
\begin{equation}
    Q = \frac{1}{4\pi} \int \mathbf{n} \cdot 
    \left( 
        \frac{\partial \mathbf{n}}{\partial x} \times \frac{\partial \mathbf{n}}{\partial y} 
    \right) 
    dx\,dy.
    \label{eq:Q}
\end{equation}
where $\mathbf{n}$ is a unit vector field. Under smoothness and appropriate boundary compactification conditions, the resulting topological charge is quantized to integer values.

Geometrically, $Q$ measures the number of times the field $\mathbf{n}(x,y)$ wraps the unit sphere $S^2$ as one traverses the real-space plane. The integer-valued topological charge $Q$ characterizes distinct classes of skyrmionic configurations. A configuration with $Q = +1$ corresponds to a standard single-winding skyrmion in which the vector field $\mathbf{n}(\mathbf{r})$ wraps once around the unit sphere. When $Q = -1$, the texture carries an opposite topological charge. Depending on the convention, this may arise from reversed vorticity and/or core polarity rather than a change in chirality alone. More generally, configurations with $Q = m$ ($m = 2, 3, \ldots$) represent higher-order, multi-winding skyrmions, where the field encircles the sphere $m$ times. These configurations belong to distinct topological sectors that cannot be continuously deformed into each other without leaving the configuration space of smooth normalized fields, reflecting the quantized and integer-valued nature of the second homotopy group $\pi_2(S^2) = \mathbb{Z}$.

Beyond integer-valued textures, this framework also accommodates fractionalized counterparts such as merons and bimerons under suitable boundary conditions. In particular, when the order-parameter field covers only half of the Bloch sphere due to open boundaries or incomplete compactification, a meron carries a fractional topological charge $Q=\pm\tfrac{1}{2}$. A bimeron may be regarded as a composite texture formed by two meron-like constituents. These composite configurations introduce additional internal degrees of freedom,such as their relative separation,providing a versatile platform for exploring topological excitations.

These theoretical constructs have been recently realized in engineered acoustic platforms, particularly through surface acoustic waves (SAWs). \textcite{PhysRevLett.127.144502} utilized a hexagonal acoustic metasurface to confine standing SAWs as illustrated in Fig.~\ref{fig:skyrmion}(a), generating a N\'eel-type skyrmion lattice that can be dynamically manipulated via source phases and amplitudes. Building on this, \textcite{SciAdv.9.eadf3652} observed phononic skyrmions as three-dimensional hybrid-spin configurations at solid interfaces, see Fig.~\ref{fig:skyrmion}(b), highlighting their broadband topological stability against structural defects. More recently, \textcite{ApplPhysLett.125.042204} realized acoustic meron--antimeron lattices on square metastructures supporting spoof SAWs [see Fig.~\ref{fig:skyrmion}(c)], which are similarly reconfigurable. Together, these experiments establish that engineering SAW interference allows robust imprinting and manipulation of topological spin textures in purely acoustic systems.

\begin{figure}[!htbp]
  \centering
  \includegraphics[width=\linewidth]{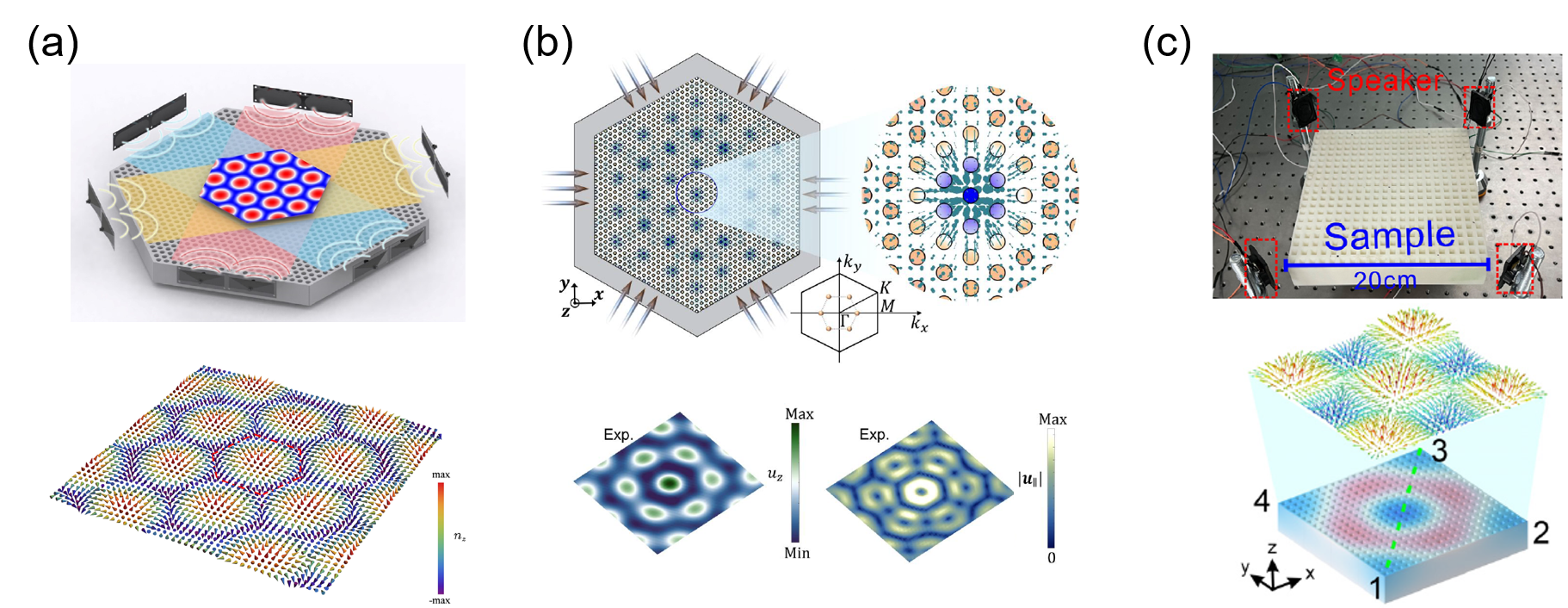}
  \caption{Experimental realization and characterization of acoustic and phononic skyrmion lattices. (a) An acoustic metasurface of skyrmion lattice, with the experimentally measured acoustic velocity field revealing a N\'eel-type skyrmion pattern. (b) A phononic skyrmion lattice in a hexagonal elastic metaplate, as experimentally measured through the axial and transverse displacement fields. (c) Schematic diagram of the acoustic meron experimental setup and the experimental measurement of sound pressure distribution of the meron lattice.
 (a) Adapted from~\citetext{\citealp{PhysRevLett.127.144502}}. (b) Adapted from~\citetext{\citealp{SciAdv.9.eadf3652}}. (c) Adapted from~\citetext{\citealp{ApplPhysLett.125.042204}}.}
  \label{fig:skyrmion}
\end{figure}

Beyond hosting purely acoustic topological states, SAWs offer a non-contact, low-power and programmable method for the dynamic manipulation of magnetic skyrmions and merons via magnetoelastic coupling. In thin films and multilayers, periodic strain from Rayleigh or leaky SAWs couples via magnetoelastic interactions to produce torques that nucleate and annihilate Skyrmion–antiskyrmion pairs, yielding stable N\'eel-type Skyrmions~\citetext{\citealp{NatNanotechnol.15.361}}. By matching SAW parameters to Skyrmion size, dense lattices can be formed, with number and position precisely tunable for synaptic or logic functionalities~\citetext{\citealp{Nanotechnology.33.115205}}. SAWs also drive Skyrmion and Skyrmionium motion without current~\citetext{\citealp{ApplPhysLett.121.242406}; \citealp{NatCommun.15.1018}}. Acoustic strain gradients and spin waves create forces allowing smooth drift along the SAW propagation, with velocity and trajectory tunable via amplitude, frequency, and damping~\citetext{\citealp{JPhysD.56.084002}; \citealp{ApplPhysLett.124.202407}}. Orthogonal standing SAWs can trap individual Skyrmions in quadrupole potentials and control sub-10-nm translations or curved paths~\citetext{\citealp{SciRep.13.1922}}. Moreover, standing SAWs in ferroelectric/ferromagnetic heterostructures can suppress Skyrmion Hall motion by counteracting the Magnus force, enabling rectilinear, multi-channel transport~\citetext{\citealp{NatCommun.14.4427}; \citealp{JApplPhys.133.203904}} [see Fig.~\ref{fig:control magnet skyrmion}(a,b)]. Overall, SAWs provide a unified platform for programmable, reconfigurable control of topological spin textures, paving the way for spin-wave logic, nonvolatile memory, and topological computing~\citetext{\citealp{AdvPhysRes.4.2400206}}, as illustrated in Fig.~\ref{fig:control magnet skyrmion}(c).

\begin{figure}[!htbp]
  \centering
  \includegraphics[width=\linewidth]{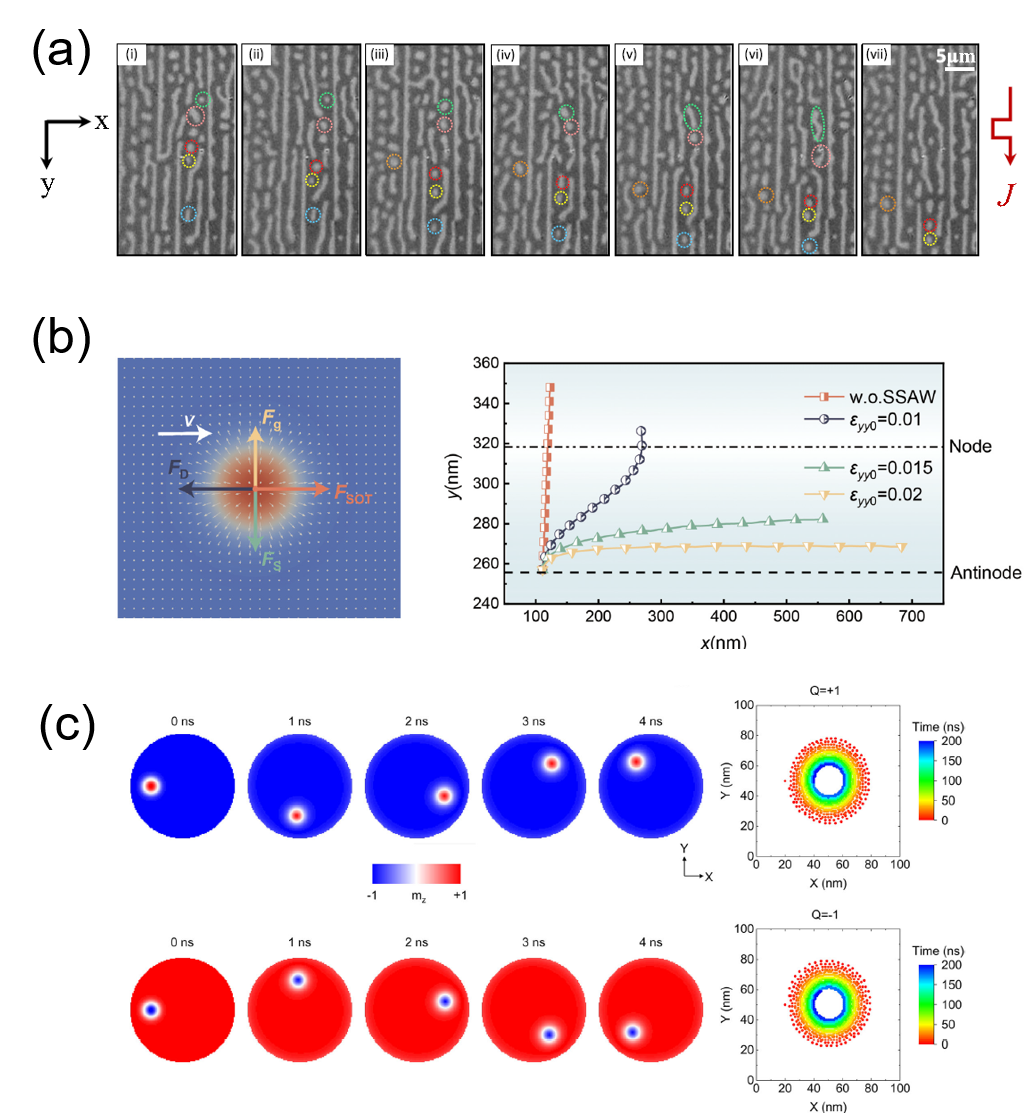}
  \caption{SAWs-driven motion and trajectory control of skyrmions. (a) Sequential images of skyrmion Hall effect with the excitation of SAWs. (b) Suppressed trajectories of skyrmion motion via different amplitudes of standing SAWs. (c) The motion trajectory of a magnetic skyrmion with a topological charge $Q = \pm 1$ driven by the shear horizontal wave.
 (a) Adapted from~\citetext{\citealp{NatCommun.14.4427}}. (b) Adapted from~\citetext{\citealp{JApplPhys.133.203904}}. (c) Adapted from~\citetext{\citealp{AdvPhysRes.4.2400206}}.}
  \label{fig:control magnet skyrmion}
\end{figure}

Recent work demonstrates that Skyrmion-like structures can intrinsically emerge as eigenmodes in tailored acoustic architectures, enabling mode-level topological engineering. Specifically, \textcite{ApplPhysLett.122.022201} showed that a single Archimedean spiral resonator naturally supports multiple localized Skyrmionic eigenmodes at distinct frequencies, formed by its velocity field distributions without requiring external interference. Building on this, \textcite{AdvSci.11.2401370} investigated coupled spiral metastructures, revealing that these eigenmodes hybridize into bonding and antibonding states whose guided propagation exhibits frequency-dependent digital-sequence patterns. In a complementary approach, \textcite{PhysRevB.111.174302} exploited higher-order multipolar eigenmodes to realize N\'eel-type Skyrmions and meron–antimeron lattices on the surface of subwavelength acoustic Mie resonators. Furthermore, \textcite{SciBull.69.1653} directly observed meronic configurations in an acoustic kagome lattice (a topological Euler insulator), linking these local textures to non-Abelian topological invariants.

In summary, topological spin textures---such as skyrmions, merons, and bimerons---provide a versatile framework for engineering complex velocity fields in acoustic systems. Rigorously defined by the continuous vector field $\mathbf{n}(\mathbf{r})$, these textures have been realized across diverse platforms, including Archimedean spirals, Mie resonators, kagome lattices, and programmable surface acoustic wave devices. Emerging either as intrinsic eigenmodes or through wave interference, these textures establish acoustic metastructures as a versatile platform for bridging fundamental theory with practical applications in topological mode engineering, information processing, and hybrid wave-matter interactions.

\subsection{\label{sec:topological mechanics}Topological Mechanical Metamaterial at Zero-Frequency Limit}

Building upon the remarkable progress in topological phononic metamaterials, where finite-frequency acoustic waves have revealed rich topological phases, it is instructive to recognize a complementary class of metamaterials that operate in the zero-frequency limit~\citetext{\citealp{NatPhys.10.39}}. These mechanical frames, known as topological isostatic structures, are defined by a balance between degrees of freedom and constraints. Positioned at the verge of mechanical instability, they exhibit floppy modes localized at system boundaries. These soft modes are zero‑energy mechanisms that govern mechanical failure and remain topologically robust against structural disorder within the bulk. The topological attributes of floppy modes are captured by the polarization vector, determined by winding numbers of the mechanical bands, which dictate the boundary localization of floppy modes. Consequently, the boundary hosting the floppy modes is mechanically soft, whereas the opposite boundary remains rigid.

Topological mechanical metamaterials have been realized in the one‑dimensional mechanical analog of the Su–Schrieffer–Heeger chain~\citetext{\citealp{NatPhys.10.39}; \citealp{ProcNatlAcadSciUSA.111.13004}}, the two‑dimensional generalized kagome lattice~\citetext{\citealp{NatCommun.8.14201}}, and the three‑dimensional deformed pyrochlore lattice~\citetext{\citealp{PhysRevLett.133.106101}}, all exhibiting protected boundary floppy modes. Notably, topological floppiness has also been uncovered in fully disordered isostatic networks, including fiber networks and quasicrystals~\citetext{\citealp{PhysRevLett.120.068003}; \citealp{PhysRevX.9.021054}}, demonstrating that crystalline symmetry is not a prerequisite. Furthermore, nonlinear isostatic structures enrich this field by hosting zero‑energy Guest–Hutchinson modes that drive topological phase transitions in mechanical attributes~\citetext{\citealp{PhysRevX.6.041029}; \citealp{PhysRevLett.131.046101}}. Finally, the robustness of topological mechanical softness proves more general than expected: even when systems are slightly driven away from the isostatic point, their topological characteristics persist~\citetext{\citealp{PhysRevLett.122.248002}; \citealp{ExtMechLett.57.101911}}, enabling phenomena such as topologically protected asymmetric wave transmission~\citetext{\citealp{PhysRevLett.121.094301}}.

Recent developments in topological mechanical metamaterials highlight their interplay with soft condensed matter physics, providing new insights through topology‑guided fracturing~\citetext{\citealp{NewJPhys.20.063034}; \citealp{PhysRevLett.135.148202}; \citealp{NatCommun.17.2420}}, selective buckling~\citetext{\citealp{PNAS.112.7639}}, topological origami~\citetext{\citealp{PhysRevLett.116.135501}}, and dislocation‑bounded topological rigidity~\citetext{\citealp{NatPhys.11.153}}. Importantly, because topological floppy modes reside at zero frequency, they are immune to damping losses, greatly enhancing the prospects of topological isostatic structures in applications such as non‑Abelian computing~\citetext{\citealp{Nature.577.636}}, mechanical isolation~\citetext{\citealp{Nature.542.461}}, and topological wheels~\citetext{\citealp{ExtMechLett.46.101344}}. 

 \subsection{\label{sec:topological_optomechanics}Topological Optomechanical Systems}

A natural extension of topological mechanical metamaterials arises when mechanical motion is coupled to optical cavities, thereby entering the realm of topological optomechanical systems. In contrast to purely mechanical realizations, where topology is chiefly encoded by geometry and elastic couplings, optomechanical platforms introduce an additional control layer in which optical and mechanical degrees of freedom are coupled and jointly driven~\citetext{\citealp{RevModPhys.86.1391}}. This extension is especially appealing at the nanoscale, where on-chip nanomechanical and nanoelectromechanical systems have already established high-frequency topological phonon transport~\citetext{\citealp{ProcNatlAcadSci.114.E3390}; \citealp{Nature.564.229}; \citealp{NatNanotechnol.16.576}}. Optomechanics enriches this setting further by providing active, in situ control together with sensitive optical readout of the same mechanical excitations.

The natural starting point is therefore not a purely mechanical lattice, but a hybrid photon--phonon system in which optical cavity modes \(a_i\) and mechanical modes \(b_i\) are coupled through radiation pressure. At the lattice level, this may be described schematically by

\begin{equation}
\begin{split}
   H
=\sum_i \omega_{c,i} a_i^\dagger a_i + \sum_i \Omega_i b_i^\dagger b_i
-\sum_i g_i a_i^\dagger a_i (b_i+b_i^\dagger) \\
+H_{\mathrm{hop}}
+H_{\mathrm{drive}}, 
\end{split}
\end{equation}

where \(H_{\mathrm{hop}}\) denotes intersite optical and mechanical hopping and \(H_{\mathrm{drive}}\) the external optical driving. In the driven and linearized regime, the radiation-pressure interaction gives rise to enhanced photon-phonon couplings \(G_i\), so that topology may be encoded either in hybridized photon-phonon polaritons or, after elimination of the optical sector in an appropriate parameter regime, in an optically programmed mechanical model. In the latter case one obtains an effective Hamiltonian of the form
\begin{equation}
H_{\mathrm{eff}}^{(m)}
=
\sum_i \tilde{\omega}_i\, b_i^\dagger b_i
+
\sum_{ij}
\left(
J_{ij}^{\mathrm{eff}} e^{i\phi_{ij}} b_i^\dagger b_j
+\mathrm{h.c.}
\right),
\end{equation}
where the renormalized on-site frequencies \(\tilde{\omega}_i\), coupling amplitudes \(J_{ij}^{\mathrm{eff}}\), and hopping phases \(\phi_{ij}\) are controlled by optical detuning, drive strength, and multimode interference. In this way, cavity optomechanics provides direct access to the same ingredients that underpin programmable topological phononics more broadly: tunable on-site terms, complex couplings, synthetic gauge fields, and nonreciprocal transport. This perspective was articulated in the seminal proposal of topological phases of sound and light in optomechanical arrays, where radiation-pressure coupling was shown to support chiral phonon transport together with topological photonic and hybridized excitations~\citetext{\citealp{PhysRevX.5.031011}}.

\begin{figure}[!htbp]
  \centering
  \includegraphics[width=\linewidth]{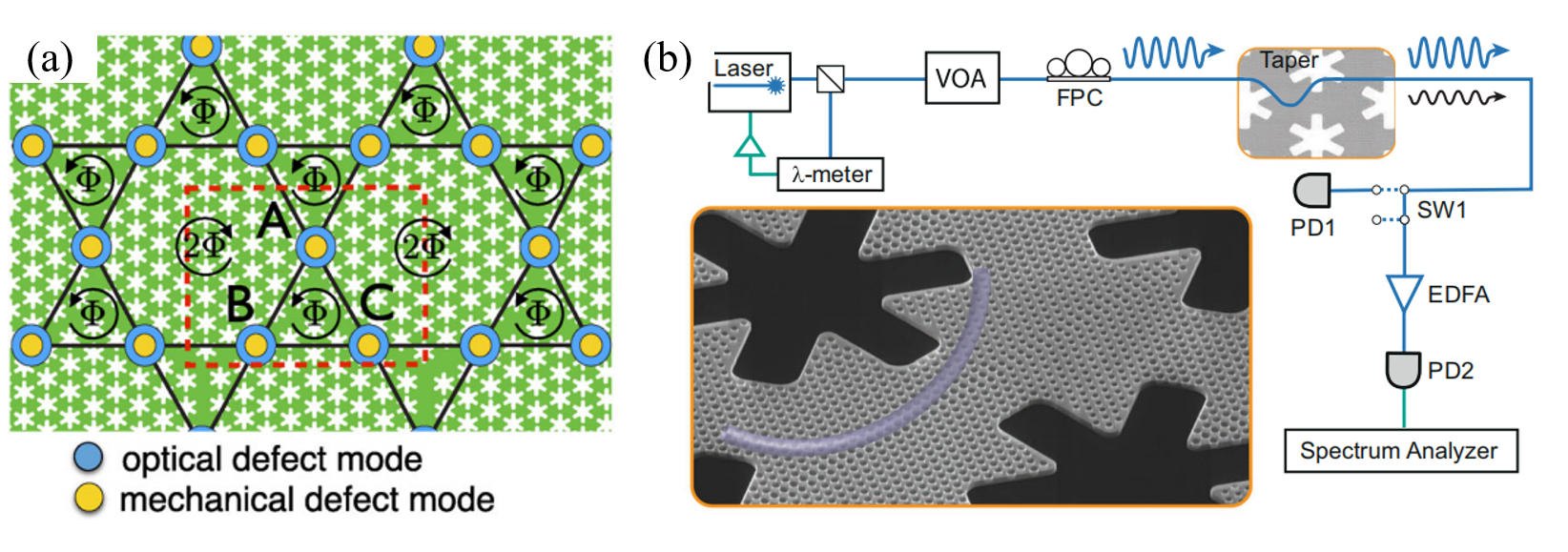}
  \caption{Sketch and Experimental  characterization of optomechanical arrays. (a) A kagome optomechanical lattice supporting optical and vibrational modes, The effective magnetic fluxes (indicated) add up to zero, realizing a Chern insulator. (b) A schematic of  optomechanical readout setup used to measure the phononic properties of the topological cavity structure.
 (a) Adapted from~\citetext{\citealp{PhysRevX.5.031011}}. (b) Adapted from~\citetext{\citealp{NatCommun.13.3476}}.}
  \label{fig:topological_optomechanics}
\end{figure}

A distinctive feature of optomechanical systems is the dynamic programmability of these effective parameters. Experimentally, this was illustrated by the realization of synthetic gauge fields for phonon transport in a nano-optomechanical system, where multimode optomechanical interactions produced a controllable nonreciprocal phase for phonon transfer between nanomechanical modes, thereby emulating an Aharonov--Bohm-type response for vibrations~\citetext{\citealp{NatNanotechnol.15.198}}. On the theoretical side, related studies of optomechanical arrays further showed that the interplay of optomechanical interaction and photonic spin-orbit coupling can induce robust and tunable nonreciprocal topological phononics in lattice settings~\citetext{\citealp{PhysRevB.101.085108}}. Taken together, these developments show that light is not merely a probe of topological phonon transport, but also a means of imprinting directionality and gauge structure directly onto the mechanical dynamics.

Recent experiments have brought topological optomechanics into the regime of large-scale, on-chip phononic transport. In particular, topological phonon propagation has been observed in a multiscale optomechanical crystal~\citetext{\citealp{NatCommun.13.3476}}, shown in Fig.~\ref{fig:topological_optomechanics}(b). More generally, such progress places topological phonon transport in a nanostructured optomechanical architecture where preparation, routing, and detection can be integrated within the same optical platform. From the perspective of artificial topological phononics, it illustrates how mechanically encoded topology can be extended by driven light--matter interactions that provide direct access to the generation, manipulation, and readout of topological mechanical excitations.

Looking ahead, this platform also points naturally toward quantum interfaces. Because cavity optomechanics provides coherent conversion channels between mechanical motion and optical fields, and can be further combined with localized quantum emitters or spin qubits, topological acoustic edge channels in optomechanical arrays have been proposed as robust carriers for long-distance quantum state transfer~\citetext{\citealp{NewJPhys.21.113030}}. Topological optomechanical systems thus delineate a broader frontier in which topology, driven control, nonreciprocity, and optical accessibility are combined within a single platform, with prospects ranging from nonreciprocal phononic circuitry and hybrid photon--phonon state engineering to quantum-coherent control of topological mechanical excitations. A conceptually complementary development is the demonstration of topological energy transfer in a cryogenic optomechanical device, where adiabatic cycles around an exceptional point produce nonreciprocal transfer between two vibrational modes~\citetext{\citealp{Nature.537.80}}. Although the topology involved there belongs to the parameter-space structure of a non-Hermitian spectrum, rather than to band topology in real or synthetic lattices, it provides a natural bridge to the non-Hermitian topological phonons discussed in the following chapter.

%
% ****** End of file apssamp.tex ******

% ****** Start of file apssamp.tex ******
%
%   This file is part of the APS files in the REVTeX 4.2 distribution.
%   Version 4.2a of REVTeX, December 2014
%
%   Copyright (c) 2014 The American Physical Society.
%
%   See the REVTeX 4 README file for restrictions and more information.
%
% TeX'ing this file requires that you have AMS-LaTeX 2.0 installed
% as well as the rest of the prerequisites for REVTeX 4.2
%
% See the REVTeX 4 README file
% It also requires running BibTeX. The commands are as follows:
%
%  1)  latex apssamp.tex
%  2)  bibtex apssamp
%  3)  latex apssamp.tex
%  4)  latex apssamp.tex
%
%\documentclass[%
%reprint,
%superscriptaddress,
%groupedaddress,
%unsortedaddress,
%runinaddress,
%frontmatterverbose, 
%preprint,
%preprintnumbers,
%nofootinbib,
%nobibnotes,
%bibnotes,
%amsmath,amssymb,
%aps,
%pra,
%prb,
%rmp,
%prstab,
%prstper,
%floatfix,
%]{revtex4-2}

%\usepackage{graphicx}% Include figure files
%\usepackage{dcolumn}% Align table columns on decimal point
%\usepackage{bm}% bold math
%\usepackage{hyperref}% add hypertext capabilities
%\usepackage[mathlines]{lineno}% Enable numbering of text and display math
%\linenumbers\relax % Commence numbering lines

%\usepackage[showframe,%Uncomment any one of the following lines to test 
%%scale=0.7, marginratio={1:1, 2:3}, ignoreall,% default settings
%%text={7in,10in},centering,
%%margin=1.5in,
%%total={6.5in,8.75in}, top=1.2in, left=0.9in, includefoot,
%%height=10in,a5paper,hmargin={3cm,0.8in},
%]{geometry}

%\preprint{APS/123-QED}
%\section{\label{sec:level1}Non-Hermitian topological phonon}
\section{\label{sec:level1}Non-Hermitian topological phonons}
%\thanks{A footnote to the article title}%

%\author{Ann Author}
%\altaffiliation[Also at ]{Physics Department, XYZ University.}%Lines break automatically or can be forced with \\
%\author{Second Author}%
%\email{Second.Author@institution.edu}
%\affiliation{%
%Authors' institution and/or address\\
%This line break forced with \textbackslash\textbackslash
%}%

%\collaboration{MUSO Collaboration}%\noaffiliation

%\author{Charlie Author}
%\homepage{http://www.Second.institution.edu/~Charlie.Author}
%\affiliation{
%Second institution and/or address\\
%This line break forced% with \\
%}%
%\affiliation{
%Third institution, the second for Charlie Author
%}%
%\author{Delta Author}
%\affiliation{%
%Authors' institution and/or address\\
%This line break forced with \textbackslash\textbackslash
%}%

%\collaboration{CLEO Collaboration}%\noaffiliation

%\date{\today}% It is always \today, today,
             %  but any date may be explicitly specified

%\begin{abstract}
%An article usually includes an abstract, a concise summary of the work
%covered at length in the main body of the article. 
%\begin{description}
%\item[Usage]
%Secondary publications and information retrieval purposes.
%\item[Structure]
%You may use the \texttt{description} environment to structure your abstract;
%use the optional argument of the \verb+\item+ command to give the category of each item. 
%\end{description}
%\end{abstract}

%\keywords{Suggested keywords}%Use showkeys class option if keyword
                              %display desired
%\maketitle

%\tableofcontents

%\section{\label{sec:level2}Introduction}
\subsection{\label{sec:level2}Introduction}

   Non-Hermitian systems, characterized by energy exchange with the environment, constitute a paradigm that extends conventional Hermitian physics and affords enhanced control over wave dynamics and topology. Their unique features, such as complex eigenfrequencies and skewed eigenvectors~\citetext{\citealp{PhysRevLett.80.5243}; \citealp{PhysRevA.68.062111}; \citealp{JPhysAMathTheor.47.035305}; \citealp{RevModPhys.96.045002}}, have led to several breakthroughs that have reshaped our understanding of topological physics. In experimental contexts, the realization of non-Hermitian systems often requires microscopic control over local dissipation, gain, and non-reciprocal interactions among elementary building blocks. Phononic platforms are particularly well-suited for this task due to the relative ease of engineering non-Hermitian parameters~\citetext{\citealp{NatPhys.14.11}; \citealp{NatPhys.17.9}; \citealp{NatRevPhys.6.11}}. This section offers a bird's-eye view of the core concepts of non-Hermitian topology and surveys key achievements in phononic systems. Readers seeking a more in-depth and comprehensive understanding of the topic should consider the following dedicated reviews:~\citetext{\citealp{NatPhys.14.11}; \citealp{Science.363.6422}; \citealp{NatMater.18.783}; \citealp{NatNanotechnol.18.706}; \citealp{NatRevPhys.6.11}; \citealp{AdvMater.37.2307998}} summarize the developments of non-Hermitian physics in different experimental systems.~\citetext{\citealp{AdvPhys.69.249}; \citealp{RevModPhys.93.015005}} are comprehensive reviews that lean more toward theoretical aspects of non-Hermitian systems.~\citetext{\citealp{NatRevPhys.4.745}} primarily focuses on topology of exceptional points. The non-Hermitian skin effect is discussed in~\citetext{\citealp{AnnuRevCondensMatterPhys.14.83}; \citealp{FrontPhys.18.53605}}.
   
\subsection{\label{sec:level2}Non-Hermitian topology: spectrum and wavefunctions}

    The breaking of Hermiticity enriches topological physics by introducing two pivotal features: the emergence of complex energy spectra and the non-orthogonality of eigenvectors. These features, in turn, give rise to two distinct but related topological frameworks that have no Hermitian counterparts. The first is a spectral topology originating from the structure of the complex eigenvalues, while the second is a geometric topology encoded in the evolution of the non-orthogonal eigenvectors.
   
   In Hermitian systems, where the eigenvalues lie on the real axis, the eigenvalue manifold is homeomorphic to the parameter space. Consequently, the topology arises solely from the evolution of eigenvectors. In non-Hermitian systems, however, the eigenvalue manifold differs topologically from the parameter space manifold and can exhibit a richer spectral topological landscape due to complex eigenvalues. A closed loop in the parameter space, such as the Brillouin zone, typically maps to an area-enclosing loop in the complex energy plane. This distinction is quantified by the spectral winding number ($\mathcal{W}_E$)~\citetext{\citealp{PhysRevX.9.041015}; \citealp{PhysRevLett.123.066405}; \citealp{PhysRevLett.126.086401}}:
    \begin{equation}
    \mathcal{W}_E = \frac{1}{2\pi i} \oint_{C_\lambda} d\lambda \cdot \nabla_{\lambda} \ln \det \left( H(\lambda) - E_r \mathbb{I} \right)
    \end{equation}
   \noindent where $C_\lambda$ is a closed path in the parameter space, $E_r$ is an arbitrary reference energy, and $\mathbb{I}$ is the identity matrix. A non-zero winding number indicates that the path $C_\lambda$ encloses a spectral singularity, a feature with no Hermitian analog. This spectral topology redefines the concept of a bandgap, generalizing it into two distinct types: point gaps and line gaps~\citetext{\citealp{PhysRevLett.120.146402}}. A system possesses a point gap around a reference energy $ E_r \in \mathbb{C}$ if $E_r$ is not an eigenvalue for any wavevector $k$ in the Brillouin zone, i.e., $\det\left[ H(k) - E_r\mathbb{I} \right] \neq 0$ for all $k$, which signifies intrinsic non-Hermitian topology and phenomena like the non-Hermitian skin effect (NHSE). In contrast, a line gap exists if the spectrum is partitioned by a continuous line in the complex plane with spectral bands residing on either side of this line, which is more analogous to a conventional bandgap~\citetext{\citealp{PhysRevX.8.031079}}. These will be discussed in a later section.

    \begin{figure}[!htbp]
    \centering
    \includegraphics[width=\linewidth]{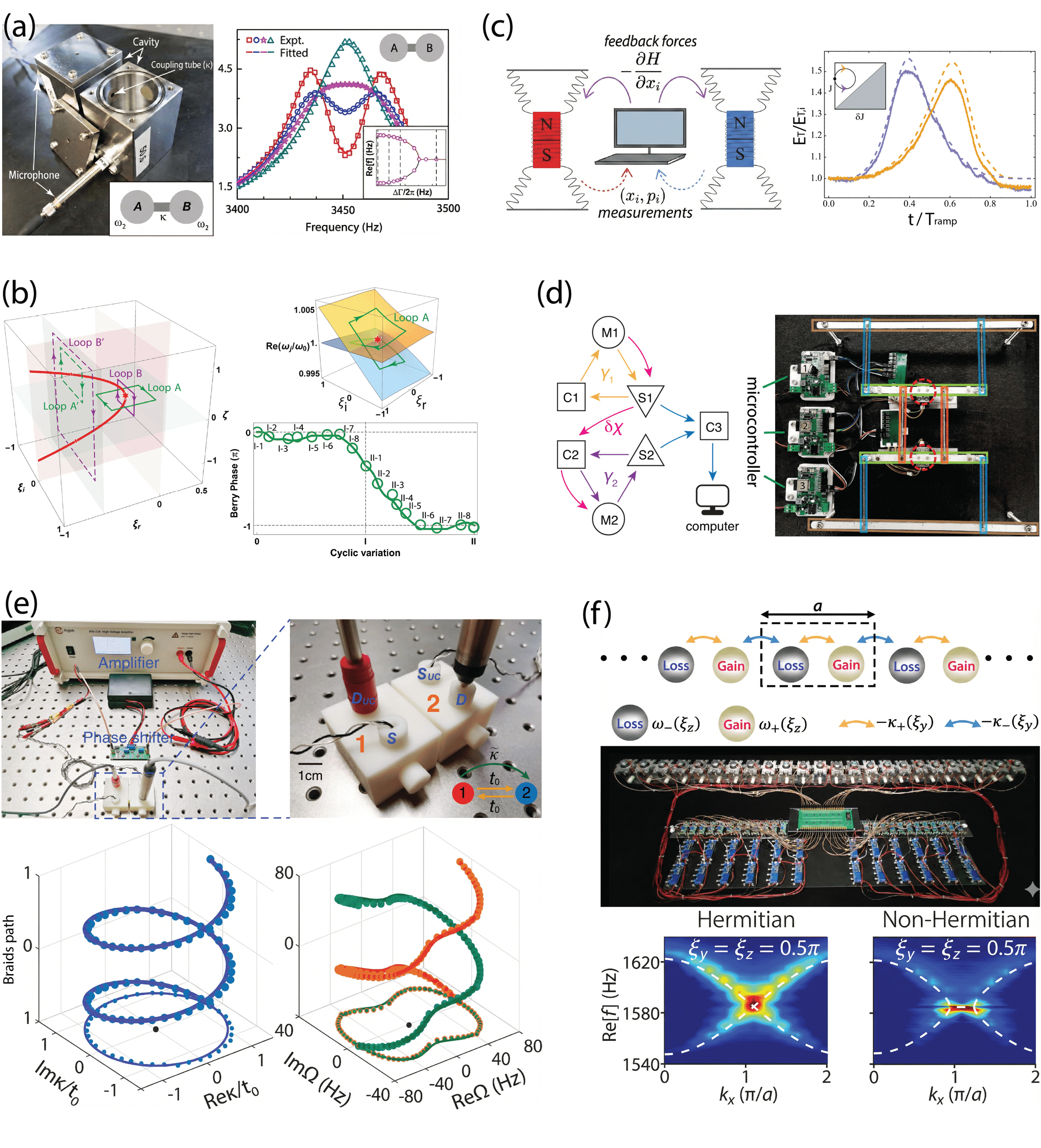}
    \caption{Experimental realizations of EP and its topological characteristic. (a) Second-order EP in coupled acoustic cavities; peak coalescence is achieved via dissipation tuning. (b) Topology of an exceptional parabola: a state encircling an EP (red star) requires two cycles to return, accumulating a $\pi$ Berry phase across a branch cut. (c) Probing imaginary part of Berry phases in a feedback-controlled dimer, corresponding to the total energy amplification/damping during the non-Hermitian dynamics evolution. (d) Active mechanical platform verifying exceptional chains stabilized by non-Hermitian latent symmetries. (e) Non-Hermitian Bloch braids in a binary cavity system, synthesized via momentum-resolved unidirectional feedback. (f) Weyl exceptional rings and synthetic Fermi arcs observed in a 1D non-Hermitian phononic crystal. (a) Adapted from~\citetext{\citealp{PhysRevX.6.021007}}. (b) Adapted from~\citetext{\citealp{PhysRevLett.127.034301}}. (c) Adapted from~\citetext{\citealp{PhysRevResearch.5.L032026}}. (d) Adapted from~\citetext{\citealp{PhysRevLett.131.237201}}. (e) Adapted from~\citetext{\citealp{PhysRevLett.130.017201}}. (f) Adapted from~\citetext{\citealp{PhysRevLett.129.084301}}.} 
    \label{fig:EP-Topology}\end{figure}

    While the spectral topology concerns eigenvalues, the eigenvector topology in non-Hermitian systems requires a fundamental rethinking of the basis states. Unlike Hermitian operators where left and right eigenvectors are related by the conjugate transpose, non-Hermitian Hamiltonians necessitate a biorthogonal formalism. The right ($|\psi_n^R\rangle$) and left ($|\psi_n^L\rangle$) eigenvectors are defined as distinct entities satisfying $H |\psi_n^R\rangle = E_n |\psi_n^R\rangle$ and $H^\dagger |\psi_n^L\rangle = E_n^* |\psi_n^L\rangle$. The complete basis satisfies the biorthogonality condition $\langle \psi_n^L | \psi_m^R \rangle = \delta_{nm}$, rather than the conventional orthonormality of Hermitian systems. The relationship between left and right eigenvectors was recently experimentally characterized in acoustic systems~\citetext{\citealp{FrontPhys.20.054202}}. To quantify the deviations from Hermiticity, one introduces the phase rigidity $r_n$, defined as the overlap of the right and left eigenvectors: 
    \begin{equation}
    r_n = \frac{\langle \psi_n^L | \psi_n^R \rangle}{\sqrt{\langle \psi_n^L | \psi_n^L \rangle \langle \psi_n^R | \psi_n^R \rangle}}
    \end{equation}
    In the Hermitian limit, the left and right eigenvectors are identical, yielding $r_n = 1$. However, in non-Hermitian systems, the eigenvectors become skewed (non-orthogonal), causing $r_n$ to drop below unity. This skewness reaches its extreme limit at singular points---typically exceptional points, where eigenvectors coalesce and $r_n$ vanishes.

   Correspondingly, eigenvectors form fiber bundles over eigenvalue manifolds, and the evolution of skewed eigenvectors in non-Hermitian Hamiltonians encodes essential eigenvector topology. A fundamental manifestation of this eigenvector topology is the geometric (Berry) phase, a gauge-invariant phase factor acquired by a state (or a unitary matrix for multiple states) during an adiabatic cyclic evolution in parameter space. The non-orthogonality of eigenvectors in non-Hermitian systems leads to a Berry phase that manifests a richer set of physical phenomena. The non-Hermitian Berry connection is defined as 
    \begin{equation}
    A_n^{\alpha\beta}(\lambda) = \langle \psi_n^\alpha | \nabla_{\lambda} \psi_n^\beta \rangle
    \end{equation}
       with $\alpha, \beta \in \{L, R\}$. For a closed curve $C$, the corresponding non-Hermitian Berry phase
    \begin{equation}
    \gamma_n^{\alpha\beta} = \oint_{C} A_n^{\alpha\beta} d\lambda
    \end{equation}
   reveals the topological properties associated with the singularities enclosed. In contrast to its Hermitian counterpart, which is strictly real, the non-Hermitian Berry phase is generally complex-valued. The real part corresponds to the conventional geometric phase, while the imaginary part quantifies the cumulative non-unitary amplification or attenuation along the cycle~\citetext{\citealp{PhysRevA.87.012118}}. This complex-valued geometric phase is not merely a theoretical construct; its physical significance has been confirmed through direct measurements in phononic platforms, such as coupled mechanical oscillators [Fig. \ref{fig:EP-Topology}(c)]~\citetext{\citealp{PhysRevResearch.5.L032026}}.

\subsection{\label{sec:level2}Exceptional Points and their topology}

    The parametric evolution of eigenvalues on the complex energy plane generates a rich set of phenomena intimately tied to spectral topology. Perhaps the most striking feature is the emergence of non-Hermitian degeneracies known as EPs~\citetext{\citealp{JPhysAMathGen.37.2455}; \citealp{CzechJPhys.54.1039}}. Mathematically, EPs are branch-point singularities where two or more eigenvalue manifolds coalesce. Unlike Hermitian degeneracies (diabolic points) where eigenvectors remain orthogonal, at an EP, the Hamiltonian becomes defective and can only be brought into a Jordan canonical form. This implies that the algebraic multiplicity of the eigenvalue exceeds its geometric multiplicity, rendering the Hamiltonian non-diagonalizable~\citetext{\citealp{Meyer2008matrix}}. Critically, at this singularity, the right and left eigenvectors themselves coalesce, satisfying the self-orthogonality condition $\langle \psi_{\mathrm{EP}}^L | \psi_{\mathrm{EP}}^R \rangle = 0$. This collapse manifests as a singularity in the non-Hermitian Berry connection and lies at the heart of non-Hermitian topology. 
   
   The topology of EPs is fundamentally encoded in the global structure of the complex energy spectrum. Typically, a closed loop encircling an EP in parameter space smoothly connects eigenvalues from distinct Riemann sheets, resulting in the permutation of states. This process underpins the non-trivial spectral topology, which can be characterized by the discriminant number or eigenvalue vorticity~\citetext{\citealp{PhysRevLett.118.040401}; \citealp{PhysRevLett.123.066405}; \citealp{PhysRevLett.126.086401}}. More formally, it is a fundamental manifestation of the fact that the first homotopy group of the space of non-Hermitian Hamiltonians is the braid group~\citetext{\citealp{PhysRevB.101.205417}; \citealp{PhysRevB.103.155129}; \citealp{RepProgPhys.87.078002}}. Consequently, through the suitable design of the parameter space and encircling paths, diverse braid and knot structures can be realized in the complex energy spectrum, as illustrated in Fig. \ref{fig:EP-Topology}(e)~\citetext{\citealp{PhysRevLett.127.034301}; \citealp{PhysRevLett.130.017201}; \citealp{PhysRevResearch.5.023038}; \citealp{PhysRevLett.134.126603}}. Phononic crystals have provided an excellent platform to explore EPs and their associated topology. For instance, in acoustic systems, coupled networks of lossy cavities---featuring independently tunable on-site resonances, inter-cavity couplings, and dissipation---offer a versatile means of realizing high-order or anisotropic EPs and probing their unique spectral properties [see Fig. \ref{fig:EP-Topology}(a, b)]~\citetext{\citealp{PhysRevX.6.021007}; \citealp{PhysRevLett.121.085702}}.

   Mathematically, the vanishing of the system's characteristic polynomial discriminant provides a robust framework for locating the emergence of EPs. A second-order EP imposes two constraints (the real and imaginary parts of a Hamiltonian's discriminant are equal to zero), defining a codimension-2 manifold. Consequently, EPs typically trace 1D curves, such as exceptional rings (closed loops) or arcs (open lines) in a 3D parameter space. The imposition of physical symmetries can reduce this codimension, enabling EPs to stably exist as 2D structures (e.g., surfaces). This geometric principle enables the engineering of complex EP structures and underpins novel topological phases, such as non-Hermitian Weyl semimetals that host exotic features like exceptional Fermi rings and hinge arcs~\citetext{\citealp{PhysRevB.91.125438}; \citealp{PhysRevB.97.075128}; \citealp{PhysRevB.99.081102}; \citealp{PhysRevLett.127.196801}}. For example, targeted gain-loss patterns can convert Hermitian nodal lines into intrinsically non-Hermitian degeneracies like exceptional lines and exceptional rings, further expanding the landscape of non-Hermitian topological phononics, as depicted in Fig. \ref{fig:EP-Topology}(f)~\citetext{\citealp{PhysRevLett.129.084301}; \citealp{PhysRevResearch.5.023020}; \citealp{PhysRevLett.134.116606}}. Alternatively, generalizing concepts like latent symmetries to non-Hermitian scenarios has led to the realization of stable exceptional chains in mechanical oscillators [Fig. \ref{fig:EP-Topology}(d)]~\citetext{\citealp{PhysRevLett.131.237201}}.
    
   Beyond two-band systems, the landscape of non-Hermitian topology becomes even more complex and diverse, as the coalescence of multiple bands can give rise to higher-order EPs with hybrid topological characteristics. For example, the exceptional nexus (both a higher-order exceptional point and the cusp where multiple exceptional arcs meet) has been experimentally identified through measurements of critical wavefunctions, and its topology is captured by a hybrid invariant comprising Berry phases accumulated along loops in the parameter space of the non-Hermitian system~\citetext{\citealp{Science.370.6520}}. Other advanced characterization schemes have also been demonstrated, such as experimentally tracking a parabolic EP trajectory while simultaneously measuring the discriminant number (from eigenvalues) and the Berry phase (from eigenstates) to fully define the system's topology~\citetext{\citealp{PhysRevLett.127.034301}}. In 3D periodic synthetic momentum space, a resultant-based winding invariant was introduced that selectively detects order-3 exceptional lines while remaining insensitive to order-2 exceptional surfaces, thereby diagnosing the associated topological currents and predicting their evolution under perturbations~\citetext{\citealp{NatCommun.14.6660}}. Such a complex topological landscape inherently supports a rich braid group topology governing the global connectivity of the eigenstates. The permutation of eigenstates across different Riemann sheets, induced by branch cuts and EPs, realizes non-Abelian state exchanges, which have been demonstrated by the sequential encircling of distinct exceptional arcs in three-state phononic systems~\citetext{\citealp{NatlSciRev.9.nwac010}}. Collectively, these theoretical and experimental advances provide a unified operational framework for detecting, characterizing, and harnessing the rich topology of EPs in phononic systems.

\subsection{\label{sec:level2}Band topology in non-Hermitian phononic systems}

    Non-Hermitian physics also introduces more profound and diverse modifications to band topology. We first consider line-gapped systems, typically characterized by spatially modulated gain and loss. The topology associated with line gaps does not essentially differ from that of their Hermitian counterparts, because any line-gapped spectrum can be smoothly deformed back to the real axis. In this case, non-Hermiticity modifies the transport characteristics of the bulk and boundary modes via amplification or attenuation~\citetext{\citealp{PhysRevLett.118.040401}; \citealp{PhysRevLett.120.146402}}. For instance, in artificial acoustic boron nitride crystals supporting valley states, sublattice-selective gain and loss preserve the line-gap topology and valley character intact while rendering acoustic valley states either amplified or damped~\citetext{\citealp{PhysRevLett.120.246601}}. By tuning a critical loss that controls the splitting of topological boundary modes, a topologically protected EP has been realized, enabling unidirectional reflectionless propagation and robustness against geometric disorder~\citetext{\citealp{PhysRevLett.121.124501}; \citealp{PhysRevApplied.15.014025}; \citealp{PhysRevApplied.22.014046}}. In higher-order topological phononic crystals, the spatial patterning of non-Hermiticity allows the corner and hinge states to be selectively tuned into amplifying or decaying modes, as shown in Fig. \ref{fig:NH-Band}(b)~\citetext{\citealp{PhysRevLett.122.195501}}. An elegant experimental platform for realizing these phenomena consists of phononic crystals composed of thermoplastic rods coated with carbon nanotube films, which act as tunable gain media via the electro-thermoacoustic effect~\citetext{\citealp{AdvMater.36.2406567}}. By further engineering non-Hermitian textures on these active elements, one can break the chiral symmetry of whispering-gallery modes, facilitating the directional out-coupling and the generation of topological single modes with selectable handedness~\citetext{\citealp{Nature.597.655}} [Fig. \ref{fig:NH-Band}(c)].

    \begin{figure}[!htbp]
    \centering
    \includegraphics[width=\linewidth]{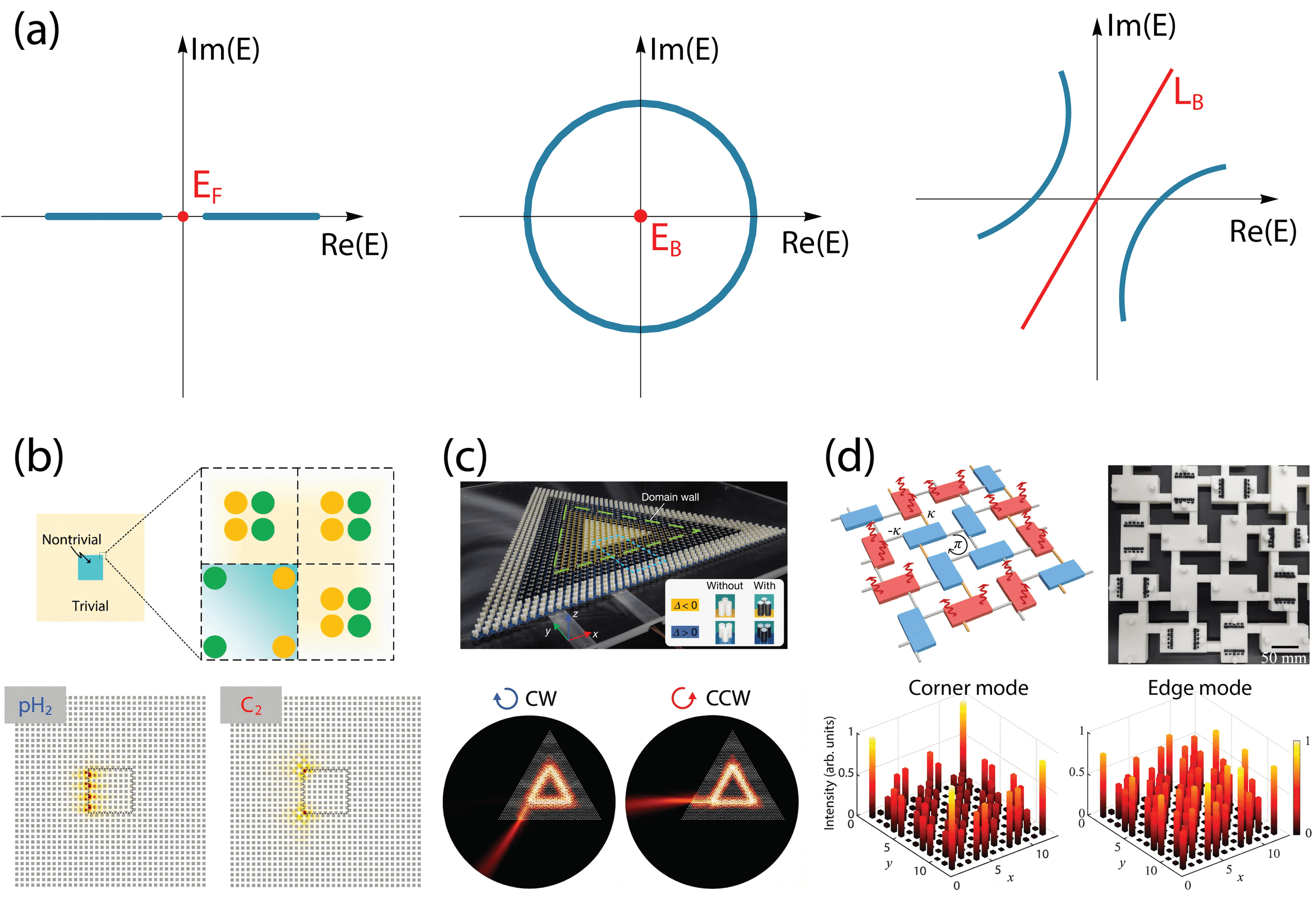}
    \caption{
    (a) Hermitian and non-Hermitian bandgaps: a bandgap in Hermitian systems (left); a non-Hermitian point gap characterized by a non-trivial spectral winding around a reference point (middle); a non-Hermitian line gap (right). (b) A second-order topological sonic crystal utilized $\mathcal{PT}$-symmetric metamolecules to support topological corner states.(c) Active topological whispering-gallery insulator. Carbon-nanotube-induced electro-thermoacoustic coupling (insets) breaks chiral symmetry, enabling directional mode out-coupling along the triangular domain wall. (d) Non-Hermiticity routes the acoustic quadrupole topological insulator via sealing resonators with absorbing materials. (b) Adapted from~\citetext{\citealp{PhysRevLett.122.195501}}. (c) Adapted from~\citetext{\citealp{Nature.597.655}}. (d) Adapted from~\citetext{\citealp{NatCommun.12.1888}}.
    }\label{fig:NH-Band}\end{figure}
   
   Beyond modifying existing states, non-Hermiticity can serve as a key mechanism for inducing topological phase transitions~\citetext{\citealp{PhysRevLett.115.040402}}. By controlling on-site gain and loss, a topologically trivial phononic crystal can be driven across a phase transition into a non-Hermitian higher-order topological phase, marked by the emergence of topological corner states~\citetext{\citealp{PhysRevLett.123.073601}}. The resulting bulk topology is characterized by biorthogonal nested Wilson loops or edge polarizations. Such topological boundary states, emerging solely from non-Hermiticity, have been experimentally realized in both 1D~\citetext{\citealp{PhysRevB.101.180303}} and 2D~\citetext{\citealp{NatCommun.12.1888}} acoustic crystals, as depicted in Fig. \ref{fig:NH-Band}(d). This approach offers remarkable reconfigurability, enabling topological modes to be engineered at arbitrary positions within the lattice~\citetext{\citealp{PhysRevB.107.L201108}}.

\subsection{\label{sec:level2}Point gap topology and non-Hermitian skin effect}

    More fascinating topological phenomena emerge when the Bloch bands exhibit point gaps. A point gap implies that the bands, as complex functions of Bloch wavevector k, enclose a nonvanishing area in the complex energy plane~\citetext{\citealp{PhysRevX.8.031079}; \citealp{PhysRevLett.123.066405}}. Physically, such non-trivial spectral winding originates from the spatial asymmetry in hopping amplitudes, clearly breaking the reciprocity of the system. This is prototypically illustrated by the Hatano-Nelson model, where the hopping rate in the forward direction ($t_R$) differs from that in the backward direction ($t_L$). Under periodic boundary conditions (PBC), this non-reciprocity forces the eigenvalues to trace a loop $E(k) = t_R e^{-ik} + t_L e^{ik}$ in the complex plane, creating a point gap that encloses an interior region.
    Unlike line gaps, point-gapped spectra cannot be deformed onto the real axis via rotation and stretching. Consequently, point-gapped bands give rise to intriguing and unique non-Hermitian phenomena. The most prominent of these is the NHSE~\citetext{\citealp{PhysRevLett.116.133903}; \citealp{PhysRevB.97.121401}; \citealp{PhysRevA.99.062112}}, wherein the bulk modes from this band are skin modes localized at one end of the lattice under open boundary conditions (OBCs). This failure of the conventional bulk-boundary correspondence (BBC) arises because the Bloch eigenmodes for wavevectors $k$ and $-k$ are no longer degenerate in energy, preventing the formation of standing waves in a finite system [see Fig. \ref{fig:NHSE}(a)]. 
   
   To understand the localization mechanism, one can employ an imaginary gauge transformation $k \to k - i\kappa$ (or equivalently, a similar transformation of the basis $\psi_n \to e^{-\kappa n} \psi_n$). For 1D single-band models, a specific choice of $\kappa = \ln\sqrt{|t_R/t_L|}$ can map the non-reciprocal non-Hermitian Hamiltonian to a reference Hermitian Hamiltonian with reciprocity. While this transformation creates an exponentially localized basis that restores the Hermitian bulk spectrum under OBC, it is forbidden under PBC as it violates the single-valuedness of the wavefunction. This singularity in the similarity transformation mathematically elucidates the extreme sensitivity of the spectrum to boundary conditions. The NHSE fundamentally stems from the breakdown of the Bloch theorem. Under OBCs, standing-wave solutions cannot form because the two Bloch eigenmodes possessing opposite wavevectors have different complex energies within the point-gapped band~\citetext{\citealp{PhysRevLett.125.126402}}. A wavepacket injected into the lattice undergoes unidirectional propagation and accumulates at the boundary. In the absence of loss, the wavepacket grows to extraordinary amplitudes at the boundary and exponentially decays throughout the lattice bulk. The NHSE has an elegant connection to Hermitian band topology, revealed through a bijective mapping of the non-Hermitian model $H(k)$ to an auxiliary chiral-symmetric Hermitian Hamiltonian:
    \begin{equation}
    \tilde{H}(k) = \begin{bmatrix} 0 & H(k) - E_r \\ H^\dagger(k) - E_r^* & 0 \end{bmatrix}
    \end{equation}
   \noindent where $E_r$ is the reference complex energy. The auxiliary Hamiltonian $\tilde{H}(k)$ is topologically non-trivial if $E_r$ lies within the point gap of $H(k)$. As such, the skin modes can be traced to the zero-energy topological modes of the semi-infinite system associated with $\tilde{H}(k)$~\citetext{\citealp{PhysRevLett.124.086801}}.
   
   The existence of the NHSE invalidates foundational principles that, in Hermitian systems, rely on conventional Bloch band theory. Notably, the BBC, a cornerstone of Hermitian topological band theory that relates topological invariants of the bulk under PBC to the existence of topological boundary states in a finite system, is no longer applicable. This breakdown has motivated the non-Bloch band theory~\citetext{\citealp{PhysRevLett.121.086803}; \citealp{PhysRevLett.123.066404}; \citealp{ProgTheorExpPhys.2020.12A102}}. The theory generalizes the Bloch wavevector $k \in [-\pi, \pi) \subset \mathbb{R} $ to a non-Bloch wavevector $\beta \in \mathbb{C} $. 
   Instead of solving the characteristic equation $\det[H(e^{ik}) - E] = 0$ on the unit circle $|e^{ik}|=1$, one solves $\det[H(\beta) - E] = 0$ on a generalized trajectory known as the Generalized Brillouin Zone (GBZ), where $|\beta|$ generally deviates from unity. As a result, the concept of the Brillouin zone is extended to an arbitrary closed loop [see Fig. \ref{fig:NHSE}(b)]. Crucially, obtaining the proper GBZ requires matching both the OBC spectrum and boundary conditions for the wavefunctions. The non-Bloch theory is powerful in that it not only yields the correct complex wavevectors that match OBC eigenmodes influenced by the NHSE but also provides a generalized scheme to compute topological invariants~\citetext{\citealp{PhysRevLett.121.136802}; \citealp{PhysRevLett.121.086803}; \citealp{PhysRevLett.123.066404}; \citealp{PhysRevB.102.085151}; \citealp{PhysRevLett.125.226402}; \citealp{ProgTheorExpPhys.2020.12A102}}, thereby restoring the BBC. However, obtaining the GBZ is relatively straightforward for 1D systems but remains exceedingly difficult in higher-dimensional systems. Several intriguing schemes have been proposed, such as the open-bulk winding number~\citetext{\citealp{PhysRevLett.123.246801}} and the doubled Hamiltonian method~\citetext{\citealp{PhysRevLett.124.056802}}. A commonly adopted alternative strategy bypasses the explicit construction of the GBZ by relying directly on the open-boundary spectrum. This approach relates topological phase transitions to delocalization transitions of biorthogonal boundary states via biorthogonal polarization~\citetext{\citealp{PhysRevLett.121.026808}; \citealp{PhysRevB.99.081302}; \citealp{PhysRevResearch.2.043046}}. 

    \begin{figure}[!htbp]
    \centering
    \includegraphics[width=\linewidth]{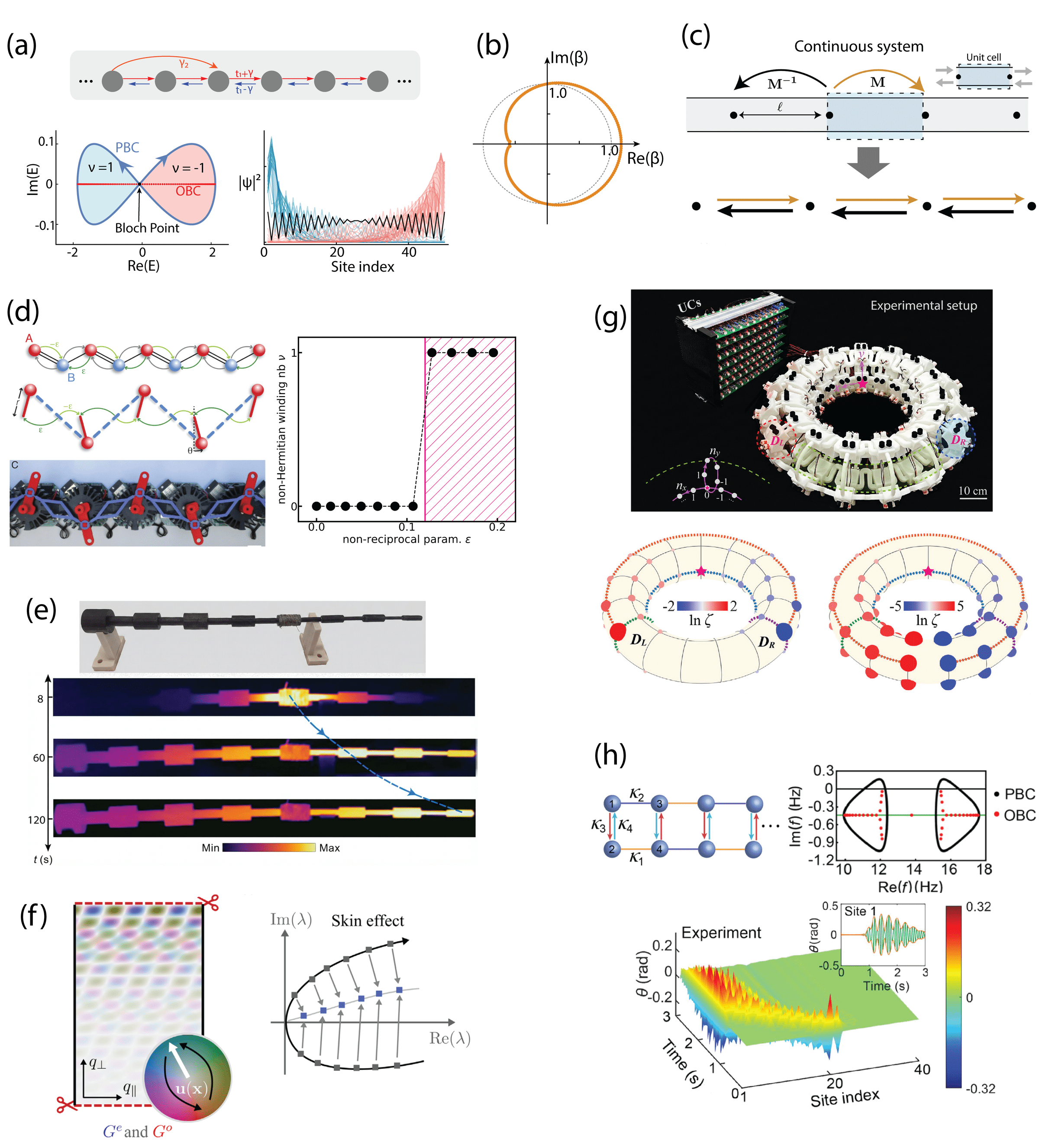}
    \caption{ Experimental observation of NHSE in phononic systems. (a, b) 1D non-reciprocal lattice with long-range coupling. PBC spectra exhibit complex-energy windings, while OBC modes display the NHSE. At the Bloch point, the skin effect vanishes. (b) depicts the corresponding GBZ. (c) Acoustic analog of the Hatano-Nelson model in the active feedback waveguide. (d) Non-reciprocal Kane-Lubensky chain as a mechanical analog of the non-Hermitian SSH model, showing winding number transitions.  (e) Diffusion-based NHSE in a thermal system. (f) Active elastic media with odd elastic moduli. Left: Eigenmodes of an odd elastic slab when the horizontal boundaries (red) are open and the vertical boundaries (black) are periodic. Right: Non-Hermitian spectral arcs (black solid) deform to a degeneracy line (blue squares) when a horizontal boundary is introduced. (g) Experimental demonstration of the dislocation-induced NHSE, showing dramatic accumulation of sound energy density at the dislocation site. (h) Dynamic evolution in glide-time-symmetric NH-SSH chains. (c) Adapted from~\citetext{\citealp{PhysRevResearch.6.L012061}}. (d) Adapted from~\citetext{\citealp{PNAS.117.29561}}. (e) Adapted from~\citetext{\citealp{SciBull.69.1228}}. (f) Adapted from~\citetext{\citealp{PhysRevLett.125.118001}}. (g) Adapted from~\citetext{\citealp{AdvMater.38.e14101}}. (h) Adapted from~\citetext{\citealp{NatCommun.15.6544}}.
    }\label{fig:NHSE}\end{figure}

   Since its theoretical prediction, the NHSE has been demonstrated across a remarkable variety of phononic platforms. Early successes were achieved in coupled mechanical oscillator systems, where local feedback was used to engineer the requisite non-reciprocity~\citetext{\citealp{NatCommun.10.4608}; \citealp{NewJPhys.22.053004}; \citealp{PNAS.117.29561}}. Specifically, by introducing non-reciprocal couplings into an active mechanical metamaterial, researchers realized a non-reciprocal Kane-Lubensky chain---the classical analog of the non-Hermitian SSH model---enabling the explicit observation of non-Hermitian topology and validating BBC [see Fig. \ref{fig:NHSE}(d)]~\citetext{\citealp{PNAS.117.29561}}. In acoustics, the implementation of asymmetric, long-range couplings has enabled the realization of complex spectral windings and the bipolar NHSE~\citetext{\citealp{NatCommun.12.6297}}. Departing from the coupling of discrete resonant modes, the integration of active feedback control within acoustic waveguides has further facilitated the synthesis of asymmetric transfer matrices, thereby extending the NHSE into the realm of continuous media [shown in Fig. \ref{fig:NHSE}(c)]. NHSEs have been further diversified across various platforms, ranging from elastic metamaterials~\citetext{\citealp{ApplPhysLett.121.022202}; \citealp{PhysRevResearch.2.023173}; \citealp{NatCommun.12.5935}}, to thermal diffusion systems~\citetext{\citealp{CommunPhys.4.230}; \citealp{SciBull.69.1228}} [Fig. \ref{fig:NHSE}(e)]. 
   
   The $\mathbb{Z}_2$ NHSE is fundamentally distinguished by a vanishing net winding number and is stabilized by TRS. A hallmark of this phase is the emergence of Kramers pairs in the OBC spectrum, which manifest as a spin-dependent NHSE where eigenmodes undergo boundary-selective localization according to their spin polarization~\citetext{\citealp{PhysRevLett.124.086801}}. This topological phenomenon was initially demonstrated in acoustic ring lattices, where a TRS-invariant $\mathbb{Z}_2$ skin effect was achieved by tailoring biased dissipation patterns in pseudospin modes, proving that global non-reciprocity is not a prerequisite~\citetext{\citealp{NatCommun.12.5377}}. Building on these advances, the $\mathbb{Z}_2$ NHSE was subsequently realized in spinless bilayer acoustic architectures through the sophisticated engineering of projective mirror symmetries~\citetext{\citealp{PhysRevLett.136.026601}; \citealp{AdvMater.37.2506739}}. A distinctive manifestation of the NHSE has also been identified in passive, static mechanical metamaterials, where it appears as boundary-localized deformations that attenuate toward the lattice interior~\citetext{\citealp{SciAdv.9.eadf7299}}. The implementation of this effect relies on a spatio-temporal mapping protocol that projects the temporal evolution of a dynamical system onto a specific spatial axis. By substituting the time variable with an additional spatial dimension, a formal mathematical correspondence is established between the static Rayleigh model and wave dynamics.

    The theoretical framework of the NHSE naturally generalizes to higher dimensions~\citetext{\citealp{PhysRevB.102.241202}; \citealp{NatCommun.13.2496}; \citealp{PhysRevX.14.021011}; \citealp{SciBull.70.51}}. In these systems, non-Hermiticity can confine modes not just to 1D edges but also to 0D corners, giving rise to higher-order skin effects protected by lattice spatial symmetries~\citetext{\citealp{NatCommun.12.4691}; \citealp{ApplPhysLett.121.041701}; \citealp{AdvMater.36.2403108}; \citealp{PhysRevB.111.014314}}. A sophisticated method to induce such non-Hermiticity in 2D elastic metamaterials involves the integration of piezoelectric feedback loops; by rendering the effective elastic tensor asymmetric, these systems facilitate precisely controlled directional wave amplification~\citetext{\citealp{PhysRevLett.125.118001}; \citealp{PNAS.120.e2209829120}} [shown in Fig. \ref{fig:NHSE}(f)]. Another profound development in this regime is the geometry-dependent skin effect (GDSE), where skin modes emerge only along boundaries whose macroscopic symmetry is incompatible with the intrinsic lattice symmetry~\citetext{\citealp{NatCommun.13.2496}}. This subtle interplay among point-gap topology, lattice geometry, and boundary termination has been experimentally verified in both acoustic~\citetext{\citealp{SciBull.68.2330}; \citealp{NatCommun.14.4569}} and mechanical~\citetext{\citealp{PhysRevLett.131.207201}} platforms, revealing a localization mechanism distinctive to higher dimensions. In 3D architectures, by tailoring dissipative coupling in waveguides, a pristine nodal ring can be bifurcated into a pair of exceptional rings, effectively realizing exceptional-line semimetals~\citetext{\citealp{PhysRevLett.134.116606}}. Within this framework, a non-vanishing winding number, defined along the direction perpendicular to the lateral boundaries of a 3D square-frustum acoustic crystal, underpins the emergence of the 3D GDSE on the crystal's side facets.

   The NHSE also manifests at internal crystalline defects (such as dislocations)~\citetext{\citealp{PhysRevB.104.L161106}; \citealp{PhysRevB.104.L241402}; \citealp{PhysRevB.106.L041302}; \citealp{PhysRevB.109.035425}}. These defects act as effective internal boundaries, trapping a vast number of modes at their cores. This "dislocation NHSE" has been directly visualized in active acoustic metamaterials, where states accumulate at dislocations with a particular Burgers vector and are depleted from those with the opposite, creating localized hotspots of acoustic energy~\citetext{\citealp{AdvMater.38.e15496}; \citealp{AdvMater.38.e14101}}, as shown in Fig. \ref{fig:NHSE}(g). This phenomenon highlights a robust interplay between point-gap topology and the topology of lattice defects, providing a pathway for defect-engineered phononic devices.
   
   Beyond static localization, the NHSE profoundly influences the dynamical behavior. In dissipative systems, where dynamics are governed by the imaginary part of the spectrum, the closure of the imaginary gap can trigger sharp, transient "edge bursts", a rapid amplification of energy at the boundary, even for bulk excitations. This phenomenon has been demonstrated in acoustic crystals incorporating unidirectional amplifiers~\citetext{\citealp{PhysRevLett.128.120401}; \citealp{AdvMater.38.e15529}}. The GBZ framework is also essential for predicting the rich dynamical phase diagrams of non-Hermitian systems, which encompass phenomena such as bulk amplification, unidirectional transport, and boundary wave trapping~\citetext{\citealp{NatCommun.15.6544}} [see Fig. \ref{fig:NHSE}(h)]. Moreover, the localization features of the NHSE can be observed even in passive systems by applying exponentially modulated (complex-frequency) drive to emulate a "virtual gain," enabling direct visualization of skin modes in otherwise passive acoustic structures~\citetext{\citealp{NatCommun.13.7668}}.

\subsection{\label{sec:level2}Non-Hermitian Topological Modes}

    Recent theoretical developments have clarified not only how non-Hermiticity reshapes topological phases but also how it can be harnessed for engineering and reconfiguring topological modes. A central insight is the delicate interplay between the NHSE and lattice symmetries: at a critical balance, the NHSE can compete with the localization of topological modes, driving their delocalization into the bulk or reconfiguring their spatial profile entirely. This principle is exemplified in the non-Hermitian Su-Schrieffer-Heeger model, where non-reciprocity can transform a topological zero-energy state into a delocalized bulk state~\citetext{\citealp{PhysRevLett.125.206402}; \citealp{PhysRevB.103.195414}; \citealp{Nature.608.50}; \citealp{AAPPSBull.33.23}}. When extending this concept to higher-dimensional phononic crystals, the NHSE can similarly alter the localization characteristics of topological modes~\citetext{\citealp{Nature.608.50}; \citealp{SciAdv.9.eadf7299}; \citealp{PNAS.121.e2408843121}} [see Fig. \ref{fig:NHSE-TMs}(a, c)]. In Kagome crystals, zero-energy corner states---which are typically bound states in the continuum (BICs)---can be transformed into extended states residing within a localized continuum, a phenomenon described as an “inverse BIC” transition~\citetext{\citealp{PhysRevLett.129.264301}} [see Fig. \ref{fig:NHSE-TMs}(b)]. 
    
    Furthermore, introducing non-Hermiticity into Chern insulators yields hybrid skin effects, characterized by the coexistence of localized topological boundary states and extended bulk states. This phenomenon arises from the interplay between the NHSE and topological chiral edge states, whereby these chiral edge states coherently accumulate at lower-dimensional boundaries (e.g., corners)~\citetext{\citealp{PhysRevB.98.165148}; \citealp{PhysRevLett.125.186802}; \citealp{NatCommun.11.5745}; \citealp{PhysRevB.102.121405}; \citealp{PhysRevB.106.035425}; \citealp{PhysRevB.105.075128}; \citealp{PhysRevB.108.075122}}. Experimentally, such phenomena have been realized in phononic crystals through spatially staggered loss engineering, where tailored dissipation profiles break reciprocity~\citetext{\citealp{PhysRevLett.133.126601}; \citealp{PhysRevLett.134.176601}} [see Fig. \ref{fig:NHSE-TMs}(d)].
    
    \begin{figure}[!htbp]
    \centering
    \includegraphics[width=\linewidth]{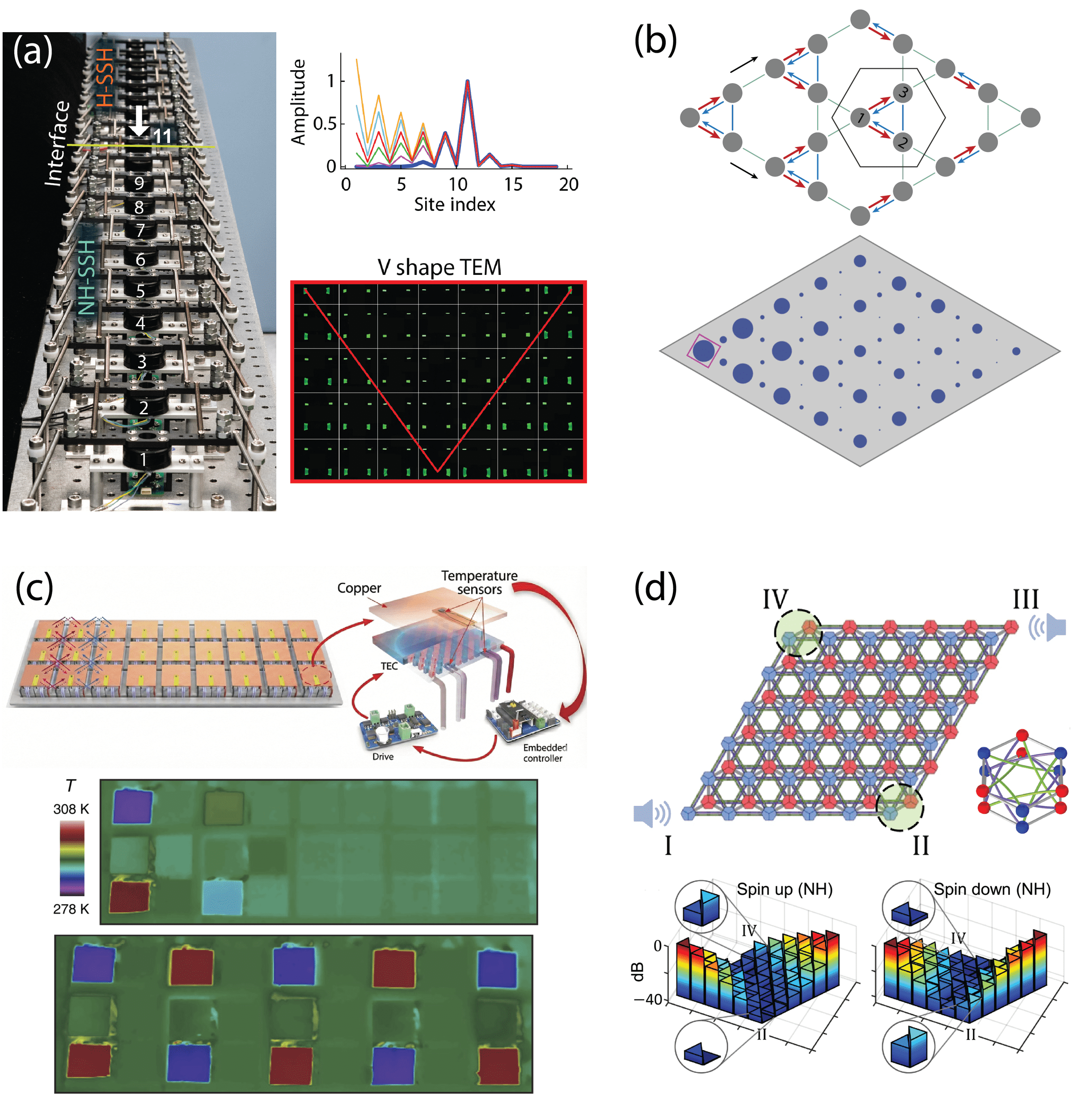}
    \caption{ Advanced non-Hermitian topological states in phononic systems. (a) Non-Hermitian SSH mechanical resonators. (Left) Experimental setup. (Top right) Transition of topological interface states from localized to extended modes by tuning intracell non-reciprocity. (Bottom right) 2D stacked NH-SSH structures with engineered non-Hermitian distributions, manifesting V-shaped topological modes. (b) Kagome mechanical lattice with non-reciprocal gauge fields. Demonstration of fully extended states embedded within a continuum of localized states. (c) Topological zero modes in thermal diffusion. Mapping an anti-Hermitian Hamiltonian to a Hermitian-like form via auxiliary-dimension engineering, realizing delocalized topological mode in a diffusive system. (d) Acoustic Kane-Mele model. Implementation of hybrid skin-topological states exhibiting spin-dependent localization. 
    (a) Adapted from~\citetext{\citealp{Nature.608.50}}. (b) Adapted from~\citetext{\citealp{PhysRevLett.129.264301}}. (c) Adapted from~\citetext{\citealp{PNAS.121.e2408843121}}. (d) Adapted from~\citetext{\citealp{PhysRevLett.133.126601}}.}\label{fig:NHSE-TMs}\end{figure}

%\section{\label{sec:level1}Interactions of non-Hermitian topology with other mechanisms}
\subsection{\label{sec:level2}Interactions of non-Hermitian topology with other mechanisms}
     The introduction of disorder into non-Hermitian systems gives rise to new localization and delocalization phenomena. On one hand, non-Hermitian disorder (e.g., random on-site gain/loss) can induce novel topological phases, such as the non-Hermitian higher-order topological Anderson insulator~\citetext{\citealp{SciChina-PhysMechAstron.63.267062}; \citealp{PhysRevB.103.224203}}. These are characterized by real-space invariants within a biorthogonal framework and have been realized in phononic platforms~\citetext{\citealp{PhysRevApplied.18.064079}; \citealp{SciChina-PhysMechAstron.66.294311}}. On the other hand, disorder brings new twists to the physics of point-gap topology. For example, a 1D non-reciprocal ring can host a mixture of localized and extended states, which are associated with real and complex eigenvalues, respectively~\citetext{\citealp{PhysRevLett.77.570}}. The emergence of such a mobility edge marks a sharp departure from Hermitian systems, where the scaling theory of localization forbids such a transition. It was also shown that more complicated winding in the spectrum, e.g., introduced by long-range hoppings, can produce localized modes with complex eigenvalues in such systems~\citetext{\citealp{PhysRevLett.134.066301}}, as illustrated in Fig. \ref{fig:NH-OtherMech}(a). A real-space winding number can remain quantized even in the presence of strong disorder, thereby preserving the NHSE~\citetext{\citealp{PhysRevB.103.L140201}; \citealp{PhysRevB.106.014207}}. Experiments in acoustic crystals have demonstrated that increasing disorder strength can trigger a “boundary-switching” skin effect, wherein wavefunctions undergo relocation from one edge to the opposite~\citetext{\citealp{PNAS.122.e2422154122}}, as shown in Fig. \ref{fig:NH-OtherMech}(b).

    \begin{figure}[!htbp]
    \centering
    \includegraphics[width=\linewidth]{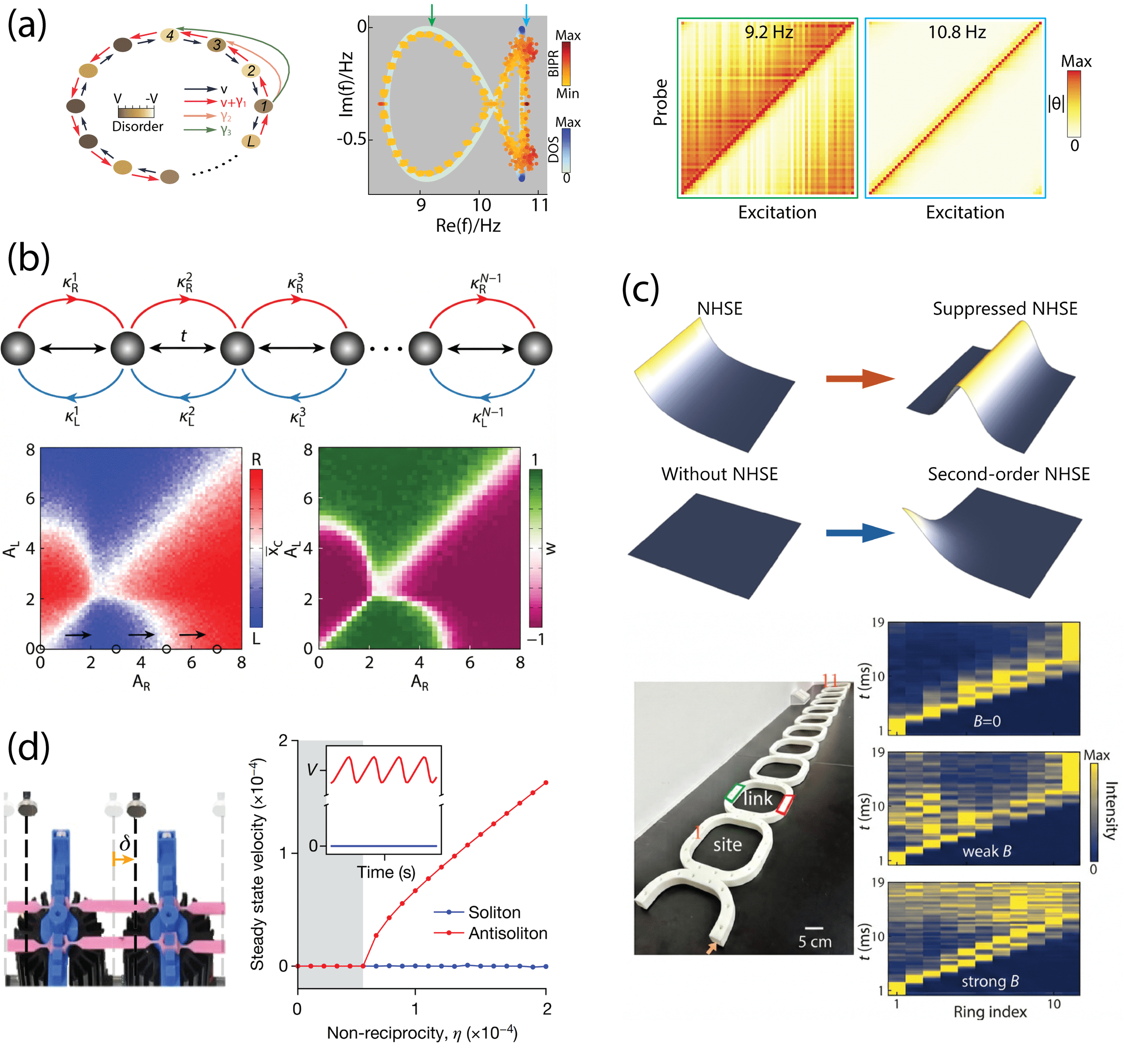}
    \caption{ Interplay of disorder, synthetic fields, and nonlinearity in non-Hermitian systems. (a) A non-Hermitian disordered ring with long-range non-reciprocity. Color scales denote the density of states (DOS, blue) and biorthogonal inverse participation ratio (BIIPR, yellow). (Right) Response maps showing extended and localized modes at specific frequencies. (b) Non-reciprocity disorder induced localization transition in a 1D non-Hermitian acoustic crystal with disordered non-reciprocity. (c) Schematic of the interaction between the NHSE and synthetic magnetic fields, leading to field-tunable skin effects. (d) The plot shows the steady-state velocity of solitons (blue) and antisolitons (red) in a bistable potential as a function of non-reciprocity strength. (a) Adapted from~\citetext{\citealp{PhysRevLett.134.066301}}. (b) Adapted from~\citetext{\citealp{PNAS.122.e2422154122}}. (c) Adapted from~\citetext{\citealp{ApplPhysRev.11.031410}}. (d) Adapted from~\citetext{\citealp{Nature.627.528}}.}\label{fig:NH-OtherMech}\end{figure}
    
    Synthetic magnetic fields represent another mechanism that can compete with the NHSE. In Hermitian systems, electrons under a magnetic field undergo cyclotron motions, resulting in the well-known Landau quantization where energy bands become flat. In non-Hermitian systems, this localization provides a mechanism to counteract the boundary-pinning tendency of the NHSE~\citetext{\citealp{PhysRevLett.127.256402}; \citealp{PhysRevLett.130.103602}; \citealp{SciBull.69.1667}; \citealp{PhysRevB.111.024201}}. As the effective (pseudo) magnetic field strength is increased, skin modes are observed to delocalize from the boundary and transition back into the bulk. This process is accompanied by a contraction of the spectral area in the complex plane and the restoration of quantized Landau levels. The competition between the NHSE and Landau quantization offers a powerful knob for manipulating wave localization~\citetext{\citealp{ApplPhysRev.11.031410}} [shown in Fig. \ref{fig:NH-OtherMech}(c)].
   
   The interplay between nonlinearity and non-Hermiticity has recently catalyzed the development of transformative wave-control paradigms~\citetext{\citealp{PhysLettA.408.127484}; \citealp{ChaosSolitonFract.172.113545}; \citealp{PNAS.120.e2217928120}}. By leveraging the interplay between NHSE and the bistable characteristics of nonlinear on-site potentials, researchers have demonstrated sophisticated manipulation of solitonic and anti-solitonic excitations [Fig. \ref{fig:NH-OtherMech}(d)]. The inherent self-localization of solitons serves to decouple non-reciprocal wave packets from the characteristic boundary sensitivity of the NHSE; consequently, in strongly nonlinear regimes, the velocity and stability of these wave packets can become invariant to fluctuations in interfacial conditions~\citetext{\citealp{Nature.627.528}}. Furthermore, it has been established that by rigorously maintaining chiral symmetry within nonlinear acoustic architectures, topological edge states can be pinned to zero energy across an expansive range of nonlinear magnitudes, effectively preventing the symmetry-breaking frequency shifts typically encountered in nonlinear topological systems~\citetext{\citealp{SciPostPhys.18.1034}}. Furthermore, active mechanical metamaterials exploiting non-reciprocal interactions can harness the nonlinear NHSE for directional energy amplification, facilitating autonomous and adaptive locomotion modes driven by environmental feedback loops~\citetext{\citealp{Nature.639.935}}.\\

\section{\label{sec:level1}Topological phononic devices}

Topological phononic metamaterials exhibit unconventional wave phenomena and distinctive physical signatures, such as suppressed backscattering, defect- and disorder-robust transport, low-loss propagation, unidirectional boundary transport, and emergent synthetic degrees of freedom associated with boundaries. These features provide a powerful foundation for functional devices with enhanced robustness, directional controllability, and compact integrability. To date, such topological functionalities have been realized in a broad range of platforms, including Fano-resonant structures, Su-Schrieffer-Heeger (SSH) lattices, pseudospin and valley-pseudospin systems, on-chip topological components, acousto-optic devices, topological surface acoustic wave (SAW) platforms, and integrated chip-scale as well as acoustofluidic architectures.

\subsection{\label{sec:level2}Topological phononic devices in airborne and elastic systems}

Conventional phononic devices are often hindered by signal crosstalk and limited operating efficiency. Topological phononic metamaterials provide an alternative framework for realizing robust phononic functionalities, owing to their suppressed backscattering and directional wave transport. A paradigmatic example is the one-dimensional SSH model, in which midgap interface states appear at domain walls between phases with distinct topological invariants. This mechanism has been exploited to realize a disorder-robust topological solver for second-order differential equations~\cite{NatCommun.10.2058}; see Fig.~\ref{fig:1}(a).

Fano resonance, a generic wave-scattering phenomenon characterized by an asymmetric and ultrasharp spectral profile~\citetext{\citealp{NatPhotonics.11.543}; ~\citealp{AdvOptPhoton.13.703}}, is of particular interest for sensing because of its sensitivity to structural parameters and environmental perturbations. In a one-dimensional acoustic system, a topological Fano resonance was proposed that combines robustness against geometric disorder with sensitivity to environmental variation~\cite{PhysRevLett.122.014301}; see Figs.~\ref{fig:1}(b) and \ref{fig:1}(c). Such a mechanism points to a route toward sensing devices that integrate topological protection with enhanced spectral selectivity.

Beyond that, acoustic pseudospin---closely analogous to electronic spin---can be engineered in metamaterials through double-Dirac-cone and zone-folding mechanisms~\citetext{\citealp{NatPhys.12.1124}; \citealp{PhysRevLett.118.084303}; \citealp{PhysRevLett.122.086804}; \citealp{NatCommun.11.3768}}. On this basis, an elastic topological whispering-gallery resonator was demonstrated, providing a route toward integrated topological phononic circuits with substantially reduced scattering loss~\cite{NatCommun.9.3072}; see Fig.~\ref{fig:1}(d). Along similar lines, topological switches and logic gates were realized by combining trivial and nontrivial domains~\cite{PhysRevLett.128.015501}; see Fig.~\ref{fig:1}(e). In addition, a programmable acoustic topological insulator composed of two digital units, ``0'' and ``1'', was proposed using honeycomb-lattice sonic crystals formed by cylindrical rods of different diameters. Based on this platform, a single-pole double-throw switch was experimentally implemented~\cite{AdvMater.30.1805002}; see Fig.~\ref{fig:1}(f).

\begin{figure}
    \centering
    \hspace*{-0.2cm}
    \vspace*{-1.6cm}  
    \includegraphics[
        scale=0.35,
        trim=100 20 20 80, 
        clip
    ]{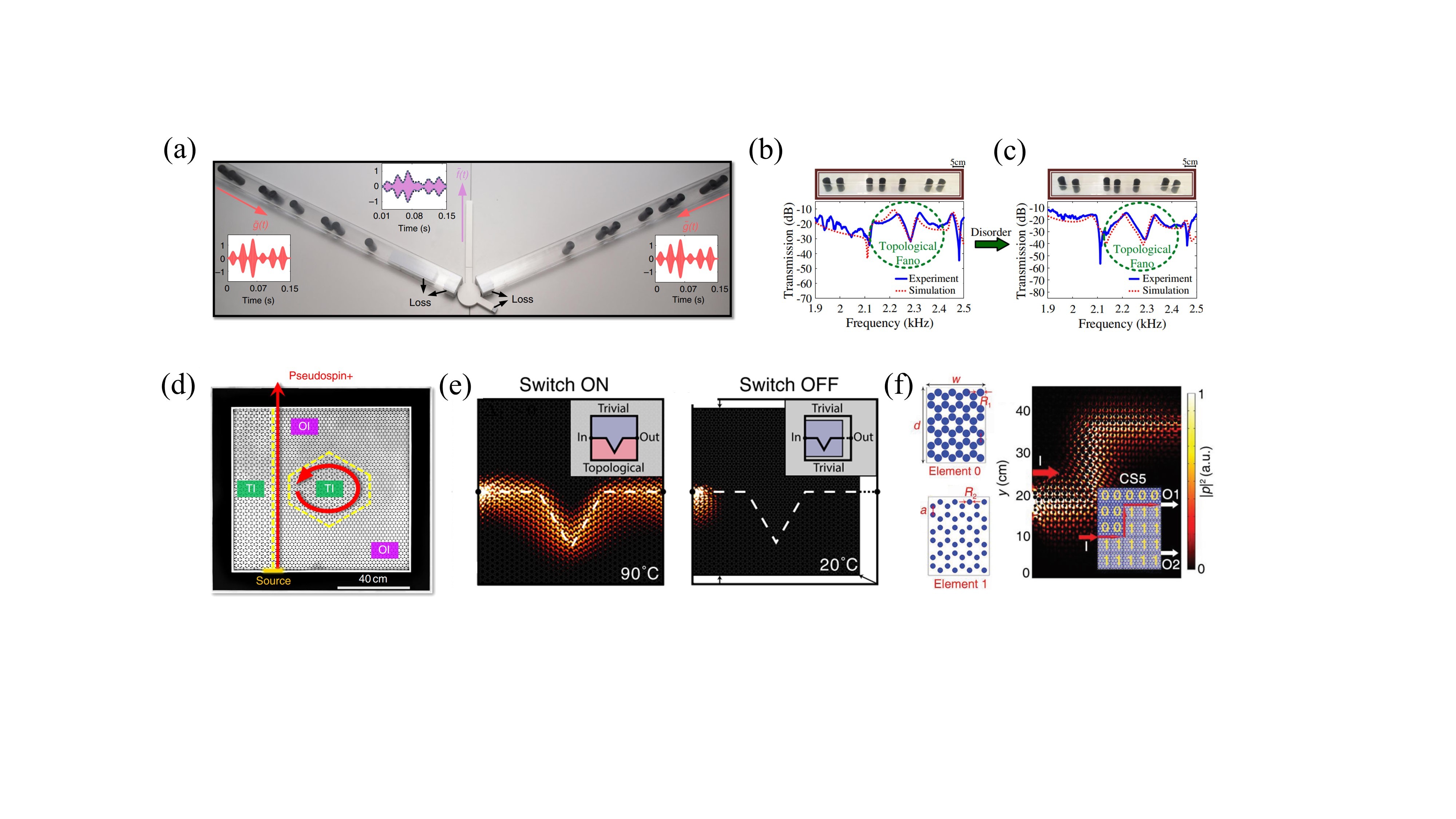}
    \caption{Functional devices in topological phononics in air and   elastic regime. (a) Experimental realization of the second-order differential equation solver. The measured output envelope (\textit{f}(\textit{t}),purple) agrees well with both the numerical simulation (gray) and the exact solution (dashed). (b),(c) Experimental demonstration of topological Fano resonances and their robustness. Nylon rods embedded in a transparent square acoustic waveguide produce a topological Fano resonance as shown in the lower panel. (d) The elastic topological whisper-gallery resonator. (e) Topological waveguide supporting robust transport despite channel bending,which can serve as a switch. (f) Left: digital elements of “0” and “1”. Right: distributions of acoustic intensity field in programmable acoustic topological insulators. (a) Adapted from~\citetext{\citealp{NatCommun.10.2058}}. (b),(c) Adapted from~\citetext{\citealp{PhysRevLett.122.014301}}. (d) Adapted from~\citetext{\citealp{NatCommun.9.3072}}. (e) Adapted from~\citetext{\citealp{PhysRevLett.128.015501}}. (f) Adapted from~\citetext{\citealp{AdvMater.30.1805002}}.}\label{fig:1}
    \vspace*{0cm}
    
\end{figure}

In parallel with the development of pseudospin-based platforms, the valley degree of freedom has provided another effective framework for realizing nontrivial topological phases and phononic functionalities, analogous to its role in two-dimensional gapped valleytronic materials~\citetext{\citealp{NatPhys.13.369}; \citealp{SciAdv.9.eadi8500}; \citealp{AdvMater.36.2311611}; \citealp{PhysRevLett.135.126603}}. Inspired by the biosonar of dolphins, a topological acoustic antenna was proposed~\cite{AdvMater.30.1803229}; see Fig.~\ref{fig:2}(a). This design supports superdirective needle-like sound radiation with a beamwidth below $10^\circ$, while robust sound reception is achieved through acoustic valley transport; see Fig.~\ref{fig:2}(b). By introducing a heterostructure composed of three valley sonic crystals, a topological waveguide supporting valley-locked edge states was demonstrated, together with reflection-free beam splitting and beam convergence associated with valley locking~\cite{NatCommun.11.3000}; see Fig.~\ref{fig:2}(c). Owing to the high flexibility of phononic crystals in design and fabrication, a range of valley-based acoustic devices, including filters, valves, and diverters, were subsequently realized~\cite{AdvMater.37.2500757}; see Fig.~\ref{fig:2}(d). Moreover, by extending the valley concept from two to three dimensions and incorporating valley pseudospin, three-dimensional topological refraction induced by valley kink states, as well as a new type of topological cavity, were demonstrated~\cite{SciAdv.10.eadp0377}; see Fig.~\ref{fig:2}(e).

\begin{figure}
    \centering
    \hspace*{-0.05cm}
    \vspace*{0cm}  
    \includegraphics[
        scale=0.35,
        clip
    ]{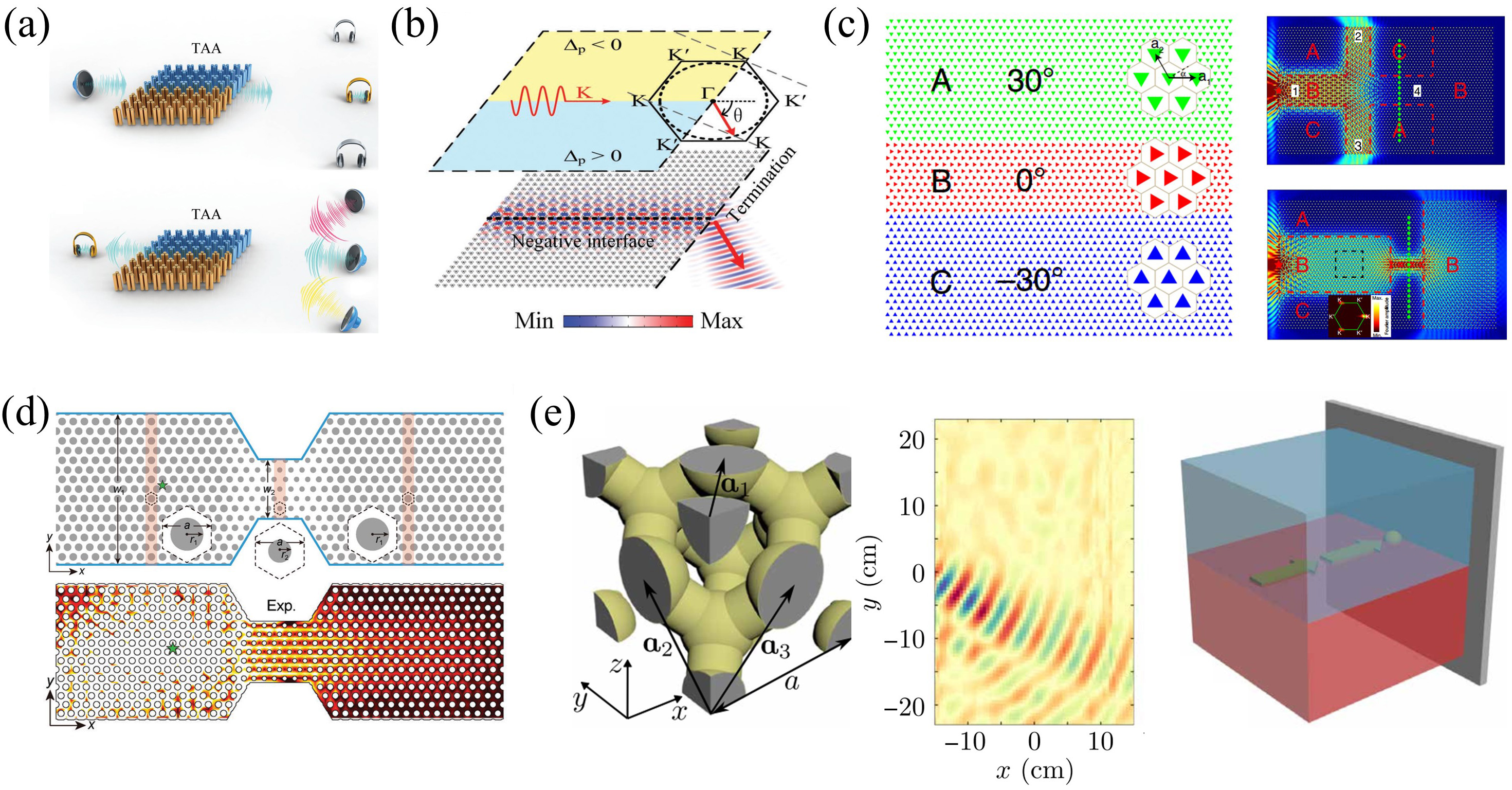}
    \caption{Applications in air and elastic media via valley degree of freedom. (a) Schematic and operating principle of the topological acoustic antenna. 
(b) Top: \textit{k}-space analysis of the out-coupling of K-projected edge modes at the negative interface. The black solid hexagon denotes the first Brillouin zone, and the black dashed circle indicates the dispersion contour. Bottom: simulated acoustic pressure field. (c) Left: schematic of a heterostructure composed of three sonic-crystal domains, A, B, and C. Right: splitting and converging. (d) Top: schematic of the valley-based acoustic device, consisting of input and output ports connected by a narrow channel. Bottom: experimental demonstration. (e) Left: acoustic cubic unit cell. Middle: topological refraction when the out-coupling plane is normal to the propagation direction of the kink states. Right: a novel cavity formed by blocking the out-coupling channel. (a),(b) Adapted from~\citetext{\citealp{AdvMater.30.1803229}}. (c) Adapted from~\citetext{\citealp{NatCommun.11.3000}}. (d) Adapted from~\citetext{\citealp{AdvMater.37.2500757}}. (e) Adapted from~\citetext{\citealp{SciAdv.10.eadp0377}}. \label{fig:2} }
     \vspace*{0cm}
\end{figure}

Additionally, for device applications, topological-charge (TC) manipulation offers an effective route for expanding the functionality of topological phononic crystals and artificial acoustic devices. Metasurfaces based on TC multiplexing can generate multiple acoustic vortex singularities and enable their independent encoding and remote dynamic control, providing a versatile platform for multichannel information transfer, structured acoustic fields, and acoustofluidic manipulation~\cite{Sci.Adv.11.eadw1701}. Moreover, combining conjugate TC pairing with orbital meta-atom design establishes robust interference channels in mode space, enabling high- and super-resolution acoustic displacement metrology~\cite{Nat.Commun.15.8391} and enhancing sensing performance in complex media. These advances show that TC-based sound-field control is moving acoustic metasurfaces and topological artificial structures beyond wavefront shaping toward high-performance functional devices.

\subsection{\label{sec:level4}On-chip topological phononic devices}
To advance compact phononic integrated circuits and device miniaturization, topological concepts have been extended to high-frequency platforms, including microelectromechanical, on-chip, and nanoelectromechanical metamaterials~\cite{AdvMater.33.e2006521}. In particular, by exploiting the valley-pseudospin degree of freedom, valley topological phases have been realized in microfabricated phononic crystals on silicon chips~\cite{NatMater.17.993}, where reflection-immune Z-shaped waveguides and pseudospin edge-state partitioning were demonstrated; see Fig.~\ref{fig:5}(a). This mechanism has subsequently been extended to ultrahigh frequencies, enabling the realization of the GHz-scale topological valley Hall effect in nanoelectromechanical phononic crystals and the demonstration of a topological beam splitter~\cite{NatElectron.5.157}; see Fig.~\ref{fig:5}(b). These devices are of considerable interest for applications in wireless communication and sensing. 

Pseudospin, alongside the valley degree of freedom, plays a central role in on-chip phononic topological devices. Directional topological acoustic transport and filtering were successfully achieved using silicon nitride nanomembranes.~\cite{Nature.564.229}; see Fig.~\ref{fig:5}(c). An alternative route to nanomechanical topological insulators is provided by introducing an auxiliary orbital degree of freedom\citetext{\citealp{NatNanotechnol.16.576}; \citealp{NatCommun.13.6597}; \citealp{NatCommun.14.8162}; \citealp{AdvMater.36.e2312421}; \citealp{SciChinaPhysMech.68.214312};\citealp{SciChinaPhysMech.69.254311}}, through which topological bending waveguides have been realized; see Fig.~\ref{fig:5}(d).

Higher-order topological insulators constitute a distinct class of topological phases governed by a generalized bulk-boundary correspondence. They host unconventional topologically protected states and thus offer new opportunities for acoustic positioning and sensing. The first on-chip MHz-scale higher-order topological corner states were demonstrated in micromechanical metamaterials based on patterned etched structures~\cite{SciBull.66.1959}; see Fig.~\ref{fig:5}(e). These states can be exploited in high-$Q$ resonators and integrated waveguides. Subsequently, higher-order monolithic topology was realized in nanomechanical metamaterials operating at ultrahigh frequencies. The corresponding topological nanomechanical corner states exhibit low mode volume, high $Q$ factors, and operating frequencies approaching the GHz regime, thereby opening new possibilities for vibration control and energy harvesting~\cite{NanoLett.24.15421}; see Fig.~\ref{fig:5}(f). 

Moreover, synthetic dimensions provide an additional route for realizing topological-state-based devices. For example, on-chip chiral zeroth Landau levels were demonstrated, providing a robust bulk transport channel for wave guiding. In the same system, 0D edge states were also observed when the surface Fermi arc states were subjected to a pseudomagnetic field~\cite{PhysRevLett.133.256602}; see Fig.~\ref{fig:5}(g).

\begin{figure}
    \centering
    \hspace*{-0.25cm}
    \includegraphics[scale=0.1]{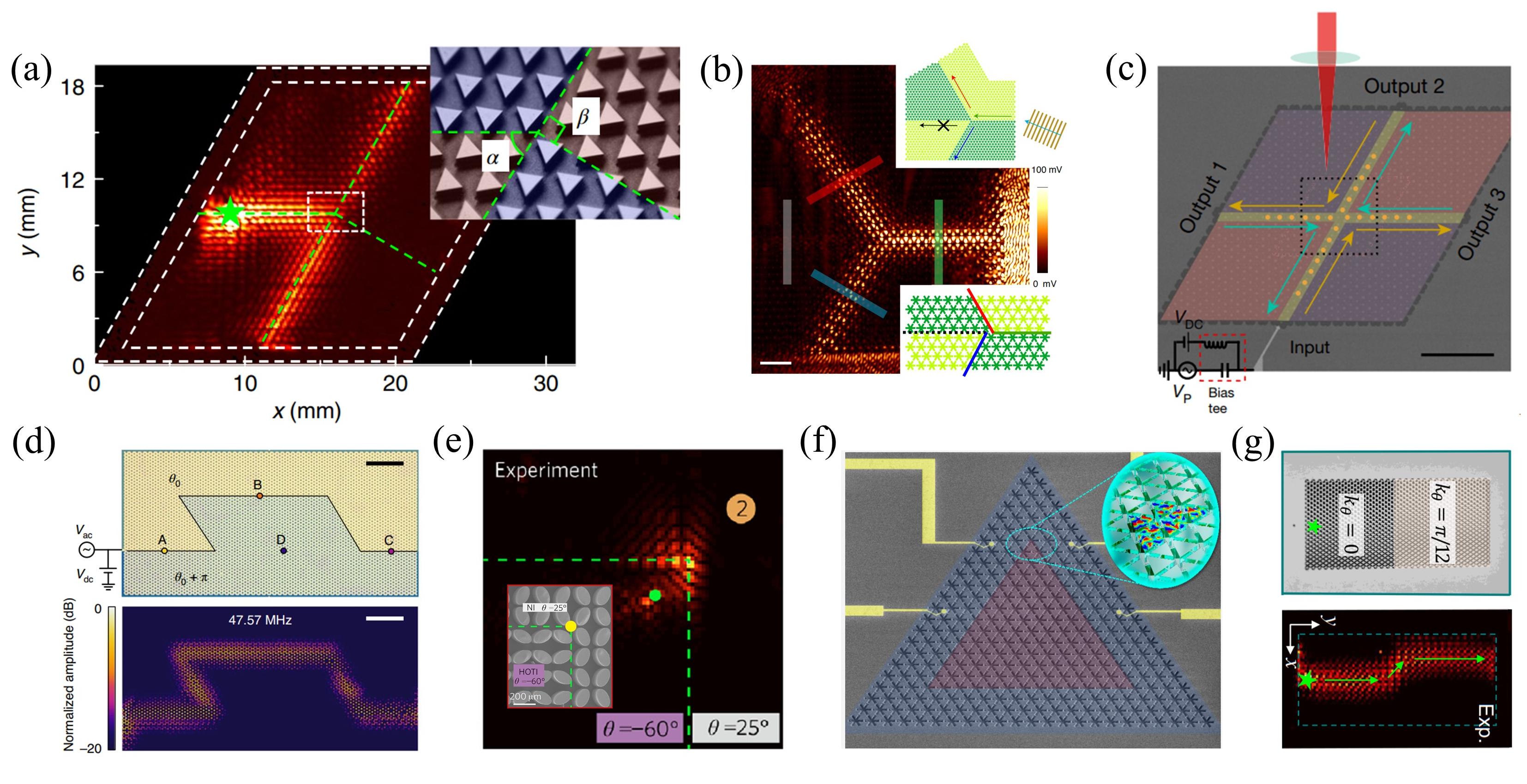}
    \caption{On-chip and high-frequency topological phononic devices. (a) On-chip partition of edge states at topological channel intersections. 
(b) Gigahertz topological beam splitting enabled by the valley Hall effect. (c) SEM image of a pseudospin-filter configuration. (d) Topological bending waveguide realized on a nanomechanical platform with an auxiliary orbital degree of freedom. (e) Photograph of a mechanical higher-order topological insulator on a silicon chip, together with measured corner displacement profiles. 
(f) SEM image of a LiNbO$_3$-based near-GHz higher-order topological insulator, providing a platform for energy harvesting. (g) Topological waveguiding enabled by a synthetic dimension. 
(a) Adapted from~\citetext{\citealp{NatMater.17.993}}. 
(b) Adapted from~\citetext{\citealp{NatElectron.5.157}}. 
(c) Adapted from~\citetext{\citealp{Nature.597.655}}. 
(d) Adapted from~\citetext{\citealp{Nature.564.229}}. 
(e) Adapted from~\citetext{\citealp{NatNanotechnol.16.576}}. 
(f) Adapted from~\citetext{\citealp{SciBull.66.1959}}. 
(g) Adapted from~\citetext{\citealp{NanoLett.24.15421}}.}
    \label{fig:5}
\end{figure}

\subsection{\label{sec:level5}Topological SAW devices}
 SAW is essential information carrier in microscale devices and are widely used in modern wireless communication and sensing systems. However, conventional SAW devices are often limited by scattering and crosstalk under high-frequency and high-sensitivity operating conditions, restricting their applications in emerging areas such as integrated circuits, biosensing, and millimeter-wave communications. Topological insulators have therefore attracted considerable interest because their robustness against defects and disorder, together with suppressed backscattering, can markedly improve device stability and operational tolerance in complex environments. In addition, unidirectional transport, such as pseudospin-dependent wave propagation, can reduce energy dissipation, making topological platforms attractive for low-power and high-bandwidth applications. By combining SAW technology with topological physics, topological SAW devices integrate interference resistance, efficient directional transport, miniaturization, and chip-scale compatibility, thus providing a promising route toward high-frequency and high-density devices for 6G communications and the Internet of Things.

In one-dimensional systems, resonator-based devices operating in the strong-coupling regime are often constrained by their sensitivity to fabrication imperfections. Using nanoscale grooves etched on a lithium niobate(LN) surface, a gigahertz SAW resonator with topological strong coupling was demonstrated~\cite{AdvMater.36.2312861}. Based on this platform, a dense wavelength-division multiplexing device was further realized through the coupling of multiple topological interface states; see Fig.~\ref{fig:6}(a). Synthetic dimensions provide another distinct route toward topological SAW devices. By using nanoscale translational deformation as an effective synthetic dimension, a monolithic gigahertz SAW topological rainbow was realized~\cite{PhysRevLett.133.267001}. This approach enabled on-chip single-mode rainbow filters over a broad operating-frequency range, as well as rainbow-type resonators; see Fig.~\ref{fig:6}(b).

Furthermore,  using the zone-folding mechanism, a Rayleigh-type SAW topological insulator was demonstrated on an interdigital-transducer (IDT)-based platform operating at frequencies of several tens of megahertz~\cite{PhysRevAppl.16.044008}. Alternatively, valley-locked SAW edge transport was realized in miniaturized phononic crystals on a semi-infinite substrate~\cite{NatCommun.13.1324}. Notably, the original one-dimensional edge transport was extended into a quasi-two-dimensional form by introducing SAW Dirac ``semimetal'' layers into the domains. Based on this mechanism, finite-width SAW waveguides and splitters were realized, providing new opportunities for hybrid phononic circuits; see Fig.~\ref{fig:6}(c).

Since most acoustic devices are passive, and phononic integrated circuits remain limited by the lack of low-loss and scalable control of acoustic waves, electrical control of gigahertz traveling acoustic waves at room and millikelvin temperatures has recently been proposed~\cite{NatElectron.5.348}; see Figs.~\ref{fig:6}(d)-\ref{fig:6}(f). In this scheme, a phase modulator was realized by electrically tuning the elasticity of a lithium niobate acoustic waveguide through the electroacoustic effect, and an electroacoustic amplitude modulator was further implemented in a Mach--Zehnder interferometer configuration.

Recently, the exploration of topological SAW devices has been extended to higher-order topological phases. Based on a two-dimensional generalized SSH model and a primitive unit composed of nickel (Ni) pillars grown on a square aluminum nitride (AlN) film, ultrastable and versatile topological signal processing was achieved within an ultracompact footprint~\cite{rs.3.rs-6176660/v1}; see Figs.~\ref{fig:6}(g) and \ref{fig:6}(h).

\begin{figure}[h]
    \centering
    \hspace*{0cm}
    \vspace*{0cm}  
    \includegraphics[
        scale=0.35, 
        clip
    ]{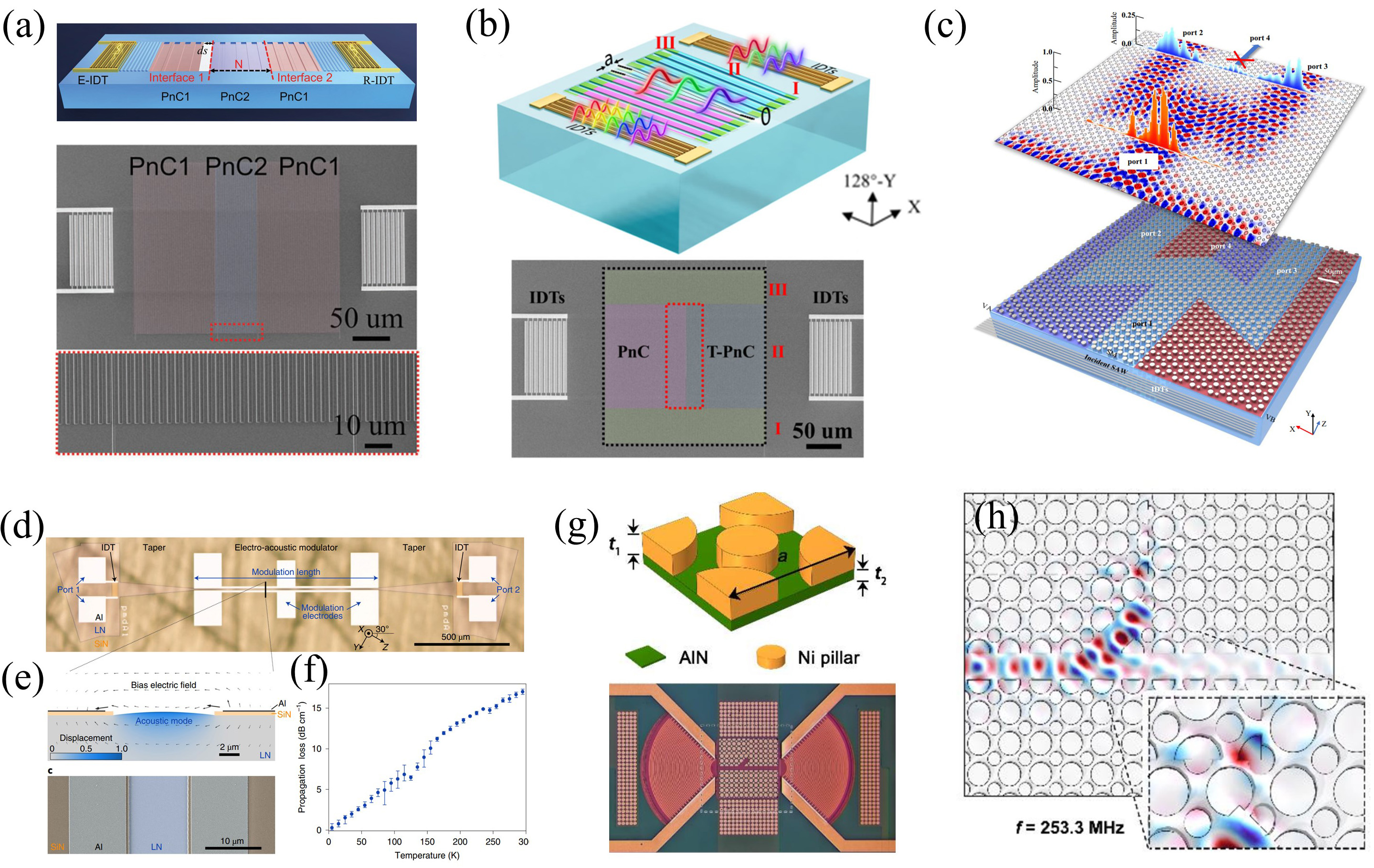}
    \caption{Topological SAW resonators, waveguides, and signal-processing devices. (a) Top: design of a coupled topological gigahertz SAW resonator. Bottom: SEM image of the fabricated device. 
(b) Top: schematic of on-chip topological rainbow resonators. Bottom: top-view SEM image of the phononic crystal with a Born--von Karman boundary. 
(c) High-throughput SAW splitter based on valley-locked states. (d) LN electroacoustic platform. (e) Top: cross section of the acoustic waveguide in the modulation region. The normalized displacement-field intensity, shown by the blue shaded region, represents the simulated fundamental mode. Bottom: false-colored SEM image of the acoustic waveguide. 
(f) Measured propagation loss of the acoustic waveguide as a function of temperature. 
(g) Top: schematic of primitive cell formed by Ni pillars on a square AlN film. Bottom: optical images of the topological phononic crystal. 
(h) Higher-order topological SAW devices with 6\% deformation of the Ni pillars. 
(a) Adapted from~\citetext{\citealp{AdvMater.36.2312861}}. 
(b) Adapted from~\citetext{\citealp{PhysRevLett.133.267001}}. 
(c) Adapted from~\citetext{\citealp{NatCommun.13.1324}}. 
(d)-(f) Adapted from~\citetext{\citealp{NatElectron.5.348}}. 
(g),(h) Adapted from~\citetext{\citealp{rs.3.rs-6176660/v1}}.}
    \vspace*{0.5cm}
    \label{fig:6}
\end{figure}

\subsection{\label{sec:level7}Topological phononic chips and acoustofluidic devices} 
Because phonons have much shorter wavelengths than photons at the same frequency, they offer intrinsic advantages for miniaturized and highly integrated devices. In addition, compared with electrons, phonons can carry information with extremely low energy dissipation and reduced heat generation. Combined with the robust wave control enabled by topology, these features make topological phononic chips---which use high-frequency sound waves or phonons as information carriers---a promising platform for addressing bottlenecks in modern information technology in selected application scenarios and for enabling new paradigms of on-chip information processing. In this context, reconfigurable topological phononic circuits operating in the GHz regime were demonstrated, together with a distinctive Mach--Zehnder interferometer (MZI) featuring intensity modulation and broad bandwidth, thereby enabling high-speed modulation of topological phonon transport pathways~\cite{NatElectron.8.689}; see Figs.~\ref{fig:8}(a)-\ref{fig:8}(d). This platform also exhibits strong compatibility with integrated photonics.

 Ingeniously coupling topological wave materials with hydrodynamics creates new opportunities for both visualizing topological physics directly and advancing practical device development. In particular, topological states in aqueous environments may enable a new generation of biomedical devices for advanced bioparticle manipulation. Along this direction, topological acoustofluidic chips were designed and experimentally realized~\cite{NatMater.24.707}; see Fig.~\ref{fig:8}(e), revealing a rich interplay between elastic valley pseudospin and nonlinear fluid dynamics. Valley streaming vortices were demonstrated, and based on them, chiral swirling patterns with backscattering-immune particle manipulation were achieved~\cite{NatMater.24.707}; see Fig.~\ref{fig:8}(f), thereby advancing acoustic tweezers and noncontact high-precision particle control. Furthermore, extraordinary topological pressure wells were observed in fluids, generating nanoscale trapping fields for DNA manipulation; see Figs.~\ref{fig:8}(g) and \ref{fig:8}(h), and opening new possibilities for molecular screening.

\begin{figure}
    \centering
    \hspace*{-0.25cm}
    \includegraphics[scale=0.33]{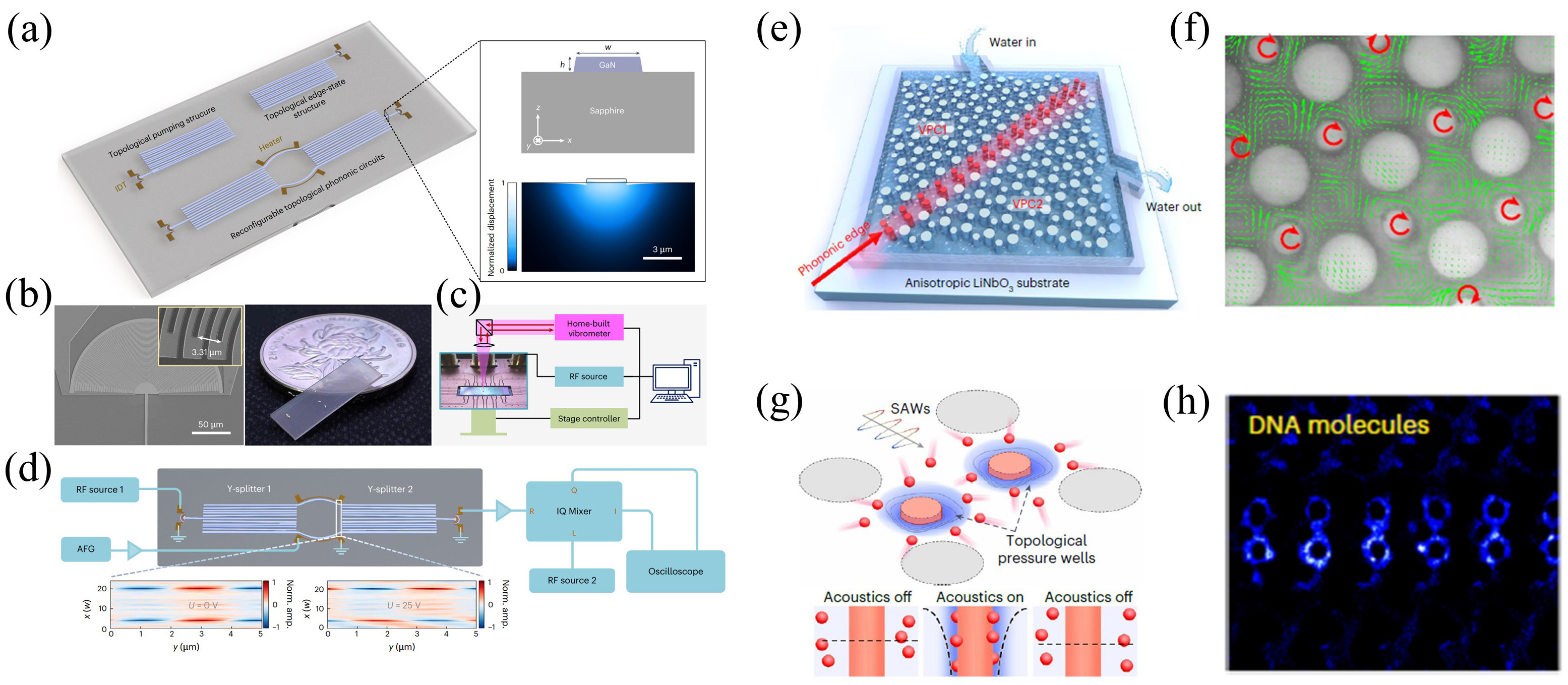}
    \caption{Integrated topological phononic and acoustofluidic platforms. (a) Left: schematic of the integrated topological phononic device. Top right: cross-sectional view of the GaN-on-sapphire phononic waveguide. Bottom right: simulated displacement field of the fundamental quasi-Rayleigh waveguide mode. 
(b) Left: SEM image of the aluminum IDT for excitation. Right: photograph of the fabricated chip. (c),(d) Schematic and experimental setup for the topological MZI. (e) Schematic of the topological acoustofluidic device on a 128$^\circ$ Y-cut LiNbO$_3$ substrate. 
(f) Valley acoustofluidic synchronous rotation of 1 \textmu m particles. (g) Schematic of topological pressure wells along the edge for particle concentration. (h) Experimental manipulation of DNA molecules at the frequency of topological edge states in a device with lattice constant \textit{a} = 50 \textmu m.  
(a)-(d) Adapted from~\citetext{\citealp{NatCommun.16.8116}}.
(e)-(h) Adapted from~\citetext{\citealp{NatMater.24.707}}.}
    \vspace*{0.16cm}
    \label{fig:8}
\end{figure}

\subsection{\label{sec:level8}Outlooks }
The milestone impact of topological phononic devices on both fundamental physics and technological innovation underscores the importance of interdisciplinary research. Future advances in this field will increasingly depend on the integration of condensed-matter physics, materials science, quantum acoustics, acousto-optics, nonlinear science, biomedicine, and communication technologies. Such a convergence is expected not only to deepen the understanding of topological wave phenomena across multiple physical settings, but also to drive the development of miniaturized, multifunctional, and intelligent phononic devices. In particular, the combination of topology with active tunability, low-loss on-chip integration, hybrid signal processing, and emerging biomedical functionalities is likely to open new directions for next-generation information and sensing technologies.

\section{Summary and Future Research Prospects}

Over the past decade, topological phononics has become a vibrant interdisciplinary field, integrating topological physics with lattice dynamics and classical wave physics to redefine phonon control---from atomic vibrations in solids to mechanical waves in metamaterials. This review surveys the field’s major development, starting with its core theoretical framework: reformulating phonon dynamics as a Schr\"{o}dinger-like eigen-problem to apply topological invariants and bulk-boundary correspondence, and extended to non-Hermitian and Floquet frameworks that expand topological landscapes beyond the conventional equilibrium Hermitian limits. We cover two key interconnected areas: natural solids where crystalline symmetry dictates Weyl, Dirac, and other topological phonons, and artificial metamaterials which enable topological phonons with engineered symmetry, synthetic gauge fields, and non-Hermitian effects. Experimental breakthroughs (via inelastic neutron/X-ray scattering, momentum-resolved EELS, acoustic pump-probe spectroscopy, etc.) have validated theoretical predictions, from Weyl phonons in FeSi to higher-order topological states in acoustic metamaterials. Non-Hermitian topological phononics introduces phenomena like exceptional nodes and the non-Hermitian skin effect, while Floquet engineering unlocks dynamical topological phases via temporal modulations.

Current frontiers include direct probing of unconventional topological invariants (e.g., entanglement entropy, entanglement spectrum~\citetext{\citealp{NatCommun.15.1601}; \citealp{SCPM.68.124311}; \citealp{SCPM.69.234311}}, and quantum metric~\citetext{\citealp{PRL.136.116602}}), synthetic dimensions, acoustic topological textures (skyrmions, merons), and nonlinear topological phonons, etc. Technologically, it enables robust devices (e.g., on-chip phonon waveguides, SAW resonators, acoustofluidic tweezers) leveraging topological robustness. A unified picture is already there: topology endows phonons with disorder-resistant effects, addressing a key phononic engineering challenge and valuable for understanding properties of genuine disordered interfaces. This review bridges natural/artificial phononic systems, Hermitian/non-Hermitian frameworks, and fundamental/applied aspects, positioning topological phononics as a transformative paradigm.

Topological phononics is poised for further rapid growth, with key emerging trends. Quantum topological phononics~\citetext{\citealp{NatPhys.11.37}; \citealp{Science.364.368}; \citealp{APLMater.13.101107}} merges phonons with quantum effects, building on optomechanical systems and quantum materials (e.g., cold atomic and molecular gases). Non-Hermitian topological phononics, a dynamic subfield, has redefined topological paradigms by incorporating gain, loss, and nonreciprocity. Future work will explore higher-order and hybrid non-Hermitian topology, such as exceptional links and knots, the interplay of non-Hermitian effects with disorder and nonlinearity~\citetext{\citealp{NatMater.23.1386}}, enriching biorthogonal bulk-boundary correspondence, and experimental advances in on-chip non-Hermitian phononic platforms operating at GHz frequencies.

Floquet engineering via periodic spatiotemporal modulation has become a powerful tool to create topological phases unavailable in equilibrium. Future directions include merging periodic driving with non-Hermitian effects to realize anomalous Floquet skin effect and Floquet exceptional points, exploring nonadiabatic topological pumping beyond adiabatic limits, and developing Floquet topological phonon lasers with enhanced coherence and directionality. Real-space topological textures and higher-order topology are also expanding, with research focusing on the dynamics of acoustic skyrmions and merons, the extension of topological bulk-defect correspondence to non-Hermitian systems, and their technological applications.

Interdisciplinary integration is driving scientific and technological innovation in topological phononics. For instance, topological phononics is generalized to biological mechanical structures~\citetext{\citealp{PhysRevE.83.021913}; \citealp{PhysRevLett.103.248101}} and active soft matter~\citetext{\citealp{Nat.Phys.13.1091}} where robust topological phonon and sound edge modes are revealed across diverse physical systems. In biomedicine, topological acoustofluidics is advancing with applications like topological acoustic tweezers for manipulation and sorting of cells and tissues. Hybrid systems combining topological phononics with photonics enable efficient acousto-optic conversion and quantum state transfer, while extensions to geophysics may offer improved earthquake protection structure design.

Fundamental challenges remain, including the direct experimental observation of topological surface phonons in solids, and pushing topological phononics to high-frequency and nanoscale regimes (also with high quality factors and low loss) for on-chip applications. Data-driven discovery and design are accelerating the field, with machine learning being used to predict topological phases and optimize device performance, and inverse design tools enabling the rapid prototyping of metamaterials with targeted topological properties and functions.

Topological phononics has evolved from a theoretical curiosity to an impressive emergent field with profound implications for both fundamental physics and technology. As the field continuously advances, it will unlock further new phenomena, effects, and technological applications. By addressing remaining fundamental questions and leveraging experimental and computational advances, topological phononics is about to become a model interdisciplinary field fusing fundamental scientific exploration with cutting-edge technological developments.

% \nocite{*}

%\bibliographystyle{unsrtnat}
\bibliography{ref}% Produces the bibliography via BibTeX.

\section*{Acknowledgments}
J.-H.J was supported by the National Key R\&D Program of China (2022YFA1404400), National Natural Science Foundation of China (grant No. 12125504), the CAS Pioneer Hundred Talents Program, and the Priority Academic Program Development (PAPD) of Jiangsu Higher Education Institutions. T.Z. acknowledges the support from the National Key R\&D Project (Grant Nos. 2023YFA1407400 and 2024YFA1409200) and the National Natural Science Foundation of China (Grant Nos. 12374165 and 12047503). G. M. was supported by the National Natural Science Foundation of China under grant No. T2525002, the National Key R\&D Program (2022YFA1404400), the Hong Kong Research Grants Council (RFS2223-2S01, 12301822, 12300925), and the Hong Kong Baptist University (RC-RSRG/23-24/SCI/01, RC-SFCRG/23-24/R2/SCI/12). X. W. acknowledges the support of Hong Kong Research Grants Council (JRFS2526-2S07). F. G. acknowledges the support of Jiangsu Funding Program for Excellent Postdoctoral Talent No. 2025ZB161, the Postdoctoral Fellowship Program (Grade C) of China Postdoctoral Science Foundation No. GZC20252196, Basic Research Program of Jiangsu (Grant No. BK20250791), China Postdoctoral Science Foundation (2025M773337), and the Basic Science Research Project of Higher Education Institutions in Jiangsu Province (Natural Science) (25KJB140015). Z. C. acknowledge the support from the National Natural Science Foundation of China (Grant No. 12744154) and the National Key R\&D Project (Grant No. 2023YFA1406900). X. Z. acknowledge the support from the National Key R\&D Program of China (Grants No. 2023YFA1407700).
\end{document}